# Improving Privacy Protection in the area of Behavioural Targeting

ACADEMISCH PROEFSCHRIFT

ter verkrijging van de graad van doctor
aan de Universiteit van Amsterdam
op gezag van de Rector Magnificus
prof. dr. D.C. van den Boom
ten overstaan van een door het college voor promoties
ingestelde commissie, in het openbaar te verdedigen in de Agnietenkapel op
woensdag 17 december 2014, te 14:00 uur
door

**Frederik Johannes Zuiderveen Borgesius**

geboren te Ede

| | |
|---|---|
| Promotor: | Prof. dr. N.A.N.M. van Eijk |
| Overige Leden: | Prof. dr. S. Gutwirth |
| | Prof. dr. N. Helberger |
| | Prof. dr. M. Hildebrandt |
| | Prof. mr. J. J.C. Kabel |
| | Dr. A. M. McDonald |
| | Dr. J.V. J. Van Hoboken |

Faculteit der Rechtsgeleerdheid

**Concise table of contents**

# Table of contents

# Acknowledgements

I thank everybody who has helped and supported me during the writing of this study. In particular, I am grateful to my supervisor Nico van Eijk, and my daily supervisor Joris van Hoboken. Their guidance was superb. I thank Serge Gutwirth, Natali Helberger, Mireille Hildebrant, Jan Kabel, and Aleecia McDonald for taking part in the reading committee for the thesis. I thank Annabel Brody for her help with the English language.

The IViR Institute for Information Law is an inspiring work environment – and a fun workplace. My colleagues feel more like friends. My study has benefited especially from discussions about research ideas – and research troubles – with Bernt Hugenholtz, Bodó Balázs, Ot van Daalen, Mireille van Eechoud, Stef van Gompel, Lucie Guibault, Christian Handke, Hielke Hijmans, Kristina Irion, Esther Janssen, Thomas Margoni, Tarlach McGonagle, Manon Oostveen, Joost Poort, Ana Ramalho, João Quintais, Joan-Josep Vallbé, Christiaan Alberdingk Thijm, and Bart van der Sloot. I have good memories of following the IViR research master program with, and writing articles with, David Korteweg and Stefan Kulk. Discussions with my IViR office mate Axel Arnbak are always stimulating – and usually hilarious too. At the IViR, Margriet Pauws-Huisink and Anja Dobbelsteen deserve special thanks.

I have benefited enormously from the assistance of librarians at the University of Amsterdam, the IViR, the NYU School of Law library, the Peace Palace library, the Koninklijke Bibliotheek, and the University of Hong Kong law library. Special thanks go out to Rosanne de Waal and Pascal Braak for their help finding obscure publications, and for their patience.

I am still surprised and grateful that so many people made time in their busy schedules to discuss research ideas with me. At the University of Amsterdam I have benefited from discussions with Joep Sonnemans, Beate Rössler, Candida Leone, Kati Cseres, Giuseppe Dari-Mattiacci, Corette Ploem, Marieke Oderkerk, Lonneke van der Velden, Chantal Mak, Marco Loos, Martijn Hesselink, and Peter Neijens. At the University of Amsterdam, Arnout Nieuwenhuis, Janne Nijman and Edgar du Perron have helped to shape an excellent research climate for PhD candidates. The meetings of the Amsterdam Platform for Privacy Research provide much inspiration, and I thank all the participants.

Parts of this study were written during a stay at New York University School of Law. The university and the Privacy Research Group offered an inspiring research environment. I had thought-provoking discussions with Sasha Romanosky, Malte Ziewitz, Sophie Hood, danah boyd, Heather Patterson, Luke Stark, Florencia Marotta-Wurgler, Oren Bar-Gill, Fernando Gomez, and Lewis Kornhauser. Special thanks go out to Helen Nissenbaum and Solon Barocas for stimulating discussions, and for making me feel so welcome.

In the US, I thank Milton Mueller, Felix Wu, Malavika Jayaram, Martin French, Julia Angwin, Jennifer Valentino-DeVries, Arvind Narayanan, Ashkan Soltani, Balachander Krishnamurthy, Jasmine McNealy, Joel Reidenberg, Joseph Turow, Neil Richards and Dennis Hirsch. Chris Hoofnagle is always extremely nice and helpful: thanks Chris! Discussions with Seda Gürses are always refreshing, and funny. In the UK, my thanks go to Douwe Korff, Bendert Zevenbergen, Paul Bernal, Monica Horten, Andy McStay, and especially Ian Brown. In Norway, I thank Ann Rudinow Sætnan, Dag Wiese Schartum and Jon Bing. In Canada, I thank Ann Cavoukian and Ian Kerr. In Germany, I thank Jan Schallaboeck, Hansjürgen Garstka, and Theo Röhle. At the Vrije Universiteit Brussels, I thank the LSTS team, and in particular Paul de Hert, Rocco Bellanova, Irina Baraliuc, and Dariusz Kloza. In Israel, I thank Tal Zarsky. At Hong Kong University, the people at the Law and Technology Centre deserve my gratitude, and especially Anne Cheung and Marcelo Thompson.

In the Netherlands, I thank Feer Verkade, Peter Hustinx, Jacob Kohnstamm, and the team at Bits of Freedom. At Leiden University, I thank the eLaw team, and especially Gerrit-Jan Zwenne, Simone van der Hof, Bart Schermer, and Esther Keymolen. I thank Michel van Eeten, Engin Bozdag, Hadi Asghari, Shirin Tabatabaie, and Jeroen van den Hoven at Delft University. At the Radboud University Nijmegen, I thank Bart Jacobs, Jaap-Henk Hoepman, and Merel Koning. At Tilburg University I thank the TILT team, and especially Eleni Kosta, Ronald Leenes, Bert-Jaap Koops and Nadezha Purtova. At the Vrije Universiteit van Amsterdam, I thank Evelien Brouwer. At the Open University, I thank Marianne Idenburg-Amorison and Titia Zwalve.

Regarding Do Not Track and the technology behind behavioural targeting, I have learnt a lot from Matthijs Koot, Rob van Eijk, Joe Hall, Rigo Wenning, Jonathan Mayer, Justin Brookman, and the people at the World Wide Web Consortium.

The chapter on behavioural economics has benefited from discussions with Alberto Alemanno, Anne-Lise Sibony, Shara Monteleone, Omer Tene, Alessio Pacces, Alessandro Acquisti, and workshop participants at the Privacy Law Scholars Conference, at the NYU Privacy Research Group, at the Erasmus School of Law in Rotterdam, and at the Nudging in Europe conference in Liège.

My students at the IViR and elsewhere deserve thanks for their curiosity and their sometimes challenging questions.

I thank the Prins Bernhard Fonds for a contribution for my research semester at New York University. Grants by the following organisations have made my research master and my stay at Hong Kong University possible: the Stichting Algemeen studiefonds, the Stichting Neeltje Buis, the Stichting Noorthey, the Stichting Max Cohen Fonds, the Stichting Talent Support, and the Amsterdam University Fund.

Special thanks are reserved to all my friends and family – I have not given you enough attention during the writing of the study. Sekander Raisani deserves extra thanks for encouraging me to study law, and for the many adventures while touring the world together.

In hope that I didn’t forget to thank people who deserve my gratitude – but I seriously doubt it. Please accept my sincere apologies if I did forget you. (This PhD project has made me more absent-minded and more forgetful.)

Finally I thank my parents, my two brothers, and Sjakie. Much love and thanks go to Sjoera. Thank you all!

# 1 Introduction

Behavioural targeting involves the monitoring of people's online behaviour. It uses the collected information to show people individually targeted advertisements. To give a simplified example, a person who frequently visits websites about cars and soccer might be profiled as a male sports enthusiast. If that same person books a flight to Amsterdam on a website, advertising for tickets for a game of the local football club, Ajax, may be shown.

Behavioural targeting could benefit firms and individuals. Advertising funds an astonishing amount of internet services. Without paying with money, people can use online translation tools, access online newspapers, use email accounts, watch videos, and listen to music. Many people prefer targeted ads to random ads, and appreciate the book recommendations by online bookstores based on earlier interactions with the store. But behavioural targeting also raises privacy concerns. For instance, data collection can cause chilling effects. Using cookies or other technologies, firms compile detailed profiles based on what internet users read, what videos they watch, what they search for, etc. Profiles can be enriched with up-to-date location data of users of mobile devices, and other data that are gathered on and off line. People have little control in relation to what happens to information concerning them. Many different types of firms are involved in behavioural targeting, which results in a complicated system where information about people is combined, analysed, and auctioned off in almost real time. Furthermore, behavioural targeting enables discriminatory practices. A firm can exclude people from its advertising campaign, based on their individual profiles. Ads and websites can be personalised for each visitor.

This study examines ways in which the law could improve privacy protection in this area. Broadly speaking, this study explores two primary ways in which privacy can be defended. The first focuses on *empowering* the individual, for example by requiring firms to obtain the individual's consent before they collect data. This empowerment approach is present in current data protection law. Under the data protection regime, personal data "must be processed fairly for specified purposes and on the basis of the consent of the person concerned or some other legitimate basis laid down by law."[1] The phrase "on the basis of the consent of the person concerned" appears to be a strong requirement, but is undermined in practice, due to the fact that many people click "I agree" to any statement that is shown on the web. Behavioural studies cast doubt on the potential of informed consent as a means to defend privacy.

The second approach focuses on *protecting* rather than empowering the individual. This approach is also present in data protection law. Many data protection rules can protect privacy in the area of behavioural targeting, even if people agree to consent requests. For instance, firms must always secure the data they process, and can't use data for new purposes at will. Such requirements should mitigate the chance that personal information may be used in unexpected ways that harm people. But the data protection regime should be supplemented with additional rules. The study concludes with recommendations on how to improve privacy protection.

## 1.1 Research question

This study aims to answer to following question.

> How could European law improve privacy protection in the area of behavioural targeting, without being unduly prescriptive?

[1] Article 8 of the EU Charter of Fundamental Rights. See for the full titles of the legal texts: Legal texts, at the end of this study.

Some remarks about the research question and the terminology. In this study, behavioural targeting is analysed by distinguishing five phases: (1) data collection, (2) data storage, (3) data analysis, (4) data disclosure, and (5) the use of data for targeted advertising.

The study focuses in particular on three privacy problems regarding behavioural targeting: chilling effects, the lack of individual control over personal information, and the risk of discriminatory or manipulative practices.[2] Privacy is notoriously difficult to define. Three privacy perspectives are distinguished in this study, as discussed in the next section.

The phrase European law as used in the research question, refers to regulation by the European Union. The study also takes the norms into account that follow from the European Convention on Human Rights and the related case law.[3] To protect privacy in the area of behavioural targeting, the EU lawmaker mainly relies on the e-Privacy Directive,[4] and the Data Protection Directive.[5] The e-Privacy Directive requires firms to ask the user's consent for using tracking cookies and similar technologies.

Data protection law can be seen as a means, a legal instrument, to protect privacy, fairness, and related interests. This study agrees with De Hert & Gutwirth, who characterise the legal right to privacy as an "opacity tool", and data protection law as a "transparency tool."[6] They say that the right to privacy in the European Convention on Human Rights prohibits intrusions into the private sphere.[7] This right aims to give the individual the chance to remain shielded, or to remain opaque. This prohibition isn't absolute. Exceptions to the prohibition are possible under strictly defined

---

[2] This study uses the phrases "data" and "information" interchangeably, and uses "risk" and "uncertainty" interchangeably.
[3] The norms that follow from the European Convention on Human Rights and the related case law form an integral part of the general principles of law for the EU (see e.g. CJEU, C-131/12, Google Spain, 13 May 2014, par. 68).
[4] Directive 2002/58/EC (amended by Directive 2009/136/EC). This study uses the consolidated text.
[5] Directive 95/46/EC.
[6] De Hert & Gutwirth 2006.
[7] Article 8(1) of the European Convention on Human rights.

conditions, for instance for national security, or for the protection of the rights of others.[8]

Data protection law takes a different approach than the legal right to privacy. In principle data protection law allows data processing, if the data controller complies with a number of requirements. Data protection law aims to ensure fairness by requiring firms to be transparent about personal data processing. It's a legal tool that aims to ensure that the processing of personal data happens fairly and transparently.[9]

In January 2012 the European Commission presented a proposal for a Data Protection Regulation,[10] which should replace the Data Protection Directive from 1995. At the time of writing, it's unclear whether the proposal will be adopted. The most optimistic view seems to be that the Regulation could be adopted in 2015.[11] The proposed Regulation is based on the same principles as the Directive.

The study looks for regulatory responses to protect privacy, "without being unduly prescriptive." The lawmaker shouldn't take measures that are excessive or unreasonable. For this study, the "without being unduly prescriptive" requirement implies that the lawmaker must respect certain boundaries. First, in line with positive law, the study assumes that some legal paternalism is acceptable, but that the lawmaker should stay away from boundless paternalism. Second, regulation shouldn't impose unreasonable costs on society. The use of the word unduly illustrates that law isn't an exact science.

### 1.2 Methodology

This is a legal study, which is situated in the field of information law. Information law is "the law relating to the production, marketing, distribution and use of information goods and services. Information law comprises a wide set of legal issues at the crossroads of intellectual property, media law, telecommunications law, freedom of expression and right to privacy."[12] More specifically, this is a study in the field of data protection law, as Europeans might say, or information privacy law, as Americans might say.[13]

[8] Article 8(2) of the European Convention on Human rights.
[9] See in more detail on De Hert & Gutwirth 2006, opacity tools and transparency tools: chapter 4, section 3, and chapter 9, section 2.
[10] European Commission proposal for a Data Protection Regulation (2012).
[11] See: European Council 2014, p. 2.

The study contains normative and descriptive research. The research question is normative, and concerns what the law ought to be, rather than what the law is. One of the goals of European data protection law is to "protect the fundamental rights and freedoms of natural persons, and in particular their right to privacy with respect to the processing of personal data."[14] This study agrees with the argument that data protection law doesn't achieve this goal in the area of behavioural targeting.[15] Parts of this study take a primarily descriptive approach, and provide an analysis of current law.[16]

The study considers different options, and maps out strengths and weaknesses of different regulatory strategies. According to Rubin, "[c]ontemporary legal scholars are now generally aware that their work consists of recommendations addressed to legal decision-makers, recommendations that are ultimately derived from value judgments rather than objective truth."[17] Whether or not this is true for legal scholars in general, this study proceeds on that basis.[18] Answering the research question necessarily entails making "legal-political choices."[19]

---

[12] Institute for Information Law 2013.
[13] Information privacy "concerns the collection, use and disclosure of personal information" (Schwartz & Solove 2009, p. 1). Chapter 5 shows that the concept of "personal information" raises questions in the area of behavioural targeting.
[14] Article 1(1) of the Data Protection Directive.
[15] See the references in chapter 8, section 1. See for an evaluation of the current regime also chapter 7.
[16] See for instance, chapter 3, section 2, and chapter 4, 5 and 6. See on the merits of descriptive legal research Posner 2007, p. 437.
[17] Rubin 1988, p. 1904.
[18] See for criticism on such "policy-driven research" Van Gestel & Micklitz 2013, p. 10.
[19] The phrase is borrowed from Vranken 2006, p. 123. See also Hesselink 2009.

There's no commonly agreed upon privacy definition. The European Court of Human Rights says that the right to private life, the right to privacy for short, is "a broad term not susceptible to exhaustive definition."[20] Borrowing from Gürses, this study distinguishes three privacy perspectives in order to bring some structure to the discussion.[21] The first perspective focuses on limited access to the private sphere. The second focuses on individual control over personal information. The third perspective focuses on privacy as the freedom from unreasonable constraints on identity construction. The three perspectives highlight different privacy aspects of behavioural targeting.

In order to analyse the appropriate regulatory response to behavioural targeting, the study focuses in particular on three privacy problems. First, the massive collection of information on user behaviour can cause chilling effects. People may adapt their behaviour if they suspect their behaviour is monitored. For instance, they might hesitate to look for medical information, or to read about certain political topics on the web. Second, people lack control over data concerning them. The online behaviour of hundreds of millions of people is tracked, without their knowledge or consent. The data flows behind behavioural targeting are complicated, and people have scant knowledge of what happens to their data. Third, there's a risk of unfair social sorting or discriminatory practices. And some fear behaviourally targeting could be used to manipulate people. Firms can personalise ads and other website content to each individual visitor, and personalised ads could be used to exploit people's weaknesses. In sum, this study takes a broad view of privacy related problems.[22]

Legal scholars often use several methods in one study to answer a research question.[23] This research follows that tradition. For example, this study draws inspiration from the field of consumer law and general contract law. In these fields, certain problems are comparable to those in data protection law: how should the balance be struck between empowering and protecting people? Taking inspiration from other areas of law could be seen as an internal comparative law method.[24]

---

[20] ECtHR, S. and Marper v. The United Kingdom, No. 30562/04 and 30566/04.
[21] Gürses 2010. The three perspective are based on Warren & Brandeis 1890; Westin 1970; Agre 1998. See chapter 3, section 1.
[22] See chapter 3, and especially section 3.
[23] Herweijer 2003.

Research from other disciplines provides valuable insights for this study. The study draws on law and economics and behavioural economics research. Law and economics provides a tool to analyse certain problems with informed consent to behavioural targeting.[25] Behavioural economics aims to improve economic theory by including findings from psychology and behavioural studies. Empirical research by scholars such as Acquisti, Cranor and McDonald provides information on how people make privacy choices in practice.[26] On paper, current data protection law looks better than it operates in practice. Salter & Mason might characterise the approach of this study as follows:

> Such research, from the start, expressly advocates a reform in law. (…) Such proposals will (…) be supported by evidence that changes in social patterns, lifestyles, attitudes and economic circumstances now mean that the policy underlying a particular area of legal regulation has become outdated and anachronistic, even if it fully meets the aspirations of the black-letter model.[27]

It's emphasised that this is legal research. The study primarily takes an internal legal viewpoint, rather than an external viewpoint, which would be the case in legal sociology, or in law and economics.[28] Arguments have to fit in the European legal system.[29] For example, law and economics can help to analyse problems, but in the European legal system economic arguments don't trump other arguments.[30] The economic analysis is meant as an addition to legal discourse, and doesn't aim to improve economic theory. Chapter 2, which describes the facts regarding behavioural targeting, uses literature from the fields of marketing, computer science, and media studies. In summary, this is legal research with a small degree of interdisciplinarity.[31]

---

[24] Vranken 2006, 88; Herweijer 2003.
[25] Law and economics can be described as the "economic analysis of legal rules and institutions" (Posner 2011, xxi). See chapter 7, section 2 for an introduction to the field.
[26] See chapter 7.
[27] Salter & Mason 2007, p. 162-163.
[28] See Hart 1961.

Some methods that are common in legal research are absent or less prominent in this study. For example, the systemisation and analysis of case law plays a minor role in this study, because case law on behavioural targeting is scarce.[32] External comparative law, the comparison of national legal systems, doesn't play a role in this study. While a comparison of the regulation of behavioural targeting in the United States and in Europe would be an interesting research topic, this study doesn't adopt that approach, in order to keep the scope of the study manageable. That said, the study does take inspiration from American scholars.[33]

This study relies on desk research, and uses several types of sources, such as the usual sources for legal research: regulation, case law, legislative history and legal literature. Research libraries and the usual databases were used to find literature. The study mainly refers to sources in English. If there was a choice between a source in English and a source in another language, usually the English source was chosen. Literature tips were asked from specialists in the various disciplines and fields of law that are present in this study. For this study, in addition to his education in information law, the author did coursework in behavioural economics, and in law and economics.[34] During the research, preliminary results were discussed with academics from various disciplines from many countries. Conversations with firms doing behavioural targeting, regulators, and lawyers also provided valuable insights. Any errors in the study are mine.

---

[29] Hesselink 2009, p. 39; Smits 2009, p. 54.
[30] See chapter 7, section 2.
[31] See Schrama 2011, p. 161, Smits 2009, p. 51-54. In the taxonomy of Siems, this study might be called "basic interdisciplinary research" (Siems 2009).
[32] The analysis of case law does play a role, especially in chapter 3, section 2, and chapter 6. Furthermore, the study refers to opinions of Data Protection Authorities, in a way that resembles how many studies refer to case law.
[33] See for a comparison between European and US privacy regulation Blok 2002 (Dutch and US law); Hoofnagle 2010; Purtova 2011 (focusing on property rights on personal data); Tene & Polenetsky 2012 (focusing on behavioural targeting). See also Korff, D. et al. 2010, comparing eleven countries.

This study aims to be reasonably pragmatic, and relevant for policy discussions. The study doesn't examine whether data protection law should be completely abolished, so the lawmaker can start again with a clean slate to develop a new privacy regime. This would be an interesting thought experiment, but it's unlikely that the EU would abolish data protection law.[35] More generally, the European legal system is accepted as the background for this study. The study doesn't consider solutions that would require completely reforming the legal and political system, or resigning from the European Convention on Human Rights.

The choice to take a European perspective for this study, as opposed to a Dutch perspective for instance, is also largely pragmatic. If a EU member state would adopt regulation regarding behavioural targeting, it would likely be less effective than if the EU did so. The author is aware that the choice of a European perspective is also a political choice. Another political choice is this study's implicit assumption that the harmonisation of European laws is desirable. In data protection scholarship it's relatively common to assume that harmonisation is desirable, as many legal data protection instruments aim at both protecting fundamental rights and protecting the free cross border flow of personal data.[36] The European (rather than national) focus influences this study's style. For instance, the study gives a relatively large amount of attention to the opinions of the Article 29 Working Party, an advisory body in which national Data Protection Authorities cooperate.[37] While not legally binding, the Working Party's opinions are influential. They give an idea of the views of European national Data Protection Authorities.

---

[34] At the University of Amsterdam I followed the courses "Behavioural Economics" by Prof. J. Sonnemans, and "Law and Economics" by Prof. Dari-Mattiacci. At New York University I followed the courses "Economic Analysis of Law" by Prof. L.A. Kornhauser, "Comparative Law and Economics of Contracts", by Prof. F.L. Gomez, and "Consumer Contracts" (behavioural law and economics) by Prof. O. Bar-Gill and Prof. F. Marotta-Wurgler.

[35] This is unlikely for many reasons. One reason is that the EU Charter of Fundamental Rights includes a right to the protection of personal data in article 8.

[36] See González Fuster 2014, p. 130.

The topic of this thesis is a moving target, in various ways. For example, firms develop new tracking technologies all the time. The legal landscape is also subject to change. While the thesis was in progress, proposals to amend European data protection law were presented by the European Commission and discussed and amended in Brussels. The research was concluded on 1 November 2014. Developments after that date aren't taken into account, with a few minor exceptions. Parts of the thesis build on and include parts of the earlier work of the author.[38]

## 1.3 Societal and scientific relevance

Research shows that many people worry about their online privacy, and that many find behavioural targeting to be a privacy invasion.[39] Vast amounts of information about hundreds of millions of people is collected for behavioural targeting. European regulation of cookies and proposals to amend European data protection law have been the topic of much policy discussion. There's a constant stream of articles in the academic and popular press on behavioural targeting. The idea that privacy protection in the area of behavioural targeting leaves something to be desired is widely shared in literature.[40]

No national regulator has come up with a definitive answer in relation to how to regulate behavioural targeting, a practice that started in the mid-1990s, and grew into a major business during the last decade. Everywhere in the world, behavioural targeting is a relatively new phenomenon. As scholars and regulators worldwide are struggling to come up with answers, this study might be relevant outside Europe as well. Roughly a hundred countries have laws that are based on the same principles as European data protection law.[41] Hence, certain problems, for instance regarding consent in the online environment, also arise outside Europe.

---

[37] Article 29 of the Data Protection Directive. See for more details on the Working Party: chapter 4, introduction.
[38] See in particular Van Der Sloot & Zuiderveen Borgesius 2012; Zuiderveen Borgesius 2011; Zuiderveen Borgesius 2013; Zuiderveen Borgesius 2013a; Zuiderveen Borgesius 2014.
[39] See for research on people's attitudes regarding behavioural targeting chapter 7, section 1.
[40] See chapter 7, chapter 8, section 1, chapter 9, section 1.

What does this study add to existing scholarship? This study is among the first to discuss the implications of behavioural economics research for European data protection policy.[42] The topic of whether data protection law should apply to pseudonymous data is discussed in depth in the study. The study contains a detailed analysis of the role of informed consent in data protection law. And the study gives much attention to the tension between protecting and empowering the individual within data protection law.

Legal scholars have discussed online privacy problems since the 1990s. In recent years, many authors have expressed scepticism about the potential of informed consent as a privacy protection measure.[43] The complicated data flows behind behavioural targeting make transparency and informed choices especially difficult.[44] Kosta analysed consent in European data protection law, using mainly legal-historical analysis and external comparative law.[45] This study could be seen as a next step after her thesis.

De Hert & Gutwirth characterisation of data protection law as a legal "transparency tool" influences this study.[46] The choice of the three privacy perspectives in this study is deeply influenced by the work of Gürses.[47] Surveillance scholars such as Gandy and Lyon inform the discussion on social sorting and discrimination.[48] Scholars such as Hildebrandt et al., Solove, and Zarsky distinguish different phases of personal data processing.[49] The analysis of behavioural targeting in this study builds on their work. The work by media scholars, such as Turow and Bermejo, provides information for this study's discussion of the behavioural targeting practice.[50] Computer science researchers such as Krishnamurthy, Mayer, Soltani, and Van Eijk provide insights into behavioural targeting practices.[51] Such research is sometimes referred to as "web privacy measurement."[52]

---

[41] In September 2013, Greenleaf counted 101 countries in the world with a data protection law (Greenleaf 2013a, Greenleaf 2013b).
[42] Other studies that take behavioural economics insights into account when discussing European data protection law include Brown 2011; Brown 2013; Helberger et al. 2012.
[43] See e.g. Bygrave & Schartum 2009; Blume 2012. See also the references in chapter 7, section 3 and 4.
[44] See e.g. Helberger et al. 2012; Kosta 2013a. See further chapter 7. US scholars are often even more pessimistic about "informed consent" as a privacy protection tool; see e.g. Barocas & Nissenbaum 2009; Nissenbaum 2011; Tene & Polenetsky 2012; Solove 2013.
[45] Kosta 2013a.
[46] De Hert & Gutwirth 2006.

Several authors in North America analyse consent to data processing through a law and economics lens, and consider the implications of behavioural economics for online privacy.[53] In Europe, authors such as Cserne, Howells, and Luth examine the implications of behavioural economics for consumer law, without, however, discussing privacy.[54] A couple of European authors, such as Brown, take behavioural economics into account when discussing data protection law.[55]

## 1.4 Scope of the study

The research question concerns behavioural targeting, the monitoring of people's online behaviour to use the collected information to show people individually targeted advertisements. Behavioural targeting enables more possibilities than targeted advertising, such as personalised content and personalised prices (a type of price discrimination). Such topics are mentioned in passing, but aren't the main focus of this study. In the long term behavioural targeting may decrease ad revenues for some website publishers. The changing power relations in the media landscape resulting from behavioural targeting are not an independent topic of inquiry in this study.[56]

---

[47] Gürses 2010
[48] Gandy 1993; Lyon 2002. Surveillance studies can be described as follows. "The contribution of surveillance studies is to foreground empirically, theoretically and ethically the nature, impact and effects of a fundamental social-ordering process. This process comprises the collection, usually (but not always) followed by analysis and application of information within a given domain of social, environmental, economic or political governance" (Lyon et al. 2012, p. 1).
[49] Hildebrandt et al. 2008; Solove 2006; Solove 2008; Zarsky 2004.
[50] Turow 2011.
[51] Mayer & Mitchell 2012; Gomez et al. 2009; Hoofnagle et al. 2012; Krishnamurthy & Wills 2009; Van Eijk 2011.
[52] Berkeley Law 2012.
[53] E.g. Kerr et al. 2009; Nehf 2005.
[54] Luth 2010; Cserne 2008; Howells 2005.
[55] Brown 2011; Brown 2013; See also Helberger et al. 2012.

The scope of this study is limited to the EU. When examining legal privacy protection in Europe, one automatically ends up looking at data protection law. As noted, data protection law is the main legal instrument to protect information privacy in Europe. Considering their specific relevance for behavioural targeting, a couple of themes within data protection law were selected for this study.[57] The study doesn't aim to give an overview of all data protection provisions that could be relevant for behavioural targeting. For example, the question of which national data protection law applies to firms based in or outside Europe falls outside the study's scope, as do trans-border data flows.[58] Whether special rules are needed for children isn't discussed in this study.[59] A discussion of the so-called right to be forgotten and the problematic interplay between data protection law and freedom of speech falls outside this study's scope.[60] The study doesn't examine the competence of the EU to adopt data protection rules.[61]

This isn't a handbook listing all European regulation and case law that might be relevant for behavioural targeting.[62] For example, advertising law, non-discrimination law, competition law, and consumer law are excluded from the analysis.[63] Having said that, the study does take some inspiration from consumer law.[64] Providing a new privacy definition isn't among the study's goals. This isn't a work of legal philosophy or political science.[65] The democratic deficit of the EU and the influence of lobbying on EU regulation are outside the study's scope.

---

[56] See on that topic chapter 2, section 2; chapter 7, section 2. See in more detail Turow 2011.
[57] See chapter 4, 5 and 6.
[58] On the territorial scope of data protection law, see Moerel 2011, chapter 1-4; Kuner 2010; Kuner 2010a; Piltz 2013. On transborder data flows see Moerel 2011; Kuner 2013.
[59] See on children and data protection law Van Der Hof et al. 2014.
[60] See on a right to be forgotten Ausloos et al. 2012 (mostly positive); Van Hoboken 2013 (more critical); Mayer-Schönberger 2009 (US focused). See also CJEU, C-131/12, Google Spain, 13 May 2014, and on that case Kulk & Zuiderveen Borgesius 2014.
[61] The 1995 Data Protection Directive is based on the old article 95 of the Treaty establishing the European Community, which corresponds to article 114 of the Treaty on the Functioning of the EU (consolidated version 2012). The European Commission proposal for a Data Protection Regulation (2012) is based on article 16 of the Treaty on the Functioning of the EU (consolidated version 2012).
[62] Good overview books about data protection law generally are Bygrave 2002; Bygrave 2014; Büllesbach et al. 2010; European Agency for Fundamental Rights 2014; Kuner 2007.

A distinction by Baldwin et al. can help to clarify this study's scope. They distinguish six tools that the state can use to regulate behaviour.[66] In current data protection law, three of the tools are present: to command, to inform, and to confer protected rights. The first tool is commanding. For example, the state can prohibit or require certain activities. This strategy can be found in data protection law. For example, personal data processing is only allowed if there's a legal basis for the processing. The second tool is using information for policy goals. For instance, the law can require firms to disclose certain information to help people make decisions. Data protection law partly relies on this strategy. Firms are required to be transparent about data processing towards the data subject. This study doesn't analyse other ways of informing the data subject, such as education. The third tool is conferring protected rights. The state can grant people rights, which they can enforce themselves. Tort law and property rights could be seen as examples of this approach.[67] Parts of data protection law also grant people rights they can enforce themselves. People can take a firm to court when their data protection rights are infringed.[68] Even the rights granted to individuals by constitutions and human rights treaties, such as the right to privacy, could be seen as an example of this regulatory strategy.[69] Some authors have discussed introducing property or intellectual property rights on personal information; a discussion of such proposals falls outside this study's scope.[70]

---

[63] See for a media law angle Helberger 2013. See on non-discrimination law and profiling Hildebrandt et al. 2008; Zarsky et al. 2013; De Vries et al. 2013.

[64] There isn't much literature that applies European consumer protection law to behavioural targeting. See for exceptions Van Der Sloot 2012; Centre for the Study of European Contract Law (CSECL) & Institute for Information Law (IViR) 2012; European Data Protection Supervisor 2014.

[65] See for a legal philosophical angle on behavioural targeting Hildebrandt & Gutwirth (eds.) 2008, and for a philosophy and ethics angle Nissenbaum 2011; Bozdag & Timmersmans 2011; Bozdag & Van De Poel 2013. See on data protection law through a political science lens Bennett 1992; Regan 1995; Heisenberg 2005; Newman 2008, and specifically on the regulation of cookies Kierkegaard 2005.

[66] Baldwin et al. 2011, p. 106, and generally chapter 7. Baldwin et al. work in the field of "regulation studies", see chapter 8, section 1 of this study. Lessig distinguished four "modalities" of regulation: law, architecture (technology, or "code"), social norms and the market (Lessig 2006). This study focues mostly on law.

[67] Baldwin et al. 2011, p. 126-128. See also Ogus 2004, p. 257-258.

[68] See chapter 8, section 1.

Three other types of policy tools fall outside the study's cope: deploying wealth, harnessing markets, and acting directly.[71] First, the state could use tax or subsidies to influence behaviour. For instance, many European states fund public broadcasters. Deploying wealth has been suggested as an instrument in the area of online privacy: in France there was discussion about the possibility of taxing the use of personal data.[72] Second, the state can aim to guide markets, with competition law for example.[73] Third, the state can act directly. For instance, the state can construct a bridge or a road, or organise hospitals or a pension scheme. The state played a large role in developing the internet.[74] In principle, the state could also set up, or help to set up, internet services such as websites or even search engines.

### 1.5 Outline

The outline of the thesis is detailed below. The research question was introduced in this first chapter: how could European law improve privacy protection in the area of behavioural targeting, without being unduly prescriptive?

Chapter 2 explains what behavioural targeting is, by distinguishing five phases. During the first phase of behavioural targeting, firms track people's online behaviour. Second, firms store data about individuals. Third, firms analyse the data. Fourth, firms disclose data to other parties. In the fifth phase, data are used to target ads to specific individuals.

---

[69] Human rights are typically inalienable, while, for instance, property rights usually can be transferred to others (see Calabresi & Melamed 1972).

[70] The most extensive discussion of property rights on personal data is Purtova 2011. She characterises such proposals as a means to improve data subject control, and suggests mandatory protective rules are needed as well (p. 244). See on introducing a type of intellectual property right on personal data Dommering 2010; Dommering 2012.

[71] Baldwin et al. 2011, p. 106 and further.

[72] Collin & Colin 2013.

[73] Some scholars have asked what competition law can do to protect privacy interests, but for the moment there are more questions than answers. See Geradin & Kuschewsky 2013; European Data Protection Supervisor 2014.

[74] Bing 2009.

Chapter 3 discusses the right to privacy in European law, and the privacy implications of behavioural targeting. Three privacy perspectives are distinguished in this study: privacy as limited access, privacy as control, and privacy as identity construction. The chapter discusses three main privacy problems of behavioural targeting. First, the massive collection of information on user behaviour can have a chilling effect. Second, people lack control over their information. Third, behavioural targeting enables social sorting and discriminatory practices. Also, some fear that personalised ads and other content could be manipulative, or could narrow people's horizons.

Chapter 4 introduces data protection law, Europe's main legal tool to protect information privacy. Data protection law aims to ensure that personal data processing happens fairly and transparently. The history of data protection law can help to understand its focus on informed consent and transparency. The chapter shows that there's a tension within data protection law between empowering and protecting the individual. This tension is a recurring theme in this study.

Chapter 5 concerns the material scope of data protection law. Many behavioural targeting firms say data protection law doesn't apply to them, because they only process "anonymous" data. The chapter makes two points. First, an analysis of current law shows that data protection law generally applies to behavioural targeting. Data protection law also applies if firms don't tie a name to individual profiles. Second, from a normative perspective, data protection law should apply.

Chapter 6 discusses the role of informed consent in the regulation of behavioural targeting. Current law regarding behavioural targeting places a good deal of emphasis on informed consent. The e-Privacy Directive requires firms to obtain informed consent for the use of most tracking technologies, such as cookies. Furthermore, in general data protection law, consent is one of the legal bases that a firm can rely on for personal data processing.

Chapter 7 analyses practical problems with informed consent in the area of behavioural targeting. The chapter reviews law and economics literature, behavioural economics literature, and empirical research on how people make privacy choices. The chapter shows that the potential of data protection law's informed consent requirement as a privacy protection measure is very limited. People generally ignore privacy policies, and click "I agree" to almost any online request.

Chapter 8 discusses measures to improve individual *empowerment*. Strictly enforcing and tightening data protection law would be a good start. For example, firms shouldn't be allowed to infer consent from mere inactivity from the individual, and long unreadable privacy policies shouldn't be accepted. User-friendly mechanisms should be developed to foster transparency and to enable people to express their choices. This study doesn't suggest that data subject control over personal information can be fully achieved. Nevertheless, some improvement must be possible, as now people's data are generally accumulated and used without meaningful transparency or consent.

Chapter 9 discusses measures to improve individual *protection*. Certain data protection principles could protect people, even if they consent to data processing. While the role of informed consent in data protection law is important, it's at the same time limited. People can't waive data protection law's safeguards, or contract around the rules. The protective data protection principles should be enforced more strictly; but this won't be enough. In addition to data protection law, more specific rules regarding behavioural targeting are needed. If society is better off if certain behavioural targeting practices don't happen, the lawmaker should consider banning them.

Chapter 10 summarises the main findings and answers the research question. There's no easy solution, but legal privacy protection can be improved in the area of behavioural targeting. The limited potential of informed consent as a privacy protection measure should be taken into account. Therefore, the lawmaker should focus less on empowering people, and more on protecting people.

* * *

# 2 Behavioural targeting

What is behavioural targeting, and how does it work? Behavioural targeting, also referred to as behavioural advertising or online profiling, involves monitoring people's online behaviour, and using the collected information to show people individually targeted advertisements.[75] The Interactive Advertising Bureau of the United States, a trade association for online and mobile advertising, describes behavioural targeting as follows:

> A technique used by online publishers and advertisers to increase the effectiveness of their campaigns. Behavioral targeting uses information collected on an individual's web browsing behavior such as the pages they have visited or the searches they have made to select which advertisements to be displayed to that individual. Practitioners believe this helps them deliver their online advertisements to the users who are most likely to be influenced by them.[76]

In a simplified example, an ad is shown on a website based on the inferred interests of that specific visitor: these interests can be inferred by an advertising network. An ad network is a firm that acts as an intermediary between websites and advertisers. The ad network might profile somebody who frequently visits websites about recipes as a food enthusiast. If that person visits a news website, the ad network displays advertising for restaurants or cookbooks. When visiting that same news website, somebody who reads a lot of legal blogs might see advertising for law books.

---

[75] See e.g. Federal Trade Commission 2000 ("online profiling") and McStay 2011 ("behavioural advertising").
[76] Interactive Advertising Bureau United States, Glossary.

The chapter is structured as follows. Below is a glossary of some key terms. Section 2.1 and 2.2 introduce online advertising and the technology used for behavioural targeting. Section 2.3 to 2.7 sketch the process of behavioural targeting, divided in five phases: (1) data collection, (2) data storage, (3) data analysis, (4) data disclosure, and (5) targeting.[77] Section 2.8 concludes.

[77] Other authors also distinguish different phases of data mining, profiling, and data processing. See for instance Hildebrandt et al. 2008 (3 phases); Solove 2006 (4 phases); Solove 2008 (4 phases); Zarsky 2004 (3 phases), and Cabena et al. 1998, p. 43-44 (5 phases).

**Glossary**

Advertising network company

Advertising network companies, ad networks for short, connect advertisers to website publishers, and serve ads on websites. Using cookies or other technologies, an ad network can recognise a user when she visit websites on which the ad network shows ads.[78]

Advertising exchange company

Ad exchanges are automated market places where advertisers can trade with multiple ad networks in one place. The Interactive Advertising Bureau US provides the following description. "Ad exchanges provide a sales channel to publishers and ad networks, as well as aggregated inventory to advertisers. They bring a technology platform that facilitates automated auction based pricing and buying in real-time. Ad exchanges' business models and practices may include features that are similar to those offered by ad networks."[79]

Behavioural targeting

Behavioural targeting is the monitoring of people's online behaviour, to use the collected information to show people individually targeted advertisements.

Click-through rate

"The number of click-throughs per ad impression, expressed as a percentage."[80] For instance, if 3 out of a 1000 people click on an ad, the click-through rate is 0.3 %.[81]

---

[78] See for a more detailed description Interactive Advertising Bureau United States 2010.
[79] Interactive Advertising Bureau United States 2010. See also section 6 of this chapter.
[80] American Marketing Association dictionary.

Cookie

HTTP cookies, cookies for short, are small text files that a server can send to a browser. First party cookies are set by the website publisher, and third party cookies are set by others, such as ad networks. Third party cookies enable ad networks to follow people around the web. Tracking technologies that rely on storing information on a user's device that are used for purposes similar to HTTP cookies are sometimes called super cookies.[82]

Interactive Advertising Bureau (IAB)

The Interactive Advertising Bureau (IAB) is a trade association of online marketers, with branches in many countries. According to the IAB Europe website, "IAB Europe is the voice of digital business. Its mission is to protect, prove, promote and professionalise Europe's online advertising, media, research and analytics industries. Together with its members – companies and national trade associations – IAB Europe represents over 5,500 organisations."[83] The IAB also "promotes self-regulation for online behavioral advertising."[84] The IAB of the United States says on its website that one of its "core objectives" is to "[f]end off adverse legislation and regulation."[85]

Real time bidding

Real time bidding is a process where advertisers (or their intermediaries) bid on an automated auction for the right to reach a specific user, who is identified with a cookie. Real time bidding "creates a data market where users' browsing data are sold at auctions to advertisers."[86]

---

[81] See section 1 of this chapter.
[82] See section 2 of this chapter.
[83] Interactive Advertising Bureau Europe, website.
[84] Interactive Advertising Bureau Europe, website.
[85] Interactive Advertising Bureau United States, website.
[86] Castelluccia et a. 2013, p. 14. See section 6 of this chapter.

Website publisher

Website owners are often called website publishers.[87]

## 2.1 Online advertising

Behavioural targeting can be seen as the latest development in a decades-old trend of increasingly targeted advertising at smaller audience segments. Because media audiences became more fragmented in the 1970s, marketers started to pay more attention to targeting audience segments.[88] In the 1980s and 1990s direct marketing progressed to database marketing, "the use of customer databases to enhance marketing productivity through more effective acquisition, retention, and development of customers."[89] Marketers started to compile increasing amounts of consumer data.

In the early 1990s, marketers gave little attention to segmentation on the internet. Users were mainly well-educated, had relatively high incomes, and were based in a small number of Western countries. When more people started to use the internet, and more websites were published, segmenting and targeting became more important for advertisers.[90]

The trend towards targeted and personalised advertising is summarised well by the Association of National Advertisers, a trade association in the United States. At its hundredth anniversary in 2010, it adopted a Marketers' Constitution. "Marketing must become increasingly targeted, focused and personal," says the first article. The Marketers' Constitution adds that the "exciting, controversial, but extraordinarily important world of behavioral advertising offers enormous efficiencies to marketers and immense value to consumers."[91]

---

[87] The IAB describes a publisher as "[a]n individual or organization that prepares, issues, and disseminates content for public distribution or sale via one or more media" (Interactive Advertising Bureau United States Glossary).
[88] Turow 2011; McStay 2011.
[89] Blattberg et al. 2008, p. 4.
[90] See Cannon et al. 2007; McStay 2011, p 128-132; Newell 1997, p. 191; Turow 2006; Turow 2011.

Behavioural targeting also aims to fulfill another desire of advertisers, who seek information on the audiences they reach.[92] A a famous phrase in marketing literature is: "I know half my advertising is wasted. The trouble is, I don't know which half."[93] Since the beginning of the twentieth century, measuring how many people are reached with advertising has been a continuous quest. Commercial mass media, such as newspapers and television, could be seen as providing audiences to advertisers.[94] Bermejo explains: "since the audience becomes a commodity, those who purchase it, advertisers, need to be certain that they are getting what they pay for."[95] Firms adapt the way they measure audiences if a new communication channel emerges. Different methods are applied to print, radio, television, or the web.[96]

For example, in 1914 American newspaper publishers established the Audit Bureau of Circulation. This organisation provided advertisers with figures about circulation, in order to dispel doubts that publishers were giving advertisers inflated figures.[97] Radio complicated matters. Counting the number of people who listen to a radio show is harder than counting how many newspapers are sold.[98] An audience measurement industry developed to provide advertisers with statistics about listeners.[99] Early research methods involved calling people at home to ask what they were listening to.[100] Later, firms such as Nielsen used recording devices called "audimeters" that were installed in households. Similar recording devices are still used for television ratings.[101] They are installed in the homes of a sample group of viewers, and record what television programmes are watched. Firms arrange panels to answer questions, in order to obtain demographic information about viewers of certain programs.[102]

---

[91] Association of National Advertisers 2009.
[92] See Aaltonen 2011 (chapter 2); Bermejo 2007; Bermejo 2011.
[93] McStay 2010, p. 187; Turow 2006, p. 21. The quotation is attributed to different people.
[94] Turow 2006, p. 6.
[95] Bermejo 2007, p. 25. See also McStay 2011, p. 130-132.
[96] Bermejo 2007, p. 38-39.
[97] Andrejevic 2009, p. 82. The Audit Bureau of Circulation organisation still exists, now under the name Alliance for Audited Media. <www.auditedmedia.com> accessed 14 February 2014.
[98] Andrejevic 2009, p. 84. Bermejo 2007, p. 38-39.
[99] Bermejo 2007, p. 38-41.
[100] Andrejevic 2009, p. 86.

### *Internet marketing*

Formerly, "for-profit activities" were not allowed on the internet, but this prohibition was lifted in the early 1990s.[103] In 1994 the first banner advertisement was shown on the web, on the website HotWired.[104] Banner ads, or display ads, are rectangular ads on websites. The first ads on the web were bought in a manner comparable to advertising on television or in newspapers. On television, an advertiser pays a fixed fee, based on the expected number of viewers during a certain period. In print, the advertiser pays for the expected number of readers, based on circulation figures. On the web, it was possible to count the number of "impressions": the number of times an ad was displayed. In a "cost per mille" model, an ad network counts how often it shows an ad, and the advertiser pays for a thousand impressions.[105]

In the mid 1990s, many larger advertisers were still hesitant to spend money on web advertising. In particular, advertisers complained about the lack of information about internet audiences. For instance, before cookies, a website publisher couldn't tell the difference between visitors. A 1996 paper which was presented at an advertising conference complained: "twenty hits could mean 20 different people visited the site, or just one person clicked a computer mouse on the site 20 different times."[106]

---

[101] Andrejevic 2009 p. 87; Bermejo 2007, p. 41-42.
[102] Bermejo 2007, p. 108.
[103] See Murray 2007, p. 72-73. In 1992 the National Science Foundation Network still listed "for-profit activities" as "unacceptable uses", subject to some exceptions (NSFNET Backbone Services Acceptable Use Policy 1992).
[104] Turow 2011, p. 43. McStay 2010, p. 18. A banner ad on HotWired is usually referred to as the first banner ad, but McStay mentions that a Wikipedia entry speaks of a banner ad in 1993.
[105] Turow 2011, p. 43-44.
[106] Hong & Leckenby 1996, p. 7.

Advertisers successfully pushed for a different way of paying for internet advertising: a "cost per click" model.[107] In this model, an advertiser only pays the website if somebody clicks on the ad. Advertisers often buy ads through advertising networks. These ad networks typically use a cost per click model as well. There are more payment models for online advertising. For instance, in a cost per conversion model, the advertiser pays for every person that takes a certain action, such as buying a product. According to a report by the Interactive Advertising Bureau (IAB), around two thirds of all online advertising income is paid for per click, or per conversion.[108] The IAB is a trade organisation of online marketers, with branches in many countries.[109]

Few internet users click on ads. When an ad is shown to 1,000 people, on average between one and five people click on the ad. Hence, the click-through rate is in the order of 0.1 % to 0.5 %. To increase the click-through rate, ad networks aim to target advertising precisely. This gives firms an incentive to collect increasing amounts of data about individual internet users.[110] Since the 1990s, click-through rates have been falling dramatically. Prices for advertising are decreasing as well.[111] The number of websites however, keeps growing, so advertising space on the web is also growing. As the supply of advertising space grows, the prices go down.[112] Prices depend on many factors, and it's difficult to find exact numbers. Generally, an advertiser pays between one and four euro for 1,000 ads (a cost per mille).[113] Website publishers receive about half of that amount; the other half goes to the ad network.[114]

---

[107] Turow 2011, chapter 2 and 3.
[108] Interactive Advertising Bureau 2013, p. 11. The report summarises such payment models as "performance-based pricing." Around 65% of the 2013 revenues in the US were priced on a performance basis. Around 33% of the revenues were priced on a cost per mille model.
[109] The website of the European branch says: "IAB Europe is the voice of digital business. Its mission is to protect, prove, promote and professionalise Europe's online advertising, media, research and analytics industries. Together with its members – companies and national trade associations – IAB Europe represents over 5,500 organisations" (Interactive Advertising Bureau Europe, website). The IAB of the US says on its website that one of its "core objectives" is to "[f]end off adverse legislation and regulation" (Interactive Advertising Bureau United States, website).
[110] Turow 2011; Strandburg 2013, p. 127.
[111] See e.g. Glaser 2014.
[112] Launder 2014.

There's "surprisingly scant research" on how effective or how expensive behaviourally targeted ads are, when compared to contextual ads.[115] A few papers suggest that behavioural targeting leads to an increase of advertising income for website publishers, but each of these papers is criticised for its methods.[116] For instance, a paper by Beales, sponsored by the Interactive Advertising Bureau US, says that behaviourally targeted ads are about 2.7 times as expensive for advertisers than ads sold in a "run of network" model. A "run of network" means that ads are presented completely randomly, usually on websites with the cheapest advertising rates. However, Beales doesn't compare behavioural targeting with contextual advertising. Contextual advertising concerns, for instance, ads for cars on websites about cars. Contextual ads are probably more expensive than completely random ads.[117]

***Power relations in online media***

On the internet it's possible to present a different ad to each individual. Therefore, the ads on a webpage aren't necessarily related to the content of that page. In print media, by contrast, groups of readers see the same ad.[118] By way of illustration, a printed newspaper with many golf players among its readers could be a good place for a golf club manufacturer to advertise. The newspaper assembles an audience, and provides the advertiser access to this audience.[119] The price of an ad is based, among other things, on the number of readers. The newspaper tells advertisers that it sells 100,000 copies, and shows research that says that 70 % of its readers play golf. With behavioural targeting, an ad network can show a golf ad anywhere on the web to a person whose profile suggests that he or she likes golf. An ad network doesn't have to buy expensive ad space on a large professional news website to advertise to an individual. The ad network can reach that individual when he or she visits an unknown website, where advertising space is cheaper.

---

[113] Turow 2011, p. 78. Mitchell reports on an average price for thousand viewers of $2.66 for an online banner ad (Mitchell 2012). Beales mentions a price for thousand viewers of $4 for a behaviourally targeted ad (Beales 2010, p. 3).
[114] Turow 2011, p. 78.
[115] Mayer & Mitchell 2012, p. 8.
[116] See e.g. Strandburg 2013, p. 100-105; Mayer & Mitchell, 2012 p. 8.
[117] See Mayer & Mitchell 2012, p. 8; Strandburg 2013, p. 100-105.
[118] Not all readers see the same ad in print media. Some print magazines and newspapers adapt advertising to regions. In one case, the cover of an US magazine showed a map on which the subscriber's address was circled (Carr 2004).
[119] See Bermejo 2007.

Turow explains that publishers have less power in the online media environment than they had in the print environment. He quotes a digital marketing firm that says: "advertisers want to pay to reach the target audiences. They don't want to pay for the creation of content."[120] Advertising intermediaries and advertisers have more power than two decades ago. Hence, in the long run behavioural targeting may decrease ad revenues for some website publishers. Publishers that produce expensive content, such as online newspapers, might be better off with ads that aren't targeted at individuals. The editor of The Atlantic complains about the effects of behavioural targeting on the media: "[t]hen the digital transition came. The ad market, on which we all depend, started going haywire. Advertisers didn't have to buy The Atlantic. They could buy ads on networks that had dropped a cookie on people visiting The Atlantic. They could snatch our audience right out from underneath us."[121]

In addition to the advertisers' wish to segment audiences and to obtain information about the audience they reach, a third factor can help to understand the rise of behavioural targeting: the development of technologies that make behavioural targeting possible. Online advertising technology is discussed in the next section.

---

[120] Turow, 2011, p. 117.
[121] Madrigal 2013.

## 2.2 Advertising technology

In 1990, Berners-Lee invented the world wide web, an application that runs on the internet.[122] We use the web when we visit a website with our browser. The web consists of millions of web pages that are connected through hypertext.[123] Hypertext transfer protocol, or HTTP, is the network protocol that was developed for the web. The protocol enables communication between web browsers and web servers.[124] A web browser is software for users to browse the web, such as Chrome, Firefox, Internet Explorer, or Safari. A web server is a computer that holds the data of a web page. The hypertext transfer protocol includes the kinds of requests that a browser can ask to a server, and the different kinds of responses a server can send back to the browser. If somebody enters the webpage address (a URL, or uniform resource locator) in the browser, the browser sends that request to a server. The server sends back the requested documents, such as text or images. The server can record information about the computer that makes a request. Such "web logs" can include the time and date of the request, the IP address of the computer that makes the request, and information about that computer, such as the browser type and the operating system.[125]

The hypertext transfer protocol is stateless. This means that a web server sees each visit to a webpage as the web browser's first visit. After the browser has received the documents it requested, it breaks off the connection. When the user clicks a link, the browser contacts a server again. In short, a stateless system has "amnesia."[126] Statelessness wasn't a problem the first years after the web was invented, but in the early 1990s firms started thinking about online commerce. However, it was difficult to build virtual shopping carts for a web shop. In the web's stateless system, the web shop would see each browser request as coming from a new visitor.

---

[122] The internet is "an electronic network that parcels application information into packets and ships them among computers over wires and wireless media, according to simple protocols (rules) known by various acronyms." Berners-Lee 2010, p. 83.
[123] A website is a collection of web pages.
[124] See generally Gillies & Cailliau 2000.
[125] Kaushik 2007, p. 26-27.
[126] Schwartz uses the phrase "amnesia" in this context (Schwartz 2001).

***Cookies***

Cookies were invented to solve the problem of statelessness on the web.[127] One of the first popular web browsers was Netscape Navigator. In 1994, a 24-year old programmer at Netscape called Lou Montulli aimed to solve the problem of statelessness, to enable firms to build shopping carts for their websites. He invented cookies to give the web a memory.[128] Netscape quickly implemented cookie technology in its browser in 1994. Netscape didn't inform the browser users, and the browser didn't enable users to manage or refuse cookies.[129]

Cookies are small text files that a server can send to a browser. The browser saves that cookie. If the browser contacts that same server again, it sends back the cookie with its request. Like this, the server can recognise the browser. This is useful to remember the contents of a virtual shopping cart, language preferences chosen by a user, or to remember that a user is logged in. Session cookies are deleted when the browser is closed. Persistent cookies remain stored if the browser is closed and when the computer is turned off.

If a server places a cookie on a computer, in principle only servers from that same domain can read that cookie.[130] In brief, website X cannot read the cookies that website Y placed. If a user visits www.bookstore.com, that website may place a cookie on his or her computer.[131] Only servers from the same domain, such as bookstore.com or accounting.bookstore.com, can read that cookie. If the user later visits www.email.com, the servers from email.com can't read the cookies of bookstore.com.

---

[127] See generally on cookies St. Laurent 1998; Elmer 2004, chapter 6; Kesan & Shah 2003; Kristol 2001.
[128] The phrase "giving the web a memory" is borrowed from Schwartz 2001.
[129] Kesan & Shah 2003, p. 300; Turow 2011, p. 47-48.
[130] However, as explained below firms have found ways to work around this.
[131] The websites in the text are examples and aren't meant to refer to real websites.

Ad networks, however, have found a way to use cookies to track people around the web. “Third party cookies” are cookies that aren’t placed by the website publisher, but by a third party. If a user visits a website, say www.news.com, it seems that all elements on the screen are presented by news.com. But different parts of a website often come from different servers. For instance, a website might have a section, or “widget”, with weather information. The widget could be sent to the visitor’s browser from widgets.com. Social network site buttons on websites, such as the Facebook Like Button, are usually loaded from third party servers as well. Likewise, ads are usually sent to the visitor’s computer by a third party, for example from the domain advertising.com. This process is invisible for the visitor, who directed his or her browser to www.news.com. (When speaking of “third party cookies”, this study refers to cookies which aren’t operated by the website publisher, but by a third party, such as an ad network.[132])

To recognise internet users, ad networks also drop and read cookies on computers. In principle, such third party cookies are the same kind of cookies as the first party cookies that are used for digital shopping carts. But if a user first visits www.news.com, and then visits www.sports.com, an ad network that serves advertising on both sites can read its own cookies. By reading its cookies, an ad network can track internet users over all websites where it serves advertising, and can compile a list of websites somebody visits. “Cookies are used in behavioural advertising to identify users who share a particular interest so that they can be served more relevant adverts,” explains the Interactive Advertising Bureau UK.[133] Tracking people over various websites is sometimes called cross-domain tracking. Tracking within one website is also possible. An online store such as Amazon can recommend books based on a visitor's earlier browsing behaviour within the site.[134] This can be called "on site" or "first party" behavioural targeting.

---

[132] This study doesn’t use “third party” in the sense of the Data Protection Directive, which defines “third party” in article 2(f).

[133] Interactive Advertising Bureau of the United Kingdom 2009, p. 4.

The distinction between first party cookies and third party cookies is somewhat fuzzy. Firms can also use first party cookies for cross-domain tracking. For instance, firms can synchronise their own cookies with those of other firms. This way, a cookie that was installed as a first party cookie, can be used for cross-domain tracking.[135]

---

[134] Amazon has an ad network (Amazon 2014). If Amazon used data gathered through its ad network for its recommendations, this wouldn't be first party behavioural targeting.

[135] See section 6 of this chapter. See also Krux 2010; Tene Polonetsy 2012, p. 7; Castelluccia et al. 2013; Hoepman 2013.

**Illustration 1. An example of a cookie**

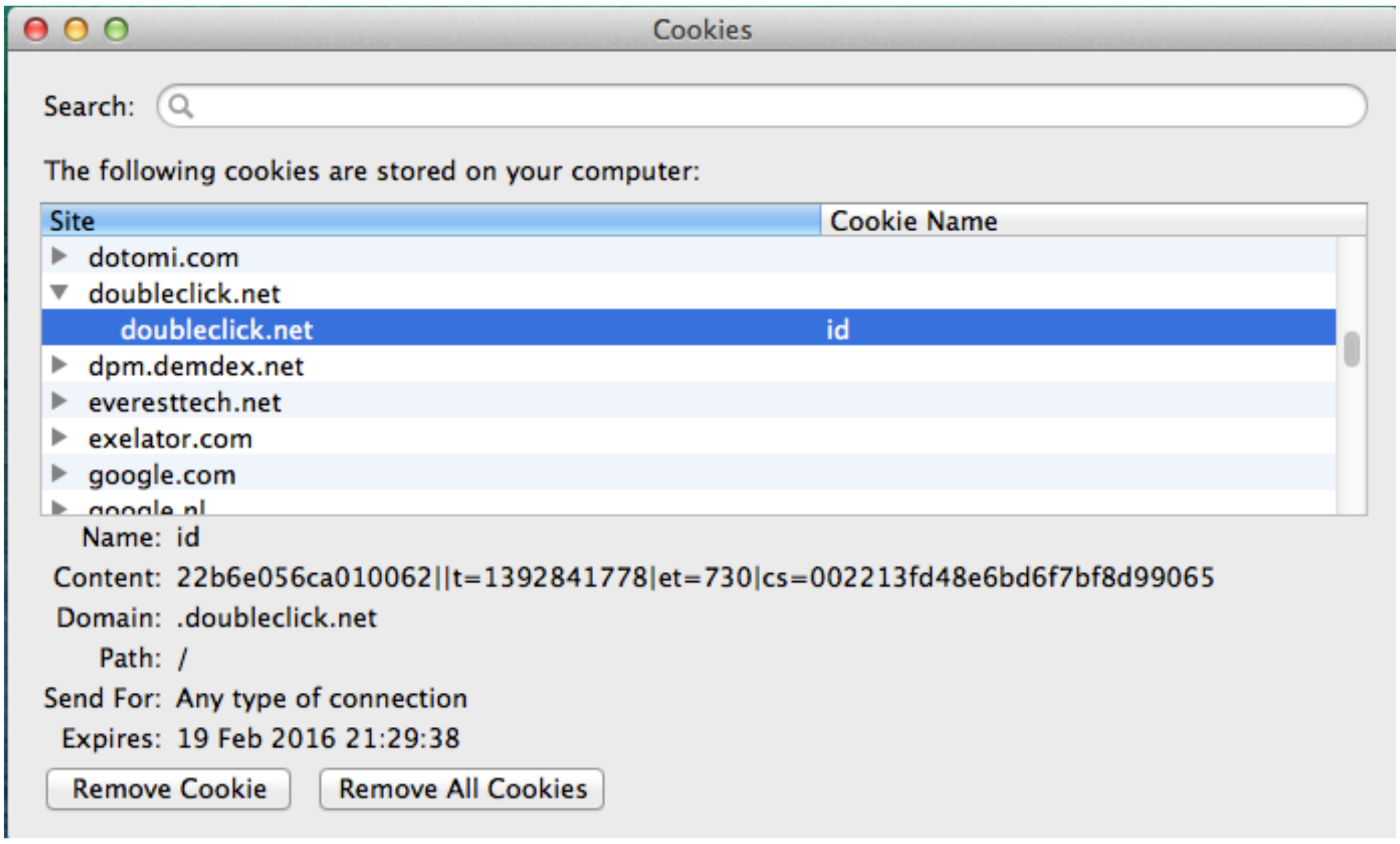

<u>Name</u>

In this case, the name of the cookie is "ID."

<u>Content</u>

The content is the unique number of the cookie. A firm doesn't have to allocate a unique number to a cookie. For instance, a cookie that is used to remember a website's laguage setting could say "en" or "nl."

<u>Domain</u>

The cookie is sent to the computer from the domain .doubleclick.net. In principle, only servers from the domain .doubleclick.net can access the cookie. In practice, firms have found ways to work around this.

Path

In this case, the path is set to “/”. In short, this implies that the cookie can be accessed from, for instance, doubleclick.net/a, and from doubleclick.net/b.[136]

Send for

In this case, the cookie says “any type of connection.” This means the cookie can be sent over an unsecured internet connection. Some cookies say “secure.” That means they can only be sent over an encrypted connection, which would make it harder for other parties to intercept and read the cookies.

Expires

The cookie expires on 16 February 2016. (The screenshot was made on 19 February 2014.) However, the cookie and the expiry date can be renewed when the user encounters DoubleClick ads on the web. If no expiry date is set for a cookie, the cookie is deleted at the end of the session: a session cookie.[137]

***Browsers***

In 1995 the Internet Engineering Task Force (IETF), an international standards body, also started discussing ways to solve the problem of statelessness.[138] For a while the IETF considered more privacy-friendly standards than the Netscape cookie standard. But as the popular Netscape browser already supported cookies, the IETF rejected the idea of developing an alternative from scratch. Therefore the IETF set out to build on Montulli's work and to improve the Netscape standard.[139]

[136] Internet Engineering Task Force 2000, RFC 2965, article 3.2.2.
[137] Internet Engineering Task Force 2000, RFC 2965, article 3.2.2.
[138] The IETF’s website says “the mission of the IETF is to make the Internet work better by producing high quality, relevant technical documents that influence the way people design, use, and manage the Internet” (Internet Engineering Task Force website).

In 1996, the IETF started worrying about a "potential privacy threat in 'third party cookies'".[140] The IETF feared that cookies could be used to track people around the web. Therefore, the IETF wanted browsers to block third party cookies by default. "We added wording to the specification that either outright prohibits a browser from accepting third-party cookies ('cookies in unverifiable transactions'), or that permits a browser to accept them, provided they are controlled by a user-controlled option whose default value is to reject them."[141] Kristol, one of the authors of the IETF standard, says that while the IETF saw the theoretical possibility for cross-domain tracking, it didn't realise that ad networks were already doing this:[142]

> Strangely enough, when we added the words about "unverifiable transactions" [i.e. third party cookies] to the [the draft for the standard], our direct motivation was not advertising networks (which at best we were only dimly aware of at that time). Instead, [IETF member] Koen Holtman had independently discovered the theoretical potential to use third-party cookies for profiling and persuaded members of the subgroup that Europeans, at least, would be very troubled by the potential abuse of privacy they could promote.[143]

Meanwhile, the marketing industry had realised the potential of cookies. Trade publication AdAge discussed the usefulness of cookies in 1996. "Ever since the Web gained prominence as a commercial medium, marketers and publishers have demanded some way to understand how users move through their sites. Enter the cookie, technology developed by Netscape Communications Corp."[144]

---

[139] Kesan & Shah 2003, p. 300-304.
[140] Kristol 2001, p. 159-160.
[141] Kristol 2001, p. 160.
[142] Kristol 2001, p. 166.
[143] Kristol 2001, p. 180.

The IETF released a standard in 1997. The standard "strongly encourages" browsers not to allow third parties to set cookies without the user's consent.[145] "A user agent should make every attempt to prevent the sharing of session information between hosts that are in different domains. Embedded or inlined objects [such as ads served by third parties] may cause particularly severe privacy problems if they can be used to share cookies between disparate hosts."[146] IETF's 1997 standard was met with hostility by the online marketing industry. Ad networks feared for their business model. One of the founders of DoubleClick, an early ad network, said that a default setting that doesn't allow third party cookies "is basically equivalent to not allowing them at all, because 99% of the population will see no reason to change the default."[147] Some firms said that large parts of the web are dependent on advertising.[148] Similar arguments are still used in current discussions about the regulation of cookies and behavioural targeting.

Kristol, who worked on the IETF standard, expected that browser vendors would implement privacy-friendly default settings. But the popular browser vendors, Microsoft and Netscape, basically ignored the 1997 standard and chose to allow third party cookies by default:[149]

---

[144] Carmichael 1996.
[145] Internet Engineering Task Force, RFC 2109, 1997, article 8.3. This document is a "request for comments", rather than a definitive standard.
[146] Internet Engineering Task Force, RFC 2109, 1997, article 8.3.
[147] Merriman 1997. DoubleClick was acquired by Google in 2007 (Google Investor Relations 2007). See on the power of default options chapter 7, section 4.
[148] Kristol 2001, p. 188. See on the economics of online advertising also chapter 7, section 2.
[149] In 2001 the IETF released a revised standard, RFC 2965. Again the standard emphasised that browser vendors should enable informed consent for third party cookies, and again browser vendors didn't follow the standard. Kristol notes that the paying customers for the major browser vendors weren't the browser users, but firms who profited, directly or indirectly, from third party tracking (Kristol 2001, p. 169-170). See regarding browsers also chapter 8, section 5 (on Do Not Track).

> We chose the default setting for third-party cookies because we felt it served the privacy expectations of users, especially European users, who, we inferred from European Union recommendations, might have high expectations. (…) Surely, we reasoned, [browser] vendors would choose to take such concerns into account for all users. Evidently we reasoned wrong. Vendors have steadfastly supported the advertising industry, leaving third-party cookies enabled by default.[150]

In 2014, most browsers still allow third party cookies by default. Perhaps this can partly be explained by the fact that most browser vendors have affiliated companies that carry out behavioural targeting and use third party cookies. In early 2013, Mozilla said it considered having its Firefox browser block third party cookies by default.[151] People would thereby have to change the browser settings to allow ad networks to track them.[152] The Interactive Advertising Bureau Europe was not amused, stating that "[t]he new Mozilla setting denies consumer choice, undermines industry efforts to responsibly manage control. Plus Mozilla threatens to completely undermine ad-funded content on the internet."[153] Later in 2013, Mozilla backtracked on its plans.[154]

Cookies don't offer a perfect tracking mechanism, because they identify a browser. A computer with two browsers installed (say Firefox and Safari) has two separate collections of cookies. If several people use the same browser on a computer, a cookie enables a website publisher to recognise the browser, rather than a person.[155]

---

[150] Kristol 2001, p. 166. The EU recommendation that Kristol refers to is: Article 29 Working Party 1999, WP 17, which said: "[c]ookies should, by default, not be sent or stored" (p. 3).
[151] Mozilla isn't in the behavioural targeting business. Mozilla does receive money from Google, which does behavioural targeting (see Mozilla blog 2011).
[152] Fowler 2013.
[153] Interactive Advertising Bureau Europe 2013. The remark that Mozilla undermines use control probably refers to the fact that the marketing industry offers people the possibility to opt out of receiving targeted advertising, but this system uses third party cookies. Hence, a browser that blocks third party cookies could be said to hinder the industry's opt-out system. See on the industry's opt-out system chapter 6, section 3, and chapter 8, section 5.
[154] Temple 2013b.
[155] This would be different if the users have separate accounts on the computer.

Furthermore, cookies rarely work in smart phone apps, and some browsers for smart phones block third party cookies by default. For instance, the Safari browser blocks third party cookies, which makes it harder for ad networks to track people's browsing behaviour on Apple devices.[156]

***Beyond cookies***

While many firms still use cookies for behavioural targeting, there are other ways to collect data for behavioural targeting. For instance, web beacons, or web bugs, are invisible elements on a web page or in an email message. Website publishers, or third parties such as ad networks, can operate a beacon. The firm that uses a beacon can see whether the web page has been visited or whether the email message has been opened or forwarded. Through a beacon, firms can set and read cookies as well. Beacons in emails can also be used to tie an email address to a cookie-based profile.[157]

People who want to avoid being tracked on the web can block or delete third party cookies. It has been estimated that 20-25 % of all internet users delete third party cookies.[158] This doesn't mean that people manually delete or block cookies. Anti-virus software sometimes deletes third party cookies. And as noted, Apple's Safari browser blocks third party cookies by default.

But some firms work around such browser settings.[159] For instance, Google bypassed the settings of the Safari browser.[160] There are many ways for firms to circumvent cookie deletion. In 2009 researchers found that firms used "flash cookies" for tracking.[161] Flash cookies are harder to delete than conventional cookies. Flash cookies were placed through more than half of the 100 most popular websites in the United States.[162] European firms have used them as well.[163] Some firms use flash cookies to reinstall, or "re-spawn", regular cookies that were deleted by the user.[164] Trade publication Mediapost wrote in 2009 about a firm: "[w]hen Tatto began to develop its core behavioral frameworks and algorithms, it believed Flash cookies would remain the best way to slow the ability of consumers to delete cookies from their computers."[165] In sum, people deleted third party cookies to protect their privacy, and many firms re-installed those tracking cookies, on purpose, to circumvent people's privacy preferences.

---

[156] Felten 2012.
[157] Kaushik 2007, p. 28-30.
[158] Kaushik 2009, p. 129.
[159] Krishnamurthy & Wills 2009.
[160] Felten 2012; Mayer 2012. In the US, Google paid 22.5 million dollar to the Federal Trade Commission to settle charges it misrepresented privacy assurances to Safari users (FTC 2012, with further references). Also in the US, Google entered a 17 million dollar settlement agreement with multiple states in 2013 (Schneiderman 2013). At the time of writing, there's an on-going court case in the United Kingdom regarding the same matter (See High Court 16 January 2014, Vidal-Hall & Ors v Google Inc [2014] EWHC 13 (QB))
[161] Gomez et al 2009. See also Cranor & McDonald 2011; Ayenson et al. 2011; Hoofnagle et al. 2012.

There are more ways to re-install third party cookies that users have deleted. Computer researcher Kamkar shows that an identifier can be placed in fourteen different locations on a computer. He invented the "evercookie" that is stored in all these locations. It's therefore difficult to delete. The evercookie makes it possible to track an internet user when he or she uses different browsers on one device.[166] Identifiers that are used for purposes similar to third party cookies are sometimes called super cookies or zombie cookies.[167] "The entire point of new tracking methods," conclude Hoofnagle et al., "seems to be to ensure that users are ignorant of them."[168]

Another way to track people is by passive device fingerprinting. This technique involves recognising a device by looking at information it transmits, without first placing a cookie or similar identifier. A computer's browser can be recognised by looking at characteristics such as its settings, plug-ins and installed fonts. A device fingerprint is "a set of system attributes that, for each device, take a combination of values that is, with high likelihood, unique, and can thus function as a device identifier.”[169] Researchers have fingerprinted smart phones by looking at the accelerometer, the sensor that measures vibration or acceleration.[170] Some firms use device fingerprinting for behavioural targeting. One firm claims to have fingerprinted 1.5 billion devices.[171] While some savvy users may know how to delete flash cookies and other identifiers, it’s very difficult to prevent one’s device being recognised by its fingerprint.[172]

---

162 Soltani et al. 2009.
163 Helberger et al. 2011. See also Helberger et al. 2012.
164 Soltani et al. 2009.
165 Sullivan 2009.
166 Kamkar 2010. See also Ayenson et al. 2011.
167 Olsen 2011.
168 Hoofnagle et al. 2012, p. 291. Hoofnagle adds about tracking: "in recent years, the methods have started to look more like computer hacking" (quoted in Temple 2011).

People’s behaviour can also be tracked by installing software on their devices. Such software is called adware if it displays advertising. If people don’t like adware, they tend to call it spyware.[173] Adware is usually bundled with software installed by a user, such as file sharing software,[174] a music player[175] or a browser toolbar.[176] A firm called Flurry offers analytics software that app developers can include in their apps. Flurry’s analytics software is installed on over 1.4 billion mobile devices. Flurry also enables advertisers to target mobile users. In 2014, Yahoo announced that it would acquire Flurry.[177]

Deep packet inspection takes a different approach than the above-mentioned technologies. Deep packet inspection entails opening the digital packets that are sent over the internet, to look at the contents.[178] To illustrate, a firm called Phorm contracted with internet access providers to inspect their customers’ internet traffic. In 2006 a large access provider in the United Kingdom did tests with Phorm, without informing its subscribers. After media attention and parliamentary hearings, English access providers severed their business ties with Phorm. Later Phorm focused on other regions, such as South America and Asia.[179] Mobile operators can use deep packet inspection for behavioural targeting as well.[180] Deep packet inspection enables firms to access more data than web browsing behaviour. For instance, a firm that uses deep packet inspection can read the contents of email messages.[181]

---

[169] Acar et al. 2013, p. 1. See generally on device fingerprinting Eckersley 2010; Joosen et al. 2013.
[170] Temple 2013a; Dey at al. 2014.
[171] Iovation 2013.
[172] Acar et al. 2013. See on device fingerprinting and EU law chapter 8, section 4.
[173] See on spyware and adware Federal Trade Commission 2005, p. 3-4.
[174] The popular file sharing software Kazaa was bundled with adware by 121 Media, which would later change its name to Phorm. McStay 2011, p. 20.
[175] Realplayer included spyware: Smith 1999. SONY included spyware on music CDs (Russinovich 2005; Federal Trade Commission 2007).
[176] For example, the firm Dollarrevenue enticed people to install a toolbar that also collected information. The Dutch telecommunications regulator fined the company one million euro based on the Dutch implementation of article 5(3) of the 2002 e-Privacy Directive, but the fine was overturned in appeal (College van Beroep voor het bedrijfsleven (Trade and Industry Appeals Tribunal), 20 June 2013, ECLI:NL:CBB:2013:CA3716 (Dollarrevenue/Autoriteit Consument en Markt). See in English: Libbenga 2007).
[177] Yahoo 2014 (Flurry).
[178] See generally Asghari et al. 2012; Kuehn & Mueller 2012; Kuehn 2013; Parsons 2013.

Behavioural targeting isn't limited to the world wide web. For instance, providers of smart phone apps often enable ad networks to do behavioural targeting. Apps make use of the internet, but not necessarily of the web.[182] Many types of firms are interested in behavioural targeting income. For example, Akamai, an internet infrastructure provider that can see up to 30% of all internet traffic, is reported to inspect traffic for behavioural targeting.[183] In 2013, a Dutch firm in the smart TV business was found to track people's viewing behaviour. The firm had plans to use the data for behavioural targeting.[184]

***Recent developments***

In around 2007 the online marketing industry had recovered from the Dotcom crash of 2000. Since then, the online marketing industry is becoming increasingly centralised. Scale is important for behavioural targeting.[185] An ad network that can follow people over only a dozen websites may not be able to compile profiles that are detailed enough to improve the click-through rate on ads. Research shows that an increasingly small number of parties collects increasing amounts of data.[186] Large players such as Google, Yahoo, Microsoft and Facebook often buy smaller marketing firms.[187] In 2012, 70% of all online advertising revenue in the United States went to the top 10 marketing firms, according to a report by the Interactive Advertising Bureau.[188] 89% of the revenue went to the top 50.[189] Another report says five firms, Facebook, Google, Yahoo, Microsoft and AOL, collected 51% of all income from display advertising in the US in 2013.[190] In 2009, that share was 38%.[191] By one estimate, Facebook and Google accounted for two thirds of all mobile advertising income worldwide in 2013.[192] In sum, there's increasing consolidation in the online marketing industry.

---

[179] See on Phorm McStay 2011, p. 15-42; Bernal 2011; European Commission 2009. See also chapter 6, section 3.
[180] Center for Democracy & Technology 2013, p. 6; Cisco 2014. See also Verizon 2014.
[181] This wouldn't work if the emails were encrypted.
[182] Berners-Lee 2010.
[183] Angwin 2010.
[184] College bescherming persoonsgegevens (Dutch DPA) 2013 (TP Vision).
[185] Brown et al. 2010, p. 74; Evans 2008; Evans 2009.
[186] Krishnamurthy & Wills 2009.

In autumn 2013, Microsoft and Google presented plans for their own proprietary tracking identifiers.[193] Apple already had a similar technology in place.[194] Such developments could lead to less competition in the behavioural targeting business. Rotenberg warns that people would have less control than with cookies: "the problem is about to get much worse – tracking techniques will become more deeply embedded and a much smaller number of companies will control advertising data."[195] For example, a smart phone manufacturer could decide to block tracking technologies of competitors on its phones. If an advertiser wanted to reach users of such phones, it couldn't choose any ad network, but would have to work with the phone developer.[196]

Currently behavioural targeting happens mostly when people use a computer or a smart phone. But the borders between offline and online are melting away.[197] Phrases such as ubiquitous computing, the Internet of Things, and ambient intelligence have been used to describe – or promote – such developments.[198] When the new version of IP addresses is implemented (IPV6), there will be so many IP addresses that every object could have its own IP address.[199] If objects are connected to the internet, firms could use the data processed through those objects for behavioural targeting.[200] For example, a fridge that's connected to the internet could order groceries. Firms could analyse consumption patterns for marketing purposes.[201]

---

[187] Evans 2009; Angwin 2014, p. 31.
[188] Interactive Advertising Bureau 2013, p. 11.
[189] Interactive Advertising Bureau 2013, p. 11.
[190] Pew Research Center 2014. See also Pew Research Center's Project for Excellence in Journalism 2013.
[191] Pew Research Center 2014. See also Pew Research Center's Project for Excellence in Journalism 2013.
[192] Emarketer 2014. Pew Research Center 2014 says "nearly three quarters (73%) of (…) mobile display dollars [in the US] are collected by five companies – Facebook, Google, Pandora, Twitter and Apple."
[193] Soltani 2013 (about Google); Peterson 2013 (about Microsoft).
[194] Arnott 2013.
[195] Quoted in Tate 2013.
[196] The Interactive Advertising Bureau is actively discussing the future of behavioural targeting. For instance, it has a working group examining "privacy and tracking in a post-cookie world" (Interactive Advertising Bureau United States 2014).
[197] Hildebrandt 2011, p. 11. See also Greenfield 2006.

Some recent developments remind one of behavioural targeting in the physical space, like in the film the Minority Report.[202] For instance, an Italian firm sells mannequins with built-in cameras. The firm's website says that the mannequins "would make it possible to 'observe' who is attracted by your windows and reveal important details about your customers: age range; gender; race; number of people and time spent."[203] A drinks machine in Japan uses a camera to estimate age and gender of the user, to recommend drinks.[204] There are billboards with facial recognition technology that adapt their images to the people looking at the billboard.[205] One firm summarises: "a few years from now, we and other companies could be serving ads and other content on refrigerators, car dashboards, thermostats, glasses, and watches, to name just a few possibilities."[206]

---

[198] "Ubiquitous computing has as its goal the nonintrusive availability of computers throughout the physical environment, virtually, if not effectively, invisible for the user" (Weiser 1993, p. 71). The internet of things can be described as "a dynamic global network infrastructure with self configuring capabilities based on standard and interoperable communication protocols where physical and virtual "things" have identities, physical attributes, and virtual personalities and use intelligent interfaces, and are seamlessly integrated into the information network." (Bassi et al. 2011). Ambient intelligence refers to "digital environments in which the electronics are sensitive to people's needs, personalized to their requirements, anticipatory of their behavior and responsive to their presence" (Philips Research 2014; see also Van Den Berg 2009).
[199] See Internet Engineering Task Force 1995, RFC 1883.
[200] An article in the Pervasive Computing Journal describes possibilities for targeted advertising in a ubiquitous computing environment, and calls ubiquitous advertising "the killer application for the 21st century" (Krumm 2011).
[201] See Calo 2013a.
[202] Spielberg 2002.
[203] Almax 2012.
[204] Lies 2010.
[205] Chen 2012.
[206] Google 2013 (letter to United States Securities and Exchange Commission).

### 2.3 Phase 1, data collection

As previously noted, this study analyses behavioural targeting by distinguishing five phases: (1) data collection, (2) data storage, (3) data analysis, (4) data disclosure, and (5) targeted advertising. The distinction in five phases is a tool to analyse the behavioural targeting process. The distinction helps when analysing privacy problems and when applying data protection law in later chapters. The phases don't suggest a chronological description of the behavioural targeting process. Different phases overlap. For instance, selling data to another firm falls within phase 4, data disclosure. But the buyer that obtains data is in phase 1, data collection.

During the first phase of behavioural targeting, firms collect information about people's online behaviour. People's behaviour is monitored, or, as it is often called: "tracked."[207] Slightly adapting a description by the International Working Group on Data Protection in Telecommunications,[208] tracking could be described as collecting data on user activity from a computer or other device while using the internet in order to combine and analyse the data for commercial and other purposes.[209] This study uses the word "track" in a common, non-technical sense.[210]

Data collection for behavioural targeting happens on a large scale, and ad networks have a wide reach. For instance, major news websites such as the New York Times,

---

[207] In this study, the use of the word "monitoring" isn't meant to have a particular legal meaning. Article 3(2)(b) of the European Commission proposal for a Data Protection Regulation aims to make the Regulation applicable to non-EU firms that "monitor [the] behaviour" of data subjects residing in the Union.

[208] This "Berlin Group" was founded in 1983 and consists of representatives from Data Protection Authorities and other bodies of national public administrations, international organisations and scientists from all over the world.

[209] International Working Group on Data Protection in Telecommunications (Berlin Group) 2013, p. 1. The original definition is as follows: "the collection, analysis and application of data on user activity from a computer or device while using various services of the Information Society (hereinafter: the Web) in order to combine and analyze it for different purposes, from charitable and philanthropic to commercial. We consider various forms of market research to fall within this definition of Web Tracking, for example outreach measurement (the degree to which users are served with ads across the Web), engagement measurement (the degree to which users interact with services across the Web) and audience measurement (the degree to which micro profiles can be derived from users interacting with services across the Web)" (internal footnotes omitted). See for a similar definition Van Eijk 2012.

[210] See chapter 8, section 5 about the meaning of "tracking" in the context of discussions on the Do Not Track standard.

The Guardian, and BBC news allow ad networks to track their visitors.[211] In 2009 Gomez et al. analysed 400,000 websites and found that Google would be able to track people's browsing behaviour on 88% of the tested websites.[212] In 2010, 49 out of the 50 most popular American websites used tracking technologies.[213] Hoofnagle & Good found that in October 2012, a visit to the most 100 popular websites led to receiving 5493 third party cookies, from 457 different third parties. 21 of the most popular 100 sites placed more than 100 cookies. Various kinds of "super cookies" were placed through the top 100 websites as well. Moreover, the researchers found a trend towards more tracking when compared with an earlier test.[214]

Firms can collect detailed information about people's online activities, based on, for instance, what people read, what videos they watch, what they search for, and what they post on social network sites. Firms can collect up-to-date location data of users' mobile devices, data that people submit to websites themselves, and many other types of data. A 2010 industry report discusses some of the data that are collected for advertising:

> Every Web page's individual views, every word typed in a search query box (also known as the "database of consumer intentions"), every video download, and even every word in an e-mail may create one more data point that a marketer can

---

[211] On 23 February 2014, I found multiple third parties on all three websites, using Ghostery (Ghostery 2014). Ghostery is a browser plug-in which enables the user to detect and block third party tracking on websites.
[212] Gomez et al. 2009, p. 27. However, the researchers noted in 2009 that they "are not claiming that Google aggregates information from each of these trackers into a central database, though it does possess the capability to do so. It appears that they [Google] strive to keep data in silos" (p. 27). The Dutch Data Protection Authority found Google DoubleClick ads on more than 20%, and Google Analytics on more than 65% of the 8000 most popular websites in the Netherlands (College bescherming persoonsgegevens 2013 (Google), p. 12-13).
[213] Angwin 2010. The tracking-free website was Wikipedia.
[214] Hoofnagle & Good 2012.

> leverage and use to more precisely target the audience with customized media placement and messaging.[215]

Schedule 2.3 below gives an overview of the kinds of data that can be collected for behavioural targeting. Many categories in the schedule are adapted from a report on the future of advertising by Brown et al.[216] The categories serve as illustrations and sometimes overlap. Some categories concern the content of data; other categories consider the way in which data are captured.

In 2010, Brown et al. suggested that in the future, information about people's psychological and physical state might be used for targeted advertising as well.[217] Some game computers measure the player's heart rate (an example of physical state data), but currently this information isn't used for advertising.[218] In 2013, at least one firm enables advertisers to target people who play computer games with ads during times such as "congratulatory moments", or "moments of rescue."[219]

---

[215] Landry et al. 2010, p. 1. The report borrows the phrase "database of intentions" from Battelle 2003.
[216] Brown et al. 2010, p. 30-33.
[217] Brown et al. 2010, p. 39.
[218] Brown et al. 2010, p. 32.
[219] MediaBrix.

**Schedule 2.3. Types of data processed for behavioural targeting**

Web browsing data

A simple version of behavioural targeting concerns the collection of browsing behaviour, by an ad network for example. The data can reveal a lot about a person's interests. Information on a person's surfing behaviour can be seen as a category of media consumption data.

Media consumption data

Behavioural targeting can also use other types of media consumption data. For instance, a firm that offers video content on the web or smart TV could register what a person watches.[220] In some cases, software to play music or video files sent information back to the vendor.

Search data

Major search engine providers, such as Bing and Google, store all search queries of their users. The providers personalise the search results based on earlier behaviour of the user. The search queries can be used for behavioural targeting.

[220] Brown et al. 2010, p. 31.

Other intentional data

Search data can be seen as a category of intentional data: information that shows people's intentions. Firms can also infer intentional data in other ways. For instance, a person who uses an online mortgage calculator might be interested in obtaining a new mortgage.[221] And users of price comparison sites are likely to be interested in buying the product of which they compare prices.

Transaction data and pre-transaction data

Transaction data relate to what people have bought or rented.[222] Online shops can use such data for behavioural targeting. Banks and credit card firms have access to transaction data as well, but in Europe they don't seem to share such data for behavioural targeting.[223] An example of pre-transaction data is information about which products a person views in an online shop.

Demographic data

Demographic data concern for instance a person's gender or age. A book on database marketing gives the following examples: "age, sex, family size, family life cycle, income, occupation, education, religion, race, nationality."[224]

Psychographic data

These are data about a person's character. Lifestyle, social class, and personality are examples from marketing literature.[225]

---

[221] Business Wire 2012.
[222] Brown et al. 2010, p. 31.
[223] In the US, credit card companies often share data about customer purchases with direct marketers. See e.g. Dwyer v. American Express Co. 625 N.E.2d. 1351 (Ill. App. 1995).
[224] Newell 1997, p. 150.
[225] Newell 1997, p. 150.

Communication contents

People's communications can also be analysed for behavioural targeting. Some email providers analyse the contents of email messages for marketing purposes. A well-known example is Google's Gmail service.[226] Social network site providers can also analyse the contents of messages.[227]

Social data

Social data concern relationships between people.[228] People with friends that drive a Toyota may be interested in a Toyota too. Social network sites such as Facebook and LinkedIn, email service providers, and mobile operators, have access to social data.[229] Some firms automatically scan the web, searching for information about people's relationships on social network sites, or to extract information from blog post, tweets, etc.[230] Marketing firm 33Across specialises in social data, and says that it reaches "over 1.25 billion users."[231]

Self-provided data

Website publishers can ask people to provide information. It's often reasonably clear when a firm requests data for marketing purposes, for instance when a website asks for information before a visitor can download something. But sometimes people might not realise that data will be used for marketing, for example when a firm uses a game or a quiz to entice people to disclose information.[232] Search data can also be seen as a category of self-provided data.

---

226 Yahoo also scans the messages in its email service for advertising (Gallagher 2012).
227 Soltani &Valentino-DeVries 2012.
228 Brown et al. 2010, p. 31.
229 See for a definition of social network sites boyd & Ellison 2007.
230 McStay 2011, p. 5. See also Gürses 2010, p. 100-101.
231 33 Across 2012.
232 See e.g. College bescherming persoonsgegevens (Dutch DPA) 2009 (Advance Concepts).

Subscription data

Firms can ask people to provide information when they sign up for a service. Subscription data are a category of self-provided data.

Location data, fixed

Examples of fixed location data are an address and a ZIP code. An IP address often gives a rough indication of a computer's location. A location could give information about a person's environment, for instance whether he or she lives in a suburban area, a city centre, or a rural area.[233]

Location data, mobile

Mobile location data refer to mobile devices, such as phones or tablets. Mobile location data can show where a person is in almost real-time.[234] Various parties have access to such location data. Smart phone apps sometimes send location data to the app provider, or to ad networks.[235] Some in the industry have high hopes for advertising on mobile devices.[236] If a person's profile suggests that he or she likes Italian food, a pizzeria might advertise a deal when he or she is in the area around lunchtime. Some firms track people's movements in shops by analysing signals emitted by people's phones, such as Bluetooth and Wi-Fi signals.[237]

---

[233] Center for Democracy & Technology 2009, p. 16.
[234] Center for Democracy & Technology 2009, p. 16.
[235] Thurm & Iwatani Kane 2010.
[236] Peterson 2012.
[237] See Future of Privacy forum 2013.

Contextual data

Contextual data refers to data about content.[238] For instance, contextual data can concern the language and the subject matter of a web page. If a car manufacturer buys advertising space on a website about cars, that would be called contextual advertising. Ads can be matched automatically to a site's content, by having software analyse the website's text.[239] Many behavioural targeting firms aim to take the website content into account as well. A cruise operator probably doesn't want its ads to be shown next to news about a ship disaster.

Environmental data

Environmental data concern for example local conditions such as the weather.[240]

Time-related data

Many firms adapt advertising to the time of day.[241] For example, advertising for restaurants might be shown around dinnertime.

Offline data

Offline data is a catch-all phrase for data that are collected from sources other than the internet. For instance, supermarkets use loyalty card programs to collect transaction data.[242] There are various ways of tying such data to online profiles.[243] The offline/online distinction is becoming less relevant, as more devices are being connected to the internet.

---

[238] Brown et al. 2010, p. 32.
[239] See for instance Google Adsense 2014.
[240] Brown et al. 2010, p. 32. This study uses the phrase contextual data for data about content. Brown et al. use the phrase contextual data differently as the overarching term for data that aren't about a person but about the environment.
[241] Brown et al. 2010, p. 32. See also McStay 2010, p. 44; McStay 2011, p. 5.
[242] See Pridmore 2008.
[243] In the United Kingdom, marketing firm Yahoo has enriched online profiles with data obtained from loyalty cards (Charlton 2010).

### 2.4 Phase 2, data storage

In the second behavioural targeting phase, firms store the data, tied to a unique identifier such as a cookie. For instance, a profile of a person might contain a list of websites that somebody visited. Or a profile might contain a person's interest categories, such as "cooking & recipes" or "mountain & ski resorts."[244] A profile is a "set of correlated data that identifies and represents a data subject."[245] An individual profile generally refers to a single person.[246] For ease of reading, this study also uses the word "profile", rather than "individual profile." Instead of individual profile, phrases such as "data double",[247] "data shadow",[248] or "digital dossier" are also used in literature.[249]

In computer science, nameless individual profiles are referred to as pseudonymous. "A pseudonym is an identifier of a subject other than one of the subject's real names." [250] Firms using behavioural targeting often call individual profiles "anonymous", when they don't tie a name to the profiles. Group profiles don't contain information about a specific person, but about a group or a category.[251] Unlike individual profiles, group profiles can be anonymous. "Anonymity of a subject means that the subject is not distinguishable from the other subjects within a set of subjects."[252] Chapter 5 returns to the topic of pseudonymous data, and shows that data protection law generally applies to such data.

Some firms have individual profiles on hundreds of millions of people. For instance, Facebook had over 1 billion monthly active users in 2014.[253] Google says its "Display

---

[244] The examples taken from Google Ad Settings 2014.
[245] Hildebrandt & Backhouse 2005, p. 106.
[246] Hildebrandt & Backhouse 2005, p. 106. In the context of this study, a pseudonymous profile may contain information about multiple users of one computer.
[247] Haggerty & Ericson 2000.
[248] Garfinkel 2000, p. 70. Garfinkel says the phrase "data shadow" was coined by Alan Westin in the 1960s.
[249] Solove 2004, chapter 2.
[250] Pfitzmann & Hansen 2010, par. 9.
[251] Hildebrandt & Backhouse 2005, p. 106.
[252] Definition taken, and slightly adapted, from par. 3 and footnote 18 of Pfitzmann & Hansen 2010.
[253] Facebook says it had "1.35 billion monthly active users as of September 30, 2014" (Facebook 2014).

Network reaches 83% of unique Internet users around the world."[254] But some lesser-known firms also have information about many people, such as the Rubicon Project ("600 million monthly unique users"),[255] and AddThis ("1.7 unique users worldwide").[256]

Firms can enrich individual profiles by tying data sets together. For instance, some firms can tie a name or an email address to a profile. Providers of social network sites such as Facebook know the names of many users. An email provider that uses behavioural targeting could tie an email address to many of its profiles. If a firm knows the name behind a profile, it could use the name to add more data to the profile.[257] Behavioural targeting profiles can be detailed. The ValueClick firm tells its advertising customers: "our database stores an average of 204 attributes for 97% of all online users."[258]

Some firms tie data collected on one device to data collected on another device: "cross device targeting." Somebody who searched for a car on his or her computer might be targeted with related ads on his or her phone.[259] If somebody uses the same email or social network account on both devices it's easy to link the devices to one person. Another way to link a person to multiple devices is looking at the IP address. If somebody uses his or her smart phone and laptop at home, both devices may use the same IP address every night. It's also possible to follow somebody while he or she uses various devices by analysing that person's browsing behaviour. "Users have very specific browsing patterns," explains Hoepman. "Everyone has his personal list of favourite websites (recall that your top five favourite movies are quite identifying). In fact, your browsing history becomes unique after a few visited websites. And people read their favourites in a fixed order."[260] A firm called Drawbridge uses this technique, and claims it has connected "over 1 billion customers across devices."[261] The firm claims it can also recognise different users of one device by analysing their behaviour.[262]

---

[254] Google AdWords 2014. "The Display Network is a collection of partner websites and specific Google websites – including Google Finance, Gmail, Blogger, and YouTube – that show AdWords ads. This network also includes mobile sites and apps."
[255] Rubicon Project.
[256] AddThis 2014.
[257] See e.g. Charlton 2010.
[258] Elsewhere on the website, ValueClick adds that the number concerns all online users "tracked by ComScore in the US" (ValueClick). In 2014 the firm merged with other firms to form Conversant Media, which claims to be able to target 263 million people (Conversant 2014).
[259] See e.g. Harper 2011.

Some firms add data gathered offline to online profiles.[263] Even when the name of the person behind a profile isn't known, it may be possible to do this. For instance, a firm that knows in which neighbourhood a computer's IP address is located, could add information about the average housing price in that neighbourhood to a profile. One American firm uses the location of IP addresses to infer "120 demographic variables" about people, including information such as "life stage, affluence, home ownership, auto interests, political affiliation, and social connectivity."[264]

---

[260] Hoepman 2014. For a majority of users, the browsing history is unique (Castelluccia et al. 2013a). Regarding identifying people by their favourite movies, Hoepman refers to Narayanan & Shmatikov 2008. See also Sivaramakrishna 2012; Cain, Miller & Sengupta 2013.
[261] Drawbridge 2014.
[262] Cain Miller & Sengupta 2013.
[263] Combining offline and online data is sometimes called "onboarding" (Federal Trade Commission 2014, p. 27).
[264] Semcasting.

**Schedule 2.4. Examples of individual profiles**

*- The person with ID xyz on his or her computer, that uses IP address 146.50.68.36, visited the following 2000 websites:*

*(1) hockey.com,*

*(2) basketball.com,*

*(3) soccer.com,*

*(…)*

*(1998) redrunningshoes.com,*

*(1999) blackrunningshoes.com,*

*(2000) bluerunningshoes.com.”*

*- The person with ID xyz on his or her computer likes sports and running shoes.*

## 2.5 Phase 3, data analysis

In phase 3, firms analyse the data. Somebody who reads a lot of legal blogs could be profiled as a person who is interested in the law. A firm may or may not delete the data it has collected about somebody's online behaviour after deducing that person's interests.[265]

Data can be analysed in various ways. For instance, data mining is the process of finding new information in data sets. Data mining can be described as "the nontrivial extraction of implicit, previously unknown, and potentially useful information from data."[266] Data mining doesn't have to begin with a hypothesis. Software is used to analyse the data in order to find correlations, and these correlations can be unexpected.[267] One firm found that customers who buy certain accessories for their cars often default on their credit. As the New York Times reports, "[a]nyone who purchased a chrome-skull car accessory or a 'Mega Thruster Exhaust System' was pretty likely to miss paying his bill eventually."[268] Conversely, people who bought felt pads for under the feet of their furniture to prevent scratches on the floor, almost always repaid their credit without problems.[269]

Firms may also construct predictive models.[270] For example, a firm might find the following model. *If a person visits website A, B, C and D, there's a 0.4 % chance that the person clicks on ads for product E.* Siegel defines a predictive model as follows.

> A mechanism that predicts a behavior of an individual, such as click, buy, lie, or die. It takes characteristics of the individual as input, and provides a *predictive score* as output. The higher the score, the more likely it is that the individual will exhibit the behavior.[271]

---

[265] Schunter & Swire 2013, p. 10-16.
[266] Frawley et al 1992, p. 58. See on data mining Custers 2004; Barocas 2010; Barocas 2014; Zarsky et al. 2013.
[267] Siegel 2013, p. 98. See in more detail: Barocas 2014, p. 54-56.
[268] Duhigg 2009. See also Brunton & Nissenbaum 2011.
[269] Duhigg 2009.
[270] Predictive models are roughly comparable with non-distributive group profiles (see Hildebrandt 2008).
[271] Siegel 2013, p. 26 (emphasis original). Predictive models can also be used to predict when somebody lies, or to predict how old somebody is likely to become: to predict behaviour "such as click, buy, lie, or die."

Siegel gives an example of a predictive model that was used for online advertising. A publisher of a website where people could search for scholarships wanted to improve the click-through rate on the site's ads. The following predictive model was found.

> IF the individual
>
> > is still in high school
> >
> > AND
> >
> > expects to graduate college within three years
> >
> > AND
> >
> > indicates certain military interest
> >
> > AND
> >
> > has not been shown this ad yet
>
> THEN the probability of clicking on the ad for the Art Institute is 13.5 percent.[272]

In brief the model says: if a website visitor fits in four categories (the input), there's a 13.5 % chance that he or she clicks on an ad for the Art Institute (the output). 13.5% might not seem like a high number, but the probability that a random website visitor clicked the ad was only 2.7%.[273] Siegel says it's unclear why people who expressed an interested in the military are more likely to click on the ad. He adds that causation is irrelevant: "it's important not to assume there is a causal relationship."[274] Whether people who expressed interest in the military see the ad as relevant is of little interest for the website; what is of interest is whether or not people are likely to click. Likewise, an ad network doesn't need causal relations. The goal is improving the chance that a person will click on an ad. As an aside, this implies that claims that behavioural targeting leads to "more relevant" ads should perhaps be taken with a grain of salt.[275]

---

[272] Siegel 2013, p. 26.
[273] In general, click-through rates are much lower. See section 1 of this chapter.
[274] Siegel 2013, p. 27.

By definition, predictive models aren't always accurate when applied to individuals. To illustrate, when a predictive model says that there's a 60% chance that people who visit sports websites also like running shoes, it's still possible that a person who visits sports websites doesn't like running shoes. And a person with an IP address from a neighbourhood with expensive real estate might be a poor student, renting a small room in an expensive villa. A book on data mining and marketing explains that predictions don't have to be accurate to increase profit.

> The fact is that, to take a typical application of data mining to direct marketing, *95 percent of the people picked by data mining to be likely responders to an offer will not respond*. In other words, at the level of individual consumers, *data mining predictions are nearly always wrong*. (…)
>
> The reason that data mining is valuable, despite being so very inaccurate, is that although only 5 percent of the people predicted to respond actually do so, that may be a significantly higher number than would have responded if no data mining model had been used. The ability of data mining to identify a population within which we can expect a 5 percent response rate, instead of the 2.5 percent response rate we could achieve

---

[275] Google says its behavioural targeting system makes ads "more relevant" and "more interesting" (Wojcicki 2009). See also Interactive Advertising Bureau Europe - Youronlinechoices (about).

> without data mining, makes it worthwhile from a business point of view.[276]

In short, accuracy isn't needed for behavioural targeting to be a good business decision. A firm doesn't have to predict accurately to improve profits. "Predicting better than pure guesswork, even if not accurately, delivers real value," notes Siegel.[277] Any improvement to the click-through rate is welcome. Say the chance that random internet users click on an ad for chairs is 0.1 %. An ad network could improve the click-through rate on the ad if it had the following predictive model: *If a person visits more than 10 websites about furniture every week, there's a 0.4 % chance that the person clicks on ads for chairs*. Hence, the predictive model, while not very accurate in predicting people's interests, can lead to a 400% improvement of the return on investment.

Behavioural targeting typically involves profiling. Hildebrandt offers the following definition.

> Profiling is the process of "discovering" correlations between data in databases that can be used to identify and represent a subject and/or the application of profiles (sets of correlated data) to individuate and represent a subject or to identify a subject as a member of a group or category. In the case of group profiling the subject is a group (which can be a category or a community of persons).[278]

---

[276] Berry & Linoff, p. 20 (emphasis original). See also Danna & Gandy 2002, p. 379.
[277] Siegel 2013, p. 11.
[278] Hildebrandt et al. 2008, p. 241, which refers to Hildebrandt 2008, p. 19. See in detail on defining profiling Bosco et al. 2013.

With profiling, information about a group of people can be applied to a person who isn't part of that group. To illustrate, the American retail store Target wanted to reach people with advertising during moments in life when they're more likely to change their shopping habits, as usually it's hard to make people change their habits. Therefore, Target wanted to know when female customers were going to give birth. "We knew that if we could identify them in their second trimester, there's a good chance we could capture them for years."[279] By analysing the shopping behaviour of customers, Target was able to construct a "pregnancy prediction" score, based on 25 products. If a woman buys those products, Target can predict with reasonable accuracy that she's pregnant.[280] Hence, Target uses data from a group of people to predict something about a person who wasn't part of that group.

Calo suggests firms might soon be able to analyse large amounts of data in order to find the characteristics and weaknesses of individuals. Is a person easier to persude with an ad in orange colours, or on rainy afternoons? "Firms will increasingly be able to trigger irrationality or vulnerability in consumers," says Calo.[281] "A firm with the resources and inclination will be in a position to surface and exploit how consumers tend to deviate from rational decision making on a previously unimaginable scale."[282] A press release of a marketing firm suggests that Calo's worries may not be completely unfounded. "New beauty study reveals days, times and occasions when US women feel least attractive."[283] The firm suggests advertising beauty products on Mondays. A Dutch firm is doing research on "persuasion profiling", which "lets you gain insight into your customer's psychological patterns (…)."[284] A firm could add to a profile what kinds of arguments convince a person to buy a product, rather than the person's interests. Does he or she react to discounts, or to phrases such as "special offer, only today"?

---

[279] Duhigg 2012, quoting the statistician of Target.
[280] Duhigg 2012. See on the Target case also Siegel 2013, chapter 2.
[281] Calo 2013, p. 5.
[282] Calo 2013, p. 22. See on the risk of manipulation through behavioural targeting chapter 3, section 3. See on "biases", deviations from rational decision making chapter 7, section 4.
[283] PHD Media 2013.
[284] PersuasionAPI. See also Kaptein 2011; Kaptein 2012; Kaptein & Eckles 2010; Groot 2012.

**Schedule 2.5. Examples of predictive models**

*- There is a 0.4% chance that a person who visits websites about consumer electronics, clicks on ads for phones.*

*- There's an 80% chance that a person who lives in neighbourhood X, has an income that is lower than 1500 euro a month.*

### 2.6 Phase 4, data disclosure

The fourth behavioural targeting phase concerns data disclosure. Firms make data available to other firms. Two kinds of data disclosure can be distinguished. First, a firm might sell copies of data to other firms.[285] For example, data brokerage is a large industry in the US. Data brokers are "companies that collect consumers' personal information and resell or share that information with others."[286] Firms can buy data to tie them to online profiles. For instance, a firm called Collective enriches online profiles with off-line consumer data from more than "35 world-class data providers such as Polk, Nielsen and eXelate, integrated into profiles representing the most desirable segments of the US online audience."[287] The American firm CampaignGrid merges data from its database with registered voters with cookie-based profiles for political campaigns. The firm deletes the name from the profile after it merges the different data sets, and suggests that this makes the profiles "de-identified."[288] However, chapter 5 shows that European data protection law usually applies to pseudonymous individual profiles. Deleting the name from a profile is not enough to remain outside the scope of data protection law, and is not enough to make information anonymous.[289]

[285] From a legal perspective, data may not be goods that can be "sold." We'll leave this issue aside.
[286] Federal Trade Commission 2014, p. 1. See also Federal Trade Commission 2013a.
[287] Collective 2011.

A second type of data disclosure doesn't involve selling copies of the data. For example, an ad network can allow an advertiser to target individuals based on their characteristics. The ad network shows the ad on behalf of the advertiser. The advertiser usually doesn't receive a copy of the data in a profile. This type of data disclosure could be seen as a modern version of list rental. With list rental, a list broker sends leaflets to a set of people, based on what it knows about those people. The advertiser doesn't receive a copy of the list.[290]

Another example of data disclosure is cookie matching, or cookie synching, "linking the profiles of a single user in databases of two independent companies."[291] Cookie synching happens routinely. For instance, researchers found that the cookies of Google's DoubleClick ad network are synched with cookies of at least 125 other firms.[292] Depending on the design of the system, cookie synching may or may not involve disclosing copies of data to others.

***Real time bidding***

Ad networks can bid on automated auctions for the chance to show an ad to a person, a process which is referred to as "real time bidding", "audience selling", or "audience buying."[293] Ad exchanges are automated market places where advertisers can trade with multiple ad networks in one place.[294] Ad exchanges owned by Google, Yahoo, and Microsoft are among the largest.[295] Real time bidding "creates a data market where users' browsing data are sold at auctions to advertisers."[296]

---

[288] CampaignGrid 2012. See also Kreiss 2012.
[289] See section 4 of this chapter.
[290] Under data protection law, list rental should probably be seen as a type of data disclosure. See chapter 6, section 2.
[291] Castellucia et al. 2013, p. 1.
[292] Castellucia et al. 2013, p. 7.
[293] See for example Pubmatic 2011.

In brief, real time bidding works as follows. A website has an empty spot for a banner ad. Somebody visits the website. An ad network that works with the website recognises this person as the cookie with, for instance, number 22be6e056ca010062||t=1392841778|cs=002213fd48e6bd6f7bf8d99065. For ease of reading, this study speaks of ID *xyz*. When the website is loaded in the user's browser, the ad network offers the empty banner spot on the advertising exchange (the auction). The ad network can include information about the person behind ID *xyz*, such as the person's inferred interests and location, and the time of day.

Other ad networks bid to reach a person who is, for instance, interested in cars, just visited a website with information about loans, and as been visiting websites with reviews of a certain car type for the past three weeks. The ad network that submits the highest bid obtains the right to target an ad to this specific group. Then, the winning bidder (for instance another ad network) can display an ad on the website for an advertiser. This process happens automatically and within a few milliseconds. (For more information on targeted advertising, see the next section, on phase 5 of the behavioural targeting process.) Researchers conclude that "user's browsing history elements are routinely being sold off for less than $0.0005."[297] Billions of such auctions take place per day.[298] "We are not buying content as a proxy for audience", explains one marketing firm. "We are just buying who the audience is."[299]

---

[294] Turow 2011 p. 79; Evans 2009. The Interactive Advertising Bureau provides a definition of advertising exchanges (Interactive Advertising Bureau United States 2010).
[295] Turow 2011, p. 79. In 2007, Google, Microsoft and Yahoo each acquired a firm running an ad exchange (Google 2011, p. 3).
[296] Castellucia et al. 2013, p. 14.
[297] Castellucia et al. 2013, p. 1.
[298] Econsultancy 2011, p. 6; Turow 2012, p. 69.
[299] Quoted in Singer 2012.

If a website publisher contracts with an ad network, and that ad network sells part of its inventory through an advertising exchange, the publisher doesn't always know in advance who will display the ads on its site. Therefore, sometimes publishers don't know which firms are collecting data on their websites.[300] "As a publisher we feel we've been raided by the ad industry," says the chairman of the Association of Online Publishers. "We've done site audits and been flabbergasted by how many third party cookies have been dropped on our site by commercial partners – they were stealing our data."[301] Some firms offer a service that wesbite publishers can use to monitor their own websites, to reduce such "data theft."[302]

The Interactive Advertising Bureau US claims that "virtually every publisher site, advertiser, ad network, or analytics firm collects or shares data with other parties in order to make the digital economy work."[303] Behavioural targeting can seem more complicated than it is, because firms tend to introduce new phrases, such as "data driven marketing,"[304] and "programmatic buying."[305] Notwithstanding this, the data flows behind a behaviourally targeted ad can be extremely complicated. Many different types of firms are involved in serving a behaviourally targeted ad, and many firms disclose information to each other. LUMA Partners, an investment bank for the media and technology sector, provides an infographic with an overview of the types of players involved in display advertising, which includes many types of firms, such as "demand side platforms", "agency trading desks", "data suppliers", and firms involved in "tag management", and "measurement and analytics."[306] It would go beyond the scope of this study to discuss each type.

[300] See for instance Martijn 2013, who interviews Dutch publishers who say they don't know what happens on their sites.
[301] Barnes, chairman of the Association of Online Publishers, quoted in Hall 2013.
[302] See for instance Krux 2014. See also Vascellaro 2010.
[303] Zaneis 2012.
[304] Data-Driven Marketing Institute 2014.
[305] Interactive Advertising Bureau United States 2014a
[306] Luma Partners 2014.

## 2.7 Phase 5, targeting

In the fifth phase of behavioural targeting, a firm targets a person with an ad, based on information about that person. Any kind of digital advertising can be based on behavioural profiles, such as display ads, ads shown by search engines, and marketing emails. Two people simultaneously visiting a website may each see a different ad, because they have different profiles. Firms can adapt ads in real time, and can serve each user a unique personalised ad.[307] The advertiser's goal is to "reach the right person with the right message at the right time."[308] A firm might also refrain from showing an ad to certain people, based on their profile.

Advertising can be defined as a "paid, mediated form of communication from an identifiable source, designed to persuade the receiver to take some action, now or in the future."[309] On the internet, the boundaries between brand advertising and direct-response advertising are blurry, because most ads enable people to click on ads to interact with advertisers.[310] The lines between behavioural targeting and other types of online advertising are blurry as well.[311] For example, nowadays ads that are shown by a search engine are often behaviourally targeted. In principle search ads don't have to be based on analysing people's behaviour over time. To illustrate, until around 2009

---

[307] Personalisation can be defined as the "use of information about a particular user that provides tailored or personalized services for the user" (Serino et al. 2005, p. 1). Some authors distinguish "system-initiated personalisation" from "user-initiated customisation" (Marathe & Sundar 2010, p. 300).
[308] TRUSTe (Drawbridge) 2013.
[309] Curran & Richards 2002, p. 74. The word mediated in this definition means "conveyed to an audience through print, electronics, or any method other than direct person-to-person contact." See for a EU legal definition of advertising: article 2(a) of Directive 2006/114/EC on misleading and comparative advertising. "Commercial communication" is defined in article 2(f) of the e-Commerce Directive 2000/31/EC.
[310] See McStay 2009, p. 7. Brand advertising aims at making a brand more famous, rather than at enticing the recipient to take action immediately. "Direct response advertising" is "[a]n approach to the advertising message that includes a method of response such as an address or telephone number whereby members of the audience can respond directly to the advertiser in order to purchase a product or service offered in the advertising message (…)" (American Marketing Association dictionary).
[311] Strandburg 2013, p. 99.

Google had refrained from behavioural targeting.[312] Now Google ties the profile of a searcher to the other data it has about that person.[313]

A category of behavioural targeting that is particularly notable for users is retargeting. Sometimes ads for a product appear to follow somebody around the web. Retargeting allows a firm to show potential customers personalised ads, based on earlier behaviour that the firm interprets as an intention to buy. Retargeted ads aim to remind the potential customer of a product. Google explains retargeting as follows to advertisers:

> Let's say you're a basketball team with tickets that you want to sell. You can put a piece of code on the tickets page of your website, which will let you later show relevant ticket ads (such as last minute discounts) to everyone who has visited that page, as they subsequently browse sites in the Google Content Network. In addition to your own site, you can also remarket to users who visited your YouTube brand channel or clicked your YouTube homepage ad.[314]

Retargeting is easy to notice. If somebody looks at red shoes in an online shop, and keeps seeing ads for those same shoes elsewhere on the web, it's obvious that the ads are tailored to the individual. Other kinds of behaviourally targeted ads can be harder to recognise. For somebody who visits the literature section of an online newspaper, it's not always clear whether an ad for a book is based on his or her earlier surfing behaviour or not. A behaviourally targeted ad might be mistaken for a contextual ad, or vice versa.

---

[312] Hoofnagle 2009.
[313] See Article 29 Working Party 2013 (Google letter), and chapter 8, section 1.
[314] Weinberg 2010.

In principle, few data are needed for retargeting, because there's no need to build a detailed profile of somebody's tastes and behaviour. A firm drops a cookie on a user's device, and the firm only needs to store the information that the person behind ID *xyz* looked at a certain product. In practice, a firm might also store the user's IP address, and the list of all websites where the firm showed the user the retargeted ad.

Behavioural targeting can also be used for political advertising. A firm gives an example of the possibilities: "targeting fathers aged 35-44 in Texas who frequent gun enthusiast websites."[315] Messages can be tailored to the profile of the recipient. In 2012, campaigners for Obama divided an email list into 26 segments, in order to be able to send each segment a different message.[316] Political behavioural targeting firm CampaignGrid claims that it reaches 90 % of American internet users. The firm enables politicians to target people with ads on LinkedIn, Facebook, and elsewhere on the web.[317] An article in the magazine *Campaigns & Elections* discusses the possibilities of digital TV for political campaigns.

> While there's plenty of potential for political campaigns in set-top box targeting, mining data from television set-top boxes and pairing it to the voter file is a good starting point this [election] cycle, according to NCC Media's Tim Kay. "It's no longer hoping you're hitting the person," says Kay, the company's political director. "Now it's about knowing whether you're hitting the person and knowing how to hit the person."[318]

---

[315] Retargeter Blog 2012.
[316] Judd 2012.
[317] CampaignGrid 2014.
[318] Williams 2014.

Not only ads, but also other content can be personalised. Major search engines personalise search results.[319] And two people visiting the same website at the same time may see a different front page.[320] To illustrate, Yahoo shows more than thirteen million different versions of its news page each day.[321] Yahoo shows the news selection that keeps visitors on the website for as long as possible, in order to show them more advertising. Yahoo doesn't ask visitors whether they want to receive personalised news. The line between content and ads can be fuzzy on the web. For instance, advertorials and "native ads" are ads that resemble editorial content.[322]

Some firms specialise in website personalisation. A company called Personyze says: "[s]egment your visitors in real-time and serve them personalized and optimized content based on their demographic, behavioural and historical characteristics."[323] Personalisation can be "based on demographics, keywords searched, referring affiliate website, articles read, favorite categories and more."[324] A website's design can also be adapted to the visitor, called morphing. "Morphing involves automatically matching the basic 'look and feel' of a website, not just the content, to cognitive styles."[325] Research suggests that website morphing could increase online sales with approximately 20%.[326] At present, website morphing doesn't seem to be widely used. As behavioural targeting makes it possible to show each person personalised ads and other content and services, Hildebrandt has called behavioural targeting an early example of ambient intelligence, technology that senses and anticipates people's behaviour to adapt the environment to their inferred needs.[327]

---

[319] See Hannak et al. 2013.
[320] Turow 2011; Pariser 2011.
[321] Yahoo 2012.
[322] See Federal Trade Commission 2013. See on the blurry line between advertising and other content Van Hoboken 2012 (chapter 10, section 3).
[323] Personyze 2014.
[324] Personyze 2014b
[325] Hauser et al. 2009, p. 202.
[326] Hauser et al. 2009.
[327] Hildebrandt 2010. See also Hildebrandt 2011, p. 12. See on ambient intelligence: Philips Research 2014; see also Van Den Berg 2009.

Behavioural targeting offers more possibilities beyond personalised advertising. For instance, firms could personalise prices based on group or individual profiles – also referred to as price discrimination.[328] A user whose profile suggests that he or she often buys expensive goods without first looking for the cheapest price online could be profiled as a "big spender."[329] A Harvard Business Review article explains that a shop could charge higher prices to some people. "Just as it's easy for customers to compare prices on the Internet, so is it easy for companies to track customers' behavior and adjust prices accordingly."[330]

It's unclear to what extent firms adapt prices to people's online profiles. Perhaps firms are hesitant to personalise prices because they fear consumer backlash.[331] However, in the US, firms have adapted credit card offers to the cookie profile of website visitors, based on a person's inferred income for instance.[332] And in 2012, Soltani et al. found that the online shop Staples charged visitors from certain areas (based on their IP address) different prices than people from other areas. This had the effect, likely unintentional, that people from high-income areas tended to pay less.[333] Opinions differ on the question of whether personalised pricing is desirable. From an economic perspective, price discrimination is a good thing, under certain assumptions.[334] On the other hand, many regard price discrimination as unfair or manipulative.[335]

---

[328] See generally Turow 2012, p. 108-110.
[329] Bluekai 2010. Marketers can buy access to "high spenders", "suburban spenders" or "big spenders" (p. 6-8). Bluekai says the profiles are "anonymous" (Bluekai 2012).
[330] Baker et al. 2001, p. 123.
[331] The English Office of Fair Trade examined whether firms raised prices based on people's online behaviour, but didn't find any evidence. The office did find that firms offer discounts based on people's profiles some cases (Office of Fair Trading 2010; Office of Fair Trading 2012). A discount for one person is a type of price discrimination, which could also be seen as a higher price for the others.
[332] Steel & Angwin 2010.
[333] Valentino-Devries et al. 2012; Angwin 2014, p. 16.

### 2.8 Conclusion

This chapter described what behavioural targeting is, and how it works. Different factors can help to understand the rise of behavioural targeting. Technology has made behavioural targeting possible. Behavioural targeting fits into a trend of increasingly targeted advertising at ever-smaller audience segments. Furthermore, advertisers have always wanted information on how many people they reached with an ad, and on what kind of people they reach. Behavioural targeting provides such information, at the individual level.

Behavioural targeting is the monitoring of people's online behaviour in order to use the collected information to show people individually targeted advertisements. In this study, the behavioural targeting process is analysed by dividing it into five phases: (1) data collection, (2) data storage, (3) data analysis, (4) data disclosure, and (5) the use of data for targeted advertising.

In phase 1, firms collect information about people's online activities. People's behaviour is monitored, or tracked. In phase 2, firms store the information about individuals, usually tied to identifiers contained within cookies, or via similar technology. As discussed later in this study, article 5(3) of the e-Privacy Directive requires consent for the use of many tracking technologies, but some tracking

---

[334] In economics, price discrimination, or price differentiation, is used in a broader sense than personalised pricing: "the practice of a seller charging different prices to different customers, either for exactly the same good or for slightly different versions of the same good. (…) [P]rice differentiation includes not only charging different prices to different customers for the same product (group pricing), but also the less controversial strategies of product versioning, regional pricing, and channel pricing" (Phillips 2005, p. 74). See generally on price differentiation Phillips 2005, chapter 4; Shapiro & Varian 1999, chapter 2 and 3. See generally on price discrimination and behavioural targeting Zarsky 2002; Odlyzko 2003; Turow 2011; Calo 2013; Narayanan 2013; Strandburg 2013; p. 138-141; Odlyzko 2014; Miller 2014.

[335] For instance, in a nationally representative survey, Turow et al. 2005 "found that they [US adults] overwhelmingly object to most forms of behavioral targeting and all forms of price discrimination as ethically wrong" (p. 4). Klock 2002 argues (not focusing on behavioural targeting): "[a] sound policy would prohibit firms from charging different prices based solely on the identity of the customer" (p. 367).

technologies, such as passive device fingerprinting, may fall outside the scope of that provision.[336]

In phase 3 the data are analysed. For instance, a firm can construct a predictive model, along the following lines: if a person visits website A, B, C and D, there's a 0.5 % chance the person will click on ads for product E. For behavioural targeting to be useful, a predictive model doesn't have to be accurate when applied to an individual. Chapter 5 shows that predictive models are outside the scope of data protection law, as a predictive model doesn't refer to an identifiable person.[337]

Phase 4 concerns data disclosure. Firms make data available data to advertisers or other firms. For example, an ad network can enable advertisers to target individuals with ads, based on their behavioural profiles. Or a firm can sell copies of data to other firms. Many types of firms are involved in behavioural targeting, and the resulting data flows are complicated. The complicated data flows make it difficult to explain to people what happens to information about them (see chapter 7).[338]

In phase 5 data are used to target ads to specific individuals. Behavioural targeting enables advertisers to reach a user, wherever he or she is on the web. A website publisher often doesn't know in advance who will serve ads on its website. Firms can personalise ads and other website content for each visitor.

Website publishers can increase their income by allowing ad networks to track their visitors and to display behaviourally targeted ads. But in the long term behavioural targeting may decrease ad revenues for some website publishers. For example, an ad network doesn't have to buy expensive ad space on a large professional news website to advertise to an individual. The ad network can show an ad to that person when he or she visits an unknown website, where advertising space is cheaper. Chapter 7 returns to the topic of the economics of behavioural targeting.[339] But first we turn to the privacy implications of behavioural targeting, in the next chapter.

[336] See chapter 6, section 4, and chapter 8, section 4.
[337] See chaoter 5, section 2.
[338] See in particular chapter 7, section 3 and 4,

* * *

[339] See in particular: chapter 7, section 2.

# 3 Privacy

What are the privacy implications of behavioural targeting? To answer this question, this chapter distinguishes three perspectives on privacy in section 3.1: privacy as limited access, privacy as control over personal information, and privacy as the freedom from unreasonable constraints on identity construction. The three perspectives highlight different concerns during the behavioural targeting process.[340]

Section 3.2 discusses the right to privacy in European law, and the privacy case law of the European Court of Human Rights. The European Court of Human Rights interprets the right to privacy generously, and refuses to define the scope of protection. Section 3.3 discusses three privacy concerns regarding behavioural targeting. First: chilling effects relating to massive data collection on user behaviour. Second: the lack of individual control over personal information. Third: social sorting and the risk of manipulation. Section 3.4 concludes.

## 3.1 Three privacy perspectives

Many people have no trouble thinking of an example of a privacy violation.[341] Countless civil rights organisations aim to defend privacy, and judges have to apply the concept.[342] But after more than a century of attempts by scholars from various disciplines, it has been impossible to agree on a definition. Privacy has been called "elusive and ill-defined",[343] "a concept in disarray",[344] and a "messy, complicated, and rather vague concept."[345] Looking for a privacy definition in literature "we find chaos",[346] "nobody seems to have any very clear idea what the right to privacy is",[347] and "even its pronunciation is uncertain."[348]

---

[340] As noted, this study uses "privacy", and "private life" interchangeably. Article 7 of the EU Charter of Fundemantal Rights and article 8 of the European Convention on Human Rights use the phrase "respect for private and family life". See in detail on the difference between "private life" and "privacy" González Fuster 2014, p. 82-84; p. 255. This study also uses "fundamental rights" and "human rights" interchangeably (see on these terms González Fuster 2014, p. 164-166).
[341] See Nippert-Eng 2010.
[342] See Bennet 2008 for an overview.

As noted, in this study three privacy perspectives are distinguished: privacy as limited access, privacy as control over personal information, and privacy as the freedom from unreasonable constraints on identity construction.[349] The classification is based on work by Gürses, who discusses three privacy research paradigms in the field of software engineering.[350]

The classification helps to structure discussions about privacy. However, there are no clear borders between the three privacy perspectives, which overlap in different ways. Furthermore, none of the three privacy perspectives is meant as absolute. Privacy as limited access doesn't suggest that people want to be completely alone. Privacy as control doesn't suggest that people should have full control over data concerning them. And privacy as the freedom from unreasonable constraints on identity construction doesn't suggest that people should be allowed to lie to everyone to improve their image.

---

[343] Posner 1978, p. 393.
[344] Solove 2009, p. 1.
[345] Boyd 2011, p. 497.
[346] Inness 1996, p. 3.
[347] Thomson 1975, p. 312.
[348] Marshall 1975, p. 242.
[349] Many other classifications are possible. For instance, Solove distinguishes 6 perspectives (Solove 2002), and Rössler distinguishes three perspectives (Rössler 2005, p. 6). Another possible distinction is that between relational and informational privacy (see e.g. Dommering & Asscher 2000; Kabel 2003).
[350] Gürses uses a slightly different terminology and distinguishes (i) "privacy as confidentiality: hiding", (ii) "privacy as control: informational self-determination", and (iii) "privacy as practice: identity construction" (Gürses 2010, p. 24-32) See for a similar taxonomy Berendt 2012.

***Privacy as limited access***

In the late 19th century, the invention of the snap camera by Kodak enabled people to create photos on the spot. Until then, people needed to be still for a picture, so people had to cooperate when a picture was taken of them. But the new cameras made it possible for the paparazzi to take photos of people without being noticed.[351] In 1890 this led two US authors, Warren & Brandeis, to write an influential article: "The right to privacy."[352]

> Recent inventions and business methods call attention to the next step which must be taken for the protection of the person, and for securing to the individual what Judge Cooley calls the right "to be let alone." Instantaneous photographs and newspaper enterprise have invaded the sacred precincts of private and domestic life; and numerous mechanical devices threaten to make good the prediction that "what is whispered in the closet shall be proclaimed from the house-tops."[353]

Warren & Brandeis argued for legal protection of privacy, to safeguard "the right to be let alone."[354] They suggested that the common law implicitly recognised a right to privacy already, citing precedents on, for example, breach of confidence, copyright, and defamation. Ever since, scholars, judges, and lawmakers have tried to adapt the concept of privacy to cope with new developments and new technologies.[355]

This study categorises Warren & Brandeis in the group of the first privacy perspective: privacy as limited access to the private sphere. The privacy as limited access perspective is categorised together with privacy as secrecy,[356] confidentiality,[357] solitude,[358] seclusion,[359] and as a right not to be annoyed.[360] Privacy as limited access emphasises the freedom from interference by the state or others. Privacy as limited access is about a personal sphere, where people can remain out of sight and in peace. Gavison describes the limited access perspective well.

---

[351] Solove 2009, p. 15.
[352] Warren & Brandeis 1890.
[353] Warren & Brandeis 1890, p. 195, internal footnote omitted.
[354] Warren & Brandeis don't actually define privacy as the right to be let alone (see Gavison 1980, p. 437).
[355] See e.g. ECtHR, Von Hannover v. Germany, No. 59320/00, 24 September 2004, par. 74. See also Gassman & Pipe 1974, p. 12.

> Our interest in privacy (…) is related to our concern over our accessibility to others: the extent to which we are known to others, the extent to which others have physical access to us, and the extent to which we are the subject of others' attention.[361]

Roughly, two categories within privacy as limited access can be distinguished.[362] First: privacy as confidentiality. When others access information that a person wishes to keep for him- or herself, there's a privacy interference. Second, privacy interferences can occur when people are disturbed, or interrupted, for instance by telemarketers. Varian speaks of privacy as a "right not to be annoyed."[363]

Privacy as limited access aptly describes many privacy infringements. Seeing privacy as limited access implies that too much access to one's private sphere interferes with privacy. A classic example is privacy violations by paparazzi that intrude on private affairs. Section 3.3 discusses how tracking people's activities for behavioural targeting can interfere with privacy as limited access.

---

[356] Posner 1978. See Solove 2002, p. 1105.
[357] Gürses 2010, p. 24.
[358] Westin 1970, p. 31.
[359] American Law Institute 1977.
[360] Varian, p. 102.
[361] Gavison 1980, p. 423.
[362] Posner 1981, p. 31, note 7. See also Solove 2009, p. 21-24.
[363] Varian 2009, p. 102. The European Court of Human Rights says that receiving unwanted or offensive spam amounted to an interference with a person's right to respect for his private life. But the Court didn't find that Italy should have done more to comply with its positive obligations (ECtHR, Muscio v. Italy, No. 31358/03, 13 November 2007 (inadmissible)).

While too much access to a person fittingly describes many privacy violations, the perspective also has weaknesses. In some ways, the privacy as limited access perspective is too narrow. For example, people often want to disclose information about themselves to others, but still have expectations of privacy. Disclosing personal information is an important part of building relationships, and of functioning in society.[364] Hence, the social dimension of privacy seems to receive insufficient attention under the privacy as limited access perspective. And sometimes people want to disclose information to firms to receive personalised service. Solove notes that privacy as secrecy is problematic as well, as many situations that people would describe as a privacy infringement don't concern information that is secret.[365] Private matters such as a person's debts can hardly be described as a secret.[366] In sum, many aspects of privacy seem to be outside the scope of privacy as limited access.

Privacy as limited access is also too broad, according to Solove. The right to be let alone is a great slogan, but as a definition it's too vague. "A punch in the nose would be a privacy invasion as much as a peep in the bathroom," says Allen.[367] Solove adds that privacy as limited access doesn't explain which aspects of one's life are so private that access shouldn't be permitted.

> The theory provides no understanding as to the degree of access necessary to constitute a privacy violation. In the continuum between absolutely no access to the self and total access, the important question is where the lines should be drawn – that is, what degree of access should we recognize as reasonable?[368]

---

[364] See Solove 2009, p. 23; Rouvroy 2008, p. 25.
[365] Solove 2009, p. 24.
[366] Solove 2009, p. 24.
[367] Allen 1988, p. 7.
[368] Solove 2009, p. 20.

Among others, Nissenbaum says that in a modern society it's hard to define what is private.[369] "Despite a superficial elegance," adds Bennet, "one cannot restrict privacy rights and claims to the domain of the 'private' because contemporary socio-technical systems have blown away these clear distinctions."[370] What should be seen as private when discussing social network sites?[371]

In conclusion, while the privacy as limited access perspective has weaknesses, the perspective fits well when discussing many privacy infringements.

***Privacy as control***

At the end of the 1960s several books, sometimes called "the literature of alarm",[372] discussed the threats of the increasing amount of personal information that the state and other organisations gathered, often using computers.[373] In his book Privacy and Freedom, Westin introduced a privacy definition that would become very influential:

> Privacy is the claim of individuals, groups or institutions to determine when, how and to what extent information about them is communicated to others.[374]

This can be summarised as privacy as control. Around the late 1960s many feared that state agencies or other large organisations were amassing information about people. The use of computers for data processing added to the worries. Some feared that computers would make decisions about people.[375] Westin summarises the anxieties

369 Nissenbaum 2010, chapter 6.
370 Bennet 2011, p. 541-542.
371 See on privacy management by young people on social network sites boyd 2014.
372 Gassman & Pipe 1974, p. 12.
373 See e.g. Packard 1966; Westin 1970; Miller 1971; Sieghart 1976. See for more references Blok 2002, p. 243-247, Regan 1995, p. 13-14.
374 Westin 1970 (reprint of 1967). Warren & Brandeis made a similar remark: "The common law secures to each individual the right of determining, ordinarily, to what extent his thoughts, sentiments, and emotions shall be communicated to others" (Warren & Brandeis 1890, p. 198).
375 Bennett 1992, in particular p. 118-123; Mayer-Schönberger 1997, p. 221.

well. "You do not find computers in street corners or in free nature; you find them in big, powerful organisations."[376] (Nowadays computers and smart phones are everywhere, but often data still flow towards large organisations.)

A 1972 UNESCO report warned that digital information about a person "may be used as the basis for passing judgment on him, a secret judgement from which there can be no appeal and which, because it is based on a computer, is thought to be objective and infallible." The report adds that such decisions could be based on information that's wrong, irrelevant, or taken out of context: "in fact the information used may be inexact, or out of date or of no real significance, with the result that the final conclusion amounts to a 'scientific sophism'."[377]

A 1974 report by the Organisation for Economic Co-operation and Development (OECD) said that the idea of privacy was – or should be – shifting from the limited access approach to the control approach.[378] The report suggests that, if people fear that organisations make decisions about them without the possibility of having a say in the decision process, the answer doesn't have to lie in ensuring that information isn't collected. Having control over information concerning oneself may be at least as important.

> The concept of privacy in the sense of data surveillance is undergoing adaptation to the modern setting. The earlier notion that privacy is the ability of an individual to withhold information about himself, a "right to be left alone", is changing to a more practical current view required of man in a complex social environment. The concept is therefore shifting from the right of preventing the extraction or collection of

---

[376] Quoted in Bing 2007, p. 78, who relies on notes from a symposium in Paris around 1972.
[377] UNESCO 1972, p. 429.
[378] See for criticism on OECD's "fair information practices" Clarke 2000; Clarke 2002; Bonner & Chiasson 2005.

> personal facts, to the extension of control over information recorded on an individual in a personal register. The new definition emphasizes the conditions placed on information content, and the control over dissemination and use of personal data.[379]

Seeing privacy as control over information implies that a lack of control, or losing control, over personal information interferes with privacy. As Gürses notes, two categories of privacy harm can be distinguished: experienced harm, and expected harm.[380] Experienced harms are adverse effects that result from data processing. Calo calls this objective harm, "the unanticipated or coerced use of information concerning a person against that person."[381] A loss of control over information can indeed lead to harm. For example, if a firm used somebody's personal information to charge that person higher prices, the lack of control leads to quantifiable harm for the person. A profile that suggests somebody is a terrorist could cause delays at a border control, or worse. A dossier that says somebody is a troublemaker could wreck a career.[382]

Another aspect of lack of control is the *feeling* of lost control, which could be called expected harm,[383] or subjective harm, "the perception of loss of control that results in fear or discomfort."[384] Many people are uncomfortable with organisations processing large amounts of information about them – including when no human ever looks at the data.[385] People vaguely know that data about them are being collected and stored, but don't know how these data will be used. Solove compares the feeling of helplessness with the situation in Kafka's *The Trial*.[386] The main problem is "not knowing what is happening, having no say or ability to exercise meaningful control over the process."[387]

---

[379] Gassman & Pipe 1974, p. 12-13 (emphasis original). In the US, a similar suggestion was made to redefine privacy as control over personal information (United States Department of Health, Education, and Welfare 1973, p. 38-41). That same report introduces a version of the fair information principles (p. 41); see chapter 4, section 1.
[380] Gürses 2010, p. 87-89. She doesn't limit her discussion to harms resulting from a lack of control over personal information, but discusses privacy concerns in general.
[381] Calo 2011, p. 1133 (see specifically about marketing: p. 1148).
[382] Ohm speaks of a "database of ruin" (Ohm 2010, p. 1748). To illustrate, in the United Kingdom construction companies used a secret black list to deny jobs to construction workers that were deemed troublesome (Boffey 2012).
[383] Gurses 2010, p. 87-89.
[384] Calo 2011, p. 1143.

Privacy as control emphasises people's freedom to decide what should happen with information concerning them. Seeing privacy as control has the advantage of respecting people's individual preferences. Furthermore, privacy as control covers situations where one wants to share information with some, but not with others. The privacy as control perspective accommodates that people have different privacy wishes.

The privacy as control perspective can be recognised in legal practice. For instance, in 1982 the German Bundesverfassungsgericht formulated the right to informational self-determination: "the right of the individual to determine for himself whether his personal data should be divulged or utilized."[388] The Court doesn't suggest that people should have full control over data concerning them; in a modern society it's often necessary to process personal data.[389] Privacy as control has deeply influenced European data protection law.[390]

Westin's control definition also impacted scholarship.[391] Many authors use similar descriptions, such as Fried, who writes privacy "is not simply an absence of information about us in the minds of others; rather it is the *control* we have over information about ourselves."[392] Similar descriptions have been used in literature on privacy on the internet.[393] Schwartz concludes that the control perspective has become "the traditional liberal understanding of information privacy."[394] He adds that "[t]he weight of the consensus about the centrality of privacy-control is staggering."[395]

---

385 See International Working Group on Data Protection in Telecommunications (Berlin Group) 2013, p. 2-3. Some suggest there can't be a privacy interference if no human looks at the information (see e.g. Posner 2008, p. 254; Van Der Sloot 2011, p. 66).
386 Solove 2004, p. 38.
387 Solove 2004, p. 38.
388 Bundesverfassungsgericht 25 March 1982, BGBl.I 369 (1982), (Volks-, Berufs-, Wohnungs- und Arbeitsstättenzählung (Volkszählungsgesetz)), translation by Riedel, E.H., Human Rights Law Journal 1984, vol. 5, no 1, p. 94, p. 101, paragraph II.
389 Idem, p. 101, paragraph II. See also González Fuster 2014, p. 176-177.
390 See e.g. Mayer-Schönberger 1997; Bennett 1992, p. 14.
391 See generally De Graaf 1997, Blok 2002; Regan 1995.

Approaching privacy as control over information also has its weaknesses. According to Solove, privacy as control is too broad a definition, because it's unclear what "control" means. "We are frequently seen and heard by others without perceiving this as even the slightest invasion of privacy."[396] Furthermore, the definition seems to promise too much. In a modern society people must often disclose personal information to the state and other organisations.[397] If people had full control over their data, the tax office wouldn't be very successful. On the other hand, the control perspective doesn't imply that people should have full control over personal information; the right to privacy isn't absolute. The definition of privacy as control over personal information can also be criticised for being too narrow, says Solove. For instance, some privacy violations aren't covered by the definition, like being annoyed or disturbed during quiet times.[398]

Furthermore, the privacy as control perspective receives criticism because it puts too much emphasis on individual interests.[399] Many scholars argue that privacy is an important value for society, rather than merely an individual interest.[400] "Privacy has a value beyond its usefulness in helping the individual maintain his or her dignity or develop personal relationships",[401] says Regan. She adds: "society is better off if privacy exists."[402] Another problem with seeing privacy as control is that control is hard to achieve in practice. Approaching privacy as control leads to a focus on informed consent, like in data protection law.[403] Chapter 7 discusses practical problems with informed consent in the area of behavioural targeting.

---

[392] Fried 1968, p. 482. See also Miller, who describes privacy as "the ability to control the circulation of information relating to him" (Miller 1971, p. 25).
[393] Kang writes "control is at the heart of information privacy" (Kang 1998, p. 1266). Froomkin describes privacy as "the ability to control the acquisition or release of information about oneself" (Froonkin 2000, p. 1463).
[394] Schwartz 1999, p. 1613.
[395] Schwartz 2000, p. 820.
[396] Solove 2009, p. 25.
[397] See Schwartz 1999, p. 1663-1664; Blume 2012, p. 29.
[398] Solove 2009, p. 25-29.
[399] See e.g. Allen 1999; Solove 2009, p. 25-29.
[400] See e.g. Simitis 1987; Regan 1995; Schwartz 1999; Schwartz 2000; Westin 2003; Rouvroy & Poullet 2009; De Hert & Gutwirth 2006; Allen 2011; Van der Sloot 2012.
[401] Regan 1995, p. 221.

***Privacy as identity construction***

Recently, a third perspective on privacy has become popular among scholars: privacy as the freedom from unreasonable constraints on identity construction. In 1998, three decades after Westin's book, Agre discussed the privacy implications of new developments such as networked computing. He notes that "control over personal information is control over an aspect of the identity one projects to the world (…)."[404] He adds:

> Privacy is the freedom from unreasonable constraints on the construction of one's identity.[405]

This perspective, privacy as identity construction for short, is popular among European legal scholars discussing profiling.[406] Hildebrandt says the definition emphasises the link between privacy and developing one's identity. Furthermore, the definition shows that one's identity isn't something static, as it speaks of identity construction. People aren't born with an identity that stays the same their whole life. A person's identity develops, and that person can try to influence how others see him or her.

---

[402] Regan 1995, p. 221.
[403] Mayer-Schönberger 1997; Hoofnagle & Urban 2014.
[404] Agre 1998, p. 7.
[405] Agre 1998, p. 7 (capitalisation adapted).
[406] See e.g. Rouvroy 2008; Gürses 2010; Hildebrandt 2010; Hildebrandt 2011a; Roosendaal 2013. See for criticism on the identity construction perspective De Andrade 2011.

Arguably, privacy as identity construction includes privacy as limited access. Sometimes, people need to be free from interference to develop their personality, an aspect of their identity.[407] The Human Rights Committee of the United Nations says "privacy refers to the sphere of a person's life in which he or she can freely express his or her identity, be it by entering into relationships with others or alone."[408]

Privacy isn't only about keeping others at a distance or keeping things confidential. Privacy also concerns how people present themselves, how they manage their image – for instance by disclosing or withholding information. Hence, the identity construction perspective includes privacy as control over personal information. Furthermore, privacy as identity construction highlights the social dimension of privacy, and captures the relevance of context.[409] "Privacy is also implicated in users' ability to control impressions and manage social contexts," say boyd and Ellison.[410] Gürses agrees, and speaks of "privacy as practice."[411] She adds that under this perspective, privacy can be "seen as the negotiation of social boundaries through a set of actions that users collectively or individually take with respect to disclosure, identity and temporality in environments that are mediated by technology."[412]

Privacy isn't merely about control. Privacy is about not *being* controlled.[413] "The difficulty with privacy-control in the information age," says Schwartz, "is that individual self-determination is itself shaped by the processing of personal data."[414] Privacy as identity construction concerns protection against unreasonable steering or manipulation – by humans or by technology. If the environment unreasonably manipulates somebody, privacy may be violated. The environment includes technology, and could include personalised ads or other information. Many fear that too much personalised information could surreptitiously steer people's choices. For example, if a person's cookie profile suggests that he or she is conservative, a website could show that person primarily conservative content. Such personalisation might influence that person's political views, without him or her being aware. Hence, content personalisation could lead to a constraint on the construction of one's identity, and possibly an unreasonable constraint.[415] Section 3.3 discusses behavioural targeting and the risk of manipulation.

---

[407] Hildebrandt 2011a, p. 381. See also Hildebrandt et al. 2008a, p. 11.
[408] Human Rights Committee, Coeriel et al. v. The Netherlands, Communication No. 453/1991, U.N. Doc. CCPR/C/52/D/453/1991 (1994).
[409] Hildebrandt 2011a, p. 381-382. See generally on the importance of context for privacy Nissenbaum 2010.
[410] boyd & Ellison 2007, p. 222.
[411] Gürses 2010, p. 31.
[412] Gürses 2014, p. 22.
[413] Thanks to Aleecia M. McDonald. I borrow this phrase from her.
[414] Schwartz 1999, p. 1661 (capitalisation adapted).

The identity construction perspective raises the question of what identity means. There's a huge body of literature from various disciplines on the term identity.[416] FIDIS, an interdisciplinary research project on the Future of Identity in the Information Society, distinguishes two aspects of identity. First, there's a person's identity or image, as seen by others: a set of attributes. This is identity from a third person perspective. FIDIS speaks of the "common sense meaning identity."[417] A second aspect of one's identity is how a person sees him- or herself, from a first-person perspective. This could also be called somebody's individual identity, or self-identity.[418]

Like every privacy perspective, privacy as identity construction has weaknesses. For instance, it could be criticised for being too broad. Many kinds of influences could be seen as "unreasonable constraints" on identity construction. But perhaps not all these situations are best described as privacy violations.[419]

---

[415] Hildebrandt 2011a, p. 381. Westin, who sees privacy primarily as control, discussed the risk of unreasonable manipulation through subliminal advertising, "tampering with the unconscious" (Westin 1970, chapter 11).

[416] See for introductory texts on identity, with references to various disciplines Kerr et al. 2009a; Roosendaal 2013; Hildebrandt et al. 2008a.

[417] Hildebrandt et al. 2008a, p. 47.

[418] Hildebrandt et al. 2008a, p. 47. They also speak of the "relational notion" of identity.

[419] For instance, let's assume that photoshopped pictures in the media convey beauty ideals that deeply influence some people. If the altered pictures influence how people perceive themselves (too fat, too thin…), there may be an unreasonable constraint on the construction of their identity. This would bring the situation within the scope of privacy as identity construction. On the other hand, it could also be argued that such constraints aren't "unreasonable." Following that reasoning, there wouldn't be a privacy interference.

In conclusion, three groups of privacy perspectives can be distinguished: privacy as limited access, privacy as control over personal information, and privacy as the freedom of unreasonable constrains on identity construction. Each privacy perspective has strengths and weaknesses. Each perspective could be criticised for its scope, or for its vagueness. But in this study, the focus isn't on the exact scope of a definition that follows from a privacy perspective. This study doesn't argue that one privacy perspective is better than the other. The three perspectives highlight different aspects of privacy. Using one privacy perspective to discuss a problem doesn't imply that the other perspectives are irrelevant.

### 3.2 The right to privacy in European law

This section discusses the right to privacy in European law, and begins with an historical introduction. An early example of a rule that protects privacy interests, among other interests, is legal protection of the home against intrusions by the state or others. Protection of the home was granted in English case law from the sixteenth century,[420] and in the French Constitution of 1791.[421] Privacy-related interests also play an implicit role in court decisions prohibiting the publication of confidential letters from the eighteenth century.[422] Continental European law grants authors the *droit de divulgation*, that lets authors decide whether their work may be published.[423] Among the interests protected by this right are privacy-related interests.[424]

Legal protection of privacy-related interests in the area of press publications dates back centuries as well. The French Constitution of 1791 protected the freedom of the press, but also included protection against "[c]alumnies and insults against any persons whomsoever relative to their private life."[425] The law has provided protection against the use of one's image for a long time. In 1889 a German court ordered the destruction of photos of Otto van Bismarck on his deathbed, which were taken without his family's consent.[426] A French court handed down a similar judgement in 1858 regarding a portrait of an actress on her deathbed.[427]

[420] King's Bench 2 November 1765, Entick v. Carrington [1765] EWHC KB J98 95 ER 807. See on such early case law Cuddihy 2009, p. ixi.
[421] Title IV, article 9 of the French constitution of 1791.
[422] See for instance the case Chancery Court, Pope v. Curl [1741] 2 Atk. 342.
[423] See e.g. the European Copyright Code, article 3.2 (The Wittem Project 2010); Hugenholtz 2012, p. 347-348.
[424] Mayer-Schönberger 2010, p. 1864-1865.

Confidentiality of communications is another privacy-related right with a long history. King Louis XI of France nationalised the postal service in 1464. Soon the state organised mail delivery in many European countries. This gave the state the opportunity to read the letters, which, for example, happened systematically in France. In response to such practices, many states in Europe included a right to the confidentiality of correspondence in their constitutions during the nineteenth century. Hence, it was the introduction of a new communication channel (the postal service) that eventually led to the introduction of a new fundamental right.[428] In the twentieth century, the right to confidentiality of correspondence was extended to a general right to confidentiality of communications in Europe.[429] To the modern eye, legal protection of the home, legal protection against excesses of the press, and the right to confidentiality of correspondence are examples of the protection of privacy-related interests. Since the end of the nineteenth century, scholars have focused on privacy as the common feature of these different interests.[430]

The legal protection of privacy at international level blossomed after the Second World War. The Universal Declaration of Human Rights from 1948 contains a provision that protects privacy.[431] The International Covenant on Civil and Political Rights also protects privacy:

> 1. No one shall be subjected to arbitrary or unlawful interference with his privacy, family, home or correspondence, nor to unlawful attacks on his honour and reputation.
>
> 2. Everyone has the right to the protection of the law against such interference or attacks.[432]

---

[425] French Constitution of 1791 (3 September, 1791), chapter V, par. 17. See Whitman 2004, p. 1172.
[426] Zweigert & Kötz 1987, p. 688.
[427] Tribunal civil de la Seine, 16 June 1858, D.P. 1858, III, p. 62 (Rachel). See Prins 2009. See on the question of whether privacy rights do – or should – continue after death McCallig 2013; Harbinja 2013; Edwards 2013; Korteweg & Zuiderveen Borgesius 2009.
[428] See on the history of the legal protection of confidentiality of communications Steenbruggen 2009, p. 11; Hofman 1995, p. 23 and further; Ruiz 1997, p. 64-70.
[429] See for example article 5(1) of the e-Privacy Directive, and article 8 of the EU Charter of Fundamental Rights.
[430] Schoeman 1984, p. 1. See the discussion of Warren & Brandeis in the previous section.

***European Convention on Human Rights***

The right to privacy is set out in the European Convention on Human Rights, a treaty of the Council of Europe that entered into force in 1953.[433] The Council of Europe is the most important human rights organisation in Europe. It's based in Strasburg and has 47 member states, including all EU member states. All Council of Europe member states have signed up to the European Convention on Human Rights.[434] Article 8 of the European Convention on Human Rights contains the right to respect for private and family life, one's home and correspondence. Hence, it protects the right to privacy and other interests.[435]

Article 8 of the Convention is structured as follows: paragraph 1 prohibits interferences with the right to private life. Paragraph 2 shows that this prohibition isn't absolute. In many cases the right to privacy can be limited in the view of other

---

[431] Article 12 of the Universal Declaration of Human Rights.
[432] Article 17 of the International Covenant on Civil and Political Rights.
[433] The official title is: European Convention for the Protection of Human Rights and Fundamental Freedoms, as amended by Protocols Nos. 11 and 14.
[434] See the website of the Council of Europe: <www.coe.int/en/web/portal/country-profiles> accessed 14 May 2014.
[435] The European Court of Human Rights uses the phrase "private life" rather than privacy, but as noted, this study uses the phrases interchangeably. See on the distinction González Fuster 2014, p. 255.

interests, such as the prevention of crime, or the rights of others.[436] Article 8 reads as follows:

> European Convention on Human Rights
>
> Article 8, Right to respect for private and family life
>
> 1. Everyone has the right to respect for his private and family life, his home and his correspondence.
>
> 2. There shall be no interference by a public authority with the exercise of this right except such as is in accordance with the law and is necessary in a democratic society in the interests of national security, public safety or the economic well-being of the country, for the prevention of disorder or crime, for the protection of health or morals, or for the protection of the rights and freedoms of others.

### *Charter of Fundamental Rights of the European Union*

The Charter of Fundamental Rights of the European Union is a document listing the fundamental rights and freedoms recognised by the EU. The Charter was adopted in 2000, and was made a legally binding instrument by the Lisbon Treaty of 2009.[437] The Charter copies the right to private life almost verbatim from the European Convention on Human Rights. But the Charter uses the more modern and technology neutral term "communications" instead of "correspondence." The article reads as follows:

---

[436] Using a phrase from the last section, "reasonable" constraints on the freedom of identity construction don't violate privacy.

[437] See article 6.1 of the Treaty on EU (consolidated version 2012). The institutions of the EU must comply with the Charter. The Member States are also bound to comply with the Charter, when implementing EU law (article 51 of the Charter).

> Charter of Fundamental Rights of the European Union
>
> Article 7, Respect for private and family life
>
> Everyone has the right to respect for his or her private and family life, home and communications.

It follows from the EU Charter of Fundamental Rights that its article 7 offers at least the same protection as article 8 of the European Convention on Human Rights. The Charter has a separate provision that lists the limitations that may be imposed on its rights.[438] Regarding the right to private life, the limitations are similar to those listed in the second paragraph of article 8 of the European Convention on Human Rights.[439] In addition to the right to privacy, the Charter contains a separate right to the protection of personal data.[440] That right is discussed in the next chapter of this study, which introduces data protection law.[441]

The European Court of Justice says the right to privacy in the Charter and the Convention must be interpreted identically.[442] "Article 7 of the Charter must (…) be given the same meaning and the same scope as Article 8(1) of the ECHR, as interpreted by the case-law of the European Court of Human Rights (…)."[443] The privacy related case law of the European Court of Human Rights receives most attention in this study, because it's more developed than that of the European Court of Justice.

---

438 Article 52 of the EU Charter Of Fundamental Rights; Note from the Praesidium, comments on article 7 (Praesidium 2000).
439 See on the difference between article 52 of the Charter and article 8(2) of the Convention González Fuster 2014, p. 201.
440 Article 8 of the EU Charter Of Fundamental Rights.
441 See chapter 4, section 1.
442 For brevity, the "Court of Justice of the European Union" is referred to as European Court of Justice in this study. See article 19(1) of the Treaty on EU (consolidated version 2012).
443 CJEU, C-400/10, J. McB. v L. E., 5 October 2010, par. 53.

*Living instrument doctrine*

While scholars sometimes deplore the privacy's vagueness, the European Court of Human Rights uses the vagueness as an advantage. This way, the Court can apply the right to private life to unforeseen situations. The European Court of Human Rights interprets the rights granted in article 8 generously, and refuses to define the ambit of the article. The Court "does not consider it possible or necessary to attempt an exhaustive definition of the notion of private life."[444] The Court says it takes "a pragmatic, common-sense approach rather than a formalistic or purely legal one."[445] This allows the Court to adapt the protection of article 8 to new circumstances, such as technological developments. The Court's dynamic approach has been called the "living instrument doctrine."[446] The Court puts it as follows. "That the Convention is a living instrument which must be interpreted in the light of present-day conditions is firmly rooted in the Court's case-law."[447] The Court uses a "dynamic and evolutive" interpretation of the Convention, and states that "the term 'private life' must not be interpreted restrictively."[448]

> It is of crucial importance that the Convention is interpreted and applied in a manner which renders its rights practical and effective, not theoretical and illusory. A failure by the Court to maintain a dynamic and evolutive approach would indeed risk rendering it a bar to reform or improvement (…).[449]

---

[444] See e.g. ECtHR, Niemietz v. Germany, No. 13710/88, 16 December 1992, par. 29. The Court consistently confirms this approach. See e.g. ECtHR, Pretty v. United Kingdom, No. 2346/02, 29 April 2002, par. 61; ECtHR, S. and Marper v. United Kingdom, No. 30562/04 and 30566/04. 4 December 2008, par. 66.

[445] ECtHR, Botta v. Italy (153/1996/772/973), 24 February 1998, par. 27.

[446] Mowbray 2005.

[447] ECtHR, Matthews v. United Kingdom, No. 24833/94, 18 February 1999, par. 39. The Court started the "living instrument" approach in ECtHR, Tyrer v. United Kingdom, No. 5856/72, 25 April 1978, par. 31.

[448] Christine Goodwin v. United Kingdom, No. 28957/95, 11 July 2002, par 74; ECtHR, Amann v. Switzerland, No. 27798/95, 16 February 2000, par. 65.

[449] Christine Goodwin v. United Kingdom, No. 28957/95, 11 July 2002, par 74. See also ECtHR, Armonas v. Lithuania, No. 36919/02, 25 November 2008, par. 38.

The Court's dynamic approach is evident in the privacy case law. In 1978 for instance, the Court brought telephone calls under the scope of article 8, although the Convention speaks of private life and correspondence.[450] In 2004 the Court said: "increased vigilance in protecting private life is necessary to contend with new communication technologies which make it possible to store and reproduce personal data."[451] In the 2007 Copland case, the Court brought internet use under the protection of article 8. After repeating that phone calls are protected, the Court simply said that "[i]t follows logically that e-mails sent from work should be similarly protected under article 8, as should information derived from the monitoring of personal internet usage."[452] The Court adds that people have reasonable expectations of privacy regarding their use of the internet.[453]

The right to private life protects many aspects of personal development. In the 2008 Marper case, concerning storage of DNA samples in a police database, the Court lists some aspects of private life that it has brought under the scope of article 8.

> The Court recalls that the concept of "private life" is a broad term not susceptible to exhaustive definition. It covers the physical and psychological integrity of a person. It can therefore embrace multiple aspects of the person's physical and social identity. Elements such as, for example, gender identification, name and sexual orientation and sexual life fall within the personal sphere protected by article 8. Beyond a person's name, his or her private and family life may include

---

[450] ECtHR, Klass and others v. Germany, No. 5029/71, 6 September 1978, par. 41.
[451] ECtHR, Von Hannover v. Germany (I), No. 59320/00, 24 September 2004, par 70.
[452] ECtHR, Copland v. United Kingdom, No. 62617/00, 3 April 2007, par. 41 (capitalisation adapted, internal citations and numbering deleted).
[453] ECtHR, Copland v. United Kingdom, No. 62617/00, 3 April 2007, par 42. The European Court of Human Rights doesn't apply the same "reasonable expectation of privacy" test as US Courts. The European Court says: "A person's reasonable expectations as to privacy is a significant though not necessarily conclusive factor" (ECtHR, Perry v. United Kingdom, No. 63737/00, 17 July 2003, par. 37). See on the US Schwartz & Solove 2009, p. 106-137

> other means of personal identification and of linking to a family. Information about the person's health is an important element of private life. The Court furthermore considers that an individual's ethnic identity must be regarded as another such element. Article 8 protects in addition a right to personal development, and the right to establish and develop relationships with other human beings and the outside world.[454]

### *Horizontal effect*

The Convention was originally envisioned to protect people against the state. The state has a negative duty not to interfere too much in people's lives. But the Court also derives positive duties for states from the Convention. Hence, sometimes the state has to take action to protect people from interferences by other private actors. The Court summarises this as follows.

> Although the object of article 8 is essentially that of protecting the individual against arbitrary interference by the public authorities, it does not merely compel the State to abstain from such interference: in addition to this primarily negative undertaking, there may be positive obligations inherent in an effective respect for private or family life (…). These obligations may involve the adoption of measures designed to

[454] ECtHR, S. and Marper v. United Kingdom, No. 30562/04 and 30566/04. 4 December 2008, par. 66 (internal citations omitted; capitalisation adapted).

> secure respect for private life even in the sphere of the relations of individuals between themselves (…).[455]

People can't sue another private party under the European Convention on Human Rights.[456] But people can complain to the Court if the state doesn't adequately protect their rights against infringements by other non-state actors. This way, the Convention's privacy right has a horizontal effect.[457] The Court says it "does not consider it desirable, let alone necessary, to elaborate a general theory concerning the extent to which the Convention guarantees should be extended to relations between private individuals *inter se*."[458]

The positive obligations can be far-reaching.[459] The Court requires states to *effectively* protect the Convention rights: "Article 8, like any other provision of the Convention or its Protocols, must be interpreted in such a way as to guarantee not rights that are theoretical or illusory but rights that are practical and effective."[460] A state can fail in its positive obligations to ensure effective protection of the right to private life if non-state actors handle personal data carelessly. For instance, having a data protection law that allows people to claim for damages after a data breach isn't always sufficient.[461]

Some commentators are sceptical of the horizontal effect of human rights.[462] Others say it's "self-evident" that human rights have horizontal effect.[463] Gutwirth argues that protecting a public interest is a more acceptable reason to interfere with privacy than aiming for profit.

---

455 ECtHR, Z v. Finland, No. 22009/93, 25 February 1997, par. 36 (capitalisation adapted). See also ECtHR, Mosley v. United Kingdom, 48009/08, 10 May 2011, par 106.

456 Article 34 of the European Convention on Human Rights.

457 See generally Akandji-Kombe 2007; De Hert 2011; Verhey 1992, Verhey 2009.

458 ECtHR, VGT Verein Gegen Tierfabriken v. Switzerland, No. 24699/94, 28 June 2001, par. 46.

459 See generally on positive requirements following from article 8 of the European Convention on Human Rights in the field of data protection De Hert 2011. To what extent the EU Charter of Fundamental Rights has horizontal effect is unclear (see Kokott & Sobotta 2014, p. 225).

460 ECtHR, Biriuk v. Lithuania, No. 23373/03, 25 November 2008, par. 37. See also ECtHR, Airey v. Ireland, No. 6289/73, 9 October 1979, par. 24-25.

461 ECtHR, I. v. Finland, No. 20511/03, 17 July 2008, par. 47.

462 See e.g. De Vos 2010.

463 Gutwirth 2002, p. 38.

> If privacy is protected against acts of the public authorities, should it "a fortiori" not be protected against individual acts, too? After all, the government acts on behalf of the public interest, which seems to be a more legitimate reason for an invasion of privacy than, for example, personal profit seeking of a businessman.[464]

***Three privacy perspectives in case law***

The above-mentioned three privacy perspectives – privacy as limited access, privacy as control, and privacy as identity construction – can be recognised in the case law of the European Court of Human Rights, although the Court doesn't use this taxonomy.[465] Privacy as limited access lies at the core of article 8: "the essential object and purpose of Article 8, [is] to protect the individual against arbitrary interference by the public authorities."[466] But the Court also emphasises privacy as limited access in cases where non-state actors interfere with privacy. "The right to privacy consists essentially in the right to live one's own life with a minimum of interference."[467]

The Court mentions keeping personal information confidential as well. "The concept of private life covers personal information which individuals can legitimately expect should not be published without their consent (…)."[468] In some judgments, the reasoning of the Court reminds one of the perspective of privacy as a right to be let alone. In a 2004 case, the Court took into account that paparazzi harassed the Princess of Monaco.[469] In sum, article 8 comprises privacy as limited access.

---

[464] Gutwirth 2002, p. 38.
[465] See for an overview of the article 8 case law, using other taxonomies Harris et al. 2009, p. 361-424; Heringa & Zwaak 2006.
[466] ECtHR, Niemietz V. Germany, No. 13710/88, 16 December 1992, par. 31. Harris et al. also see privacy as limited access, "a private space into no-one is entitled to enter", as the core of the concept of private life (Harris et al. 2009, p. 367).
[467] This definition of privacy is taken from Parliamentary Assembly, Resolution 428 (1970) containing a declaration on mass communication media and human rights. The Court cited the definition in several cases, including cases where non-state actors infringed on privacy. See ECtHR, Von Hannover v. Germany (I), No. 59320/00, 24 September 2004, par 42; ECtHR, Von Hannover v. Germany (II), Nos. 40660/08 and 60641/08, 7 February 2012, par. 71; ECtHR, Mosley v. United Kingdom, 48009/08, 10 May 2011, par. 56.
[468] ECtHR, Flinkkilä and others v. Finland, No. 25576/04, 6 April 2010, par. 75.

Privacy as control is also present in the case law of the European Court of Human Rights. The Court says "it would be too restrictive to limit the notion [of private life] to an 'inner circle' in which the individual may live his own personal life as he chooses and to exclude therefrom entirely the outside world not encompassed within that circle."[470] In several cases, the Court cites a Resolution of the Parliamentary Assembly of the Council of Europe on the right to privacy.[471] "In view of the new communication technologies which make it possible to store and use personal data, the right to control one's own data should be added to this definition."[472] In a case where a picture was taken without consent, the Court says it's a problem if "the person concerned would have no control over any subsequent use of the image."[473] Privacy as control can also be recognised in cases where the Court accepts a right for people to access[474] or to correct[475] personal data regarding them.

The Court has established that storing personal data can interfere with privacy, regardless of how those data are used.[476] "The mere storing of data relating to the private life of an individual amounts to an interference within the meaning of article 8 (…). The subsequent use of the stored information has no bearing on that finding."[477] However, the Court said this in a case where the state stored personal data that are particularly sensitive (DNA data). In some cases where private parties store personal data, the Court also says that the mere storage interferes with privacy, but again the data were rather sensitive.[478] In some other cases the Court didn't see personal data processing as a privacy interference. Hence, for the Court some personal data processing activities don't interfere with privacy.[479] Sometimes the European Court of Human Rights also applies data protection principles (see the next chapter).[480] The Court has cited the Data Protection Convention,[481] and the Data Protection Directive.[482]

---

[469] ECtHR, Von Hannover v. Germany (I), No. 59320/00, 24 September 2004. In principle, offensive spam email interferes with privacy (ECtHR, Muscio v. Italy, No. 31358/03, 13 November 2007 (inadmissible)).
[470] ECtHR, Niemietz v. Germany, No. 13710/88, 16 December 1992, par. 29. The Court also stresses control over personal information in ECtHR, Von Hannover v. Germany (II), Nos. 40660/08 and 60641/08, 7 February 2012, par. 96.
[471] See e.g. ECtHR, Von Hannover v. Germany (I), No. 59320/00, 24 September 2004, par 72; ECtHR, Von Hannover v. Germany (II), Nos. 40660/08 and 60641/08, 7 February 2012, par. 71.
[472] Parliamentary Assembly, Resolution 1165 (1998), on the right to privacy.
[473] ECtHR, Reklos and Davourlis v. Greece, No. 1234/05, 15 January 2009, par 40. See also par. 42-43 for a control perspective on privacy. See also ECtHR, Von Hannover v. Germany (II), Nos. 40660/08 and 60641/08, 7 February 2012, par. 96.
[474] See e.g. ECtHR, Gaskin v. United Kingdom, Application no. 10454/83, 7 July 1989, par. 49; ECtHR, McMichael v. United Kingdom, No. 16424/90, 24 February 1995, par 92; ECtHR, Mcginley and Egan v. United Kingdom (10/1997/794/995-996), 9 June 1998. par 97.
[475] See e.g. ECtHR, Rotaru v. Romania, No. 28341/95, 4 May 2000, par. 46; Christine Goodwin v. United Kingdom, No. 28957/95, 11 July 2002, par. 93; ECtHR, Segerstedt-Wiberg and others v. Sweden, No. 62332/00, 6 June 2006, par. 99; ECtHR, Cemalettin Canli v. Turkey, No. 22427/04, 18 November 2008, par 41-43; ECtHR, Ciubotaru V. Moldova, No. 27138/04, 27 April 2010, par. 51, par. 59.
[476] See e.g. ECtHR, Leander v. Sweden, No. 9248/81, 26 March 1987, par. 48; ECtHR, Amann v. Switzerland, No. 27798/95, 16 February 2000, par. 69; ECtHR, Copland v. United Kingdom, No. 62617/00, 3 April 2007, par. 43-44; ECtHR, S. and Marper v. United Kingdom, No. 30562/04 and 30566/04. 4 December 2008, par. 67, par 121.

The other important European Court, the European Court of Justice, says that privacy is threatened by any personal data processing – and doesn't limit its remarks to sensitive data.[483] This is in line with the EU Charter of Fundamental Rights, which requires fair processing for any kind of personal data. The Court says about the right to privacy and data protection: "as a general rule, any processing of personal data by a third party may constitute a threat to those rights."[484] As the Data Protection Directive requires, the Court does differentiate between non-special personal data and "special categories of data", such as data regarding health, religion or race.[485] In sum, privacy as control can be recognised in the case law of the European Court of Human Rights and the European Court of Justice.[486]

---

477 ECtHR, S. and Marper v. United Kingdom, No. 30562/04 and 30566/04. 4 December 2008, par. 67 (capitalisaion adapted).

478 In a case where a private party held photographic material, the mere retention of that personal information interfered with private life (ECtHR, Reklos and Davourlis v. Greece, No. 1234/05, 15 January 2009, par. 42). See along similar lines (regarding video surveillance by a private party) ECtHR, Köpke v. Germany, No. 420/07 (inadmissible), 5 October 2010.

479 See e.g. ECtHR, Perry v. United Kingdom, No. 63737/00, 17 July 2003, par 40: "the normal use of security cameras *per se* whether in the public street or on premises, such as shopping centres or police stations where they serve a legitimate and foreseeable purpose, do not raise issues under Article 8 § 1 of the Convention." See De Hert & Gutwirth 2009, p. 24-26; Kranenborg 2007, p. 311-312; Kokott & Sobotta 2014, p. 223-224; González Fuster 2014, p. 101.

480 See on the data protection principles chapter 4, section 2.

481 See for an early case ECtHR, Z v. Finland, No. 22009/93, 25 February 1997, par. 95.

482 Examples of cases where the Court mentions the Data protection Directive include ECtHR, Romet v. The Netherlands, No. 7094/06, 14 February 2012; ECtHR, M.M. v. United Kingdom, No. 24029/07, 13 November 2012; ECtHR, S. and Marper v. United Kingdom, No. 30562/04 and 30566/04. 4 December 2008; ECtHR, Mosley v. United Kingdom, 48009/08, 10 May 2011.

483 CJEU, C-291/12, Schwartz v. Stadt Bochum, 17 October 2013, par. 25. See also the judgment on the Data Retention Directive CJEU, C-293/12 and C-594/12, Digital Rights Ireland Ltd, 8 April 2014, par. 29.

484 CJEU, C-291/12, Schwartz v. Stadt Bochum, 17 October 2013, par. 25. See also the judgment on the Data Retention Directive CJEU, C-293/12 and C-594/12, Digital Rights Ireland Ltd, 8 April 2014, par. 29.

The third privacy perspective, the freedom from unreasonable constraints on identity construction, can be recognised in the case law as well. For example, in a case regarding privacy infringements by the press, the Court emphasises privacy's function for the construction of one's personality. "As to respect for the individual's private life, the Court reiterates the fundamental importance of its protection in order to ensure the development of every human being's personality. That protection extends beyond the private family circle to include a social dimension."[487]

The Court says the right to private life should enable a person to "freely pursue the development and fulfilment of his personality."[488] The right to private life also includes a social dimension and "comprises the right to establish and develop relationships with other human beings."[489] In a 2012 judgment concerning Princess Caroline of Monaco, who complained about privacy violations by the press, the reasoning of the Court relates to the privacy as identity construction perspective.

> The Court reiterates that the concept of private life extends to aspects relating to personal identity, such as a person's name, photo, or physical and moral integrity; the guarantee afforded by article 8 of the Convention is primarily intended to ensure

---

[485] See e.g. CJEU, C-101/01, Lindqvist, 6 November 2003.

[486] The European Court of Justice isn't very explicit on the question of whether it sees privacy as control over personal information. However, the Court's reasoning does remind one of privacy as control sometimes. For instance, in the Data Retention case the Court says that the "fact that data are retained and subsequently used without the subscriber or registered user being informed", entails a "particularly serious" interference with the right to privacy (CJEU, C-293/12 and C-594/12, Digital Rights Ireland Ltd, 8 April 2014, par. 37). The Google Spain case, emphasising the right to request erasure of data (possibly too much), also fits the privacy as control perspective (CJEU, C-131/12, Google Spain, 13 May 2014).

[487] ECtHR, Biriuk v. Lithuania, No. 23373/03, 25 November 2008, par. 38.

[488] ECtHR, Shtukaturov v. Russia, No. 44009/05, 27 March 2008, par. 83.

[489] ECtHR, Amann v. Switzerland, No. 27798/95, 16 February 2000, par. 65; ECtHR, Perry v. United Kingdom, No. 63737/00, 17 July 2003, par. 65.

> the development, without outside interference, of the personality of each individual in his relations with other human beings.[490]

In conclusion, judges and lawmakers try to adapt the right to privacy to new developments and technologies. The right to privacy is laid down in the European Convention on Human Rights, and in the EU Charter of Fundamental Rights. The European Court of Human Rights interprets the right to privacy generously, and refuses to pin itself down to one definition. Each of the three privacy perspectives that was discussed in section 3.1 can be recognised in the case law of the European Court of Human Rights.

## 3.3 Privacy implications of behavioural targeting

There are many privacy problems with behavioural targeting.[491] This study focuses in particular on three problems. First, the massive collection of data about user behaviour can lead to chilling effects. A second problem is the lack of individual control over personal information. A third problem is social sorting and the risk of manipulation.[492] The problems are related and partly overlap.

***Chilling effects relating to massive data collection on user behaviour***

Many people find data collection for behavioural targeting creepy or invasive.[493] The tracking for behavioural targeting has often been compared with following somebody on the streets.[494] People use the internet for many things, including things that they would prefer to keep confidential.[495] As Berners-Lee notes, browsing behaviour can reveal a lot about a person:

> The URLs which people use reveal a huge amount about their lives, loves, hates, and fears. This is extremely sensitive material. People use the web in crisis, when wondering whether they have STDs, or cancer, when wondering whether they are homosexual and whether to talk about it, to discuss political views which may to some may be abhorrent, and so on.[496]

---

[490] ECtHR, Von Hannover v. Germany (II), Nos. 40660/08 and 60641/08, 7 February 2012, par 95 (capitalisation adapted). See also ECtHR, Niemietz v. Germany, No. 13710/88, 16 December 1992, par 29. Arguably, the privacy as identity construction perspective could also be recognised in the Google Spain judgment of the European Court of Justice, although the Court based its reasoning mostly on data protection law. People could try to shape their identity by influencing search results regarding their name (CJEU, C-131/12, Google Spain, 13 May 2014).
[491] See generally on privacy (and related) problems regarding behavioural targeting Turow 2011; Castelluccia & Narayanan 2012; Federal Trade Commission 2012. See also Hildebrandt & Gutwirth (eds.) 2008, on profiling, and Richards 2013, on surveillance, and the references therein.
[492] Van Der Sloot gives a similar analysis of privacy problems resulting from data collection in the area of behavioural targeting. But he argues that the problems are better conceptualised as data protection problems, rather than as privacy problems (Van Der Sloot 2011).

For example, many websites about health problems allow third parties to track their visitors. People might search for information about unwanted pregnancies, drugs, suicidal tendencies, or HIV. Medical problems can be embarrassing or simply personal. People may have an individual privacy interest in keeping confidential that they read about such topics. But if a chilling effect occurred, the problem would go beyond individual interests. People with questions about health might refrain from looking for information if they fear being tracked.[497] It would be detrimental for society if a person failed to seek treatment for a contaguous disease.

People also use the internet to read about news and politics. Third party tracking happens on the websites of most newspapers. But people could feel uneasy when firms monitor their reading habits. A person's political opinion could be inferred from his or her reading habits. People may want to read a communist, Christian, or Muslim news site. And a political opinion that is uncontroversial now, could become suspicious in the future.[498] Many conclusions could be drawn from people's browsing behaviour – the right or the wrong conclusions.[499] Somebody might be looking for information about cancer for a friend. And somebody who reads about bombing airports isn't necessarily a terrorist.

---

[493] See chapter 7, section 1 for research on people's attitude regarding behavioural targeting.
[494] See e.g. Kang 1998, par 1198-1199; Kristol 2001, p. 180; Chester 2007, p. 134; International Working Group on Data Protection in Telecommunications (Berlin Group) 2013, p. 2-3.
[495] As Richards puts it, a record of somebody's browsing behaviour is "in a very real sense a partial transcript of the operation of a human mind" (Richards 2008, p. 436).
[496] Berners-Lee 2009. He discusses behavioural targeting that relies on deep packet inspection, but his remark is relevant for behavioural targeting in general.
[497] See United Nations High Commissioner for Human Rights 2014, p. 5; Castelluccia & Narayanan 2012, p. 9.

People have individual interests in keeping their reading habits confidential, but it's also in the interest of society that people don't fear surveillance. Frank La Rue, the Special Rapporteur on the promotion and protection of the right to freedom of opinion and expression for the United Nations, says privacy is essential in order to enjoy the right to seek and receive information.

> States cannot ensure that individuals are able to freely seek and receive information or express themselves without respecting, protecting and promoting their right to privacy. Privacy and freedom of expression are interlinked and mutually dependent; an infringement upon one can be both the cause and consequence of an infringement upon the other.[500]

Behavioural targeting could be seen as a form of surveillance, as defined by Lyon: "any collection and processing of personal data, whether identifiable or not, for the purposes of influencing or managing those whose data have been garnered."[501] The goal of data processing for behavioural targeting is influencing people with advertising. Lyon stresses that the word surveillance doesn't imply that a practice is sinister. But he adds that surveillance always implies "power relations."[502]

---

[498] Berners-Lee 2009 makes a similar point. See also Turow et al. 2012.
[499] Van Hoboken 2012, p. 323; Purtova 2011, p. 44-46.
[500] La Rue 2013, p. 20.
[501] Lyon 2001, p. 2. A United Nations report speaks of "communications surveillance" (La Rue 2013, p. 3).
[502] Lyon 2001, p. 16. See also Clarke, who speaks of dataveillance, "the systematic use of personal data systems in the investigation or monitoring of the actions or communications of one or more persons" (Clarke 1999).

The chilling effect of surveillance can be illustrated by the Panopticon, a circular prison designed by Bentham.[503] The prison has a watchtower in the middle, and the guards can watch the prisoners at all times. The prisoners can always see the watchtower, so they're reminded that they could be being watched at any given time. But the prisoners can't see whether they are being watched. Therefore, they will adapt their behaviour.[504]

Behavioural targeting fits Lyon's definition of surveillance, but there's no threat of punishment. However, as the German Bundesverfassungsgericht notes, not knowing how personal information will be used can cause a chilling effect as well. "If someone is uncertain whether deviant behaviour is noted down and stored permanently as information, or is applied or passed on, he will try not to attract attention by such behaviour."[505] Unfettered surveillance could lead to self-censorship. The Court adds that this threatens society as a whole. "This would not only impair [the individual's] chances of development but would also impair the common good, because self-determination is an elementary functional condition of a free democratic community based on its citizens' capacity to act and to cooperate."[506]

It has been suggested that online tracking doesn't merely influence people's behaviour, but also their thoughts. In the US, Richards argues that surveillance threatens the possibility to "develop ideas and beliefs away from the unwanted gaze or interference of others." Therefore, he says, the first amendment (that protects freedom of speech) should be interpreted in such a way that it safeguards intellectual privacy. "Intellectual privacy is protection from surveillance or interference when we are engaged in the processes of generating ideas – thinking, reading, and speaking with confidantes before our ideas are ready for public consumption."[507] Similarly, Cohen argues for a "right to read anonymously."[508]

---

[503] Foucault 1977.
[504] It has been suggested that behavioural targeting is worse than a Panopticon, as firms can store all the information they gather (International Working Group on Data Protection in Telecommunications (Berlin Group) 2013).
[505] Bundesverfassungsgericht 25 March 1982, BGBl.I 369 (1982), (Volks-, Berufs-, Wohnungs- und Arbeitsstättenzählung (Volkszählungsgesetz)), translation by Riedel, E.H., Human Rights Law Journal 1984, vol. 5, no 1, p. 94, p. 100, paragraph II.
[506] Bundesverfassungsgericht 25 March 1982, BGBl.I 369 (1982), (Volks-, Berufs-, Wohnungs- und Arbeitsstättenzählung (Volkszählungsgesetz)), translation by Riedel, E.H., Human Rights Law Journal 1984, vol. 5, no 1, p. 94, p. 100, paragraph II.

In Europe, Van Hoboken suggests that privacy is necessary to enjoy the right to impart and receive information.

> It can be argued that the user's privacy is a precondition for the fundamental right to search, access and receive information and ideas freely. Free information-seeking behavior can be quite negatively affected if the main available options to find information online entail comprehensive surveillance and storage of end-users behavior without appropriate guarantees in view of intellectual freedom.[509]

The chilling effect could be greater if communications, such as email messages, are also monitored. The European Court of Human Rights says that the mere threat of surveillance threatens fundamental rights. In a case regarding a German law that empowered the authorities to inspect mail and to listen to telephone conversations, the Court warns that the "menace of surveillance can be claimed in itself to restrict free communication."[510] In another case, the Court states that such a "threat necessarily strikes at freedom of communication between users of the telecommunications services and thereby amounts in itself to an interference with the exercise of the

---

[507] Richards 2014. See Richards 2008; Richards 2013.
[508] Cohen 1995. See also Kang 1998, p. 1260.
[509] Van Hoboken 2012, p. 226, internal footnote omitted. While he discusses surveillance by search engines, his remarks are also relevant for behavioural targeting.
[510] ECtHR, Klass and others v. Germany, No. 5029/71, 6 September 1978, par. 37.

applicants' rights under article 8, irrespective of any measures actually taken against them."[511]

The European Court of Human Rights says that monitoring traffic data (sometimes called metadata), rather than the content of communications, also interferes with the right to privacy.[512] According to the Bundesverfassungsgericht, the retention of traffic data by telecommunications companies for law enforcement can invoke a "feeling of permanent control", because people feel a "diffuse threat."[513]

> [A] preventive general retention of all telecommunications traffic data (…) is, among other reasons, also to be considered as such a heavy infringement because it can evoke a sense of being watched permanently (…). The individual does not know which state official knows what about him or her, but the individual does know that it is very possible that the official does know a lot, possibly also highly intimate matters about him or her.[514]

Along the same lines, the European Court of Justice states that storing traffic data by telecommunications companies for law enforcement purposes "is likely to generate in the minds of the persons concerned the feeling that their private lives are the subject of constant surveillance."[515] The cases concern surveillance for law enforcement, but similar conclusions can be drawn about behavioural targeting.[516] Once private parties hold personal data, law enforcement bodies can, and indeed often do, access those data. In a case regarding monitoring internet traffic by a private party, the Advocate General of the European Court of Justice states that such monitoring "constitutes, by its very nature, a 'restriction' (…) on the freedom of communication enshrined in article 11(1) of the Charter (…)."[517]

---

[511] ECtHR, Liberty and others v. United Kingdom, No. 58243/00, 1 July 2008, par. 56. See also par. 104-105. See similarly United Nations High Commissioner for Human Rights 2014, p. 7.
[512] ECtHR, Malone v. United Kingdom, No. 8691/79, 2 August 1984, par. 83-84; ECtHR, Copland v. United Kingdom, No. 62617/00, 3 April 2007. See also CJEU, C-293/12 and C-594/12, Digital Rights Ireland Ltd, 8 April 2014.
[513] Traffic data are, in short, data processed for the purpose of the conveyance of a communication (see article 2(b) of the e-Privacy Directive). See chapter 5, section 6.
[514] Bundesverfassungsgericht 2 March 2010, BvR 256/08 vom 2.3.2010, Absatz-Nr. (1 - 345), (Vorratsdatenspeicherung) [Data Retention]. Translation by Bellanova et al. 2011, p. 10.
[515] CJEU, C-293/12 and C-594/12, Digital Rights Ireland Ltd, 8 April 2014, par. 37.
[516] See Article 20 Working Party 2014, WP 217, p. 37.

The early history of the right to confidentiality of communications illustrates the connection between that right and the right to freedom of expression. Nowadays the right to confidentiality of communications is regarded as a privacy-related right.[518] But when it was developed in the late eighteenth century, confidentiality of correspondence was seen as an auxiliary right to safeguard freedom of expression.[519] The right to confidentiality of communications in the e-Privacy Directive also applies to web browsing behaviour.[520]

Behavioural targeting firms collect information about people's online activities, which can include information that people don't want to disclose. Privacy as limited access captures this. Moreover, some tracking practices invades people's private sphere. For instance, a smart phone's location data could disclose where a person's house is, or where that person sleeps. Tracking that involves accessing information on people's devices can also interfere with privacy as limited access. The e-Privacy Directive's preamble discusses tracking technologies such as adware and cookies, and says that people's devices are private: "[t]erminal equipment of users of electronic communications networks and any information stored on such equipment are part of the private sphere of the users requiring protection under the European Convention for the Protection of Human Rights and Fundamental Freedoms."[521] Similarly, the

---

[517] Opinion AG Cruz Villalón, 14 April 2011, par 73 (for CJEU, C-70/10, Scarlet v. Sabam, 24 November 2011, Scarlet Sabam AG) (capitalisation adapted). The Advocate General is an independent advisor to the European Court of Justice (see article 252 of the consolidated version of the Treaty on the functioning of the EU).
[518] See for instance article 7 of the EU Charter of Fundamental Rights.
[519] Ruiz 1997, p. 67. See also ECtHR, Autronic AG v. Switzerland, No. 12726/87, 22 May 1990, par. 47.
[520] See chapter 6, section 4.
[521] Recital 24 of the e-Privacy Directive.

German Bundesverfassungsgericht says people have a "right to the guarantee of the confidentiality and integrity of information technology systems."[522]

Privacy as control and privacy as identity construction are also relevant when discussing chilling effects. For instance, the lack of individual control over data processed for behavioural targeting could aggravate the chilling effect. And if surveillance indeed influenced people's thoughts, it could constrain the development of their identity.[523] Regardless of how data are used at later stages, tracking people's behaviour (phase 1 of behavioural targeting) can cause a chilling effect. But data processing in later phases can worsen the chilling effect. For instance, a firm could find new information about a person by analysing the collected data.[524]

***Lack of individual control over personal information***

A second privacy problem regarding behavioural targeting is that people lack control over information regarding them. One aspect of the lack of individual control is information asymmetry. The online behaviour of hundreds of millions of people is tracked, without them being aware.[525] A visit to a website can lead to receiving dozens of tracking cookies from firms that people have never heard about. As Cranor notes, "it is nearly impossible for website visitors to determine where their data flows, let alone exert any control over it."[526]

Furthermore, people have scant knowledge about what firms do with data about them, and what the consequences could be. Personal data is auctioned off, shared and combined, without people being aware. "Users, more often than not, do not understand the degree to which they are a commodity in each level of this marketplace."[527] If people don't even know who holds information about them, it's clear they can't exercise control over that information.

[522] Bundesverfassungsgericht, 27 February 2008, decisions, vol. 120, p. 274-350 (Online Durchsuchung).
[523] Diaz & Gürses 2012.
[524] Schermer 2007, p. 136-137; Schermer 2013, p. 139.
[525] Hoofnagle et al. 2012, p. 291.
[526] Cranor 2012, p. 1.

Firms rarely explain clearly what they do with people's data. Privacy policies often use ambiguous language, and don't help to make the complicated data flows behind behavioural targeting transparent. It's rare for people to have consented in a meaningful way to behavioural targeting. As discussed in more detail in chapter 7, people don't know enough about the complex data flows behind behavioural targeting to understand what they are being asked to consent to. And if firms ask consent, they often make using a service conditional on consent to tracking. Many people feel they must consent to behavioural targeting when encountering such take-it-or-leave-it choices.

Behavioural targeting can lead to experienced harms.[528] Data could be used in ways that harm people. For instance, a profile could be used to charge higher prices to a person. A health insurer might learn that somebody was reading about certain diseases, or about alcohol addiction. Furthermore, storing information about people is inherently risky. Data can leak, to insiders or outsiders. For instance, an employee might access the information stored by a firm. In one case, an internet firm's employee accessed information in user accounts, such as messages and contact lists.[529] Or a hacker or another outsider might obtain the data. 32 million user passwords were accessed at a firm that develops social media apps and runs an ad network.[530] And in the US, data brokers accidentally sold personal data to criminals.[531] A data breach could lead to spam, embarrassment, identity fraud, or other unpleasant surprises.[532]

---

[527] White House (Podesta J et al.) 2014, p. 41.
[528] See on experienced harms and expected harms Gürses 2010, p. 87-89; Calo 2011. See also section 3 of this chapter.
[529] Checn 2010.
[530] See about this data breach at RockYou: Hoffman 2011.
[531] For instance, US data broker Acxiom sold personal data about thousands of people to a criminal gang (Van der Meulen 2010, p. 76-77; 206-209). Experian also sold personal information to criminals (Krebs 2013).
[532] The harms can be diverse. In one case, a US data broker sold information to a stalker that used the information to locate and murder a woman (Remsburg v. Docusearch, Inc. 816 A.2d (N.H. 2003)).

Identity fraud can be costly for the victim, and for society as a whole – even without taking privacy interests into account.[533]

A general risk resulting from data storage is function creep: using data for other purposes than the original collection purpose.[534] For instance, commercial databases tend to attract the attention of law enforcement bodies.[535] People would protest a law requiring everyone to provide the police with lists of all of the websites they visit daily. But many behavioural targeting firms collect such data. And when the data are there, the police can demand access.[536] Firms like Facebook and Google, both using behavioural targeting, get many demands for police access.[537] Moreover, intelligence agencies could access data held by firms.[538] Schneier summarises: "[t]he primary business model of the Internet is built on mass surveillance, and our government's intelligence-gathering agencies have become addicted to that data."[539] The data that have been gathered for behavioural targeting can thus be used for new purposes. But also the technologies that have been developed for behavioural targeting could be used for new purposes. For instance, the National Security Agency (US) appears to have used tracking cookies of behavioural targeting firms to unmask users of the Tor anonymity service.[540] Using surveillance technologies for new purposes could be called "surveillance creep."[541]

---

[533] Van Den Hoven 1997. See on identity fraud ECtHR, Romet v. The Netherlands, No. 7094/06, 14 February 2012.
[534] Function creep can be seen as a breach of data protection law's purpose limitation principle (see chapter 4, section 3). See Dahl & Sætnan 2009.
[535] See the Data Retention Directive. See also Van Hoboken 2012, p. 324-325. Haggerty & Ericson 2000.
[536] Or, to take an example by Schneier "[i]magine the government passed a law requiring all citizens to carry a tracking device. Such a law would immediately be found unconstitutional. Yet we all carry mobile phones" (Schneier 2013a).
[537] See Google Transparency Report 2014; Facebook Government Requests Report 2014.
[538] See on state access to commercial data Soghoian 2012 (regarding the US); Brown 2012; Arnbak et al. 2013; Koning 2013. See also the special issue on systematic government access to private-sector data of the journal International Data Privacy Law, volume 4, issue 1, February 2014.
[539] Schneier 2013a. He's from the US, but his remarks are relevant for Europe too.
[540] See Reisman et al. 2014, with further references.
[541] See Marx 2005.

Apart from experienced harms, the lack of individual control over personal information can lead to the expectation of harm, or subjective harm.[542] People may vaguely realise that organisations hold data about them. Many people fear their information will be used, without their knowledge, for unexpected purposes.[543] A majority of Europeans doesn't trust internet companies such as search engines and social networks sites to protect their personal information.[544] The lack of control problem has an individual and a societal dimension. Information based harms, such as identity fraud, are costly both for victims and society as a whole.[545] For instance, the European Commission suggests that consumers' privacy anxieties hinder online business.[546]

In sum, transparency and individual control are lacking during every behavioural targeting phase. The ideal of privacy as individual control over personal information doesn't seem close to materialising in the area of behavioural targeting.

***Social sorting***

A third privacy risk resulting from behavioural targeting concerns social sorting and the risk of manipulation. Behavioural targeting enables what surveillance scholars refer to as social sorting.[547] In Lyon's words, social sorting involves "obtain[ing] personal and group data in order to classify people and populations according to varying criteria, to determine who should be targeted for special treatment, suspicion, eligibility, inclusion, access, and so on."[548] For example, an advertiser could use discounts to lure affluent people to become regular customers. But the advertiser might want to avoid poor people because they're less profitable. Or advertisers could target poor people with offers for certain products, such as predatory lending schemes.[549] Legal scholars tend to speak of discrimination when discussing social sorting.[550]

---

[542] Gürses 2010, p. 87-89; Calo 2011. See also section 3 of this chapter. The expectation of harm that results from a lack of individual control over personal data could also be called a chilling effect.
[543] European Commission 2011 (Eurobarometer), p. 146.
[544] European Commission 2011 (Eurobarometer), p. 138.
[545] The phrase "information based harms" is borrowed from Van Den Hoven 1997.
[546] European Commission proposal for a Data Protection Regulation (2012), p. 1. See also recital 5 of the e-Privacy Directive. From an economic perspective, information asymmetry is a societal problem because it's a type of market failure (see chapter 7, section 3).
[547] See chapter 1, section 3, for a description of surveillance studies.
[548] Lyon 2002a, p. 20. See on surveillance and marketing Pridmore & Lyon 2011.

Firms classify people as "targets" or "waste", says Turow. "Marketers justify these activities as encouraging *relevance*. But the unrequested nature of the new media-buying routines and the directions these activities are taking suggest that *narrowed options* and *social discrimination* might be better terms to describe what media-buyers are actually casting."[551] The Dutch Data Protection Authority expresses similar concerns: "profiling can lead to stigmatisation and discrimination and to a society in which free choice has become illusory."[552] European Data Protection Authorities add that "[t]his may perpetuate existing prejudices and stereotypes, and aggravate the problems of social exclusion and stratification."[553]

Social sorting isn't a new phenomenon. By placing billboards for expensive cars in wealthy neighbourhoods, firms can target population segments based on location. Since the 1980s database marketing allows for segmentation on the individual level.[554] A book on database marketing explains that firms shouldn't treat all customers the same:

> Successful relationship marketing forces us to look at a new marketing fact of life. The buyer-seller relationship is not a democracy. All customers are not created equal. All customers are not entitled to the same inalienable rights, privileges, and

[549] To illustrate, one US firm sells an "Online Ad Network Direct Response Buyers Mailing List": "These responsive buyers have also expressed an interest in additional promotions, and 60% of these impulse buyers had their bank cards declined. (…) This self reported age 18+, third party verified database is perfect for subprime financial or credit repair offers. Gender, DOB, homeowner, marital status, income and a variety of other demographics are also available" (Mailing List Finder 2014).

[550] See Richards 2013, p. 1957-58.

[551] Turow 2011, p. 89. See also Dixon & Gellman 2014; White House (Podesta J et al.) 2014, p. 53; Barocas 2014.

[552] College bescherming persoonsgegevens, Annual report 2011, p. 2.

[553] Article 29 Working Party 2013, WP 203, p. 45.

[554] Gandy speaks of the "panoptic sort" (Gandy 1993).

> benefits. (…) That means some customers must earn "better treatment" than others, whatever that means. If you can't accept this undemocratic fact, quit reading and close the book, right now. Database relationship marketing is not for you.[555]

With behavioural targeting, marketers don't need people's names to classify them.[556] For instance, an advertiser that seeks wealthy customers could avoid a person whose cookie profile shows that he or she visits websites about credit card debt problems, or whose IP address shows that he or she is from a poor neighbourhood. And if a cookie shows that a peerson often hunts for bargains at price comparison sites, an advertiser might conclude the person is too careful with money to be a profitable customer. An advertiser could exclude that person from campaigns. Or advertisers could target people with more money. For instance, a firm called Bluekai offers an "auction marketplace for all audience data", where marketers can buy access to pseudonymous profiles of "high spenders."[557] Behavioural targeting makes social sorting easier and more effective: firms can categorise people as targets and waste, and treat them accordingly.

***Manipulation***

Some fear that behavioural targeting could be used to manipulate people. Broadly speaking, this study summarises two risks under the heading manipulation. First, personalised advertising could become so effective that advertisers have an unfair advantage over consumers. Second, there could be a risk of "filter bubbles" or "information cocoons", especially when behavioural targeting is used to personalise not only ads, but also other content and services.[558] In brief, the idea is that personalised advertising and other content could surreptitiously steer people's choices.

---

[555] Newell 1997, p. 136.
[556] Turow 2011, chapter 4.
[557] Marketers can buy access to "high spenders", "suburban spenders" or "big spenders" (Bluekai 2010, p. 6-8). Bluekai says the profiles are "anonymous" (see e.g. Bluekai 2012). In 2014, BlueKai was acquired by Oracle (Oracle 2014).
[558] The phrases are from Pariser 2011 and Sunstein 2006.

Personalised ads could be used to exploit people's weaknesses or to charge people higher prices. Calo worries that in the future, firms could find people's weaknesses by analysing massive amounts of information about their behaviour: "digital market manipulation." With modern personalised marketing techniques, "firms can not only take advantage of a general understanding of cognitive limitations, but can uncover and even trigger consumer frailty at an individual level."[559] For example, a firm could target ads to somebody when he or she is tired, or easy to persuade for another reason. Firms could tailor messages for maximum effect. In short, firms could obtain an unfair advantage over people.[560]

Following the definition quoted in the last chapter, advertising is "designed to persuade the receiver to take some action."[561] Hence, advertising always aims to persuade or influence people. Persuading people could become unfair when targeted ads influence people too much. Zarsky gives an example of somebody who might become a vegetarian. The example is slightly adapted here. Suppose an ad network tracks the behaviour of Alice. The ad network analyses Alice's browsing behaviour, and applies a predictive model. Alice has never thought about becoming a vegetarian, but the model suggests that the person behind ID *xyz* (Alice) is statistically likely to become a vegetarian within 2 years. One firm starts targeting Alice with ads for steak restaurants. Another firm targets Alice with ads about the advantages of a vegetarian diet. Hence, firms could steer Alice's behaviour, while Alice isn't even aware of being influenced.[562] Scholars from various disciplines say that profiling changes the power balance between firms and individuals.[563] Data Protection Authorities agree.[564]

[559] Calo 2013, p. 1. See also chapter 2, section 5.
[560] See on fairness chapter 4, section 4.
[561] Curran & Richards 2002. See chapter 2, section 7.

Behavioural targeting could be used for purposes beyond advertising. The risk of manipulation is greater when firms personalise not only advertising, but also other content and services. However, as noted, the line between advertising and other content is fuzzy on the web.[565] Zarsky speaks of the autonomy trap, "the ability of content providers to influence the opinions and conceptions of individuals by providing them with tailored content based on the provider's agenda and the individual's personal traits."[566] Zarsky argues that the autonomy trap is one of the main threats resulting from data mining. He calls it "a scary concept, portraying a frightening picture of a dysfunctional society."[567] However, in 2004 he didn't think behavioural targeting practices already brought this risk.[568]

In his book "Republic.com", Sunstein discusses risks from too much customised content.[569] He's mainly concerned about people locking themselves into "information cocoons" or "echo chambers", by only reading like-minded opinions.[570] He worries about user-driven personalisation (customisation) and not about media-driven personalisation (which happens without people's deliberate input).[571] But in later work Sunstein expresses similar worries about software personalising content automatically.[572] He discusses two risks. First, citizens in a democratic society need to come across opinions that differ from their own opinions to fully develop themselves. People might drift towards more extreme viewpoints if they don't encounter opposing viewpoints. "Unplanned, unanticipated encounters are central to democracy itself."[573] Second, if everyone locked themselves in their own information cocoons, people might have fewer common experiences. But Sunstein says a diverse democratic society needs shared experiences as "social glue."[574] Along similar lines, the Council of Europe says public service media should promote "social cohesion and integration of all individuals, groups and communities."[575]

---

[562] Zarsky 2002, p. 40.
[563] See e.g. Schwartz & Solove 2009, p. 2; Gürses 2010, p. 51; Acquisti 2010a, p. 11; Purtova 2011, p. 42-43; Richards & King 2013. As noted above (under chilling effects), Lyon says surveillance always implies power relationships (Lyon 2001, p. 16).
[564] International Working Group on Data Protection in Telecommunications (Berlin Group) 2013, p. 7 (capitalisation adapted).
[565] See chapter 2 section 7. See about the distinction between editorial content and advertising Van Hoboken 2012 (chapter 10, section 3).
[566] Zarsky 2004, p. 30 (original footnote omitted). Zarsky borrows the phrase from Schwartz, but Zarsky defines it differently (see Schwartz 2002, p. 821-828).
[567] Zarsky 2002, p. 42. See on data mining chapter 2, section 5.
[568] Zarsky 2004, p. 46.
[569] Sunstein 2002.
[570] He describes "information cocoons" as "communication universes in which we hear only what we choose and only what comfort sus and pleases us" (Sunstein 2006, p. 9).
[571] The phrases user- and media-driven personalisation are used by Helberger 2013, p. 5-6. User-driven personalisation can be called customisation, and media-driven personalisation can be called personalisation (Treiblmaier et al 2004).

Pariser speaks of a filter bubble, "a unique universe of information for each of us."[576] Say a search engine personalises search results. The search engine's software learns that people who click on links to website X, are likely to click on links to website Y. Therefore, the software recommends website Y to people who click on links to website X. As a result, the search engine could mainly provide links to conservative news sites to somebody whose profile suggests that he or she is conservative. And the search engine could offer mostly results from left-leaning websites to a person categorised as progressive. If people think they see a neutral or complete picture, the search engine could narrow their horizon, without them being aware. Adverse effects of too much personalisation can occur accidentally. Hence, a filter bubble can occur when a firm doesn't aim to manipulate a person. Many authors share at least some of the concerns about filter bubbles and information cocoons.[577]

However, others are sceptical about the risks of personalisation.[578] The fear for filter bubbles leads to several questions. First, how much personalisation goes on? Research finds only limited personalisation in Google's search results.[579] Likewise, personalisation on news websites seems to be in its infancy.[580] But search engines do adapt search results to regions.[581] And one paper finds that watching extreme right videos on YouTube is likely to lead to recommendations for other extreme right videos.[582]

---

[572] Sunstein 2013.
[573] Sunstein 2002, p. 9. See also Sunstein 2006.
[574] Sunstein 2002, p. 9.
[575] Council of Europe, Committee of Ministers, Recommendation CM/Rec(2007)3 of the Committee of Ministers to member states on the remit of public service media in the information society, 31 January 2007, article I.1(a).
[576] Pariser 2011, p. 9.
[577] See for instance Hildebrandt 2008a; Bozdag & Timmersmans 2011; Castelluccia & Narayanan 2012, p. 14; Oostveen 2012; Angwin 2014, p. 14-15; Lessig 2006 (chapter 11).
[578] See for instance McGonagle 2011, p. 198; Hoboken 2012, p. 286-287; p. 301; Jenkins 2008.

A second and more difficult question concerns the long-term effects of personalisation. Does personalised content really influence people and does it really harm our democracy? So far, there's little empirical evidence.[583] However, firms can influence people's emotions. For instance, Facebook published results of an experiment, which involved manipulating the user messages ("posts") that 689,003 users saw in their news feeds. "When positive expressions were reduced, people produced fewer positive posts and more negative posts; when negative expressions were reduced, the opposite pattern occurred."[584] Hence, Facebook succeeded in influencing the emotions of users.

Third, assuming that personalisation could deeply influence people, wouldn't the many possibilities to broaden one's horizon outweigh the effects of personalisation? For example, the web offers many kinds of unexpected content. In other words: how likely is it that the possible harm materialises? It appears that people do encounter information outside their own comfort zones.[585] And before the web became popular, people could lock themselves in their own echo chambers, by only choosing newspapers and radio stations that reinforced their existing opinions. In sum, it's unclear how much we should worry about filter bubbles at present. But problems could arise in the future, whith further technological developments.[586] As previously noted, behavioural targeting could be seen as an early example of ambient intelligence: technology that senses and anticipates people's behaviour in order to adapt the environment to their inferred needs.[587]

[579] Hannak et al. 2013.
[580] Thurman & Schifferes 2012; Turow 2011, p. 195. Moreover, as discussed in chapter 2, section 5, a predictive model for behavioural targeting might predict a click-through rate of 0.1 % to 0.5 %. Such models don't seem to enable very accurate personalisation.
[581] Hoboken 2012, p. 188.
[582] O'Callaghan et al. 2013.
[583] Van Hoboken 2012, p. 286; p. 301-302.
[584] Kramer et al. 2014.
[585] See e.g. Gentzkow & Shapiro 2011; LaCour 2014.

In some contexts, undue influence would be more worrying than in others. The societal impact might be limited if behavioural targeting makes somebody buy a different brand of laundry detergent. But behavioural targeting in the context of elections raises more serious concerns. In the US, politicians use behavioural targeting. In principle, behavioural targeting would enable a political party to present each individual a personalised ad. In practice, it would make more sense to work less granularly. A political party could present itself as a one-issue party to each individual: "rhetorical redlining."[588]

By way of illustration, say a politician has a profile of Alice, identified by ID *xyz* in a cookie on her device. A predictive model says that the person behind ID *xyz* (Alice) probably dislikes immigrants. The politician shows Alice personalised ads, in which the politician promises to curtail immigration. The politician has a cookie-profile of Bob that suggests that Bob has more progressive views. The ad targeted to Bob says that the politician will fight discrimination of immigrants in the job market. The ad doesn't mention the politician's plan to limit immigration. Similarly, in ads targeted at jobless people, the politician mentions plans to increase the amount of money people on welfare receive every month. People whose profile suggests that their main concern is paying less tax, receive an ad stating that the politician will limit the maximum welfare period to six months. Hence, without technically lying, the politician could say something different to each individual. This doesn't seem to be a recipe for a healthy democracy.

---

[586] See Oostveen 2012.
[587] Hildebrandt 2010. See chapter 2, section 7.
[588] Turow et al. 2012, p. 7. See generally on behavioural targeting and profiling by politicians Barocas 2012; Bennett 2013; Kreiss 2012.

"Voter surveillance" is widespread in the US, says Bennett. He suggests that this can be partly explained by the absence of a general data protection law, and by the strong right to freedom of speech in the US. In Europe, data protection law limits the legal possibilities to obtain personal data.[589] However, it appears political parties in Europe look to the US practices: "candidates and political parties elsewhere have reportedly looked with great envy on the activities of their US counterparts and longed for similar abilities to find and target potential supporters and to ensure that they vote."[590]

The problems of unfair discrimination and manipulation surface in phase 5. A firm decides – or has software automatically decide – to show personalised ads or other content to a specific individual.[591] Other people are excluded because their profile suggests they won't become profitable customers. As long as the data aren't applied to an individual (phase 5), the sorting doesn't happen. But data analysis (phase 3) is a crucial step. For instance, a firm might discover that people who buy certain accessories for their cars are likely to default on payments. That model could be applied in phase 5, to deny someone credit.[592] Targeted advertising wouldn't be possible without collecting data. However, a firm could use data about one group of people to construct a predictive model, to apply that model to a person who isn't part of the group. Hence, while social sorting often involves processing vast amounts of information, a firm doesn't always need much information on the person to whom it applies the model.[593]

The perspective of privacy as the freedom from unreasonable constraints on identity construction fits well when discussing the risk of unfair social sorting and manipulation. If personalised ads unreasonably influenced a person's choices, that person could be constrained in building his or her personality, or constructing his or her identity. And if an ad network compiles a profile of Alice, the ad network constructs an identity of Alice (her individual profile). Hence, it's not Alice who constructs that aspect of her identity. This could be seen as a constraint on Alice's freedom to construct her identity – and possibly an unreasonable constraint.[594]

---

[589] The Data Protection Directive provides for separate rules for political parties, which are less strict than for other data controllers (article 8(2)(d); recital 30 and 36). Nevertheless, the data protection regime probably reduces the amount of personal information that is available for political parties to obtain. See generally on data protection law chapter 4, section 2 and 3, and on personal data regarding political opinions chapter 5, section 7, and chapter 9, section 6.

[590] Bennett 2013.

[591] The Data Protection Directive has a separate provision for certain types of automated decisions (article 15) see chapter 9, section 6.

[592] See chapter 2, section 6.

[593] See chapter 2, section 3 and 5 on predictive modelling. See also chapter 5, section 3, and chapter 7, section 4 (on externalities).

The privacy as control perspective is also relevant for discrimination and manipulation. With fully transparent data processing and perfect individual control over behavioural targeting data, the risk of manipulation would be reduced. And some might find targeted ads more difficult to ignore than contextual advertising, and therefore more intrusive. If that were true, targeted ads could interfere with privacy as limited access.[595]

***Social sorting as a privacy issue***

People often use the word privacy to express unease about unfair treatment involving the use of personal data.[596] "Like it or not," says Bennett, "privacy frames the way that most ordinary people see the contemporary surveillance issues."[597] Some argue that social sorting and manipulation (in phase 5 of behavioural targeting) shouldn't be conceptualised as privacy problems. Koops says it's more a question of fairness: "why not call a spade a spade and say that in this respect, it is not so much privacy that is at stake, but fair judgement and equal treatment?"[598] Likewise, surveillance scholars often suggest that privacy isn't the right frame to discuss social sorting.[599]

---

[594] Diaz & Gürses categorise the problem of discrimination under privacy as identity construction (Diaz & Gürses 2012). See also Roosendaal 2013, p. 195; International Working Group on Data Protection in Telecommunications (Berlin Group) 2013, p. 5.
[595] See Füster et al. 2010. The Council of Europe also says intrusive online direct marketing advertising interferes with privacy (Committee of Ministers, Recommendation CM/Rec(2007)16 of the Committee of Ministers to member states on measures to promote the public service value of the Internet, V, 7 November 2007).
[596] See Bennett 2011a.
[597] Bennett 2011a, p. 495. See also Richards 2014a, p. 12, p. 28.
[598] Koops 2008, p. 329-330.
[599] See Lyon 2002a. See also Van Der Sloot 2011.

For this study it's not necessary to take sides in the debate on whether social sorting and manipulation should be discussed under the topic of privacy.[600] As noted, this study includes social sorting and the risk of manipulation in the category of privacy problems. But this study doesn't argue that such problems should *always* be categorised as privacy problems. In any case, the question "is this a privacy issue?" is a different question than "is this a serious threat?" Somebody might take the risk of unfair social sorting seriously, but not see it as a privacy problem. Apart from all that, the next chapter shows that while data protection aims to protect privacy interests, it also aims for fairness more generally when personal data are processed.[601]

### 3.4 Conclusion

This chapter discussed the privacy implications of behavioural targeting. Many people dislike behavioural targeting, because they find it creepy or privacy-invasive (see chapter 7).[602] This study classifies privacy perspectives into three groups: privacy as limited access, privacy as control over personal information, and privacy as the freedom from unreasonable constraints on identity construction. The three perspectives partly overlap, and highlight different aspects of privacy.

Privacy as limited access concerns a personal sphere, where people can be free from interference. The limited access perspective is similar to approaches of privacy as confidentiality, seclusion, or a right to be let alone. This perspective implies that too much access to a person interferes with privacy. For instance, if somebody wants to keep a website visit confidential, there's a privacy interference if others learn about the visit.

A second privacy perspective focuses on the control people should have over information concerning them. The privacy as control perspective is common since the 1960s, when state bodies and other large organisations started to amass increasing amounts of information about people, often using computers. The control perspective has deeply influenced data protection law (see the next chapter).

---

[600] See on this debate Bennett 2011a, and the reactions to that article in the Surveillance & Society journal.
[601] See chapter 4, section 3 and 4. See also chapter 9, section 6.
[602] See chapter 7, section 1.

Third, privacy can be seen as the freedom from unreasonable constraints on identity construction. This privacy as identity construction perspective highlights a concern regarding modern data processing practices in the digital environment such as profiling and behavioural targeting. There could be an interference with privacy if the environment manipulates somebody. The environment can include technology.

Each of three privacy perspectives can be recognised in the case law of the European Court of Human Rights. The Court interprets the right to privacy from the European Convention on Human Rights generously, and refuses to define the scope of the right. This living instrument doctrine allows the Court to apply the right to privacy in unforeseen situations and to new developments. The Court has held that monitoring somebody's internet usage interferes with privacy.

In the area of behavioural targeting, three of the main privacy problems are chilling effects, a lack of individual control over personal information, and the risk of unfair discrimination and manipulation. First, the massive data collection on user behaviour can cause chilling effects. Data collection can cause a chilling effect, regardless of how the data are used in later phases. People may adapt their behaviour if they suspect their activities are being monitored. For example, somebody who fears surveillance might hesitate to look for medical information, or to read about certain political topics.

Second, people lack control over information concerning them, and in the area of behavioural targeting individual control over personal information is almost completely lacking. As discussed in more detail in chapter 7, people don't know what information about them is collected, how it's used, and with whom it's shared. The feeling of lost control is a privacy problem. Apart from that, large-scale personal data storage brings quantifiable risks. For instance, a data breach could occur, or data could be used for unexpected purposes, such as identity fraud.

Third, there's a risk of unfair discrimination and manipulation. Behavioural targeting enables social sorting: firms can classify people as "targets" and "waste", and treat them accordingly.[603] And some fear that behavioural targeting could be manipulative. Personalised advertising could become so effective that advertisers will have an unfair advantage over consumers. And there could be a risk of "filter bubbles" or "information cocoons", especially when behavioural targeting is used to personalise not only ads, but also other content and services.[604] The idea is that behavioural targeting could surreptitiously steer people's choices.

In sum, behavioural targeting raises privacy concerns from each of the three privacy perspectives. The next chapter turns to data protection law, the main legal instrument in the EU to protect privacy and related interests in the context of digital data processing.

* * *

[603] Turow 2011.
[604] The phrases are from Pariser 2011 and Sunstein 2006.

# 4 Data protection law, principles

The main legal instrument in the EU to protect privacy in the area of behavioural targeting is the e-Privacy Directive's consent requirement for tracking technologies, together with the Data Protection Directive. In January 2012 the European Commission presented a proposal for a Data Protection Regulation, which should replace the Data Protection Directive.[605] The proposal is based on the same principles as the Directive.[606] Data protection law is a legal tool, which aims to ensure that data processing happens fairly and transparently. Data protection law aims to ensure fairness, by imposing requirements on data controllers when they process personal data. Data protection law aims to protect privacy interests, but also other interests, such as the prevention of unfair discrimination.[607] This chapter provides an introduction to data protection law for the purposes of this study.

When discussing data protection law, this study draws in particular on opinions published by the Article 29 Working Party, an independent advisory body installed by article 29 of the Data Protection Directive. The Working Party consists of representatives of the Data Protection Authorities of the member states, the European Data Protection Supervisor, and a representative of the European Commission.[608] The Working Party has several duties, including advising the European Commission on

[605] European Commission proposal for a Data Protection Regulation (2012).
[606] See European Commission proposal for a Data Protection Regulation (2012), p. 4.
[607] Brouwer 2008, p. 194-204. Blok 2002, p. 131-132. The Impact Assessment for the proposal for a Data Protection Regulation contains a list of fundamental rights that are affected by data protection law (Impact Assessment for the proposal for a Data Protection Regulation, p. 39-40).
[608] Article 29(2) of the Data Protection Directive. The European Data Protection Supervisor (EDPS) is the supervisory authority responsible for monitoring the processing of personal data by the EU institutions and bodies (see article 41 of Regulation (EC) 45/2001 on personal data processing by the Community institutions and bodies).

EU measures affecting the rights and freedoms with regard to personal data processing.

The Working Party can also publish opinions on all matters relating to the processing of personal data.[609] Since 1997, the Working Party has published more than 200 opinions, covering topics such as the concept of personal data, consent, cookies, and behavioural targeting. The Working Party's opinions aren't legally binding, but they are influential. Although the Working Party can adopt opinions after a vote,[610] it usually makes decisions by consensus.[611] In several cases, Advocates General of the European Court of Justice have referred to the Working Party's opinions when interpretatiing data protection law.[612] Lawyers keep an eye on the Working Party's opinions when interpretatiing data protection law.[613] The European Commission proposal for a Data Protection Regulation foresees the replacement of the Working Party by a European Data Protection Board.[614] But the Working Party's opinions will remain relevant, as the Regulation uses the same terminology as the Directive.

The Working Party has been criticised, for instance, for being too extreme in its opinions.[615] And, perhaps because consensus requires compromises, the opinions aren't always crystal clear. Nevertheless, the opinions give an idea of the views of European national Data Protection Authorities.[616]

Data protection law doesn't grant property rights to data subjects. But for ease of reading this study sometimes speaks of "their data" instead of "data concerning them." The study often refers to parties that process personal data as firms or companies.[617] Data subjects are also referred to as people, persons or individuals. Personal data are also referred to as data.

---

[609] Article 29 and 30 of the Data Protection Directive.
[610] Article 29(3) of the Data Protection Directive.
[611] Gutwirth & Poullet 2008.
[612] See for instance Opinion AG (14 April 2011) for CJEU, C-70/10, Scarlet v. Sabam, 24 November 2011, par. 74-78; Opinion AG (25 June 2013) for CJEU, C-131/12, Google v. Spain (in this case the AG disagrees with the Working Party on some points).
[613] See Bamberger & Mulligan 2013.
[614] Article 64 of the European Commission proposal for a Data Protection Regulation (2012).
[615] See Interactive Advertising Bureau Europe 2010; Zwenne 2013, p. 36.
[616] Sometimes national Data Protection Authorities appear to follow a different course than the Working Party. See for instance chapter 6, section 4, on the English interpretation of consent.

The chapter is structured as follows. Section 4.1 sketches data protection law's history. Section 4.2 provides an overview of the data protection principles. Section 4.3 and 4.4 discuss data protection law's aim for transparency and for fairness. Section 4.5 considers the tension in data protection law between protecting and empowering the data subject. Section 4.6 concludes.

## 4.1 History

Data protection law's focus on making data processing transparent can be partly explained by its historical background. As noted in the previous chapter, the perspective on privacy as control over personal information became popular in the late 1960s.[618] Many people were concerned about the effects of large-scale data processing by the state and other large organisations. The use of computers for data processing added to the worries. Computers could be used to make decisions about people without human input. In this context, data protection law was developed in the early 1970s. While data protection law has evolved considerably, its underlying principles have remained largely in place.[619]

In Europe, three international organisations have been particularly important for data protection law: the Council of Europe, the Organisation for Economic Co-operation and Development, and the European Union.[620] The Council of Europe took some of the earliest steps in the field of data protection law.[621] In 1968, the Parliamentary

---

[617] This is a simplification. The Data Protection Directive distinguishes "data controllers" from "data processors", but an analysis of that distinction falls outside this study's scope. See section 2 of this chapter.
[618] See chapter 3, section 1.
[619] See generally about the early history of data protection law Hondius 1975; Sieghart 1976; De Graaf 1977; Flaherty 1989; Nugter 1990; Mayer-Schönberger 1997; Rule & Greenleaf 2010; Kosta 2013a, p. 12-82; González Fuster 2014. See for a political science angle Bennett 1992; Newman 2008.
[620] Data Protection Convention 1981, Explanatory Report, par. 14-16. See also Bennet 1992, p. 130-136; González Fuster 2014, chapter 4-8.
[621] See on the Council of Europe chapter 3, section 2.

Assembly of the Council of Europe asked the Committee of Ministers to examine whether the European Convention on Human Rights offered adequate privacy in relation to "modern science and technology."[622] The study concluded that the Convention's right to private life didn't offer sufficient protection. Therefore the Committee of Ministers adopted two resolutions in 1972 and 1973, with principles for the protection of privacy in the area of digital data processing, for the private and the public sector. In its Resolutions, which aren't legally binding, the Committee of Ministers calls for member states "to give effect to the principles," and suggests that legislation may be needed.[623]

Around the same time, in several countries similar principles were developed that should apply to data processing. The principles can be found in the world's first Data Protection Act in the German state of Hesse (1970),[624] and the first national Data Protection Act in Sweden (1973).[625] In the UK and the US, comparable principles were proposed around that time, but no comprehensive data protection laws were adopted at the time.[626] (The US did, however, adopt rules for data processing in the public sector in 1974.[627]) Legislation followed in Germany (1977), France, Austria, Norway and Denmark (1978), and later in other European countries.[628]

It's impossible to establish in which country the data protection principles were developed first, says Bennett.[629] A 1974 report of the Organisation for Economic Cooperation and Development found "a striking similarity to the independent yet correlative actions in data protection and privacy" in different countries.[630] The basic principles of data protection law are sometimes called the Fair Information Principles (FIPs), or the Fair Information Processing Principles (FIPPs).[631] Although the application of the data protection principles varies considerably, they express an almost worldwide consensus on minimum standards for fair data processing.[632]

---

622 Data Protection Convention 1981, Explanatory Report, par. 4.

623 Council of Europe, Committee of Ministers, Resolution (73)22 on the protection of the privacy of individuals *vis-à-vis* electronic data banks in the private sector, 26 September 1973; Committee of Ministers, Resolution (74)29 on the protection of the privacy of individuals *vis-à-vis* electronic data banks in the public sector, 20 September 1974.

624 Hessisches Datenschutzgesetz [Hesse Data Protection Act], Gesetz und Verordungsblatt I (1970), 625 [repealed].

625 Datalagen (Data Act), SFS (Svensk Författningssamling; Swedish Code of Statutes) 1973:289 [repealed].

626 UK: Younger Committee 1972; US: United States Department of Health, Education, and Welfare 1973.

627 The US did adopt a data protection act for the public sector (Privacy Act of 1974, Pub. L. No. 93-579, 88 Stat. 1896 (Dec. 31, 1974), codified at 5 U.S.C. 552a).

628 Mayer-Schőnberger,1997, p. 237 (footnote 3).

629 Bennet 1992, p 99. See also Hondius 1975; Mayer-Schönberger 1997, p. 221.

According to Bennett, it isn't surprising that countries developed similar principles. Computers were developing fast, were quickly being adopted, and had a "mystical or closed quality."[633] While the general public, policymakers, and academics felt uneasy about the new phenomenon of personal data processing, the threats for fundamental rights weren't exactly clear.[634] Therefore, legislation was drafted that aimed to make data processing transparent, and to make computers and databases less mysterious. "Basing legislative action on the assumption that 'the lid must be taken off' leads data protection policy to some inevitable conclusions."[635] Data protection law aims to open the black box of data processing. Making data processing transparent should help to signal problems, which could otherwise remain invisible.

Several European data protection laws from the 1970s contained restrictions on exporting personal data. This worried some countries, the US in particular. Some feared that states would use data protection acts as a disguised trade barrier.[636] Therefore, European countries and the US negotiated about more international cooperation in the Organisation for Economic Cooperation and Development (OECD). In 1980, this led to the adoption of the OECD Guidelines for the Protection of Privacy and Transborder Flows of Personal Data. The Guidelines, which aren't legally binding, are built on similar principles as current data protection law. The OECD Guidelines were updated in 2013.[637]

---

[630] Organisation for Economic Co-operation and Development, 'Developments in Data Protection and Privacy by OECD Countries', Unpublished survey from the OECD's Computer Utilization Group (Paris: OECD, Directorate for Scientific Affairs, 1975), p. 2, quoted in Bennett 1992, p. 95.

[631] Especially in US literature FIPs and FIPPs are common phrases. See for an overview of the history of the Fair Information Principles Gellman 2013, which is a document that is regularly updated.

[632] The European data protection regime goes much further than the FIPs as expressed in the OECD Guidelines (see for criticism on the OECD Guidelines Clarke 2000; Clarke 2002).

[633] Bennett 1992, p. 118-119. See also p. 21.

[634] See Van Dijk 1970, p. 34; Hondius 1975, p. 4, p. 7-8, p. 80-81; Flaherty 1989, p. 373; Bennett 1992, p. 12, p. 29.

[635] Bennett 1992, p. 121. See also González Fuster 2014, p. 126.

[636] Platten 1996, p. 15; González Fuster 2014, p. 77. Similar arguments are used in the discussion about the European Commission proposal for a Data Protection Regulation (2012).

In 1981, the Council of Europe adopted the first legally binding international instrument on data protection, the Data Protection Convention.[638] It entered into force in 1985, and contains similar principles to the OECD guidelines. The Data Protection Convention requires signatories to enact data protection provisions in their national law.[639] The Data Protection Convention has been supplemented with a number of recommendations, which aren't legally binding, regarding data processing in specific sectors.[640] For instance, there's a recommendation on personal data processing for direct marketing, on profiling, and on the protection of privacy and personal data on the internet.[641] The European Court of Human Rights sometimes cites such recommendations, although they're not legally binding.[642]

***European Union***

The European Commission had called on the European Community member states to ratify the Council of Europe's Data Protection Convention in 1981, but in 1990 only seven member states had done so.[643] Furthermore, the Data Protection Convention left possibilities for countries to raise barriers for personal data flows at the borders.[644]

---

637 Organisation for Economic Co-operation and Development, Guidelines governing the protection of privacy and transborder flows of personal data (1980, amended in 2013).
638 Data Protection Convention 1981.
639 Article 4(1) of the Data Protection Convention.
640 See on standard-setting by the Council of Europe's Committee of Ministers and Parliamentary Assembly: Nikoltchev &McGonagle 2011, p. 1-3; Nikoltchev &McGonagle 2011, p. 1-3.
641 Committee of Ministers, Recommendation (85)20 (direct marketing); Committee of Ministers, Recommendation (99)5 (privacy on the Internet), Committee of Ministers, Recommendation (2010)13 (profiling). See for an overview of the Council of Europe data protection texts: <www.coe.int/dataprotection>.
642 See for instance ECtHR, S. and Marper v. The United Kingdom, No. 30562/04 and 30566/04. 4 December 2008 (citing Recommendation (87)15); ECtHR, Von Hannover v. Germany (I), No. 59320/00, 24 September 2004; ECtHR, Von Hannover v. Germany (II), Nos. 40660/08 and 60641/08, 7 February 2012 (both citing Resolution 1165 (1998) on the right to privacy).
643 See European Commission 1981; Kuitenbrouwer 2000, p. 44; Platten 1996, p. 17-18; p. 23.
644 Article 12.3 of the Data Protection Convention allows states to derogate from the prohibition of interfering with cross border data flows, in brief because of the special nature of personal data, or to avoid circumvention of data protection law.

Many stakeholders feared that national authorities would stop the export of personal data to other European countries.[645] This led to action by the European Commission to harmonise data protection law in the EU.[646]

In 1990, the European Commission presented a proposal for a Data Protection Directive.[647] Heated discussions ensued, for instance about the proposal's rules on direct marketing.[648] Business organisations, including the European Direct Marketing Association, started lobbying intensely.[649] Marketers feared that direct mail marketing would only be allowed with the data subject's prior consent. The lobbying paid off. In 1992 the Commission presented an amended proposal, which suggests that direct mail marketing is allowed without prior consent – on an opt-out basis.[650] After the 2012 proposal for a Data Protection Regulation, history repeated itself. Marketers started lobbying heavily, in order to be able to use behavioural targeting on an opt-out basis.[651]

In 1995 the *Directive on the protection of individuals with regard to the processing of personal data and on the free movement of such data* was adopted. This Data Protection Directive has two goals. The first goal is safeguarding the free flow of personal data between member states.[652] Second, the Directive aims to "protect the fundamental rights and freedoms of natural persons, and in particular their right to privacy with respect to the processing of personal data."[653] The wording shows that the Directive protects not only privacy rights, but other rights and interests as well. The Data Protection Directive became one of the world's most influential data protection texts.[654]

---

[645] For example, the French Data Protection Authority had stopped FIAT from exporting personal data from France to Italy in 1989. The transfer was allowed after FIAT made contractual arrangements to safeguard the personal data (Schwartz 2009, p. 11). For more examples of troubles with transborder data flow within the EU see Vassilaki 1993; Stadlen 1976, p. 185-186.
[646] According to Nugter, the European Commission focused mostly on the interests of data controllers, and the European Parliament mostly on the interests of data subjects (Nugter, 1990, p. 29).
[647] European Commission 1990.
[648] See generally about the legislative history of the Data Protection Directive Simitis 1994; Platten 1996; Heisenberg 2005, chapter 3.
[649] Regan 1993, p. 266-267; Heisenberg 2005, p. 62.
[650] See chapter 6, section 2: the Directive sometimes allows processing for direct marketing without consent; namely on the basis of the balancing provision (article 7(f)), which, in short, lays down an opt-out system for direct marketing under certain circumstances.
[651] See chapter 5, section 5.
[652] Article 1(1) of the Data Protection Directive. See González Fuster 2014, p. 130.
[653] Article 1(1) of the Data Protection Directive.

Data protection law isn't just a European affair.[655] In 1990 the United Nations adopted the Guidelines for the Regulation of Computerized Personal Data Files.[656] These are essentially recommendations to national lawmakers to implement data protection principles. Many non-European countries have passed legislation inspired by the Directive. In July 2013, there were about a 100 countries in the world with a data protection law.[657] The US doesn't have a general data protection law for the private sector, which makes it a lonely exception among developed nations. Recently the Federal Trade Commission and the White House called for privacy regulation for the private sector based on a version of the fair information principles; whether these calls will lead to regulation remains to be seen.[658]

The right to data protection and the right to privacy are increasingly seen as distinct rights. In a 2012 text on the modernisation of the Council of Europe's Data Protection Convention, "[i]t is proposed to refer, in addition to the right to privacy, to the right to the protection of personal data which has acquired an autonomous meaning over the last thirty years."[659] The European Commission's 2012 proposal for a Data Protection Regulation only mentions privacy four times.[660] Article 1 of the proposal provides that the "Regulation protects the fundamental rights and freedoms of natural persons, and in particular their right to the protection of personal data." The 1995 Directive still mentions privacy in article 1.

---

[654] See Birnhack 2008.
[655] On of the first, possibly the first, versions of the fair information principles was published in the US (United States Department of Health, Education, and Welfare 1973).
[656] UN General Assembly, Guidelines for the Regulation of Computerized Personal Data Files, 14 December 1990.
[657] See Greenleaf 2013a; Greenleaf 2013b.
[658] Federal Trade Commission 2012; White House 2012.
[659] Council of Europe, The Consultative Committee Of the Convention for the Protection of Individuals with Regard to Automatic Processing of Personal Data (ets No. 108) 2012, p. 2. See also González Fuster 2014, p. 92.
[660] The four instances don't include the introduction to the proposal (European Commission proposal for a Data Protection Regulation (2012), p. 1-16). The 1995 Data Protection Directive, which is much shorter, mentions privacy thirteen times. See also González Fuster 2014, p. 242-245.

The EU Charter of Fundamental Rights, adopted in 2000 and legally binding since 2009, contains a separate right to data protection in article 8.[661] This illustrates that data protection has grown into a separate field of law in Europe.[662]

> Protection of personal data
>
> 1. Everyone has the right to the protection of personal data concerning him or her.
>
> 2. Such data must be processed fairly for specified purposes and on the basis of the consent of the person concerned or some other legitimate basis laid down by law. Everyone has the right of access to data which has been collected concerning him or her, and the right to have it rectified.
>
> 3. Compliance with these rules shall be subject to control by an independent authority.

### 4.2 Overview of the data protection principles

The Data Protection Directive lays down an omnibus regime, which applies to the private sector and the public sector (with exceptions to the latter).[663] The strength of a broadly applicable data protection law with open norms is that the law doesn't leave any gaps. Yet this regulatory approach means that the norms can't be too specific. As the European Court of Justice puts it, the Directive's "provisions are necessarily relatively general since it has to be applied to a large number of very different situations."[664]

---

[661] See also article 16 of the Treaty on the Functioning of the EU (consolidated version 2012).

[662] As González Fuster 2014 shows, a number of European countries didn't historically see data protection as a privacy-related right (chapter 2 and 3; p. 268). See generally on the "emergence of personal data protection as a fundamental right of the EU" González Fuster 2014.

[663] Some parts of the public sector are outside the scope of the Directive (see article 3(2) and article 13). Some data processing practices in the private sector are also exempted, for purely personal purposes (article 3(2). There are also exemptions for the processing for journalistic purposes (article 9). See on journalistic purposes ECJ, C-73/07, Satamedia, 16 December 2008; CJEU, C-131/12, Google Spain, 13 May 2014.

When applying data protection law, a firm has to go through a number of steps, which often require interpreting rather open norms, such as "fairly", "necessary", and "not excessive."[665] But the complicated nature of data protection law shouldn't be exaggerated. Data protection law gives a relatively objective checklist to assess the fairness of personal data processing. Data protection law can be applied without engaging in discussions about the scope or meaning of the right to privacy, a notoriously slippery concept. Imagine how difficult it would be for firms if the only guidance was: don't infringe on privacy and other fundamental rights when you process personal data.[666]

The core of data protection law can be summarised in nine principles. This study uses Bygrave's taxonomy of eight principles, but adds the transparency principle.[667] Bygrave includes this in the fair and lawful principle. The nine principles are: the fair and lawful processing principle, the transparency principle, the data subject participation and control principle, the purpose limitation principle, the data minimisation principle, the proportionality principle, the data quality principle, the security principle, and the sensitivity principle. There are no clear borders between the different principles, which overlap in different ways. Some principles consist of clusters of other principles. Some principles can be recognised in various provisions within data protection law.[668]

---

[664] ECJ, C-101/01, Lindqvist, 6 November 2003, par. 83. See also ECJ, Joined Cases C‑468/10 and C‑469/10 (ASNEF), par. 35.
[665] See article 6(1)(a), 6(1)(c), and 6(1)(c) of the Data Protection Directive.
[666] See De Hert & Gutwirth 2006, p. 94. Chapter 9, section 1 returns to the topic of general and specific rules.
[667] Bygrave has presented the taxonomy in Bygrave 2002, chapter 3 and 18, and presented an updated and more concise version in Bygrave 2014, chapter 5. I use a slightly different terminology than Bygrave.
[668] Bygrave 2002, p. 57.

The fair and lawful processing principle is the overarching norm of data protection law. Personal data have to be processed "fairly and lawfully", says the Data protection Directive.[669] The lawfulness requirement is reasonably clear: data processing has to comply with data protection law and other laws. Fairness is more vague. Among other things, it requires transparency.[670] Koops summarises that data protection law aims for "common decency."[671] Section 4 of this chapter returns to the topic of fairness.

This study sees the transparency principle as the most important principle next to the fair and lawful principle (see the next section of this chapter). Data processing must take place in a transparent manner, and secretive data collection isn't allowed (unless an exception applies, for instance for national security).[672] The European Commission proposal for a Data Protection Regulation emphasises the importance of transparency, by adding the transparency requirement to the first data protection principle: "[p]ersonal data must be (…) processed lawfully, fairly and in a transparent manner in relation to the data subject."[673]

The data subject participation and control principle aims to involve the data subject.[674] Involvement of the individual can only be achieved if he or she is aware of the processing. People derive several rights from the data subject participation and control principle. For instance, in some cases firms are only allowed to process personal data after the data subject has given consent. In many other cases, people have the right to object to data processing.[675] Data subjects have the right to obtain information from a firm about whether their data are being processed, and for what purposes.[676] The data subject also has the right to rectify or erase data,[677] and to object to certain types of automated decisions.[678] The influence of the concept of privacy as control over personal information is clear.

---

[669] Article 6(1)(a) of the Data Protection Directive. See Bygrave 2002, p. 58; Bygrave 2014, p. 146.
[670] Recital 38 of the Data Protection Directive. See also Article 29 Working Party 2006, WP 118, p. 9.
[671] Koops 2008, p. 331.
[672] See article 10 and 11 and recital 38 of the Data Protection Directive, and for exceptions article 13.
[673] Article 5(1) (a) of the European Commission proposal for a Data Protection Regulation (2012).
[674] In 2002 Bygrave spoke of the "data subject participation and control" principle; in 2014 he renamed it "data subject influence" principle (Bygrave 2002, p. 63-67; Bygrave 2014, p. 158-163).
[675] Consent: article 7(a), 8(2)(a), 26(1)(a); object: article 14 of the Data Protection Directive. See chapter 6.
[676] Article 12(a) of the Data Protection Directive.
[677] Article 12(b) of the Data Protection Directive.

According to the purpose limitation principle, personal data must be collected for specified, explicit and legitimate purposes, and must not be further processed for incompatible purposes.[679] The first requirement is sometimes called the purpose specification principle. Personal data may be processed on the basis of the consent of the person concerned or another legal basis. These other legal bases are listed exhaustively, and can only be relied upon if the processing is "necessary."[680] The purpose limitation principle and the requirement for a legal basis to process personal data are discussed in more detail below.[681]

The data minimisation principle prohibits excessive processing in relation to the processing purpose. The principle can be recognised in various provisions. For instance, a firm may not process more personal data than necessary, or store data longer than necessary.[682] Collecting personal data because "you never know, it might come in useful one day" would breach the purpose limitation principle, the data minimisation principle, and the transparency principle.[683] Chapter 9 returns to the data minimisation principle.[684]

The proportionality principle was mainly developed in case law. "One of the most striking developments over the last decade in European data privacy law", notes Bygrave, "is the emergence of a requirement of proportionality as a data protection principle in its own right."[685] Proportionality plays two roles in data protection law.[686] First, it's a general principle of data protection law. Second, proportionality is often relevant when applying data protection provisions, for instance when a provision uses the word "necessary" (see chapter 6).[687]

---

[678] Article 15 of the Data Protection Directive. See in detail about this provision chapter 9, section 6.
[679] Article 6(b) of the Data protection Directive. The principle could be summarised in the slogan "select before you collect" (Jacobs 2005, p. 1006). See Bygrave 2002, p. 61; Bygrave 2014, p. 153-157.
[680] Article 7 of the Data Protection Directive. See also article 8(2) of the EU Charter of Fundamental Rights.
[681] See section 3 of this chapter (purpose limitation) and chapter 6 (legal basis).
[682] Article 6(1)(c), 6(1)(e) of the Data Protection Directive. See Bygrave 2002, p. 59-61; Bygrave 2014, p. 151-153. See on "necessity" chapter 6, section 1 and 2.
[683] A similar phrase was used (in Dutch) during the legislative history of the Dutch Data Protection Act (Kamerstukken II 1998/99, 25 892, nr. 6, p. 34).
[684] See chapter 9, section 3.

The application of the proportionality principle can be illustrated by the data retention judgment of the European Court of Justice. The Court states that "the principle of proportionality requires that [measures] be appropriate for attaining the legitimate objectives pursued (…) and do not exceed the limits of what is appropriate and necessary in order to achieve those objectives."[688] The Court invalidates the Data Retention Directive because "the EU legislature has exceeded the limits imposed by compliance with the principle of proportionality in the light of" the right to private life and the right to data protection.[689]

The data quality principle requires an appropriate level of accuracy, completeness, and relevancy of personal data.[690] Firms must take reasonable steps to ensure they erase or rectify inaccurate data. In principle, the data controller must comply if a data subject requests to have incorrect data rectified. The data quality principle aims to reduce the chance that organisations base decisions about people on incorrect data. This corresponds with the fear of powerful organisations with opaque computers in the early 1970s. But the data quality principle remains relevant. Decisions based on incorrect data can have disastrous effects for a data subject.[691]

The security principle requires an appropriate level of security for personal data processing, and confidentiality of the data being processed. Firms that process personal data must protect the data against unauthorised disclosure or access, and other unlawful forms of processing.[692]

---

[685] Bygrave 2014, p. 147. In 2002 Bygrave didn't list the proportionality principle, but listed the "disclosure limitation principle" instead (Bygrave 2002, p. 67).
[686] Kuner 2008, p. 1616-1617.
[687] See chapter 6, section 1 and 2 (on "necessary" in article 7 of the Data Protection Directive), and chapter 9, section 3 on data minimisation and proportionality.
[688] CJEU, C-293/12 and C-594/12, Digital Rights Ireland Ltd, 8 April 2014, par. 46.
[689] CJEU, C-293/12 and C-594/12, Digital Rights Ireland Ltd, 8 April 2014, par. 69.
[690] Bygrave 2002, p. 62; Bygrave 2014, p. 163-164; article 6(1)(d) of the Data Protection Directive. The data quality principle falls outside the scope of the thesis.
[691] See for instance ECtHR, Romet v. Netherlands, No. 7094/06, 14 February 2012.

The sensitivity principle refers to the stricter regime for "special categories" of personal data. Examples are data revealing racial or ethnic origin, religious beliefs, and data concerning health or sex life.[693] Processing such special categories of data is in principle prohibited, unless a legal exception applies such as medical necessity.[694] A member state can choose to allow data subjects to override this prohibition by giving their "explicit consent."[695] Apart from the special categories of data, the nature of data is relevant when applying data protection law. More sensitive data call for stricter application of the rules.

***Additional rules***

Next to the core data protection principles, Bygrave distinguishes a second group of rules, which mainly concern enforcement of the principles.[696] For instance, compliance with data protection law is subject to control by independent Data Protection Authorities. This requirement is laid down in the EU Charter of Fundamental Rights.[697]

The Data Protection Directive distinguishes "data controllers" from "data processors." The data controller is the party that determines the purposes and means of the personal data processing.[698] The controller is responsible for compliance.[699] A data processor is a party that processes personal data on behalf of the controller.[700] The distinction between controllers and processors is difficult to make sometimes, and the distinction's usefulness has been questioned.[701] Nevertheless, the European Commission proposal for a Data Protection Regulation keeps the distinction.[702] The difficulty is apparent with behavioural targeting, because many parties can be involved in delivering an ad. The Working Party says ad networks and website publishers are often joint data controllers, as they jointly determine the purposes and means of the processing. For instance, the website publisher allows the ad network to place cookies through its site. The Working Party says a website publisher can't escape its responsibilities by saying that it doesn't know what ad networks do through its website.[703] For ease of reading, this study often refers to firms, without specifying whether a firm is the controller or the processor.

---

[692] Bygrave 2002, p. 67; Bygrave 2014, p. 164-165. See article 16 and 17 of the Data Protection Directive. See also article 4, and recitals 6, 20, 24 and 25 e-Privacy Directive. A definition of "network and information security" can be found in art. 4(c) of the ENISA Regulation (EC) 460/2004. See on communications security Arnbak 2013a.
[693] Article 8 of the Data Protection Directive. Bygrave 2002, p. 68; see also 131-132.
[694] Article 8(c) of the Data protection Directive.
[695] Article 8(2)(a) of the Data Protection Directive.
[696] Bygrave 2002, chapter 4, p. 70-83.
[697] Article 8(3) of the EU Charter of Fundamental Rights. Data Protection Authorities (DPAs) go under various names in the member states. For instance, in the United Kingdom the DPA is called the Information Commissioner's Office, and in France the Commission Nationale de l'Informatique et des Libertés (CNIL) [National Commission on Informatics and Liberty].
[698] Article 2(d) of the Data Protection Directive. The Directive also defines "third parties" and "recipients" (article 2(f) and 2(g)). This study doesn't discuss such parties.
[699] Article 6(2)(b) and 23(1) of the Data Protection Directive

In principle, the Data Protection Directive prohibits transferring personal data to countries outside the EU, if those third countries don't offer an adequate level of protection to personal data.[704] The data subject can override this prohibition by giving consent for a transfer.[705] For the US, which doesn't have the status of a country with "adequate" protection, a special "Safe Harbor" arrangement is in place. In short, firms from the US from certain sectors are deemed to offer an adequate level of protection if they agree to comply with the data protection principles.[706]

---

[700] Article 1(e) of the Data Protection Directive. The processor has mainly responsibilities regarding confidentiality (article 16).
[701] See on the roles of processors and controllers Article 29 Working Party 2010, WP 169; Van Alsenoy 2012. For criticism on the distinction Traung 2012; Purtova 2011, p. 171-174.
[702] Article 4(5) and 4(6) of the European Commission proposal for a Data Protection Regulation (2012).
[703] Article 29 Working Party 2010, WP 171, p. 11. The distinction between controllers and processors falls outside the scope of this study.
[704] Article 25 and 26 of the Data Protection Directive.
[705] Article 26(1)(a) of the Data Protection Directive.
[706] See the website about the Safe Harbor program <www.export.gov/Safeharbor>. See on the negotiations that lead to the agreement Heisenberg 2005, chapter 4. The Safe Harbor program was always controversial, but the criticism grew after the Snowden revelations about international surveillance by US Intelligence Agencies in 2013 (see LIBE Committee 2014).

The Data Protection Directive also contains rules to establish whether firms from outside the EU have to comply with the EU rules.[707] The two main rules regarding territoriality can be summarised as follows. First, European data protection law applies when processing is carried out in the context of the activities of an establishment of a firm on EU territory.[708] Second, the law applies when the firm is not established in the EU, but uses equipment situated on EU territory for personal data processing.[709] Several of the largest firms that use behavioural targeting are formally established in Europe, such as Facebook and Apple (Ireland), and Microsoft (Luxemburg). Many other non-European firms also use equipment, such as data centres, in Europe. The Working Party says, in short, that European data protection law applies to any firm that uses tracking technologies on a device in Europe, because in such cases the firm makes use of equipment (the user's device) in Europe.[710] The territorial scope of data protection law has been analysed extensively elsewhere and falls outside the scope of this study.[711] For this study the conclusion will suffice that EU data protection law often applies to firms that are usually regarded as non-European firms.

***Data Protection Regulation proposal***

After a two-year consultation period, the European Commission presented its proposal for a Data Protection Regulation in January 2012. Many scholars and civil rights organisations welcomed the proposal.[712] Others were less enthusiastic – one US scholar spoke of "more crap from the EU"[713] While based on the same principles as the Directive, the Regulation would bring significant changes. For instance, a regulation has direct effect. Unlike a directive, a regulation doesn't have to be implemented in the national laws of the member states.[714] Hence, a regulation should lead to a more harmonised regime in Europe. Less divergence between national rules should make it easier to do cross-border business.

---

[707] Article 4 of the Data Protection Directive. The Directive as such doesn't apply to firms outside the EU; rather the national provisions based on the Directive apply.

[708] Article 4(1)(a) of the Data Protection Directive. In the Google Spain case, the European Court of Justice applies this provision. In short, EU data protection law applies when a search engine operator has a subsidiary in a member state, and that subsidiary sells and promotes advertising space offered by the search engine (CJEU, C-131/12, Google Spain, 13 May 2014, dictum, 2). Regarding the territorial scope the Court follows the Advocate General, who based his reasoning, in part, on Article 29 Working Party 2008, WP 148, p. 9-12.

[709] Article 4(1)(c) of the Data Protection Directive.

[710] Article 29 Working Party 2008, WP 148, p. 9-12.

[711] On the extra-territorial reach of data protection law, see Article 29 Working Party 2010, WP 179; Moerel 2011, chapter 1-4; Kuner 2010; Kuner 2010a; Piltz 2013. The e-Privacy Directive potentioally has an even broader territorial scope than the Data Protection Directive (see Kuner 2010, p. 191-192).

[712] See for instance De Hert & Papakonstantinou 2012; EDRi (European Digital Rights) 2012. See for overview articles on the 2012 proposal Hornung 2012; Kuner 2012a; Van Der Sloot 2012a; Zanfir 2014.

With 91 provisions, the proposed Regulation is much longer than the 1995 Directive (34 provisions). There are new requirements for data controllers, such as the obligation to implement measures to ensure and demonstrate compliance.[715] In some circumstances, data controllers must undertake a data protection impact assessment before they start processing.[716] But the proposal also brings advantages for firms, such as the abolishment of the requirement to notify Data Protection Authorities of data processing practices.[717] The European Commission estimates the regulation could lead to savings for businesses of around 2.3 billion Euros per year.[718]

The proposal emphasises the ideal of data subject control. Pursuant to the preamble, "[i]ndividuals should have control of their own personal data."[719] For instance, the proposal requires consent to be "explicit" and sets out more detailed rules regarding transparency.[720] The rights to request erasure and to withdraw consent are emphasised.[721] A right to data portability is introduced, which should make it easier for people to transfer their data from one service provider to another.[722]

[713] Yakowitz 2012.
[714] Article 288 of the Treaty on the Functioning of the EU (consolidated version 2012).
[715] Chapter IV, section 1 of the European Commission proposal for a Data Protection Regulation (2012).
[716] Article 33 of the European Commission proposal for a Data Protection Regulation (2012). See on privacy and data protection impact assessments Kloza 2014, and the PIAF project (Privacy Impact Assessment Framework for data protection and privacy rights, <http://piafproject.eu>).
[717] European Commission proposal for a Data Protection Regulation (2012), p. 10 and article 28.
[718] Impact Assessment for the proposal for a Data Protection Regulation (2012), p. 3.
[719] Recital 6 (and p. 2) of the European Commission proposal for a Data Protection Regulation (2012). See also Impact Assessment for the proposal for a Data Protection Regulation (2012), p. 41.
[720] Article 4(8) and 7 of the European Commission proposal for a Data Protection Regulation (2012). See for more details chapter 6, section 3, and chapter 8, section 3.

Enforcement and the right to redress are strengthened in the European Commission proposal. In certain circumstances, Data Protection Authorities can impose fines of up to one million Euros or, in the case of an enterprise, up to 2% of its annual worldwide turnover.[723] The European Parliament has proposed fines of up to 5%.[724] Another novelty is that organisations that aim to protect data subject rights can sue a data controller that breaches data protection law.[725] The proposed Regulation also applies to the processing of personal data of people residing in the EU by a non-European firm, if the processing relates to "the monitoring of their behaviour."[726] This would apply to behavioural targeting.

The proposal has led to much debate and much lobbying.[727] Members of the European Parliament have proposed 3999 amendments.[728] In March 2014, the European Parliament adopted a compromise text ("LIBE Compromise"), which the Parliament's LIBE Committee prepared on the basis of the 3999 amendments by the members of parliament. The rules for behavioural targeting in the LIBE Compromise are less strict than those in the European Commission proposal.[729] At the time of writing, the proposed Regulation is still being discussed in Brussels. It's unclear whether the proposal will be adopted. The most optimistic view seems to be that the Regulation could be adopted in 2015.[730]

---

[721] Article 17 of the European Commission proposal for a Data Protection Regulation (2012) had the somewhat misleading title "the right to be forgotten." See on a right to be forgotten Ausloos et al. 2012 (mostly positive); Van Hoboken 2013 (more critical); Mayer-Schönberger 2009 (US focused). See also CJEU, C-131/12, Google Spain, 13 May 2014, and on that case Kulk & Zuiderveen Borgesius 2014.
[722] Article 18 and recital 55 of the European Commission proposal for a Data Protection Regulation (2012); article 15(2) of the LIBE Compromise, proposal for a Data Protection Regulation (2013).
[723] Article 79 of the European Commission proposal for a Data Protection Regulation (2012).
[724] Article 70(2a)(c) of the LIBE Compromise, proposal for a Data Protection Regulation (2013).
[725] Article 76(1) of the European Commission proposal for a Data Protection Regulation (2012).
[726] Article 3; recital 20 and 21 of the European Commission proposal for a Data Protection Regulation (2012). See also Impact Assessment for the proposal for a Data Protection Regulation (2012), p. 41-42.
[727] The website LobbyPlag shows which amendments by members of the European Parliament were copied literally from lobbyists (<http://lobbyplag.eu>). One member tabled over 150 amendments to weaken the proposal, many of which were copied from lobbyists. He later said he wasn't aware that his assistant submitted the amendments (See Nielsen 2013; Brems 2013).
[728] See LIBE Committee, Documents relating to procedure 2012/011(COD).
[729] See chapter 5, section 5; chapter 6, (the end of) section 3.

### 4.3 Transparency

A basic tenet of data protection law is that data processing should take place in a transparent manner. Following De Hert & Gutwirth, the legal right to privacy can be characterised as an "opacity tool" and data protection law as a "transparency tool."[731] Opacity tools aim "to guarantee non-interference in individual matters, or the opacity of the individual."[732] Transparency tools aim "to make the powerful transparent and accountable: they allow us 'to watch the watchdogs'"[733]

Article 8 of the European Convention on Human Rights prohibits intrusions into the private sphere. This prohibition is not absolute; privacy must often be balanced against other interests, such as the rights of others or the prevention of crime. The structure of article 8 of the Convention is as follows. In principle there's a prohibition on privacy infringements (paragraph 1): "There shall be no interference by a public authority with the exercise of this right (…)." But exceptions to this prohibition are possible under strictly defined conditions, for instance in the interests of national security, or to protect the rights and freedoms of others (paragraph 2). De Hert & Gutwirth characterise the legal right to private life as a "no, unless" rule.[734] The right aims to allow the individual to remain shielded, or to remain opaque: it's an opacity tool. Their characterisation of the legal right to privacy thus appears to be related to privacy as limited access.[735]

Data protection law takes another approach than the legal right to privacy, according to De Hert & Gutwirth. In principle, data protection law allows data processing, if the data controller complies with a number of requirements. Data protection law is mainly a regime of "yes, but."[736] Data protection law aims to manage rather than to stop data flows. "Data protection regulation does not protect us from data processing, but from unlawful and/or disproportionate data processing."[737] As Bygrave puts it, data protection law "usually posts the warning 'Proceed with care'; it rarely orders 'Stop!'"[738] One of data protection law's main tools to foster fairness is the requirement that data processing happens transparently. Data protection law aims to prevent abuse of information asymmetry.[739] "No openness, no legitimacy," says Gutwirth.[740] Hence: a transparency tool.[741]

---

[730] See European Council 2014, p. 2.
[731] De Hert & Gutwirth 2006; De Hert & Gutwirth 2008. See chapter 1, section 1, and chapter 9, section 2.
[732] De Hert & Gutwirth 2006, p. 66.
[733] De Hert & Gutwirth 2008, p. 277. See also Bennett 2011a, p. 491.
[734] De Hert & Gutwirth 2008, p. 291.
[735] See on privacy as limited access: chapter 3, section 1.

The Data Protection Directive's transparency requirements aren't a new invention. One of the first texts listing principles for fair information processing is a 1973 report from the US. The first of its five principles states: "[t]here must be no personal-data record-keeping systems whose very existence is secret." The second principle adds that "[t]here must be a way for an individual to find out what information about him is in a record and how it is used."[742] The OECD Data Protection Guidelines say that personal data "should be obtained by lawful and fair means and, where appropriate, with the knowledge or consent of the data subject."[743] The Annex to the 1980 Guidelines adds that this provision "is directed against practices which involve, for instance, the use of hidden data registration devices such as tape recorders, or deceiving data subjects to make them supply information. The knowledge or consent

---

[736] De Hert & Gutwirth 2008, p. 291.
[737] Gutwirth & De Hert 2009, p. 3. González Fuster & Gutwirth call the transparency tool interpretation of data protection law a "permissive notion", which they contrast with a "prohibitive notion" (González Fuster & Gutwirth 2013, p. 532). The permissive notion can also be recognised in Blume 2012, p. 28; Blok 2002, p. 326.
[738] Bygrave 2014, p. 122.
[739] The phrase "abuse of information asymmetry" was used in the privacy context by OrwellUpgraded 2013.
[740] Gutwirth 2002, p. 96.
[741] The analysis of De Hert & Gutwirth 2006 is widely cited. But there's also criticism; see Verbruggen 2006; Tzanou 2012; Tzanou 2013.
[742] United States Department of Health, Education, and Welfare 1973, p. 41.
[743] Collection Limitation Principle. This article 7 is phrased the same in 1980 and the 2013 version of the OECD Data Protection Guidelines.

of the data subject is as a rule essential, knowledge being the minimum requirement."[744]

Borrowing from Van Alsenoy et al., five justifications for data protection law's transparency requirements can be distinguished. First, fairness logically requires transparency: "even if one doesn't have a real say in the matter, an individual should, in principle, at least be put 'on notice' when his personal data is being processed."[745] Second, transparency is necessary to enable data subjects to exercise their rights, such as access, correction and deletion rights, and the right to opt out of data processing.[746] Third, the requirement to disclose information, for instance in a privacy policy, can nudge a firm towards reviewing its data processing practices.[747] If a firm wants to explain its data processing practices, it has to know about them.[748] Fourth, transparency fosters accountability. "If drafted properly, a privacy notice enables scrutiny of a company's data collection and use practices."[749] Transparency could also help Data Protection Authorities to obtain an overview of types and risks of processing.[750] Fifth, the transparency requirements aim to reduce the information asymmetry between data subjects and data controllers, "as a first, albeit relatively modest, step towards 'leveling the playing field' between data subjects and controllers in terms of the knowledge acquired through processing."[751] Chapter 7 shows that from an economic perspective, it's not only in the interest of the individual, but also in the public interest to reduce information asymmetry.[752]

---

[744] Annex to the Recommendation of the Council of 23rd September 1980, par. 52.
[745] Van Alsenoy et al. 2013, p. 2. See on transparency also Zarsky 2013, p. 1530-1553. US scholar Calo mentions another reason for regulators to focus on transparency requirements that should enable people to make choices: "regulators use notice to avoid having to actually regulate" (Calo 2013a, p. 795).
[746] Van Alsenoy et al. 2013, p. 2-3.
[747] Privacy policies are also called privacy notices or privacy statements.
[748] Van Alsenoy et al. 2013, p. 3. See also Solove 2013, p. 1900.
[749] Van Alsenoy et al. 2013, p. 3. See also Bennett 2011a.
[750] The obligation to notify the Data Protection Authority of (certain) processing operations can also be seen in this light (article 18 of the Data protection Directive). This obligation is abolished in the European Commission proposal for a Data Protection Regulation (2012), p. 10.
[751] Van Alsenoy et al. 2013, p. 2.
[752] Information asymmetry is a form of market failure. See chapter 7, section 3.

The Directive's article 10 and 11 concern "information to be given to the data subject." A firm must provide at least information regarding its identity and the processing purposes.[753] The firm must provide more information when necessary to guarantee fair processing. The Directive gives examples of information that could be needed to ensure fairness: the categories of data concerned, the recipients or categories of recipients, and information about the right to access and to rectify data.[754]

Article 10 applies when a firm collects data from the data subject; article 11 applies "where the data have not been obtained from the data subject." In the case of data collection for behavioural targeting on websites, the Working Party says that the website publisher is usually a joint data controller and must inform the data subject.[755] When article 10 or 11 applies can be difficult to determine, but the information that a firm must provide is the same anyway. The main difference is the moment at which the information must be given.

If article 11 applies, the firm must give the information "at the time of undertaking the recording of personal data or if a disclosure to a third party is envisaged, no later than the time when the data are first disclosed."[756] If tracking by ad networks were seen as not obtaining data from the data subject, article 11 would apply. Hence, the ad network would have to inform the data subject when it collects the data (at the time of "recording"). Or the ad network would have to inform the data subject when it allows advertisers to target people with ads, which should probably be seen as data disclosure under data protection law.[757] However, chapter 6 shows that consent is almost always required for personal data processing for behavioural targeting. If a firm seeks consent, article 10 applies, as the firm collects the data directly from the data subject.

---

[753] Büllesbach 2010, comment on article 10, p. 68.
[754] Article 10 and 11 of the data Protection Directive. The Council of Europe has given guidance for the transparency requirements for profiling (article 4 of the Profiling Recommendation (2010)13).
[755] Article 29 Working Party 2010, WP 171, p. 11.
[756] Article 11(1) of the data Protection Directive.
[757] See chapter 6, section 2 (and chapter 2, section 6).

There are some exceptions to the transparency requirement, which are discussed in chapter 8.[758]

To slightly rephrase Verhelst, a privacy policy is an instrument that a data controller can use to comply with its obligation to provide information pursuant to article 10 and 11 of the Data Protection Directive.[759] Privacy policies must be distinguished from consent requests. The Directive always requires firms to be transparent about personal data processing. Even if a firm doesn't want to rely on consent as a legal basis for processing personal data, data protection law requires transparency.

***Purpose limitation principle***

The purpose limitation principle also fosters transparency. A firm must specify the collection purposes, and personal data must not be "further processed in a way incompatible with those purposes."[760] If data subjects consent to their data being used for one goal, the purpose limitation principle should ensure that they don't have to worry that the data will be used for unrelated goals. Informed consent would be worthless if firms were free to use personal data for new purposes at will. To establish whether a new purpose is "incompatible", the collection context should be taken into account.[761]

But the purpose limitation principle isn't as strict as it might seem. First, the processing purposes must be "specified", but the law allows a firm to ask consent for many purposes, as long as these purposes are clearly described. One English author summarises: "[a]t the heart of data protection legislation is the concept that it is possible to do almost anything with personal data if the relevant consent to the relevant purpose has been obtained from the relevant individual."[762] This seems exaggerated, as consent only concerns the legal basis for processing.[763] The data subject can't waive data protection law's provisions. Nevertheless, cunningly phrased consent requests can reduce the value of the purpose limitation principle. And many people might click "yes" anyway.[764]

---

[758] See chapter 8, section 2.

[759] His definition is as follows: "a privacy statement is an instrument which the data controller can use to comply with his obligation to provide information pursuant to Articles 33 and 34 Wbp [Dutch Data Protection Act]. The data controller can formalise the content and therefore the implementation of the obligation to provide information by means of this privacy statement" (Verhelst 2012, p. 224).

[760] Article 6(1)(b) of the Data Protection Directive. The European Commission proposal for the Data Protection Regulation (2012) appears to soften the purpose limitation principle (article 6.4).

[761] See Bygrave 2014, p. 157. See generally on privacy as contextual integrity Nissenbaum 2010. The European Court of Human Rights also takes the collection context into account. See e.g. ECtHR, Niemietz v. Germany, No. 13710/88, 16 December 1992, par. 28; ECtHR, Von Hannover v. Germany (I), No. 59320/00, 24 September 2004, par. 68; ECtHR, S. and Marper v. The United Kingdom, No. 30562/04 and 30566/04. 4 December 2008, par. 67.

Second, firms have some leeway because they're allowed to process personal data for a new purpose if it's "not incompatible" with the collection purpose. While the interpretation of the phrase "not incompatible" varies by member state, it's clear that purposes that are fully unexpected for the data subject aren't allowed.[765]

Third, the Directive softens the purpose limitation principle, because "[f]urther processing of data for historical, statistical or scientific purposes shall not be considered as incompatible provided that member states provide appropriate safeguards."[766] Firms could try to claim that predictive modelling for behavioural targeting is a form of statistical analysis, which can be based on this exception.[767] Nevertheless, although the purpose limitation principle is softened somewhat, it could still protect people against unexpected uses of their data.

Other data protection provisions also aim for transparency. For example, in many circumstances firms must obtain the data subject's consent for processing. This obligation should foster transparency, as the data subject is alerted to the data processing when the firm asks for consent. Furthermore, data subjects have the right to hear from a firm what data of theirs it processes, for what purposes, and whether and to whom the data are disclosed.[768]

---

[762] Carey 2002, p. 37.
[763] Moreover, sometimes valid consent can't be obtained, because it wouldn't be voluntary. See chapter 6, section 3 and 4, and chapter 8, section 3 and 5.
[764] See chapter 7.
[765] Article 29 Working Party 2013, WP 203.
[766] Article 6(1)(b) of the Data Protection Directive (capitalisation adapted). See on statistical data also Council of Europe, Committee of Ministers (1997), Statistical Purposes Recommendation Rec(97)18; Ploem 2004, chapter 3.
[767] Firms would still need to comply with data protection law when processing personal data, for instance when collecting personal data (phase 1).

### 4.4 Fairness

The main requirement of data protection law is that data processing happens "fairly and lawfully."[769] Lawfully means that firms have to comply with data protection law and other laws. But what does fairness mean? The Data Protection Directive's preamble offers some insight. According to the preamble, fairness requires transparency: "if the processing of data is to be fair, the data subject must be in a position to learn of the existence of a processing operation and, where data are collected from him, must be given accurate and full information, bearing in mind the circumstances of the collection."[770] Furthermore, data processing should serve mankind, people's well-being, and social and economic progress.

> [D]ata-processing systems are designed to serve man; (…) they must, whatever the nationality or residence of natural persons, respect their fundamental rights and freedoms, notably the right to privacy, and contribute to economic and social progress, trade expansion and the well-being of individuals.[771]

[768] See regarding access, erasure and opt-out rights article 12 and 14 of the Data Protection Directive, and regarding access rights also article 8(2) of the EU Charter of Fundamental Rights.
[769] Article 6(1)(a) of the Data Protection Directive.
[770] Recital 38 of the Data Protection Directive. See also recital 28 of the Data Protection Directive, and European Agency for Fundamental Rights 2014, p. 76-77.
[771] Recital 2 of the data Protection Directive (punctuation adapted). Recital 2 of the European Commission proposal for a Data Protection Regulation (2012) says roughly the same. The text of these recitals resembles article 1 of the French Data Protection Act: "Information technology should be at the service of every citizen. Its development shall take place in the context of international co-operation. It shall not violate human identity, human rights, privacy, or individual or public liberties."

It's hard to disagree with this, but it might be difficult to operationalise in practice. According to Bygrave, fairness implies that "data controllers must take account of the interests and reasonable expectations of data subjects."[772] He adds that fairness, at a minimum, requires attention to proportionality. Data processing shouldn't have a disproportionate impact on the data subject. Bygrave says fairness also implies that a firm shouldn't pressure people too much into disclosing data, and shouldn't abuse monopoly-like situations.[773] As noted, fairness also logically requires transparency.

Data protection law's fairness principle can be seen as a "safety net" under the more specific requirements.[774] Usually complying with the data protection provisions should ensure that data processing happens fairly. Data protection law could be summarised as one big detailed fairness test. But the lawmaker can never foresee every situation. On rare occasions data processing that complies with all the other data protection provisions may still be illegal because it doesn't comply with the fairness requirement.[775]

For the interpretation of fairness in commercial settings such as behavioural targeting, inspiration can be drawn from European consumer law.[776] Unfair commercial practices are prohibited.[777] Under the Unfair Commercial Practices Directive, a practice is unfair when it's contrary to the requirements of professional diligence, and it's likely to distort a consumer's economic behaviour.[778] The Directive includes a list of commercial practices that are always regarded as unfair.[779] For instance, the presentation of rights given to consumers in law as a distinctive feature of the trader's offer is not allowed.[780] Advertorials that aren't clearly identified as such are prohibited.[781] It's also prohibited to describe a product as "free", if the consumer has to pay anything other than the unavoidable cost of responding to the offer and the delivery costs of the item.[782] Such requirements can be applied by analogy to consent requests for personal data processing.

---

[772] Bygrave 2002, p. 58.
[773] Bygrave 2002, p. 58-59; p. 334-337; Bygrave 2014, p. 146-147.
[774] Korff 2005, p. 37.
[775] Korff 2005, p. 37; See also Rouvroy & Poullet 2009, p. 73. See for criticism on the vagueness of the fair and lawful requirement Traung 2012, p. 40.
[776] See Article 29 Working Party 2014, WP 217, p. 44.
[777] Article 5(1) of the Unfair Commercial Practices Directive.
[778] Article 5(2) of the Unfair Commercial Practices Directive. See about the concept of unfairness in that directive Collins 2005.
[779] Article 5(1)(5) of the Unfair Commercial Practices Directive.
[780] Annex 1(10) of the Unfair Commercial Practices Directive.

Standard contract terms are unfair when they cause a significant imbalance between the rights of a consumer and a firm. In the words of the Unfair Contract Terms Directive:

> A contractual term which has not been individually negotiated shall be regarded as unfair if, contrary to the requirement of good faith, it causes a significant imbalance in the parties' rights and obligations arising under the contract, to the detriment of the consumer.[783]

Fairness and good faith have also been discussed in the context of European contract law.[784] The Draft Common Frame of Reference for European Contract Law (DCFR) is a text prepared by academics, which lays down principles, definitions, and model rules for European contract law.[785] The DCFR says: "[t]he expression 'good faith and fair dealing' refers to a standard of conduct characterised by honesty, openness and consideration for the interests of the other party to the transaction or relationship in question."[786] The European Commission proposal for a Common European Sales law

---

781 Annex 1(11) of the Unfair Commercial Practices Directive.

782 Annex 1(20) of the Unfair Commercial Practices Directive. See about this prohibition CJEU, C-428/11, 18 October 2012, Purely Creative. See also European Commission 2014a (about apps and games).

783 Article 3(1) of the Unfair Contract Terms Directive.

784 For international contract law, see article 1.7 of the UNIDROIT principles. "Each party must act in accordance with good faith and fair dealing in international trade." See also article 7(1) of the United Nations Convention on Contracts for the International Sale of Goods, which says regard is to be had to the observance of good faith when interpreting the Convention.

785 The Draft Common Frame of Reference was prepared by the Study Group on a European Civil Code (<www.sgecc.net>) and the European Research Group on Existing EC Private Law (<www.acquis-group.jura.uni-osnabrueck.de>).

786 Article I-1:103 of the Draft Common Frame of Reference; Principles, Definitions and Model Rules of European Private Law. The Principles of European Contract Law (PECL) contain similar provisions. The PECL require requires parties to act in accordance with good faith and fair dealing, which is defined almost identically as in the DCFR.[787]

The DCFR also provides rules to assess the fairness of standard terms (that haven't been individually negotiated) in business to consumer contracts. For instance, a standard term is unfair "if it is supplied by the business and if it significantly disadvantages the consumer, contrary to good faith and fair dealing."[788] The DCFR adds that a standard term that isn't drafted in plain, intelligible language may on that ground alone be considered unfair.[789] The good faith requirement in European contract law provides a tool for judges to invalidate unfair contracts. National legal systems in Europe offer judges comparable possibilities.[790] The fairness requirement in data protection law could serve a similar function.

## 4.5 Protecting and empowering the individual

Law is messy.[791] This also applies to data protection law. The Data Protection Directive is a compromise text that combines elements from earlier national data protection laws in Europe.[792] As is often the case with law, data protection law aims to strike a balance between conflicting interests, and embodies inherent tensions.[793]

For instance, data protection law aims to protect fundamental rights and to foster the internal market at the same time. The titles of the Data Protection Directive and the European Commission proposal for a Regulation reflect this. Both give rules "on the

---

parties to "act in accordance with good faith and fair dealing." A standard term is unfair "if, contrary to the requirements of good faith and fair dealing, it causes a significant imbalance in the parties' rights and obligations" (article 1:201(1) and Article 4:110(1)).

787 European Commission 2011 (proposal Common European Sales Law), article 2(b).

788 Article II. – 9:403 of the Draft Common Frame of Reference; Principles, Definitions and Model Rules of European Private Law.

789 Article II 9:402(1) requires standard contract terms to be "drafted and communicated in plain, intelligible language." (Draft Common Frame of Reference; Principles, Definitions and Model Rules of European Private Law).

790 Most national legal systems in Europe also have a good faith clause or something similar (Hesselink 2011; Korff 2005, p. 37).

791 The phrase is used by, among others, Hesselink 2009, p. 28.

792 Simitis 1994; González Fuster 2014, p. 126.

793 Bygrave 2002, p. 86; Blume 2012.

protection of individuals with regard to the processing of personal data and on the free movement of such data."[794] Business would benefit from a free flow of personal data within the EU.[795]

Data protection law also aims to balance the interests of data controllers and data subjects.[796] The law aims to protect the data subject's rights and to take the data controllers' interests into account. The law accepts that processing can be useful and necessary. For example, the state is frequently permitted to process personal data without the data subject's consent. Firms are often allowed to process personal data without consent as well.[797]

***Rules that aim for data subject control***

Another tension is between protection and empowerment of the data subject. Data protection law aims to strike a balance between protecting and empowering people.[798] On the one hand, data protection law aims to empower the data subject. The data subject participation and control principle plays an important role in European data protection law. "A core principle of data protection law," says Bygrave, "is that persons should be able to participate in, and have a measure of influence over, the processing of data on them by other individuals or organizations."[799] Data protection law is deeply influenced by the privacy as control perspective and the concept of informational self-determination.[800] Data protection law relies partly on procedural

---

[794] The 1995 Directive is an internal market directive, as it's based on the old article 95 of the Treaty establishing the European Community, which corresponds to article 114 of the Treaty on the Functioning of the EU (consolidated version 2012). The 2012 proposal is based on article 16 of the Treaty on the Functioning of the EU (consolidated version 2012).

[795] Other international data protection texts also have the dual goal of aiming for fair data processing and the free movement of personal data over borders. See e.g. the Council of Europe Data Protection Convention and the OECD Data Protection Guidelines. The Council of Europe approaches the free flow of information over boarders in the light of article 10 of the European Convention on Human Rights (see Kranenborg 2007, p. 67).

[796] Bonner & Chiasson 2005 suggest the OECD Data Protection Guidelines mainly aim to help firms and other data controllers.

[797] See the legal bases for data processing (article 7(b) and 7(f)) that are discussed in chapter 6.

[798] Blume 2012 highlights this as well, and speaks of "inherent contradictions" in data protection law.

[799] Bygrave 2002, p. 63.

[800] Mayer-Schönberger 1997, p. 232. See on informational self-determination chapter 3, section 1.

safeguards. The idea is that fair procedures regarding data processing should lead to fair outcomes.[801]

Rules that aim for data subject control can be roughly divided into two groups. First, there are rules that give the data subject a choice to allow processing or not. In some cases, firms are only allowed to process personal data after the data subject has given consent.[802] In other cases, firms are permitted to process personal data without consent, but people have a right to object on compelling legitimate grounds. This is a relative right to object. If the objection is justified, the firm must stop the processing. In the case of direct marketing, the data subject has an absolute right to object: to opt out.[803] People can also consent to the export of their data to a country without adequate legal protection of personal data. This way, a data subject can override the in-principle prohibition of transferring personal to a country outside the EU that doesn't offer "an adequate level of protection."[804] (Chapter 6 discusses the role of informed consent in the legal regime for behavioural targeting in detail.) Data protection law's transparency requirements should help data subjects to excersice their rights.[805]

A second set of rules that aim for data subject control grants the data subject rights. For instance, people have the right to access their data. People also have the right to obtain communication of data that are being processed, and of any available information regarding the source of the data. Furthermore, people have the right to obtain information regarding processing purposes, the categories of data concerned, and the recipients to whom the data are disclosed. People can also rectify, erase or block data, if the processing doesn't comply with the Directive's provisions, for instance when data are incomplete or inaccurate.[806] And if a firm breaches their data protection rights, people have the right to go to court. The Directive makes the data processor liable in case something goes wrong,[807] and gives data subjects rights that they can enforce.[808] Essentially, the Directive assigns rights and liabilities here.[809]

---

[801] Bennett 1992, p. 112; Blok 2002, p. 248-251. See for an analysis of the rights of the data subject in the context of direct marketing Korff 2005, chapter 5.
[802] Article 7(a), 8(a), 26(a) of the Data Protection Directive.
[803] Article 14 of the Data Protection Directive.
[804] Article 26(a) of the Data Protection Directive.
[805] See section 3 of this chapter on the transparency principle, chapter 7, section 3 and 4 for a critique, and chapter 8, section 2 for suggestions to improve transparency.

***Rules that aim for data subject protection***

On the other hand, many aspects of data protection law aim to protect, rather than to empower, the data subject. First, the mere existence of data protection law could be said to protect the data subject. Data protection law limits what firms can legally do with personal data. Furthermore, the data subject can't make deviating arrangements with a firm; a contract stating that data protection law doesn't apply wouldn't be enforceable.

Another example of a rule that aims to protect the individual, is the obligation for firms to secure the personal data they process.[810] This security principle protects the data subject. For instance, badly secured data could lead to data breaches, which could negatively affect the data subject. The data minimisation is another important requirement that aims to protect the individual.[811] One of the goals of minimising the amount of data processed is to mitigate risks. If fewer data are collected and stored, there are fewer data that can fall into the wrong hands. Another example of how data protection law aims to protect people is the existence of independent Data Protection Authorities that oversee compliance with the rules, as required by the EU Charter of Fundamental Rights.[812]

---

[806] Article 12(a) and 12(b) of the Data Protection Directive. See also article 8(2) of the EU Charter of Fundamental Rights.
[807] Article 23(1) of the Data Protection Directive.
[808] Article 22 of the Data Protection Directive. People rarely go to court for data protection cases. See chapter 8, section 1.
[809] See chapter 1, section 4; Baldwin et al. 2012, chapter 7.
[810] Article 17(1) of the Data Protection Directive.
[811] Article 6(1)(c) and 6(1)(e) of the Data Protection Directive.
[812] Chapter 9 returns to the topic of rules that aim to protect the data subject.

This study distinguishes protection and empowerment rules in order to structure the discussion, but it's not suggested that there's a formal legal distinction. There are no hard lines between rules that aim to protect and to empower the data subject. Some rules have a dual function. For instance, data subjects can't contract away their right to access. This limits the data subject's contractual freedom.[813] But at the same time, the prohibition of waiving one's access rights could be said to foster individual control over personal data. Data subjects would have less control over their data if they could waive their access rights. This study uses rules that aim for *data subject control* and rules that aim for *empowerment* roughly interchangeably.[814]

### 4.6 Conclusion

This chapter introduced data protection law for the purposes of this study. The key aim of data protection law is ensuring that personal data processing happens fairly and transparently. Data protection law grants people whose data are being processed rights, and imposes obligations on parties that process personal data. Personal data must be processed for specified purposes and on the basis of the consent of the person concerned or another legitimate basis provided by law. Independent Data Protection Authorities oversee compliance with the rules. Data protection law applies when "personal data" are processed. Whether data protection law applies to behavioural targeting is discussed in the next chapter.

Data protection law aims to strike a balance between protecting and empowering the data subject. On the one hand, data protection law aims to empower the data subject by fostering individual control over personal data. On the other hand, data protection law contains many safeguards that the individual can't waive. These safeguards are mainly aimed at protecting rather than empowering the individual. The tension between protection and empowerment is a recurring theme in this study.

---

[813] See in more detail on limiting contractual freedom chapter 6, section 5 and 6; chapter 9, section 2.
[814] However, chapter 9, section 2, argues that protective rules, which limit people's contractual freedom, can sometimes help to ensure real empowerment.

* * *

# 5 Data protection law, material scope

Whether data protection law applies at all to behavioural targeting is hotly debated. Many firms using behavioural targeting say they only process "anonymous" data and that data protection law, therefore, doesn't apply. For instance, the Interactive Advertising Bureau Europe states on a website on which it provides information about behavioural targeting:

> The information collected and used for this type of advertising is not personal, in that it does not identify you – the user – in the real world. No personal information, such as your name, address or email address, is used. Data about your browsing activity is collected and analysed anonymously.[815]

According to the Article 29 Working Party, however, firms usually process personal data when they use behavioural targeting. The Working Party also views behavioural targeting as personal data processing if a firm can't tie a name to the data it holds about an individual. Moreover, it's often fairly easy for a firm or another party to attach a name to the data. This chapter argues that the Working Party's view is correct. Data protection law should apply to behavioural targeting.

The chapter is structured as follows. Section 5.1 concerns the difference in scope of data protection law and the legal right to private life. Section 5.2 shows that the Working Party views behavioural targeting as personal data processing, due to the fact that a firm can use such data to "single out" a person, also when the firm can't tie a name to the data it has on an individual. Section 5.3 shows that the firm doing behavioural targeting or another party can often tie a name to data about an individual. Section 5.4 is more normative than the rest of the chapter and argues that data protection law should generally apply to behavioural targeting. Section 5.5 concerns discussions about lighter rules for pseudonymous data, triggered by the proposal for a Data Protection Regulation. Section 5.6 shows that behavioural targeting often entails the processing of special categories of data, such as data regarding health or political opinions. Section 5.7 concludes.

[815] Interactive Advertising Bureau Europe – Youronlinechoices.

### 5.1 Difference in scope of data protection law and privacy

The scope of data protection law is both broader and narrower than the right to privacy as protected by article 8 of the European Convention on Human Rights. Data protection law has a broader scope because it applies to all personal data – any information relating to an identifiable person. The scope of data protection law isn't limited to information that is sensitive or private. Hence, data protection law applies regardless of whether there's an interference with privacy.

On the other hand, the scope of data protection law is narrower than the right to privacy in article 8 of the Convention. For instance, when somebody uses binoculars to spy on a neighbour in the bathroom, there's an interference with privacy. But data protection law doesn't apply, as the spy doesn't "process" personal data.[816] Moreover, many judgments regarding the right to private life have nothing to do with personal data processing.[817]

---

[816] See for the definition of processing article 2(b) of the Data Protection Directive, and see on non-automated processing recital 15 of the Data Protection Directive. See also Pool 2014, p. 178; p. 298. Some data processing activities are outside the Directive's scope see chapter 4, section 2.

[817] See e.g. ECtHR, X and Y v. The Netherlands, No. 8978/80, 26 March 1985, on the impossibility of instituting criminal proceedings against the perpetrator of sexual assault on a mentally handicapped girl of sixteen years old. See on the difference between the right to privacy life and the right to data protection also Opinion AG (12 December 2013), C-293/12 and C-594/12, Digital Rights Ireland Ltd, par. 61; González Fuster 2014.

The European Court of Human Rights hasn't extended the protection of article 8 of the Convention to all personal data. In other words, certain data processing activities don't infringe upon privacy according to the Court.[818] If personal data processing concerns data regarding people's private life, or if data processing is extensive, the Court is likely to find that privacy is affected.[819]

Figure 5.1 illustrates the scope of the legal right to private life (article 8 of the European Convention on Human Rights) and data protection law. The scope of the right to privacy and the right to data protection partly overlap. In many cases, data protection law and the right to privacy both apply. For instance, if a firm processes personal data about a person's private life, both legal regimes apply.

Some situations are covered by the right to private life, but not by data protection law (see the left part of the figure). Somebody who spies on his or her neighbour doesn't necessarily process personal data. There may be a privacy infringement, while data protection law doesn't apply. In certain situations data protection law applies, where the right to private life doesn't (see the right section of the figure). For instance, data protection law applies to an electronic phonebook, because it includes people's names and phone numbers, which are personal data. In this instance, the right to private life doesn't necessarily apply; being listed in the phonebook doesn't have to interfere with privacy.[820]

[818] See De Hert & Gutwirth 2009, p. 24; Kranenborg 2007, chapter 4 (in Dutch) and p. 311-312 (in English).
[819] See Gellert & Gutwirth 2013, p. 526.
[820] The phonebook is merely an example. There's a special regime for subscriber directories (article 12 of the e-Privacy Directive).

**Figure 5.1**

**The scope of the legal right to privacy (article 8 of the European Convention on Human Rights) and data protection law.**

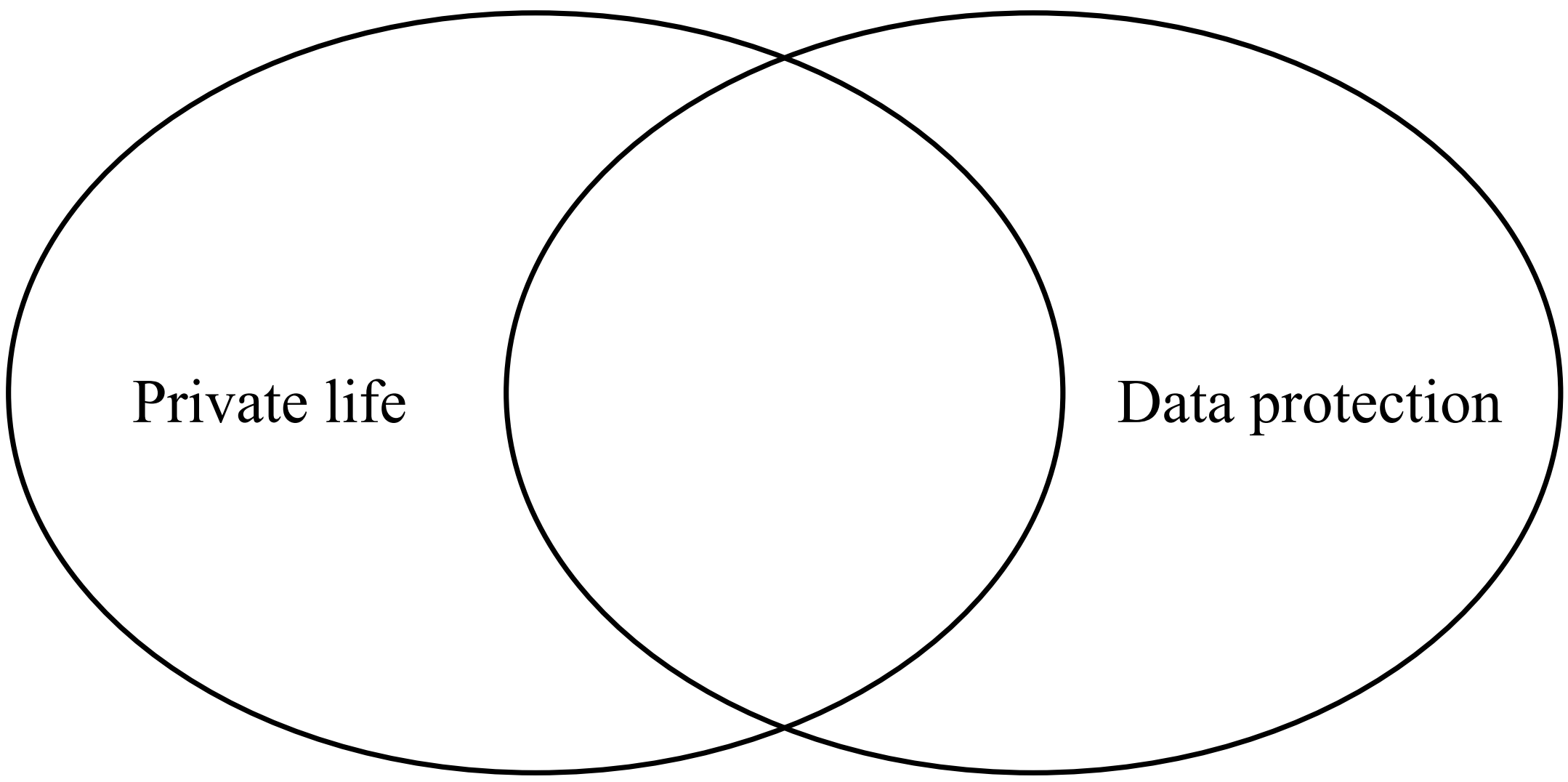

### 5.2 Data that single out a person

Data protection law only applies if "personal data" are processed. Any operation that is performed upon personal data, such as collection, storage, or analysis, falls within the definition of "processing."[821] But do firms process "personal data" when they use behavioural targeting? The personal data definition in the Data Protection Directive reads as follows:

> "Personal data" shall mean any information relating to an identified or identifiable natural person ('data subject'); an identifiable person is one who can be identified, directly or indirectly, in particular by reference to an identification number or to one or more factors specific to his physical, physiological, mental, economic, cultural or social identity.[822]

Personal data are therefore not limited to a name and address, but include all kinds of data that relate to an identifiable person. An identifiable person is someone who can be identified, directly or indirectly. The European Court of Justice has confirmed several times that information without a name can constitute personal data.[823]

In 2007 the Article 29 Working Party published a detailed opinion on the concept of personal data. The opinion is structured around four elements of the Data Protection Directive's definition of personal data: (i) any information, (ii) relating to, (iii) an identified or identifiable, and (iv) natural person. The first element is "any information." Data processed for behavioural targeting, such as digital information about a person's web browsing history, fall within the scope of "any information."[824]

---

[821] Article 2(b) of the Data Protection Directive defines processing of personal data ("processing") as: "any operation or set of operations which is performed upon personal data, whether or not by automatic means, such as collection, recording, organization, storage, adaptation or alteration, retrieval, consultation, use, disclosure by transmission, dissemination or otherwise making available, alignment or combination, blocking, erasure or destruction."

[822] Article 2(a) of the Data Protection Directive, capitalisation adapted.

[823] For the Court, personal data are "any information relating to an identified *or identifiable* individual" (CJEU, C-92/09 and C-93/09, 9 November 2010, Volker und Markus Schecke and Eifert; CJEU, C-468/10 and C 469/10, ASNEF, 24 November 2011, par. 27). See also ECJ, C-101/01, Lindqvist, 6 November 2003, par 27: "identifying [people] by name *or by other means*, for instance by giving their telephone number or information regarding their working conditions and hobbies, constitutes 'the processing of personal data (…)'" (emphasis added).

The second element is "relating to." Sometimes information relates to a person because it refers to an object, such as a computer or a car. Case law of the European Court of Justice confirms that data that relate to an object can identify a person.[825] With behavioural targeting, a firm often recognises a person's device, such as a computer or a smart phone. The Working Party explains that information may relate to a person because of one of three elements: a content element, a purpose element, or a result element.[826]

Information relates to a person because of its content when it's "about" a person. The Working Party gives the example of a patient's medical file. The information in such a file is clearly about a person, regardless of the purpose or the result of using the information.[827] When a firm holds an individual but nameless profile for behavioural targeting, that information relates to a person because of its content. For the person with ID *xyz* on his or her computer, the firm might have a list of visited websites, or a list of inferred interests. The information tied to ID *xyz* is *about* that person.

Information processed for behavioural targeting may also relate to a person because of a "result" element.[828] If a firm shows an ad to a specific person, the firm treats that person differently from others. If a firm targets an ad based on data about an individual, the data relate to that person because of the "result."

---

[824] See Article 29 Working Party 2007, WP 136, p. 6-9.
[825] See section 5.3 below, on IP addresses, in CJEU, C-70/10, Scarlet v Sabam, 24 November 2011. In Lindqvist, the Court mentions a phone number as an example of information that can identify somebody. Arguably a phone relates to an object rather than to a person (ECJ, C-101/01, Lindqvist, 6 November 2003, par 27).
[826] Article 29 Working Party 2007, WP 136, p. 9-10.
[827] Article 29 Working Party 2007, WP 136, p. 10.
[828] Article 29 Working Party 2007, WP 136, p. 11.

Information can also relate to a person because of a "purpose" element.[829] A purpose element is present if a firm uses data "with the purpose to evaluate, treat in a certain way or influence the status or behaviour of an individual."[830] If an identifier that is used for behavioural targeting is primarily linked to a device, the data attached to that identifier often "relate" to a person. If a firm processes data about an individual for behavioural targeting, the processing purpose is influencing that individual, to make that person click on an ad, or buy products. Ads are targeted to a particular device because the firm hopes that the user of that device buys something. The International Working Group on Data Protection in Telecommunications notes that advertising aims to influence people rather than devices.

> While ads may well be addressed to a machine at the technical level, it is not the machine which in the end buys the proverbial beautiful pair of red shoes – it is an individual. Thus, the claim that the processing of behavioural data for marketing is directed "only" at machines in the first place may well be seen as an attempt to blur our vision as societies on the gravity of the problem, when in reality the individual and not the machine is the only instance that can make all such tracking operations a "success" for its proponents (i.e., when the red shoes are finally being bought).[831]

Some data processing activities for behavioural targeting don't concern personal data. As previously noted, in phase 3 of behavioural targeting, a firm can use data it has to construct a predictive model: *1% of people who visit websites about sports, click on*

---

[829] See also International Working Group on Data Protection in Telecommunications (Berlin Group) 2013, p. 6.
[830] Article 29 Working Party 2007, WP 136, p. 10 (emphasis original).
[831] International Working Group on Data Protection in Telecommunications (Berlin Group) 2013, p. 3. This "Berlin Group" was founded in 1983 and consists of representatives from Data Protection Authorities and other bodies of national public administrations, international organisations and scientists from around the world.

*ads for running shoes, while 0.5% of random people clicks on such ads*.[832] Such a predictive model doesn't consist of personal data, as it doesn't relate to a specific person.

But as soon as a predictive model is applied to an individual (phase 5), the information relates to that person because of the "purpose."[833] For instance, if a person with the cookie with ID *xyz* on his or her computer visits a website, an ad network may recognise that person (ID *xyz*) as a person who visits a lot of websites about sports. The firm has a predictive model saying that people who visit websites about sports are more likely to click on ads for running shoes. Therefore, the firm shows the person advertising for shoes. At that moment, the firm applies the predictive model to a specific person, with the purpose of influencing that person.[834] Hence, the firm processes personal data. The Working Party concludes: "the information collected in the context of behavioural advertising *relates to,* (*i.e.* is about) a person's characteristics or behaviour and it is used to influence that particular person."[835] In sum, behavioural targeting often entails the processing of "information relating to a natural person."[836]

This brings us to the third element of the personal data definition. Does behavioural targeting entail the processing of data that relate to an "identifiable" person? In other words, does a firm process data that "directly or indirectly identify" a person, if it processes data about a person, and it would be hard for anybody to tie a name to the data?

Many behavioural targeting firms say they only process "anonymous" data when they don't add a name to a person's data. Therefore, the argument goes, they don't process personal data when using behavioural targeting.[837] The European concept of personal data has a broader scope than the US concept of "personally identifiable information." Although definitions in US statutes differ, the concept typically refers to information such as a name or a social security number.[838] Perhaps some US firms think that only information such as a name or a social security number makes a person identifiable.

[832] See chapter 2, section 5.
[833] Koops 2008, p. 331.
[834] See Hildebrandt et al. 2008.
[835] Article 29 Working Party 2010, WP 171, p. 9 (emphasis original).
[836] Article 2(a) of the Data Protection Directive.

Computer scientists would refer to nameless individual profiles that are used for behavioural targeting as pseudonymous data: "a pseudonym is an identifier of a subject other than one of the subject's real names."[839] A handbook on data protection law summarises: "[i]n contrast to anonymised data, pseudonymised data are personal data."[840] The Working Party concurs.[841]

The Directive's personal data definition mentions an "identification number" as an example of information that can identify a person. Cookies with unique identifiers are strings of numbers and letters. There's no reason to exclude such cookies and similar technologies from the category identification numbers.[842] A cookie or another unique identifier allows a firm to follow a person's online behaviour, and to make inferences about that person's interests. As the Interactive Advertising Bureau UK explains, "[c]ookies are used in behavioural advertising to identify users who share a particular interest so that they can be served more relevant adverts."[843]

The Working Party says that a person can be identified without knowing his or her name. In its 2007 Opinion on personal data, the Working Party says that "singling out" an individual implies identifying that individual.[844] A person is identifiable if she can be distinguished within a group.[845] A firm that aims to individualise a person wouldn't have a strong case if it argued that its aim was not to identify that person. "In fact, to argue that individuals are not identifiable, where the purpose of the processing is precisely to identify them, would be a sheer contradiction in terms."[846] In later opinions the Working Party says explicitly that cookies and similar files with a unique identifier are personal data, because they "enable data subjects to be 'singled out', even if their real names are not known."[847]

---

[837] See e.g. Interactive Advertising Bureau Europe - Youronlinechoices (about).
[838] See Schwartz & Solove 2011, with references to statutes.
[839] Pfitzmann & Hansen 2010, par. 9.
[840] European Agency for Fundamental Rights 2014, p. 36. "Data are anonymised if all identifying elements have been eliminated from a set of personal data. No element may be left in the information which could, by exercising reasonable effort, serve to re-identify the person(s) concerned. Where data have been successfully anonymised, they are no longer personal data" (internal footnote omitted), p. 45.
[841] Article 29 Working Party 2014, WP 216, p. 20.
[842] See Cuijpers et al. 2007, p. 25. See also Traung 2012, p. 37.
[843] Interactive Advertising Bureau United Kingdom 2009, p. 4. See on cookies chapter 2, section 2.
[844] Article 29 Working Party 2007, WP 136, p. 14.

> [B]ehavioural advertising involves the processing of unique identifiers be that achieved through the use of cookies, or any kind of device fingerprinting. The use of such unique identifiers allows for the tracking of users of a specific computer even when IP addresses are deleted or anonymised. In other words, such unique identifiers enable data subjects to be "singled out" for the purpose of tracking user behaviour while browsing on different websites and thus qualify as personal data.[848]

The impact assessment for the proposal for a Data Protection Regulation agrees with the Working Party about online identifiers: "[e]ven without a name or other traditional identifying attribute, it is often possible to effectively identify the individual to whom the data relates."[849] The data can be used to individuate, isolate, or individualise a person.[850] Many authors agree.[851]

---

[845] Article 29 Working Party 2007, WP 136, p. 12.
[846] Article 29 Working Party 2007, WP 136, p. 16. The Working Party doesn't make this remark in the context of behavioural targeting.
[847] Article 29 Working Party 2008, WP 148, p. 9. See also Article 29 Working Party 2010, WP 171, p. 9; Article 29 Working Party 2011, WP 188, p. 8.
[848] Article 29 Working Party 2011, WP 188, p. 8. See along similar lines CNIL 2014 (Google) (Google), p. 11-12; College bescherming persoonsgegevens (Dutch DPA) 2013 (Google), p. 50-57. The Working Party has described singling out as follows: "the possibility to isolate some or all records which identify an individual in the dataset" (Article 29 Working Party 2014, WP 216, p. 11).

The fourth element of the personal data definition says that the information must relate to a "natural person."[852] This is usually the case with behavioural targeting. However, it's possible to think of situations where behavioural targeting data tied to a unique identifier aren't personal data, because they don't refer to a person. For instance, a computer in an internet café might be used by many people.[853] An ad network that builds a profile based on a cookie placed on that computer, might compile a profile based on the surfing behaviour of a group of people. Arguably such a profile doesn't consist of personal data. Nevertheless, if a firm uses a unique identifier for behavioural targeting, it usually identifies a person.

The Working Party isn't alone in its interpretation that behavioural targeting generally entails personal data processing. The International Working Group on Data Protection in Telecommunications agrees that behavioural targeting usually entails the processing of personal data.[854] Dutch law even contains a legal presumption regarding the use of tracking cookies and similar technologies for behavioural targeting. The use of such cookies is presumed to entail the processing of personal data.[855] In a letter to Google, signed by 27 national Data Protection Authorities, the Working Party says that Google processes personal data about its "passive users."[856] These are users that are tracked by Google's ad network.[857]

---

[849] Impact Assessment for the proposal for a Data Protection Regulation (2012), Annex 2, p. 24.
[850] The phrase "individuate" is used by Hildebrandt for instance (Hildebrandt 2008, p. 19). Zwenne speaks of "isolating" (Zwenne 2013, p. 32). It must be noted that Zwenne disagrees with the Working Party about "singling out."
[851] See e.g. De Hert & Gutwirth 2008, p. 289; Leenes 2008; Traung 2010; and more hesitant: Koëter 2009. See also the references in Zwenne 2013, p. 35-36. Zwenne disagrees on this point with the Working Party: see Zwenne 2013, with references.
[852] See on the question of whether privacy rights do – or should – continue after death McCallig 2013; Harbinja 2013; Edwards 2013; Korteweg & Zuiderveen Borgesius 2009. See on the question of legal persons should be protected by data protection law Bygrave 2002, chapter 9-16.
[853] The example is taken from Article 29 Working Party 2007, WP 136, p. 17.
[854] International Working Group on Data Protection in Telecommunications (Berlin Group) 2013.
[855] Article 11.7a of the Dutch Telecommunications Act (See for a translation Zuiderveen Borgesius 2012).

Outside Europe, some regulators arrive at similar conclusions. For instance, the Privacy Commissioner of Canada says that behavioural targeting usually entails personal data processing.[858] In the US, the Federal Trade Commission (FTC) released a report in 2012 with recommendations regarding online data processing.[859] The recommendations apply to firms "that collect or use consumer data *that can be reasonably linked to a specific consumer, computer, or other device* (…)."[860] Therefore, the recommendations also apply to firms that gather data about individuals but don't tie a name to the data. However, not all regulators see behavioural targeting as the processing of personal data. For instance, the Office of the Australian Privacy Commissioner states that "[t]he information collected by online advertisers may often not be sufficient to identify you; it might just be general information about your interests and sites you have visited."[861]

In sum, according to the Working Party, a firm uses data to identify a person if the firm uses the data to single out somebody. Apart from that, we will see in the next section that firms using behavioural targeting can often attach names to the individual profiles they hold.

---

[856] Article 29 Working Party 2013 (Google letter).
[857] Article 29 Working Party 2013 (Google letter appendix), p. 2, footnote 2. Passive users are "users who does not directly request a Google service but from whom data is still collected, typically through third party ad platforms, analytics or +1 buttons." Reports by national Data Protection Authorities come to the same conclusion as the Working Party. See CNIL 2014 (Google); College bescherming persoonsgegevens (Dutch DPA) 2013 (Google)
[858] The International Working Group on Data Protection in Telecommunications also has members from outside Europe.
[859] Office of the Privacy Commissioner of Canada 2012, p. 2. See also Office of the Privacy Commissioner of Canada (Google) 2014.
[860] Federal Trade Commission 2012. p. 22 (emphasis added). In a 2014 report, the FTC includes "a persistent identifier, such as a customer number held in a "cookie" or processor serial number" in its personal data definition (Federal Trade Commission 2014, Appendix A, p. A16).
[861] Office of the Australian Privacy Commissioner 2011. The Australian Act speaks of "personal information" – not of personal data.

### 5.3 Data that identify people by name

It's often relatively easy for a firm that has an individual profile of a person, or for another party, to attach a name to that profile. To structure the analysis, this section distinguishes four situations where a firm processes data about a person.

> (i) A firm processes data about an individual, and it knows the name of the individual.
>
> (ii) A firm processes data about an individual, and it's fairly easy for the firm to tie a name to the data.
>
> (iii) A firm processes data about an individual, and it's difficult for the firm to add a name to the data, but it would be fairly easy for another party to tie a name to the data.
>
> (iv) A firm processes data about an individual, and it would be difficult for anybody to tie a name to the data.

***Situation (i)***

A firm processes data about an individual, and it has the individual's name. This firm clearly processes data about an identified person. For example, a provider of a social network site that engages in behavioural targeting often has profiles with names. Facebook requires people to register under their own name.[862]

---

[862] Facebook's Name Policy.

***Situation (ii)***

A firm processes data about an individual, and it's fairly easy for the firm to tie a name to the data. The preamble of the Data Protection Directive says: "to determine whether a person is identifiable, account should be taken of *all the means likely reasonably to be used* either by the controller or by any other person to identify the said person."[863]

The question is thus: what means can a firm that processes data about a person "reasonably likely use" to identify a person?[864] The answer to this question depends, among other things, on the state of science and technology, and on how costly it would be to identify somebody. According to the Working Party, "a mere hypothetical possibility to single out the individual is not enough to consider the person as 'identifiable'."[865]

It's often possible to identify people within an (supposedly) anonymised data set. In 2000 Sweeney found that 87% of the US population is uniquely identified by three attributes: their date of birth, gender, and ZIP code.[866] Techniques to re-identify data subjects continue to improve. Additionally, re-identification may become easier if more data sets that could be coupled with the source set become available, for instance from social network sites. Furthermore, computers keep getting faster, reducing the time needed for complicated calculations. Computer scientists summarise that de-identification of data is an "unattainable goal."[867]

Sometimes the person behind a nameless profile can be found without sophisticated data analysis. In 2006 search engine provider AOL released a data set of individual nameless search profiles, tied to a random number. Within a few days, New York Times journalists had found one of the searchers: "A face is exposed for AOL searcher no. 4417749."[868] The search queries suggested that the searcher was an elderly woman with a dog, living in a specific town. An interview confirmed that the journalists had correctly identified her.

---

[863] Recital 26 of the Data Protection Directive (emphasis added).
[864] Following the definition of "data subject" (article 4(1)) of the European Commission proposal for a Data Protection Regulation (2012), this study switches the words "likely reasonably" to "reasonably likely."
[865] Article 29 Working Party 2007, WP 136, p. 15. See also Article 29 Working Party 2014, WP 216, p. 8-9.
[866] Sweeney 2000; Sweeney 2001.
[867] Narayanan & Shmatikov 2010, p. 3. See generally on re-identification research Sweeney 2000; Sweeney 2001; Narayanan & Shmatikov 2008; Ohm 2010; Koot 2012 (chapter 2); Wu 2013; Article 29 Working Party 2014, WP 216.

A behavioural targeting firm can often tie a name to the data about an individual it processes, taking into account "the means reasonably likely to be used" by the firm.[869] For instance, some firms offer services directly to consumers. If a firm has a cookie-based profile of a user, and the same firm offers an email service to that person, it can tie the user's email address to the cookie. Most email addresses are personal data.[870] In addition, email addresses and email messages often contain the user's name. The situation is similar if a firm offers a social network site or another service where people log in.

A search engine provider that has a nameless profile, tied to a unique identifier in a cookie, can also attach a name to a profile in many circumstances, as illustrated by the AOL case discussed above. If the firm stores all search queries, tied to the cookie, it holds plenty of information about the user. The firm could identify the person based on his or her searches. If the user sometimes searches for his or her name, this would be even easier.[871] As a Google employee said in a court case, "[t]here are ways in which a search query alone may reveal personally identifying information."[872] In sum, firms that process nameless profiles can often attach a name to the data with relative ease. They process personal data.

---

[868] Barbarom & Zeller 2006. See also Van Hoboken 2012, p 318; Article 29 Working Party 2014, WP 216, p. 11.
[869] Recital 26 of the Data Protection Directive.
[870] An "info@" email address of a company might not constitute personal data, if it doesn't refer to an individual.
[871] See on such "vanity searchers" Soghoian 2007.
[872] Cutts 2006, p. 9.

***Situation (iii)***

A firm processes data about an individual. It's difficult for the firm to add a name to the data, but it would be fairly easy for *another party* to tie a name to the data. An example might be an ad network that has a cookie-based profile of a person, including an IP address. Let's assume that it's difficult for the ad network to tie a name to the profile. But for the internet access provider of the person, it's fairly easy to tie a name to the IP address. For an online shop this would be easy too, if a person orders a product and provides the shop with a name and address.

Does it matter that only another party can identify the person? According to recital 26 of the Data Protection Directive, the answer is no: "to determine whether a person is identifiable, account should be taken of all the means reasonably likely to be used either by the controller *or by any other person* to identify the said person."[873] The recital's approach is sometimes called the absolute approach. A relative approach would imply only looking at the means at the disposal of the data controller.[874]

While recital 26 suggests an absolute approach, the means at the disposal of a data processor are relevant for the purpose of determining which means are reasonably likely to be used for identification. This can be illustrated with an example from Zwenne, presented here in a slightly revised form.[875] If a random person finds some human hairs, those hairs should probably not be seen as personal data for the finder. But if the police has a hair sample and sends this to a DNA research institute to match them against a database with DNA samples, the sample should probably be regarded as personal data.

---

[873] Emphasis added. See also Bygrave 2002, p 318; Article 29 Working Party 2010, WP 171, p 9; Article 29 Working Party 2007, WP 136, p 12-21.
[874] See European Commission's Information Society and Media Directorate-General 2011, p. 18-21.
[875] Zwenne 2013, p. 26-27. Zwenne argues for a relative approach.

Sometimes, Data Protection Authorities say that personal data are identifiable for one party, and not identifiable for another party.[876] Hence, Data Protection Authorities sometimes take into account what means can be used by the firm holding the data. For instance, the English Information Commissioner's Office appears to favour the relative approach.[877] In sum, while recital 26 appears to dictate an absolute approach, the relative approach may be relevant when determining which methods are likely to be used for identification. The Working Party clearly advocates the absolute approach in a 2014 opinion on anonymisation techniques.[878]

Computer scientist Narayanan discusses various ways for ad networks to attach a name to data. For instance, many websites disclose identifying information about their visitors to ad networks – often inadvertently. Furthermore, some firms specialise in tying names to data held by ad networks. The goal of some web surveys – "Win a free iPod!" – is matching email addresses and names to data. If you provide your email address to a firm that also operates a cookie, that firm can tie the two together. If one firm has tied a name to a cookie-profile, it can provide the name to other firms that only had a nameless profile ("cookie matching"). Narayanan summarises: "[t]here is no such thing as anonymous online tracking."[879]

The discussion about behavioural targeting and the scope of data protection law resembles the discussion about IP addresses. The Working Party and many judges in Europe say that IP addresses should generally be considered to be personal data.[880] Others counter that IP addresses shouldn't be considered as personal data in all circumstances. First, some argue for a relative approach. For instance, Google says that IP addresses shouldn't be seen as personal data if the firm holding the IP address can’t tie a name to it.[881] Second, sometimes IP addresses can’t be used to identify a person.[882] For example, the country Qatar routed all internet traffic through a couple of IP addresses.[883] And some organisations access the internet though one IP address. In such cases, a mere IP address without any other information may not be enough to identify somebody.

---

[876] Impact Assessment for the proposal for a Data Protection Regulation (2012), Annex 2, p. 15-16.
[877] Information Commissioner's Office 2012, p. 21. The German situation is more complicated, but also boils down to a relative approach (see Korff 2010b, p. 4).
[878] Article 29 Working Party 2014, WP 216, p. 9.
[879] Narayanan 2011. See on cookie synching chapter 2, section 6.
[880] See about the status of IP addresses as personal data: Impact Assessment for the proposal for a Data Protection Regulation (2012), Annex 2, p. 14-16; Time.lex 2011. See for criticism on the Time.lex report: Zwenne 2013.

In the 2012 *Scarlet* case, the European Court of Justice decided that the IP addresses in that case were personal data. Copyright organisation Sabam requested internet access provider Scarlet to install a filtering system to help enforce copyrights. Scarlet refused. Prompted by the Advocate General, the European Court of Justice decided that the IP addresses are personal data. “Those addresses are protected personal data because they allow those users to be precisely identified.”[884] The Advocate General referred to opinions of the Article 29 Working Party and the European Data Protection Supervisor to support his conclusion that the IP addresses were personal data.[885]

Still, the discussion about IP addresses isn’t over. The Court uses ambiguous language, but it may have suggested a relative approach.[886] For parties that aren’t internet access providers, it’s harder to tie an IP address to a name. They may still try to argue that IP addresses are not personal data in their hands.[887] In sum, European Data Protection Authorities generally consider IP addresses to be personal data, and judges tend to take a similar position.

---

[881] See e.g. Whitten 2008.
[882] See Zwenne 2013, p. 27-28.
[883] Zittrain 2008, p. 157.
[884] CJEU, C-70/10, Scarlet v Sabam, 24 November 2011, par. 51.
[885] Opinion AG (14 April 2011) for CJEU, C-70/10, Scarlet v Sabam, 24 November 2011, par. 75-80.
[886] In an earlier publication I assumed that the Court limited its remark to IP addresses in the hands of access provider Scarlet (Kulk & Zuiderveen Borgesius 2012). Now I believe the Court may have taken an absolute approach, as the Court talks about “users”, and not about “subscribers.” See the definition of “user” (article 2(a) of the e-Privacy Directive), and of “subscriber” (article 2(k) of the Framework Directive 2002/21). A full discussion of the Scarlet Sabam case falls outside this study’s scope.
[887] In a 2013 opinion the Advocate General also sees IP addresses as personal data when they’re in the hands of Google. This suggests an absolute approach (Opinion AG (25 June 2013), C-131/12, Google Spain, par. 48). The Court has neither confirmed nor disproved this view in the subsequent judgment. In October 2014, the German Bundesgerichtshof has asked preliminary questions to the CJEU regarding the question of whether dynamic IP addresses should be seen as personal data (Bundesgerichtshof 2014; see Husovec 2014).

The case law on IP addresses is relevant for two reasons. First, the discussion resembles the discussion about behavioural targeting profiles. The case law on IP addresses confirms that nameless data that refer to a device can be personal data. But there's an important difference between IP addresses and personal profiles that are used for behavioural targeting. Individual behavioural targeting profiles contain much more information than an IP address.[888] Second, firms that process data about individuals for behavioural targeting usually tie IP addresses to the data. For instance, an ad network typically needs the IP address of the receiving device to display the ad.

To conclude, if a firm processes data about an individual for behavioural targeting, and it's fairly easy for another party to tie a name to the data, the Data Protection Directive's preamble suggests that the data sould be regarded as personal data.

***Situation (iv)***

A firm processes data about an individual, and it would be difficult for *anybody* to tie a name to the data. As it's often fairly easy for a firm to tie a name to the data it processes for behavioural targeting, the number of situations in this category is likely to be small. This category was discussed in section 2: the Working Party says it's not relevant whether a firm can attach a name to the data. If the firm uses the data to single out a person, the firm processes personal data.

### 5.4 Data protection law should apply to behavioural targeting

Many scholars say a logical interpretation of data protection law implies that data about a nameless individual should be regarded as personal data.[889] This study agrees. Some say that, if necessary, the personal data definition should be adapted to emphasise that it includes data used to single out a person. De Hert & Gutwirth have hinted at "a shift from personal data protection to data protection *tout court*."[890]

---

[888] See Van Hoboken 2012, p. 328.
[889] See for instance De Hert & Gutwirth 2008; Leenes 2008; Koëter 2009; Traung 2010; Traung 2012. But see Zwenne 2013, with references, for another view.

But, apart from an analysis of the law, why should data that are used to single out a person for behavioural targeting be regarded as personal data? First, the processing of information for behavioural targeting triggers many concerns that lie at the core of data protection law. The risks of large-scale data collection don't disappear merely because data about a person can't be tied to a name.[891] For instance, massive collection of information on user behaviour can cause a chilling effect; which remains true even if firms collect pseudonymous data. Firms compile detailed information about people, and can classify people, while the individual has little control over this process.[892] As Turow notes, "[i]f a company can follow your behaviour in the digital environment – an environment that potentially includes your mobile phone and television set – its claim that you are 'anonymous' is meaningless. (…) It matters little whether your name is John Smith, Yesh Mispar, or 3211466."[893] And a firm could discriminate against a person, for instance when the cookie says the person is "handicapped",[894] or is in the interest category "lesbian, gay, bisexual, transgender."[895]

True, certain risks are reduced when a firm doesn't attach a name to the data it holds about a person. Suppose a firm with pseudonymous profiles regarding people's browsing behaviour experiences a data breach: the firm accidentally publishes millions of browsing profiles on the web. People who see the data learn that the person behind ID *xyz* visited dirty-pictures.com, or another-embarrassing-website.com. But somebody who sees the pseudonymous browsing profile doesn't immediately learn the name of the person who visited those websites. Hence, the privacy risks are reduced, because the leak concerns pseudonymous data. There's less risk of embarrassment or other unpleasant surprises for the person behind ID *xyz*. However, the AOL search data case illustrates that it may be possible to find the person behind a pseudonymous profile.[896]

---

[890] De Hert & Gutwirth 2008, p. 289.
[891] Article 29 Working Party 2013, WP 203, p. 46.
[892] See chapter 3, section 3.
[893] Turow 2011, p. 7 (see also p. 100).
[894] Rocket Fuel, Health Related Segments 2014. All the examples are taken from US companies, but it can't be ruled out that they also operate cookies on devices in the EU.
[895] Flurry (audiences). Flurry is firm offering analytics and advertising for mobile devices. Among the demographic data that advertisers can select, Flurry lists "race" (Flurry, factual).

Privacy risks are also reduced for another reason when a firm doesn't know the name of a person behind a cookie profile. For example, say a supermarket offers a loyalty card to customers, and knows the names of those customers. If a behavioural targeting firm wanted to tie a profile based on information gathered through a supermarket loyalty card to an online profile, it would be practical if a name were linked to both profiles. Without a name, it's harder to merge data from different databases.[897]

Nevertheless, many risks remain, even when firms don't tie a name to data. The behavioural targeting industry compiles large amounts of information about people, and if data protection law didn't apply, this industry could operate largely unregulated. We will see in later chapters that applying data protection law provides more protection to internet users than only applying the e-Privacy Directive's consent requirement for cookies and similar tracking technologies.[898]

Second, a name is merely one of the identifiers that can be tied to data about a person. In some situations, the name is the most practical identifier. But for many purposes, a name isn't the most effective identifier. If the purpose is sending messages to a phone, or tracking a phone's location, a phone number or one of the ID numbers of a phone is the easiest identifier. Furthermore, a unique number is often a better identifier than a name, because names may not be unique.[899]

For an ad network that wants to track a person's browsing behaviour, or wants to target ads to a person, a cookie is a better identifier than a name. Many firms aren't interested in tying a name to data they process for behavioural targeting. When Mozilla, the firm behind the Firefox browser, considered blocking third party cookies by default, the Interactive Advertising Bureau (IAB) US reacted furiously.[900] The reaction suggests that the IAB sees the threat of not being able to use people's names for behavioural targeting as less serious than the threat that third party cookies won't work anymore.

---

[896] See section 3 of this chapter.
[897] See chapter 2, section 6.
[898] See chapter 6, 8 and 9.
[899] The Working Party notes that very common names by itself aren't always personal data, because they can't be used to identify people (Article 29 Working Party 2007, WP 136, p. 13).

Third, the goal of behavioural targeting is targeting the right person, with the right ad, at the right time. It would be odd to say that data used by a firm for individualised tracking and targeting aren't personal data. The whole point of behavioural targeting is singling people out, and targeting ads to *specific* individuals.

Seeing data that can single out a person as personal data corresponds with the rationale for the Data Protection Directive.[901] One of the Directive's goals is protecting privacy and other fundamental rights.[902] The European Court of Justice says that the Directive aims for a "high level" of protection,[903] and that fundamental rights guide the interpretation of the Directive.[904] Furthermore, "limitations in relation to the protection of personal data must apply only in so far as is strictly necessary."[905] According to the European Court of Human Rights, the right to private life is a broad term that should be applied dynamically and pragmatically.[906] This study suggests that data protection law, like the European Convention on Human Rights, should be seen as a living instrument. In the light of new developments such as behavioural targeting, it wouldn't make sense to limit the scope of data protection law to data that can identify people by name.

---

[900] Interactive Advertising Bureau Europe 2013. See chapter 2, section 2.
[901] See Korff 2010a, p. 47-48.
[902] Article 1(1) of the Data Protection Directive.
[903] ECJ, C-524/06, Huber, 16 December 2008, par. 50; CJEU, C-131/12, Google Spain, 13 May 2014, par. 66.
[904] ECJ, C-465/00, C-138/01 and C-139/01, Österreichischer Rundfunk, 20 May 2003, par. 68; CJEU, C-131/12, Google Spain, 13 May 2014, par. 68.
[905] See e.g. CJEU, C-293/12 and C-594/12, Digital Rights Ireland Ltd, 8 April 2014, par. 52; CJEU, Case C-473/12, 7 November 2013, Institut professionnel des agents immobiliers, par. 39 (with further references).
[906] See the case law mentioned in chapter 3, section 2. The Court says: "[t]hat broad interpretation [of he right to private life] corresponds with that of the Council of Europe's Convention of 28 January 1981 for the Protection of Individuals with regard to Automatic Processing of Personal Data (…)" ECtHR, Amann v. Switzerland, No. 27798/95, 16 February 2000, par. 65). See similarly ECtHR, Rotaru v. Romania, No. 28341/95, 4 May 2000, par. 43.

***Criticism on the singling out perspective***

Some authors criticise the tendency of Data Protection Authorities to interpret the personal data definition broadly and point to several disadvantages.[907] The main points are summarised here. It's concluded that the arguments aren't persuasive.

First, it has been argued that firms have less incentives for investing in pseudonymisation technology if the law covers pseudonymised data.[908] While it may be true that firms have less incentive to pseudonymise data, the law requires appropriate security measures from data controllers, and pseudonymisation can improve security. For instance, pseudonymisation can help to keep data subjects' names hidden from employees that don't need to see the names.[909] Hence, pseudonymisation can improve data security. But replacing a name with another identifier isn't enough to keep data outside the scope of data protection law.[910]

Second, some suggest applying data protection law to behavioural targeting would be bad for business and innovation.[911] This argument isn't sufficient to keep behavioural targeting outside data protection law's scope. When information is within the scope of data protection law, that doesn't mean processing is prohibited. But it does imply that firms have to comply with the data protection principles. It's certainly true that some firms would make less profit when they have to comply with data protection law. But even if fundamental rights were ignored and only economic effects were taken into account, a more relevant question is whether society as a whole wins or loses. Chapter 7 shows that it's unclear whether more or less legal privacy protection is better from an economic perspective.[912] And while innovation – a term almost as vague as privacy – is important, it doesn't trump fundamental rights. If it were good for innovation if children below eight worked in factories, we still shouldn't allow it.[913] Moreover, if regulation pushes firms towards developing new and privacy preserving technologies, this is also innovation.

---

[907] The most detailed and eloquent critique is offered by Zwenne (Zwenne 2013).
[908] Zwenne 2010, p. 336.
[909] The Data Protection Directive requires an appropriate level of security for personal data (article 17). See chapter 4, section 2 on the security principle.
[910] See European Agency for Fundamental Rights 2014, p. 45-46; Article 29 Working Party 2014, WP 216, p. 20.
[911] See e.g. Stringer 2013.

Third, some say that applying data protection law to data that identify nameless people could lead to peculiar situations. For instance, if a firm holding nameless profiles granted data subjects the right to access their data, the firm might have to ask the data subject to identify herself, which might involve asking for more personal data.[914] However, interpreting the data protection provisions in a reasonable manner can prevent absurd results.[915] As the Working Party puts it, "[i]t is a better option not to unduly restrict the interpretation of the definition of personal data but rather to note that there is considerable flexibility in the application of the rules to the data."[916]

Fourth, some say a broad interpretation of personal data implies that data protection law applies, even when there are no privacy threats. Some suggest that data protection law shouldn't be severed from the right to privacy.[917] That argument doesn't fit well with positive law, as the EU Charter of Fundamental Rights distinguishes the right to data protection and the right to privacy. Furthermore, many authors say it's an advantage that data protection law applies to all personal data, rather than just private personal data.[918]

---

[912] See chapter 7, section 2.
[913] Helen Nissenbaum made a remark among these lines at the Acatech Symposium Internet Privacy (26 March 2012, Berlin). Article 32 of the EU Charter of Fundamental rights: "The employment of children is prohibited." See for a general critique of the innovation argument Morozov 2013, and Richards 2014a, p. 28 - p. 36.
[914] Schwartz & Solove 2011, p. 1876. See also Zwenne 2013, p. 37. See on access rights and pseudonymous data: chapter 8, section 2.
[915] Like with any statue, there's also a risk that data protection law is applied in an unreasonable manner.
[916] Article 29 Working Party 2007, WP 136, p. 5.
[917] Cuijpers & Marcelis 2012.
[918] See e.g. De Hert & Gutwirth 2006, p. 94; Hildebrandt et al. 2008, p. 245. See also chapter 4, section 2, and chapter 9, section 2.

Fifth, some worry that almost everything could become personal data if data that are used to single out a person are seen as personal data. Enforcing data protection law would become too difficult. Data Protection Authorities would only be able to enforce the law against a few wrongdoers. This could lead to arbitrary decisions about enforcement, which would be bad for legal certainty. A related point is that the scope of the personal data definition becomes too uncertain. This would also be bad for legal certainty.[919]

There's merit to the point that the broad scope of data protection law makes enforcement difficult. But limiting the scope of data protection to exclude pseudonymous data wouldn't be the right reaction. Similarly, it's probably good that we have environmental law, even though it's impossible to catch everybody who breaches the law.[920] Furthermore, in legal practice the fringes of a definition can always provoke discussion. In sum, the criticism doesn't justify leaving behavioural targeting outside the scope of data protection law.

Merely ensuring that data protection law applies to behavioural targeting doesn't solve all privacy problems. But, with all its weaknesses, at least data protection law provides a framework to assess fairness when personal data are processed. And since data protection law requires firms to disclose information about their processing practices, it can help to make the processing transparent. When problems are found, this could lead to the conclusion that more regulation is needed.[921]

---

[919] This fifth point is made most convincingly by Zwenne 2013, in particular p. 33-35. Korff agrees that a broad interpretation of personal data can have drawbacks, but still argues for a broad interpretation (Korff 2010a, p. 47-48).
[920] See on privacy scholarship taking inspiration from environmental law Hirsch 2006.
[921] See also chapter 4, section 3.

### 5.5 Data protection reform and pseudonymous data

The 2012 European Commission proposal for a Data Protection Regulation led to much discussion about the law's material scope.[922] The proposal doesn't significantly alter the personal data definition. But the proposal includes "online identifiers" and "location data" in the list of examples of information that can be used to identify a data subject.[923] The preamble and the impact assessment that accompanied the proposal show that the European Commission intended data protection law to apply to behavioural targeting.[924]

The Commission's proposal chooses the absolute approach to identifiability. The definition says that the "means reasonably likely to be used by the controller *or by any other natural or legal person*" should be taken into account when determining identifiability.[925] The proposal defines personal data as "any information relating to a data subject."[926] The latter is defined as follows:

> "Data subject" means an identified natural person or a natural person who can be identified, directly or indirectly, by means reasonably likely to be used by the controller or by any other natural or legal person, in particular by reference to an *identification number*, location data, *online identifier* or to one or more factors specific to the physical, physiological, genetic,

---

[922] I took part in this debate, for instance at the Dutch and the European Parliament (Zuiderveen Borgesius 2012a).
[923] Article 4(2) of the European Commission proposal for a Data Protection Regulation (2012)
[924] See recital 20 and 46 and article 3(2)(b) of the European Commission proposal for a Data Protection Regulation (2012). See also Impact Assessment for the proposal for a Data Protection Regulation (2012), p. 31.
[925] The proposed definition incorporates parts of the old recital 26 ("the controller or by any other") in the definition of personal data.
[926] Article 4(2) of the European Commission proposal for a Data Protection Regulation (2012).

> mental, economic, cultural or social identity of that person (emphasis added).[927]

Recital 24 of the proposal discusses online tracking and elaborates on the use of "online identifiers." The recital begins with suggesting that data about a person that are attached to a unique identifier, such as a cookie, are usually personal data. A tracking cookie or another identifier can be used to profile individuals and to identify them.

> When using online services, individuals may be associated with online identifiers provided by their devices, applications, tools and protocols, such as Internet Protocol addresses or cookie identifiers. This may leave traces which, combined with unique identifiers and other information received by the servers, may be used to create profiles of the individuals and identify them. It follows that identification numbers, location data, online identifiers or other specific factors as such need not necessarily be considered as personal data in all circumstances.[928]

The recital's last sentence suggests that there may be circumstances where online identifiers shouldn't be considered as personal data.[929] It's true that in some circumstances unique identifiers might not relate to an individual, for instance when many people use the same computer. But the last sentence may create a gap in the data protection regime. Among others, the Working Party says the last sentence must be deleted, to emphasise that data protection law fully applies to unique identifiers such as tracking cookies.[930]

---

[927] Article 4(2) of the European Commission proposal for a Data Protection Regulation (2012) (capitalisation and punctuation adapted).
[928] Recital 24 of the European Commission proposal for a Data Protection Regulation (2012).
[929] During the last weeks before the European Commission proposal was presented, the last sentence was changed. An earlier version of the proposed Regulation concluded in the last sentence that the "Regulation should be applicable to processing involving such data" (European Commission proposal for a Data Protection Regulation (2012), leaked Interservice draft (2011), recital 23).

Discussion on the scope of data protection law continued after the Commission's proposal. The proposal has led to an enormous amount of lobbying, including by firms from the US[931] During the discussions about the Data Protection Regulation proposals, a new legal concept was suggested: "pseudonymous data." The Interactive Advertising Bureau, and firms such as Yahoo and Amazon, both using behavioural targeting, lobbied for amendments that would introduce a lighter regime for "pseudonymous" data.[932] At least one non-governmental organisation was in favour of a lighter regime for pseudonymous data, because that would incentivise firms to pseudonymise data, which would help data security.[933]

Some European Parliament members proposed amendments to introduce a data protection-light regime for pseudonymous data. For instance, shadow rapporteur Alvaro proposed adding a rule that would legitimise the processing of pseudonymous data. "Processing of pseudonymized data shall be lawful."[934] Other Parliament members proposed leaving pseudonymous data largely outside the scope of data protection law.[935]

***LIBE Compromise***

In January 2013, the Rapporteur for the European Parliament, Albrecht, presented his draft report. The report codifies the Working Party's view on the definition of personal data, by adding the "single out" phrase to the personal data definition.[936] Hence, any data relating to a person that "can be identified *or singled out"* are personal data. The Albrecht report thus emphasises that data protection law applies to processing data about nameless individuals. The draft report by Rapporteur Albrecht also included a definition of "pseudonymous data", as a category of personal data.[937]

---

[930] Article 29 Working Party 2012, WP 199, p. 5-6; Korff 2012, p. 32.

[931] See Albrecht 2013. Albrecht estimates that more than half of the firms that contacted him regarding the proposals are from the US (Traynor 2014). See generally on corporate lobbying in Brussels Horten 2011.

[932] See on the lobbying by the Interactive Advertising Bureau for a lighter regime for pseudonymous data Stringer 2013. Amazon proposed amendments, ready to submit (Amazon proposed amendments). See also Yahoo proposed amendments.

[933] Center for Democracy & Technology 2013a.

[934] Alvaro 2013, amendment 48, p. 31. The rule would imply that firms don't need another legal basis (such as consent) for the processing of pseudonymous data; see chapter 6.

[935] See amendment 327 by Jens Rohde & Bendt Bendtsen: "encrypted and some psydonymised [sic] data fall outside this regulation" (ITRE Amendments).

In March 2014, the European Parliament adopted a compromise text (the "LIBE Compromise"), which the Parliament's LIBE Committee prepared on the basis of the 3999 amendments by the members of parliament.[938] The LIBE Compromise defines personal data roughly the same as the 1995 Data Protection Directive.[939] But the LIBE Compromise adds location data and unique identifiers to the examples of possible identifiers. The preamble makes clear that the LIBE Compromise takes an absolute approach to identifiably. "To determine whether a person is identifiable, account should be taken of all the means reasonably likely to be used *either by the controller or by any other person* to identify or single out the individual directly or indirectly."[940]

Recital 24 of the LIBE Compromise suggests that in principle the regulation is applicable to processing unique identifiers such as cookies, IP addresses and RFID tags.[941] In other words, the Regulation seems to apply to data that can "single out" a person, including if no name is tied to the data.[942] However, the LIBE Compromise also introduces a new category of personal data: “pseudonymous data.”

---

[936] He proposed the following definition: "data subject' means an identified natural person or a natural person who can be identified *or singled out*, directly or indirectly, *alone or in combination with associated data*, by means reasonably likely to be used by the controller or by any other natural or legal person, in particular by reference to *a unique identifier*, location data, online identifier or to one or more factors specific to the physical, physiological, genetic, mental, economic, cultural, social *or gender* identity *or sexual orientation* of that person" (the emphasised words are proposed) (amendment 84, article 4(1), Draft Albrecht report).
[937] Amendment 85, article 4(2)(a), Draft Albrecht report. The draft report suggests that under certain conditions, a system like Do Not Track could be used to signify consent to the processing of such data (amendement 105, article 7(2)(a)).
[938] LIBE Compromise, proposal for a Data Protection Regulation (2013). See for a concise overview of the discussions from January 2012 to January 2014: Burton & Pateraki 2013; Kuner et al. 2014.
[939] 'Personal data' means any information relating to an identified or identifiable natural person ('data subject'); an identifiable person is one who can be identified, directly or indirectly, in particular by reference to an identifier such as a name, an identification number, location data, unique identifier or to one or more factors specific to the physical, physiological, genetic, mental, economic, cultural or social or gender identity of that person." (Article 4(2) of the LIBE Compromise, proposal for a Data Protection Regulation (2013), capitalisation and punctuation adapted).
[940] Recital 23 of the LIBE Compromise, proposal for a Data Protection Regulation (2013) (emphasis added).
[941] Recital 24 of the LIBE Compromise, proposal for a Data Protection Regulation (2013).

> “Pseudonymous data” means personal data that cannot be attributed to a specific data subject without the use of additional information, as long as such additional information is kept separately and subject to technical and organisational measures to ensure non-attribution.[943]

Such pseudonymous data are subject to a lighter regime in the LIBE Compromise. One of the main differences is that the LIBE Compromise allows processing pseudonymous data without consent in some circumstances.[944] But the introduction of the pseudonymous data category might lead to a level of protection that is too low.[945] At the time of writing of this study, the debate about the legal status of pseudonymous data is ongoing.

### 5.6 Special categories of data

Data protection law has a stricter regime for “special categories of data.” These are “data revealing racial or ethnic origin, political opinions, religious or philosophical beliefs, trade-union membership, and the processing of data concerning health or sex life.”[946] According to the European Court of Justice, data concerning health must be given a wide interpretation.[947] This suggests that "special categories of data" must be interpreted broadly.

---

[942] See also recital 23 of the LIBE Compromise, proposal for a Data Protection Regulation (2013): “To determine whether a person is identifiable, account should be taken of all the means reasonably likely to be used either by the controller or by any other person to identify or single out the individual directly or indirectly.”

[943] Article 4(2a) of the LIBE Compromise, proposal for a Data Protection Regulation (2013). The LIBE Compromise also includes a definition of encrypted data in article 4(2b). Recital 23 says the regulation doesn’t apply to anonymous data: “information that does not relate to an identified or identifiable natural person.”

[944] See chapter 6, section 2. The lighter regime for pseudonymous data has more consequences. See for instance recital 38 and 58a (on the balancing provision and profiling), health data (recital 122a and article 81(2)(a)), and article 10.

[945] European Commissioner Reding warns: “pseudonymous data must not become a Trojan horse at the heart of the Regulation, allowing the non-application of its provisions” (Reding 2014).

[946] Article 8(1) of the Data Protection Directive.

Processing special categories of data is only allowed with the data subject's "explicit consent."[948] About half of the member states require such explicit consent to be in writing. Some member states have chosen not to allow people to override the prohibition with consent.[949] There are exceptions to the in-principle processing prohibition, for instance in the medical context and for churches. These provisions aren't, however, relevant for behavioural targeting.[950]

The stricter regime for special categories of data can be explained by the wish to prevent unfair discrimination.[951] "In general, information relating to the intimate private life of persons or information which might lead to unfair discrimination should not be recorded or, if recorded, should not be disseminated," said the Council of Europe in 1972.[952] And the Directive's preamble says that special categories of data "are capable by their nature of infringing fundamental freedoms or privacy."[953] The stricter regime for special categories of data also seems to be related to privacy as limited access, or as intimacy.[954] Certain types of data are considered particularly private or sensitive.[955] Data protection instruments such as the Data Protection

---

[947] ECJ, C-101/01, Lindqvist, 6 November 2003, par. 50.
[948] See article 8(2)(a) of the Data Protection Directive. See also chapter 9, section 5.
[949] For instance: Italy and Sweden require consent to be in writing (Impact Assessment for the proposal for a Data Protection Regulation (2012), Annex 2, p. 29). See article 8(2)(a) of the Data Protection Directive.
[950] Article 8(2)-8(7) of the Data Protection Directive. There's also an exception for sensitive data that are "manifestly made public by the data subject" (article 8(2)(e). It doesn't seem plausible that firms can invoke this ground for the gathering of data for behavioural targeting. An exception might be a firm that gathers information that people publish about themselves on the web.
[951] The United Nations guidelines use the header "principle of non-discrimination" for their provision on sensitive data, article 5 (UN General Assembly, Guidelines for the Regulation of Computerized Personal Data Files, 14 December 1990).
[952] Committee of Ministers, Resolution (73)22 on the protection of the privacy of individuals *vis-à-vis* electronic data banks in the private sector, 26 September 1973, article 1.
[953] Recital 33 of the Data Protection Directive.
[954] See Bygrave 2002, p. 132.
[955] See e.g. European Union Civil Service Tribunal, Civil Service Tribunal Decision F-46/095, V & EDPS v. European Parliament, 5 July 2011, par. 163; I. v. Finland, App. No. 25011/03, 17 Jul. 2008, par. 38. See along similar lines Z v. Finland (9/1996/627/811) 25 February 1997, par. 95. See on special categories of data also chapter 9, section 5.

Convention and the United Nations Data Protection Guidelines also have stricter rules for certain types of personal data.[956]

The European Commission proposal for a Data Protection Regulation retains the stricter regime for special categories of data. The categories remain roughly the same.[957] While the proposal always requires consent to be "explicit", the distinction between special categories of data and non-special personal data remains relevant. The Regulation only allows processing of special categories of data for direct marketing and behavioural targeting after obtaining the data subject's consent.[958]

Research suggests that many people indeed regard special categories of data, such as data regarding health, as sensitive. Many people also consider data regarding their finances sensitive.[959] The European Consumer organisation says financial data should be added to the category of sensitive data.[960] However, there are cultural differences between member states. For instance, in Finland data from the tax office about people's income are publicly available.[961]

***Behavioural targeting and special categories of data***

Do firms that engage in behavioural targeting process special categories of data? Some firms do, some don't, and many operate in a grey area. There are firms that clearly process special categories of data for behavioural targeting. Some firms target advertising based on categories such as "US Hispanics",[962] "arthritis", "cardiovascular general health”, [963] “lesbian, gay, bisexual, and transgender,” [964] or “disabled/handicapped consumers.”[965] Such firms process special categories of data.

---

[956] Article 6 of the Data Protection Convention 1981; Article 5 of the UN General Assembly, Guidelines for the Regulation of Computerized Personal Data Files, 14 December 1990.

[957] Genetic data are added to the definition, and data about philosophical beliefs are deleted (article 9 of the European Commission proposal for a Data Protection Regulation (2012)). Genetic data are defined in article 4(10). See also article 33(2)(a), which suggests that processing operations involving certain kinds of data "present specific risks."

[958] See on the legal basis for processing (such as consent) chapter 6.

[959] See Leon et al. 2013. See also the Commission Regulation on Data Breaches (no. 611/2013), which gives examples of data that likely to adversely affect people's personal data or privacy in the case of a data breach. The list of examples includes "financial information (…) internet log files [and] web browsing histories (article 3(2) and recital 12).

[960] European Consumer Organisation BEUC 2013, p. 16.

[961] See ECJ, C-73/07, Satamedia, 16 December 2008.

[962] Batanga Network Inc.

It’s possible to use behavioural targeting without processing special categories of data. Suppose an ad network only works with websites about comic books. The firm puts cookies in one of three categories: people who like science fiction, people who like superheroes, and people who like other topics. Immediately after categorising people, the firm deletes the list of visited websites. In this example, the firm doesn’t process special categories of data.

But many firms using behavioural targeting operate in a grey area – perhaps most of them. Say a firm puts people (or cookies) in the category “unions and labour movement”, based on their surfing behaviour.[966] A person’s interest in the labour movement could imply a political opinion. And certain website visits could suggest a person’s sexual orientation or medical condition, even if there are no behavioural categories associated with the raw data.[967] In sum, behavioural targeting often entails the processing of data that could be considered “special categories of data.”

***e-Privacy Directive***

In 1997, two years after the Data Protection Directive, the EU adopted the Directive on personal data processing in the telecommunications sector.[968] In 2002 it was replaced by the Directive “concerning the processing of personal data and the protection of privacy in the electronic communications sector." This e-Privacy Directive was meant to be more in line with new technologies.[969]

---

[963] Yahoo! Privacy.
[964] Flurry (audiences). Flurry is firm offering analytics and advertising for mobile devices. Among the demographic data that advertisers can select, Flurry lists “race” (Flurry, factual).
[965] Rocket Fuel, Health Related Segments 2014. All the examples are taken from US companies, but it can’t be ruled out that they also operate cookies on devices within the EU. See on political behavioural targeting also chapter 2, section 7, chapter 3, section 3, and chapter 9, section 5.
[966] Google Ad Interest Categories 2014.
[967] For instance, the Office of the Privacy Commissioner of Canada finds that “Google is delivering tailored ads in respect of a sensitive category, in this case, health” (Office of the Privacy Commissioner of Canada (Google) 2014, par. 26). Of course, that report doesn’t concern EU data protection law, but the Canadian regime has similarities to the EU regime.
[968] Directive 97/66/EC (the ISDN Directive).

The e-Privacy Directive has a stricter regime that applies when certain types of firms process location data or traffic data. Such data may only be processed based on consent, or in some narrowly defined circumstances.[970] Traffic data, sometimes called metadata, are "any data processed for the purpose of the conveyance of a communication on an electronic communications network or for the billing thereof."[971] Examples of traffic data are the time of a communication, the email address of communicating partners, and the IP address used to access the internet.[972]

The Advocate General of the European Court of Justice says traffic data are "data which are in a sense more than personal."[973] Traffic data are "'special' personal data, the use of which may make it possible to create a both faithful and exhaustive map of a large portion of a person's conduct strictly forming part of his private life, or even a complete and accurate picture of his private identity."[974] With modern communication technology, the line between traffic data and communications content becomes increasingly blurry. For instance, the subject line of an email message could be seen as traffic data or as communications content, and traffic data can paint a detailed picture of a person's life.[975]

Location data are data "indicating the geographic position of the terminal equipment of a user of a publicly available electronic communications service."[976] Location data can be sensitive.[977] For example, a phone's location data can disclose visits to the hospital or a mosque, or the location of one's bed. The e-Privacy Directive's regime for traffic data and location data is similar to the regime for special categories of data in the Data Protection Directive. Unless a legal exception applies, consent is needed for the processing of traffic and location data.[978]

---

[969] Recital 4 of the e-Privacy Directive. This study refers to the consolidated version (amended in 2009), unless otherwise noted. See on the e-Privacy Directive chapter 6, section 4, chapter 8, section 4.
[970] Article 6 and 9 of the e-Privacy Directive.
[971] Article 2(b) of the e-Privacy Directive.
[972] See recital 15 of the e-Privacy Directive, and the Data Retention Directive.
[973] Opinion AG (12 December 2013) for CJEU, C-293/12 and C-594/12, Digital Rights Ireland Ltd, 8 April 2014, par. 65.
[974] Opinion AG (12 December 2013) for CJEU, C-293/12 and C-594/12, Digital Rights Ireland Ltd, 8 April 2014, par. 74)
[975] See on the blurry line between traffic data, a EU law angle: Koops & Smit 2014; Breyer 2005; United Nations High Commissioner for Human Rights 2014, p. 6-7. See for a computer science angle Felten 2013; Mayer & Mutchler 2014.
[976] Article 9 of the e-Privacy Directive.

But the scope of these provisions in the e-Privacy Directive is narrow. The requirements regarding traffic and location data only apply to providers of publicly available electronic communications services, such as internet access providers or phone operators (telecommunication providers for short).[979] The e-Privacy Directive's background as a directive regulating telecommunications companies can help to explain the narrow scope of these provisions.[980] But many firms, such as ad networks and providers of smart phone apps, process more data of a sensitive nature than telecommunications providers. However, ad networks and apps providers aren't subject to the e-Privacy Directive's rules for traffic and location data. The Working Party suggests that when applying data protection law, location and traffic data should be treated as prima facie sensitive, although they're not within the definition of "special categories of data."[981]

In the behavioural targeting area, the most relevant provision of the e-Privacy Directive is article 5(3), which requires consent for the use of most tracking technologies. The scope of article 5(3) isn't limited to telecommunications providers. That provision is discussed in the next chapter.[982]

---

[977] See Article 29 Working Party 2011, WP 185, p. 7.
[978] See article 5, 6 and 9 of the e-Privacy Directive.
[979] An "electronic communications service" is, in short, a service that consists wholly or mainly in the conveyance of signals on electronic communications networks (article 2(c) of the Framework Directive 2002/21/EC (amended in 2009)). It's thus a transmission service. See Zuiderveen Borgesius 2011a.
[980] See Arnbak 2013a, p. 9.
[981] Article 29 Working Party 2013, WP 203, p. 25; p. 66. See also the European Commission proposal for a Data Protection Regulation (2012): article 33(2)(a) suggests that certain processing operations involving location data "present specific risks." See also the Commission Regulation on Data Breaches (no. 611/2013), article 3(2) and recital 12. Financial information and web browsing histories are given as examples of data that are likely to affect privacy in case of a breach. See on the scope of the e-Privacy Directive also chapter 6, section 4, and chapter 9, section 5.

### 5.7 Conclusion

Two conclusions can be drawn from this chapter. First, an analysis of current law shows that data protection law generally applies to behavioural targeting. Second, from a normative perspective, data protection law should apply to behavioural targeting.

Personal data are "any information relating to an identified or identifiable natural person."[983] The Article 29 Working Party says that firms carrying out behavioural targeting usually process personal data; even if they don't tie a name to the data they hold about an individual. A name is not needed to identify a person. Firms process the data to single out a person. Therefore, the data processed for behavioural targeting are generally personal data. Moreover, it's often fairly easy for the firm using behavioural targeting, or for another party, to tie a name to the data.

Heated discussions about pseudonymous data ensued when the European Commission released its proposal for a new Data Protection Regulation. The debate focuses on two aspects. Should data protection law apply to pseudonymous data? And if pseudonymous data are within the scope of data protection law, should there be a lighter regime? At the time of writing of this study, the debate is ongoing.

This study argues that data protection law should apply to behavioural targeting, and argues against a lighter regime for pseudonymous data. First, many risks remain, regardless of whether firms tie a name to the information they hold about a person. For instance, surveillance can cause a chilling effect, even if firms collect pseudonymous data. And a cookie-profile that says a person is handicapped or from a bad neighbourhood could be used for unfair discrimination. Second, a name is merely one of the identifiers that can be tied to data about a person, and is not the most practical identifier for behavioural targeting. For an ad network that wants to track a person's browsing behaviour, or wants to target a person with online advertising, a cookie works better than a name. Third, the online marketing industry processes large amounts of information about people, which carries risks. If data protection law didn't apply, this industry could operate largely unregulated. For these reasons, data that are used to single out a person should be considered personal data.

---

[982] See chapter 6, section 4.
[983] Article 2(a) of the Data Protection Directive.

The fact that data protection law applies doesn't imply that processing is prohibited. It means that the firm using behavioural targeting must process the data fairly, lawfully, and transparently. Of course, merely ensuring that data protection law applies doesn't solve all privacy problems. But at least, data protection law can be used to assess the fairness of processing. The next chapter discusses the role of informed consent in data protection law.

* * *

# 6 Informed consent in data protection law

Informed consent plays a central role in the current regulatory framework for behavioural targeting. Therefore, this chapter examines the role of informed consent in data protection law. The e-Privacy Directive requires consent for the use of tracking cookies and similar technologies. And unambiguous consent is generally required as a legal basis for personal data processing for behavioural targeting.

The requirement for firms to obtain the individual's consent for certain practices is indicative of data protection law's aim to put people in control of their personal data. But while consent plays an important role in data protection law, this chapter shows the role is also limited. Data subjects can't set data protection law aside with consent.

A data controller may only process personal data on the basis of the data subject's consent, or on one of the other five legal bases. Article 7 of the Data Protection Directive lists the six possible legal bases to process personal data, starting with (a) consent. The other legal bases only allow processing when it's "necessary." Briefly stated, the other legal bases are as follows. Data processing is allowed if it's necessary (b) for the performance of a contract with the data subject, (c) to comply with a legal obligation, (d) to protect the data subject's vital interests, (e) for a task carried out in the public interest, for instance by the state, or (f) for legitimate interests of the controller that outweigh the data subjects fundamental rights.[984] This study refers to this last legal basis (f) as the balancing provision. The European Commission proposal for a Data Protection Regulation copies the same legal bases without major revisions.[985] For the private sector, the three most relevant legal bases are consent, a contract, or the balancing provision; the study focuses on these.

---

[984] Article 7 of the Data Protection Directive. See for an introduction on the legal bases Article 29 Working Party 2014, WP 207, p. 16-21. See on the six legal bases and behavioural targeting Van Der Sloot & Zuiderveen Borgesius 2011, p. 99-100.

Section 6.1 of this chapter discusses a contract with the data subject, section 6.2 the balancing provision, and section 6.3 the data subject's consent. Section 6.4 discusses the e-Privacy Directive's consent requirement for the use of tracking technologies. Section 6.5 analyses the role of consent in data protection law, and shows the role is important, but also limited. People can't set data protection provisions aside by giving consent, or by contractual agreement. Hence, data protection law limits the data subject's contractual freedom. Nevertheless, section 6.6 rejects the idea that data protection law is too paternalistic. Section 6.7 concludes.

## 6.1 Contract

A first legal basis that a firm can rely on for processing personal data is a contract. Data processing is allowed when it's "necessary for the performance of a contract to which the data subject is party (…)."[986] For example, a shop has to process certain personal data when somebody pays with a credit card. And a magazine publisher doesn't need to obtain consent to process the name and address of a subscriber, as far as these personal data are needed to deliver the magazine at the subscriber's home. The personal data are "necessary" to deliver the magazine to the subscribers and thus to fulfill the contract.[987]

Many firms can't base the processing of personal data for behavioural targeting on a contract. For instance, if an ad network collected data about people without them being aware, it's difficult to see how it could have entered a contract with those people. To illustrate, the Working Party has examined Google's privacy policy, after Google made amendments in March 2012, which allowed Google to combine user data across most Google services. According to the Working Party, Google can't rely on the legal basis contract for combining data across its various services.[988] Similarly, the Dutch Data Protection Authority rejects the idea that Google could rely on a contract to process personal data of people who Google tracks through its ad networks, because people haven't accepted any offer.

---

[985] Article 6 of the European Commission proposal for a Data Protection Regulation (2012).
[986] In some cases firms can also rely on this ground prior to entering a contract. See article 7(a) of the Data Protection Directive. See also chapter 9, section 6, on article 15 of the Data Protection Directive.
[987] See for a similar example College bescherming persoonsgegevens (Dutch DPA) 2013 (Google), p. 77.

> Passive users (…), in other words visitors to websites that use Google's (advertising) services, do not receive any proposal from Google to enter into a contract, electronically or otherwise. So they can hardly be said to have accepted an offer (since they have not even received one). Passive users will in most cases not even be aware that they have encountered or will encounter Google cookies when using third-party websites. The Terms of Service therefore certainly do not give rise to a contractual relationship with the passive users.[989]

***Necessity***

For a firm to be able to rely on the legal basis contract, the processing must be "necessary" for the performance of a contract with the data subject.[990] The *Huber* case of the European Court of Justice gives guidance for the interpretation of "necessary" in the Data Protection Directive. According to the Court, necessity "is a concept which has its own independent meaning in Community law."[991] The Court emphasises that data processing must be proportionate to the goal pursued. For instance, if anonymous data can be used to achieve the same goal, no personal data should be retained.[992] As the Advocate General explains, the word necessary sets a higher threshold than "more convenient, easier or quicker."[993] The Advocate General refers to the case law of the other European Court, the European Court of Human Rights. The latter says "[t]he adjective 'necessary' is not synonymous with 'indispensable', neither has it the flexibility of such expressions as 'admissible', 'ordinary', 'useful', 'reasonable' or 'desirable' (…)."[994] Case law of the latter court confirms that data processing must be proportionate in relation to the processing purpose.[995]

---

[988] Article 29 Working Party 2013 (Google letter). See in more detail on the investigation into Google chapter 8, section 1.
[989] College bescherming persoonsgegevens (Dutch DPA) 2013 (Google), p 85. See along similar lines CNIL 2014 (Google), p. 24-25.
[990] In some cases firms can also rely on this ground prior to entering a contract. See article 7(a) of the Data Protection Directive. See also chapter 9, section 6, on article 15 of the Data Protection Directive.
[991] ECJ, C-524/06, Huber, 16 December 2008, par. 52.

It's sometimes suggested that "necessary" in the Data Protection Directive must always be interpreted as "necessary" in the case law of the European Court of Human Rights.[996] But caution is needed when interpreting this case law from Strasbourg and Luxembourg. In the *Huber* case of the European Court of Justice, the state was the data controller. The state didn't aim to rely on the legal basis contract, but on another legal basis: data processing is "necessary for the performance of a task carried out in the public interest" (article 7(e)).[997]

An argument can be made that firms should have more leeway than the state. Some might argue that people primarily need protection against the state, rather than against other private actors. This would suggest that "necessary" must be interpreted more leniently when there is a legal basis contract (article 7(b)), than when applying article 7(e), regarding processing for public interests. On the other hand, the aim of the state should be to work for the common good, while firms aim for profit. This would suggest that a firm should have less leeway.[998] Without taking sides in this debate, it's clear that it's not enough that a firm finds it helpful or profitable to process personal data; the concept of necessity requires more.

---

[992] ECJ, C-524/06, Huber, 16 December 2008, par. 60, par. 65-68, and dictum. As noted, the proportionality is one of the core principles of data protection law. See chapter 4, section 2.
[993] Opinion AG (3 April 2008) for ECJ, C-524/06, Huber, 16 December 2008, par. 29.
[994]ECtHR, Silver and Others v. United Kingdom, No. 5947/72; 6205/73; 7052/75; 7061/75; 7107/75; 7113/75; 7136/75, 25 March 1983, par 97.
[995] ECtHR, S. and Marper v. United Kingdom, No. 30562/04 and 30566/04. 4 December 2008, par. 103. See about "necessary in a democratic society" in the article 8 case law Harris et al. 2009, p. 349-359.
[996] For instance, the Dutch legislator interprets "necessary" in the Dutch Data Protection Act the same as "necessary" in the case law of the European Court of Human Rights, and the Dutch Data Protection Authority also takes this view. (See College bescherming persoonsgegevens 2013 (Google), p. 76-77). Some commentators take a similar view (see e.g. Kranenborg & Verhey 2011, p. 84; Bygrave & Schartum 2009, p. 163). See critically on interpreting "necessary" in data protection law the same way as in article 8 of the European Convention on Human Rights: González Fuster & Gutwirth 2013, p. 538.
[997] But see Bygrave, who suggests "necessary" in other data protection law provisions should probably be interpreted the same (Bygrave 2014, p. 150).

The question of necessity can be divided into two steps: subsidiarity and proportionality.[999] The subsidiarity question concerns whether the firm could pursue its purpose in another way that's less intrusive. The relevant question is whether a lighter measure is available. That lighter measure doesn't have to perform as well as the measure in question, according to the Advocate General in the *Huber* case. "It is not necessary for the alternative system to be the *most* effective or appropriate; it is enough for it to be able to perform adequately."[1000] The second question regarding necessity is whether the data processing is proportionate. In other words, do the measures not exceed the limits of what is appropriate and necessary in order to achieve the objective?[1001]

***Necessity for the performance of a contract***

The Working Party says that the legal basis contract isn't appropriate for behavioural targeting. The processing has to be genuinely necessary for providing the service in question. According to the Working party, "it is important to determine the exact rationale of the contract, i.e. its substance and fundamental objective, as it is against this that it will be tested whether the data processing is necessary for its performance."[1002] Therefore, in general, firms can't rely on the legal basis contract for behavioural targeting.[1003]

---

[998] See Gutwirth 2002, p. 38.
[999] See for instance College bescherming persoonsgegevens (Dutch DPA) 2013 (Google), p. 76-77; p. 87-88.
[1000] Opinion AG (3 April 2008) for ECJ, C-524/06, Huber, 16 December 2008, par 16 (emphasis original).
[1001] See on the proportionality principle in data protection law: chapter 4, section 2.
[1002] Article 29 Working Party 2014, WP 217, p. 17.
[1003] The Working Party's view that behavioural targeting can be based on article 7(b) doesn't receive much criticism in the literature. Google appears to invoke the legal basis contract for behavioural targeting, but Data

> Article 7(b) is not a suitable legal ground for building a profile of the user's tastes and lifestyle choices based on his click-stream on a website and the items purchased. This is because the data controller has not been contracted to carry out profiling, but rather to deliver particular goods and services, for example. Even if these processing activities are specifically mentioned in the small print of the contract, this fact alone does not make them "necessary" for the performance of the contract.[1004]

The analysis becomes more complicated if a firm uses the same personal data for behavioural targeting and to provide its service. Suppose a firm offers an app with a personalised news service. The app analyses the user's reading habits and recommends other news articles based on the user's earlier media consumption. Processing some personal data (the user's reading habits tied to a unique identifier) is necessary for performing the contract, as the app can only offer its personalised news service by analysing the user's personal data. That processing can be based on the legal basis contract (b), because the processing is necessary for the performance of the contract. But following the Working Party's reasoning, it's not necessary for provision of the personalised news service to use the same personal data for targeted advertising. Hence, the firm must obtain consent for behavioural targeting if the firm wants to use the same data to target ads to the user.[1005]

Perhaps a firm that provides a social network site could try to argue that it can base personal data processing for behavioural targeting on a contract.[1006] A social network site provider has a direct relationship with its user. The firm would have to argue that

---

Protection Authorities in France and the Netherlands reject this idea (CNIL 2014 (Google), p. 25; College bescherming persoonsgegevens 2013 (Google), p. 85-87).

[1004] Article 29 Working Party 2014, WP 217, p. 17.

[1005] See Article 29 Working Party 2013, WP 202, p. 13.

[1006] In some cases, the user of a social network site could be seen as a data controller, but we'll leave this complication aside (see Article 29 Working Party 2009, WP 163; Helberger & Van Hoboken 2010).

it entered a contract with the user when the user opened an account. And the firm would have to argue that behavioural targeting "is necessary for the performance of a contract" with the data subject (the user). The "contract" would imply that the user discloses personal data, in exchange for the use of the service.[1007]

Indeed, European social network providers have argued that personal data processing for behavioural targeting is "part of the processing that is necessary for the performance of a contract to which the data subject is party." They add "it is absolutely necessary to provide a legal basis for denying services to users that refuse to be the subjects of targeted advertising."[1008] Facebook makes a similar argument.[1009] But the Working Party says "[t]he user should be put in a position to give free and specific consent to receiving behavioural advertising, independently of his access to the social network service."[1010] Literature also suggests that the legal basis "necessary for the performance of a contract" must be interpreted narrowly.[1011]

If a firm could rely on a contract with the data subject as a legal basis for personal data processing for behavioural targeting, the tracking and further processing would be subject to the contract. Arguably Data Protection Authorities should be more cautious when interpreting the contents of a contract, than when explaining the requirements for consent, which is a sui generis construction of data protection law. It could be argued that for the interpretation of contracts, contract law and consumer law set out the primary guidelines. For instance, under consumer law a standard contract term is unfair if, contrary to the requirement of good faith, it causes a significant imbalance to the parties' rights and obligations, to the detriment of the consumer.[1012]

---

[1007] See on such "exchanges" chapter 7, section 2.
[1008] European Social Networks 2011, p. 5.
[1009] Facebook proposed amendments 2013, p. 27. Facebook proposes the following sentence for recital 34: "Controllers should be able to make consent to the processing a condition of access to a service which may not be otherwise free." See on such take-it-or-leave-it choices section 3 and 4 of this chapter, chapter 7, section 4, and chapter 8, section 3 and 5.
[1010] Article 29 Working Party, WP 187, p. 8; p. 18.
[1011] Kuner 2007, p. 243-244.
[1012] Article 3(1) of the Unfair Contract Terms Directive. As noted in chapter 4, section 4, some EU consumer law principles could be applied to the relation between firms and data subjects.

On the other hand, even if a firm had a legal basis for processing because of a contract, the firm would still have to comply with the other data protection requirements. Therefore, the idea that Data Protection Authorities have little to say about processing that's "necessary for the performance of a contract" isn't very plausible.

There's another reason why the difference between the legal bases consent (article 7(a)) and a contract (article 7(b)) is relevant.[1013] The procedural requirements for consent in data protection law are stricter than for many contracts. In principle, any expression of will is sufficient to enter a contract, although the law sometimes requires formalities.[1014] And in general contract law, terms and conditions are often part of the contract. But as discussed below, according to the Working Party firms can't obtain consent for data processing through terms and conditions.[1015]

While the difference between the legal bases contract and consent is relevant, in some ways it doesn't matter much which of the two is the legal basis for processing. Chapter 7 discusses practical problems with informed consent to behavioural targeting. These problems would be largely the same if firms could base personal data processing for behavioural targeting on a contract.

In conclusion, the Working Party says a firm can only rely on the legal basis contract if the processing is genuinely necessary to provide the service. The Working Party's view implies that, in general, firms can't rely on this legal basis for behavioural targeting.

---

[1013] Le Métayer & Monteleone 2009 argue that consenting in data protection law shouldn't be seen as entering a contract (p. 138). See on that topic also Verhelst 2012; Van Der Sloot 2010; Traung 2012, p. 38.
[1014] Zweigert & Kötz 1987, p. 366.
[1015] See section 3 of this chapter, and chapter 8, section 3. See also Article 29 Working Party, WP 187, p. 33-34.

### 6.2 Balancing provision

A second legal basis that a firm can rely on for personal data processing is the balancing provision, also called the legitimate interest clause. In brief, a firm can rely on the balancing provision when its legitimate business interests, or those of a third party, outweigh the data subject's fundamental rights. The relevant provision is as follows.[1016]

> Member States shall provide that personal data may be processed (…) if: (…)
>
> (f) processing is necessary for the purposes of the legitimate interests pursued by the controller or by the third party or parties to whom the data are disclosed, except where such interests are overridden by the interests or fundamental rights and freedoms of the data subject which require protection under article 1(1).[1017]

The balancing provision is the appropriate ground for innocuous standard business practices.[1018] Many data processing practices happen on a small scale and bring limited risks. For instance, a bakery shop might have a list of names and addresses of regular customers on its computer, for the purpose of sending New Year's greeting cards. Within the context of an existing customer relationship, a firm can generally rely on the balancing provision for postal direct marketing for similar products (first

---

[1016] Article 7(f) of the Data Protection Directive. The official English version of the Directive says "for" ("the interests for fundamental rights"). The Directive says "or" in other languages. Therefore I assume that "for" should be read as "or." (See Korff 2005, p. 68, footnote 19; Article 29 Working Party 2014, WP 217, p. 29. The proposal for a Data Protection Regulation also uses "or").

[1017] Article 1(1) of the Data Protection Directive says: "Member States shall protect the fundamental rights and freedoms of natural persons, and in particular their right to privacy." Therefore, any interest or fundamental right of the data subject could override the interests of the data controller.

[1018] See recital 30 of the Data Protection Directive.

party direct marketing).[1019] It's often assumed that postal direct marketing to non-customers (third party direct marketing) can also be based on the balancing provision.[1020]

The balancing provision is a very open norm and national Data Protection Authorities have diverging interpretations.[1021] To foster a more harmonised approach across Europe, the Working Party released a long and detailed opinion on the balancing provision in 2014.[1022]

***Legitimate interests***

Can firms base personal data processing for behavioural targeting on the balancing provision? Let's take a simple case as an example: an ad network tracks people's browsing behaviour over thousands of websites, to compile nameless individual profiles, to single people out and target them with advertising.

A preliminary question is whether the firm has a legitimate interest.[1023] As Gutwirth notes, "the ultimate purpose of the processing should be lawful. An illegal or illegitimate interest can never be pursued by a legitimate processing operation."[1024] By way of illustration, if a controller processes personal data with the goal of unlawfully discriminating against people, the interest can't be legitimate.[1025] A legitimate interest must be lawful. The "lawfully" requirement suggests the ad network must also

---

[1019] See for instance article 23(4) of the Data Protection Act of Poland. "The legitimate interests, referred to in [the balancing provision], are considered to be: (1) direct marketing of own products or services provided by the controller (..)." See on first party direct marketing also recital 41 of the e-Privacy Directive.

[1020] The European Commission amended proposal for a Data Protection Directive (1992) says: "This balance-of-interest clause is likely to concern very different kinds of processing, such as direct-mail marketing and the use of data which are already a matter of public record; Member States are to weigh the balance of interest in accordance with procedures which they are to establish taking account in particular of the general principles [of data protection] and of the rights of data subjects" (p. 15). See also Korff 1993, p. 7-8; Korff 2005, p. 43; Carey 2002, p. 106. Recital 39b of the LIBE Compromise, proposal for a Data Protection Regulation (2013) says that postal direct marketing can generally be based on the balancing provision, even if it's not first party marketing.

[1021] Irion & Luchetta 2013, p. 53; Korff 2010a, p. 72-73; Kuner 2007, p. 245; Impact Assessment for the proposal for a Data Protection Regulation (2012), Annex 2, p. 27. Traung 2012 calls the provision "circular nonsense" (p. 41).

[1022] Article 29 Working Party 2014, WP 217, p. 7.

[1023] Article 29 Working Party 2014, WP 217, p. 24-29; Article 29 Working Party 2013, WP 203, p. 12.

[1024] Gutwirth 2002, p. 99.

[1025] See Article 29 Working Party 2013, WP 203, p. 25.

comply with other laws, such as the e-Privacy Directive's consent requirement for tracking technologies.[1026] These requirements are also relevant when a firm relies on a legal basis other than the balancing provision, but the balancing provision emphasises that the firm's interests must be legitimate.

The ad network could invoke its right to conduct a business, as protected by the EU Charter of Fundamental Rights: "[t]he freedom to conduct a business in accordance with Union law and national laws and practices is recognised."[1027] But this right isn't absolute and has to be balanced against other fundamental rights, such as the right to privacy and the right to data protection.[1028] As an aside, a firm that breached data protection provisions or other legal rules wouldn't have a strong case if it invoked its right to conduct a business. Its business wouldn't be "in accordance with Community law and national laws", as required by the Charter.[1029] This implies, for instance, that the firm must comply with the e-Privacy Directive, which requires consent for the use of most tracking technologies.[1030]

The balancing provision speaks of legitimate interests pursued by "the third party or parties to whom the data are disclosed."[1031] If an ad network allows advertisers to advertise to specific people (identified with a cookie for instance), it essentially rents out access to those people. Under the Data Protection Directive, this should probably be seen as a type of data disclosure. The definition of processing speaks of "disclosure by transmission, dissemination *or otherwise making available*."[1032] The ad network makes data available for advertisers, including when it doesn't provide them with a copy of the data. Korff notes that list rental is a type of data disclosure, and his

---

[1026] Article 29 Working Party 2014, WP 217, p. 25.
[1027] Article 16 of the EU Charter of Fundamental Rights. The Advocate General of the European Court of Justice confirms that the provision of online advertising relates to the freedom to conduct a business (Opinion AG (25 June 2013), C-131/12, Google Spain, par 95).
[1028] Article 52(3) of the EU Charter of Fundamental Rights. See also CJEU, C-70/10, Scarlet v Sabam, 24 November 2011, par. 46. The Google Spain case suggests that a firm's economic interests have less weight than the data subject's privacy rights (CJEU, C-131/12, Google Spain, 13 May 2014, par. 81, dictum, 4).
[1029] Article 16 of the EU Charter of Fundamental Rights.
[1030] See section 4 of this chapter.
[1031] The Data Protection Directive defines 'third party" in article 2(f).
[1032] Article 2(b) of the Data Protection Directive.

conclusion can be applied to ad networks by analogy.[1033] In any case, the analysis of the balancing provision remains roughly the same, regardless of whether a firm invokes its own interests, or those of third parties. Let's assume that the ad network in our example has a legitimate interest.[1034]

***Necessity***

For a firm to be able to rely on the balancing provision, having a legitimate interest is not enough; the processing must be "necessary." As noted, the question of necessity can be divided into two steps: subsidiarity and proportionality.[1035] Regarding subsidiarity: it seems questionable whether tracking people's browsing behaviour is the least intrusive manner possible for the ad network to enable advertisers to promote their products or services. There are many other types of online advertising that are less privacy-invasive, such as contextual advertising (advertising about cars on websites about cars). But an ad network that specialises in behavioural targeting could try to argue that the tracking is necessary for its business model. However, it doesn't follow that the ad network has to track people's browsing behaviour and construct detailed profiles. For the ad network, other ways of pursuing its interests may include finding a way that involves processing less personal data.[1036]

The second question regarding necessity is whether the tracking and further processing is proportionate in relation to the ad network's interests. The processing is disproportionate if it exceeds the limits of what is appropriate to pursue the ad networks business interests.[1037] For some behavioural targeting practices, which entail

---

[1033] Korff 2005, p. 63. With list rental, a list broker sends leaflets to a set of people, but the advertiser doesn't receive a copy of the list. See chapter 2, section 6.

[1034] See Article 29 Working Party 2014, WP 217, p. 25: marketing is a legitimate interest.

[1035] See for instance College bescherming persoonsgegevens (Dutch DPA) 2013 (Google), p. 76-77; p. 87-88.

[1036] Privacy enhancing technologies could help here (see Article 29 Working Party 2014, WP 217, p. 42). See on privacy preserving behavioural targeting chapter 9, section 3.

[1037] See on the proportionality principle in data protection law: chapter 4, section 2.

large-scale collection of detailed information about people, it seems questionable whether they are proportionate.[1038]

If the tracking and further processing is "necessary" for the ad network's legitimate interests, the ad network must pass another hurdle. The balancing provision requires that the ad network's interests "must not be overridden by the fundamental rights and freedoms of the data subject."[1039] The interests of the firm and the data subject must be weighed. When balancing the conflicting interests, it has to be taken into account that the right to data protection and the right to privacy are fundamental rights.[1040]

People have an interest in using the internet without being tracked. Many people find tracking and behavioural targeting intrusive.[1041] Collecting and storing data can cause a chilling effect, and large-scale data storage brings risks, such as data breaches. In some cases there could be a risk of unfair discrimination or manipulation.[1042] People have a reasonable expectation of privacy regarding their internet use, and storage of information about internet use can interfere with the right to private life, regardless of how those data are used.[1043] A Council of Europe resolution suggests that online tracking is a privacy threat:

> [P]ersonal ICT systems as well as ICT-based communications may not be accessed or manipulated if such action violates privacy or the secrecy of correspondence; access or manipulation through "cookies" or other unauthorised

[1038] See also chapter 9, section 3, and Kuner 2008.
[1039] Article 7(f) of the Data Protection Directive. This requirement could be seen as a separate, or second, balancing test. See CJEU, C-468/10 and C-469/10, ASNEF, 24 November 2011, par. 38; College bescherming persoonsgegevens (Dutch DPA) 2013 (Google), p. 88. The Working Party distinguishes more steps within the balancing provision (Article 29 Working Party 2014, WP 217).
[1040] CJEU, C-468/10 and C-469/10, ASNEF, 24 November 2011, par. 41. See also ECJ, C-465/00, C-138/01 and C-139/01, Österreichischer Rundfunk, 20 May 2003, par. 68; CJEU, C-131/12, Google Spain, 13 May 2014, par. 74.
[1041] See chapter 7, section 1, for a review of research on people's attitudes towards behavioural targeting.
[1042] See chapter 3, section 3.
[1043] ECtHR, Copland v. United Kingdom, No. 62617/00, 3 April 2007, par. 42. See the case law discussed in chapter 3, section 2.

> automated devices violate privacy, in particular where such automated access or manipulation serves other interests, especially of a commercial nature.[1044]

But the data subject's rights aren't absolute: "under certain conditions, limitations may be imposed", says the European Court of Justice. Therefore, "a fair balance [must] be struck between the various fundamental rights and freedoms protected by the EU legal order."[1045]

When balancing the opposing interests, all circumstances have to be taken into account, such as "the seriousness of the infringement of the data subject's fundamental rights."[1046] Relevant factors can include the sensitivity of the data, the scale of data collection, the reasonable expectations of the data subject, and the risks involved.[1047] For instance, mobile location data are of a rather sensitive nature. Firms can never rely on the balancing provision for processing special categories of data, such as data regarding political opinions or health, as the Data Protection Directive requires "explicit consent" for processing special categories of data for marketing purposes.[1048] The safeguards a firm has in place to protect the data subject's interests should also be taken into account. For instance, does the firm offer sufficient transparency, and does it offer a clear opt-out option?[1049]

In most cases the data subject's interests must prevail over the ad network's interests, as behavioural targeting involves collecting and processing information about personal matters such as people's browsing behaviour. Several authors have already

---

[1044] Parliamentary Assembly, Resolution 1843 (2011) The protection of privacy and personal data on the Internet and online media, 7 October 2011, par 18.6.
[1045] CJEU, C-468/10 and C-469/10, ASNEF, 24 November 2011, par. 43.
[1046] CJEU, C-468/10 and C-469/10, ASNEF, 24 November 2011, par. 44.
[1047] See Article 29 Working Party 2014, WP 217, p. 33-43.
[1048] Article 8 of the Data Protection Directive. There are exceptions for the "explicit consent" requirement, but these aren't relevant for behavioural targeting. Some member states don't accept consent as a legitimate basis for processing special categories of data. See on special categories of data chapter 5, section 6; chapter 9, section 5.
[1049] Article 29 Working Party 2014, WP 217, p. 41. See also WP 185, p. 16; Korff 2005, p. 43; College bescherming persoonsgegevens (Dutch DPA) 2013 (Google), p. 89. See on opting out below, on the right to object.

concluded that ad networks can't rely on the balancing provision for behavioural targeting that involves tracking over multiple websites.[1050] The Dutch lawmaker comes to the same conclusion.[1051] Similarly, the Working Party says that "free, specific, informed and unambiguous 'opt-in' consent (…) should be required, for example, for tracking and profiling for purposes of direct marketing, behavioural advertis[ing], data-brokering, location-based advertising or tracking-based digital market research."[1052] In sum, the most convincing view is that personal data processing for behavioural targeting that relies on following people over various internet services, can't be based on the balancing provision.

It has also been suggested that in some circumstances, firms might be able to base data processing for first party behavioural targeting on the balancing provision. For instance, perhaps an online bookstore that tracks people's behaviour within its website to provide recommendations might be able to rely on the balancing provision. Arguably people are more likely to understand what happens when they see behaviourally targeted ads, which are based on browsing behaviour within one website.[1053]

***Right to object***

The Data Protection Directive grants data subjects the right to object "on compelling legitimate grounds" to the processing of their data when firms rely on the balancing provision. If there's a "justified objection", the processing may no longer involve those data.[1054] This right is thus not an absolute right, but a qualified right to object.

---

[1050] See Koëter 2009; Traung 2010, p. 218; Antic 2012, p. 106; Moerel 2014, p. 58; Van Der Sloot 2011.
[1051] See for an English translation of the relevant remarks of the Dutch legislator: College bescherming persoonsgegevens (Dutch DPA) 2013 (Google), p. 81, footnote 294.
[1052] Article 29 Working Party 2013, WP 203, p. 46. See similarly Article 29 Working Party 2014, WP 217, p. 45; Article 29 Working Party 2013 (draft LIBE comments), p. 4; Article 29 Working Party 2013 (Google letter), Appendix, p. 4.
[1053] See Koëter 2009, p. 109-111. In the US, the Federal Trade Commission also says first party marketing could be allowed without consent, while third party marketing requires consent (Federal Trade Commission 2012, p. 44).
[1054] Article 14(a) of the Data Protection Directive.

Therefore, the data subject's reasons for objecting must be balanced against the legitimate interests of the firm.[1055]

In the case of direct marketing, the Data Protection Directive grants data subjects the right to object, without requiring the data subject to have "compelling legitimate grounds." This right to object to direct marketing must be interpreted as an absolute right to object.[1056] As Korff puts it, the Data Protection Directive "speaks of a right to 'object to' rather than a right to prevent or stop the processing in question, but it is clear that the latter is intended. If a data subject exercises the right to object to direct marketing (…), the controller in question must comply with that objection."[1057]

Behavioural targeting is a form of direct marketing, as confirmed in the code of conduct of the Federation of European Direct and Interactive Marketing for the use of personal data in direct marketing, which is approved by the Working party. "Direct marketing in the on-line environment refers to one-to-one marketing activities where individuals are targeted."[1058] The Council of Europe Recommendation on profiling confirms that people have an absolute right to object to profiling for direct marketing (in cases where the profiling doesn't require consent).[1059]

---

1055 See CJEU, C-131/12, Google Spain, 13 May 2014, par. 76.
1056 See Article 29 Working Party 2013, WP 203, p. 35.
1057 Korff 2005, p 100. Article 14 of the Data Protection Directive is somewhat difficult to read, and provides to alternative possibilities for member states to implement the right to object. Korff 2005 provides an analysis. See also Article 29 Working Party 2013, WP 203, p. 35.
1058 Capitalisation adapted. The Working Party approved the code in Article 29 Working Party 2010, WP 164. The FEDMA defines direct marketing as follows. "The communication by whatever means (including but not limited to mail, fax, telephone, on-line services etc.) of any advertising or marketing material, which is carried out by the direct marketer itself or on its behalf and which is directed to particular individuals" (code approved in Article 29 Working Party 2003, WP 77).
1059 Article 5(3) of Committee of Ministers, Recommendation (2010)13 to member states on the protection of individuals with regard to automatic processing of personal data in the context of profiling, 23 November 2010. The Recommendation applies to behavioural targeting (see the profiling definition in article 1(e)), and Polakiewicz 2013.

***Proposal for a Data Protection Regulation***

The European Commission proposal for a Data Protection Regulation duplicates the balancing provision without major changes.[1060] But the proposal requires a firm that relies on the balancing provision to provide the data subject with information about the legitimate interests pursued by the firm.[1061] The requirement to give this information could already be read in the current regime, as a firm is required to provide all information that's necessary to guarantee fair processing.[1062] But that requirement is rather vague, so it's useful that the proposal requires firms to inform the data subject about how they apply the balancing provision.[1063]

The LIBE Compromise allows firms, under certain conditions, to rely on the balancing provision for behavioural targeting with pseudonymous data.[1064] The Working Party warns that the LIBE Compromise could be misunderstood as allowing firms to base most behavioural targeting practices on the balancing provision, as long as firms use pseudonymous data.[1065]

In conclusion, under current law, personal data processing for behavioural targeting, in particular if it involves tracking an internet user over multiple websites, generally can't be based on the balancing provision. If, in rare circumstances, a firm could rely on the balancing provision for behavioural targeting, the data subject would have the right to stop the data processing: to opt out.

---

[1060] But see Purtova, who argues that the proposal tilts the balance in favour of data controllers in the new version of the balancing provision (Purtova 2014).
[1061] Article 6(f) and article 14(b) of the European Commission proposal for a Data Protection Regulation (2012).
[1062] Article 10 and 11 of the Data protection Directive. See chapter 4, section 3.
[1063] Like the Data Protection Directive, the European Commission proposal for a Data Protection Regulation (2012) uses ambiguous language to describe the right to object to the use of personal data for direct marketing (article 19(1) and 19(3)).
[1064] See article 2(a), article 6(f), and recitals 38 and 58a of the LIBE Compromise, proposal for a Data Protection Regulation (2013). The LIBE Compromise also requires a "highly visible" opt-out possibility (article 20(1); see also article 19(2)).
[1065] Article 29 Working Party 2013 (draft LIBE comments).

### 6.3 Consent for personal data processing

If firms want to process personal data, and can't base the processing on the balancing provision or another legal basis, they must ask the data subject for consent. Consent is defined as "any freely given specific and informed indication of his wishes by which the data subject signifies his agreement to personal data relating to him being processed."[1066] People can always withdraw their consent.[1067]

***Indication of wishes***

Consent must be an indication of the data subject's wishes. If there's no indication of wishes there can't be consent, so there's no need to check the other requirements for consent. The predominant view in general contract law is that an indication of wishes can be expressed in any form, and can also be implicit.[1068] Consent in data protection law can also be given in any form.[1069] For instance, dropping ones business card in a bowl with a sign saying "leave your name and address to receive our monthly newsletter" can imply consent to the processing of some personal data.[1070]

Without special circumstances, mere inactivity isn't an indication of wishes. "Consent cannot be inferred from the absolute silence of the data subject," summarises Kosta.[1071] A Council of Europe Resolution confirms that consent for online data processing "requires an expression of consent in full knowledge of such use, namely the manifestation of a free, specific and informed will, and excludes any automatic or tacit usage."[1072]

---

[1066] Article 2(h) of the Data Protection Directive.
[1067] European Commission amended proposal for a Data Protection Directive (1992), p. 2. See also Kosta 2013a, p. 251, with further references. The European Commission proposal for a Data Protection Regulation (2012) makes the right to withdraw consent explicit in article 7(2).
[1068] Zweigert & Kötz 1987, p. 688.
[1069] Kuner 2007, p. 68; Kosta 2013a, p. 386; Article 29 Working Party, WP 187, p. 11.
[1070] Article 29 Working Party, WP 187, p. 11.
[1071] Kosta 2013a, p. 256. See also Kuner 2007, p. 69.
[1072] Parliamentary Assembly, Resolution 1843 (2011) The protection of privacy and personal data on the Internet and online media, 7 October 2011, par 18(4). The Resolution is not legally binding.

In *Schecke,* the European Court of Justice says that merely informing a person that data processing will take place "thus does not seek to base the personal data processing (…) on the consent" of the data subject.[1073] The Advocate General is more explicit. "Acknowledging prior notice that publication of some kind will happen is not the same as giving 'unambiguous' consent to a particular kind of detailed publication. Nor can it properly be described as a 'freely given specific indication' of the applicants' wishes in accordance with the definition of the data subject's consent in article 2(h)."[1074] Other judgments of the European Court of Justice confirm that consent cannot easily be assumed.[1075]

In case law outside the field of data protection law, the European Court of Justice affirms that consent can't be inferred from inactivity. For instance, in two trademark cases, "implied consent (…) cannot be inferred from (…) mere silence", [1076] and "'consent' (…) must be so expressed that an intention to renounce a right is unequivocally demonstrated."[1077] In a case where the European Commission didn't initiate an infringement procedure, this inactivity "cannot be interpreted as the Commission's tacit consent."[1078]

Likewise, in general contract law, mere silence doesn't constitute an indication of will. According to the Vienna Sales Convention for instance, "[a] statement made by or other conduct of the offeree indicating assent to an offer is an acceptance. Silence or inactivity does not in itself amount to acceptance".[1079] Several proposals for international contract law use the same phrase.[1080] Indeed, it would have peculiar

---

[1073] CJEU, C-92/09 and C-93/09, 9 November 2010, Volker und Markus Schecke and Eifert, par. 63.
[1074] Opinion AG (17 June 2010) for CJEU, C-92/09 and C-93/09, 9 November 2010, Volker und Markus Schecke and Eifert, par. 79.
[1075] The Court suggests that "consent" in the Data Protection Directive requires "express" consent (CJEU, C-28/08 and T-194/04, Bavarian Lager, 29 June 2010). And the Court reads "an opportunity to determine" as requiring "prior", "free, specific and informed consent" (CJEU, C-543/09, 5 May 2011, Deutsche Telekom, par. 55-58).
[1076] ECJ, C-414/99 to C-416/99, 20 November 2001, Zino Davidoff, par. 55.
[1077] CJEU, C-482/09, 22 September 2011, Budějovický Budvar, par. 42-44.
[1078] CJEU, C-577/08, 29 June 2010, Brouwer, par. 39.
[1079] Article 18(1) of the Vienna Convention on International Sale of Goods.
[1080] The same phrase is used in article II 4:204(2) of the Draft Common Frame of Reference (Principles, Definitions and Model Rules of European Private Law), and article 34 (of Annex 1) of European Commission

results if the law allowed a seller to infer an expression of will from mere silence. A shop owner could demand payment if somebody failed to object to an offer to buy a TV.

After the European Commission presented its first proposals for a Data Protection Directive in the early 1990s, firms argued that giving people the possibility to object should suffice in order to obtain consent. The International Chamber of Commerce, a business lobbying organisation, said for instance: "[s]ince new products and services constantly emerge, it is virtually impossible for the customer or the controller (…) to foresee at the outset all the specific applications for which the customer's data could be used"[1081] If the law required unambiguous consent, "[c]ompanies would be faced with administrative burdens and potential delays in introducing new services."[1082] The International Chamber of Commerce added that opt-out systems should suffice in order to obtain consent.

> It is far more common to employ a notice or 'opt out' approach, under which individuals are informed of the use to be made of personal data and have the opportunity to object to those uses. Such an approach, or other forms of implied consent, would offer individuals an effective protection of their personal data without putting undue restrictions on all use of personal information.[1083]

The EU lawmaker didn't follow such suggestions in the final text of the 1995 Directive.[1084] The 2012 European Commission proposal for a Data Protection

---

2011 (proposal Common European Sales Law). Inertia selling, where a firm sends consumers a product and demands payment if they don't return the product, is banned in the Consumer Rights Directive (article 27).

[1081] International Chamber of Commerce 1992, p. 261.

[1082] International Chamber of Commerce 1992, p. 261. See for a similar argument regarding the 2012 proposals: Amazon proposed amendments.

[1083] International Chamber of Commerce 1992, p. 261.

[1084] Kosta 2013a, p. 83-108.

Regulation has led to comparable lobbying in favour of opt-out systems. The arguments used are still remarkably similar to those in the 1990s, although nowadays they're usually coupled with remarks about "big data."[1085]

In the UK regulators and commentators seem to be more inclined to accept a system that allows people to object – an opt-out system – as a way of obtaining "implied consent."[1086] For instance, the English Information Commissioner's Office (ICO), the regulator that oversees compliance with the e-Privacy Directive, drops cookies through its website as soon as a visitor arrives, and explains in a banner that it has done so. The ICO appears to suggest that explaining how a user can delete cookies is enough to obtain "implied consent."[1087] The English notion of implied consent has led to an infringement proceeding by the European Commission. In brief, the English implementation of the e-Privacy Directive accepted a form of implied consent as a justification to interfere with the confidentiality of communications. This became salient when a firm called Phorm assumed that people had consented to deep packet inspection for behavioural targeting. The European Commission closed the infringement proceeding after the United Kingdom amended its law.[1088]

Viewing an opt-out system as sufficient to obtain consent has been met with criticism in literature. For example, Kosta says "there is no such thing as 'opt-out consent'."[1089] She adds that "reference to 'opt-out' consent is a misnomer. An 'opt-out' regime refers to the right of a data subject to object to the processing of his personal data and does not constitute consent."[1090] The Working Party confirms that consent needs

---

[1085] See for instance Interactive Advertising Bureau United Kingdom 2012; Amazon proposed amendments; International Chamber of Commerce 2013.

[1086] Kosta 2013a, p. 192. See also Impact Assessment for the proposal for a Data Protection Regulation (2012), p. 19.

[1087] The banner says: "We have placed cookies on your computer to help make this website better. You can change your cookie settings at any time. Otherwise, we'll assume you're OK to continue" (Information Commissioner's Office 2013a)

[1088] European Commission 2009; European Commission 2012. The new law only allows interception where both the sender and recipient have consented to it (The Regulation of Investigatory Powers (Monetary Penalty Notices and Consents for Interceptions) Regulations 2011 You are here: 2011 No. 1340). See on Phorm chapter 2, section 2. See also McStay 2011, p. 15-42; Bernal 2011.

[1089] Kosta 2013a, p. 202.

[1090] Kosta 2013a, p. 387. See also Traung 2012; McStay 2012.

affirmative action. “There are in principle no limits as to the form consent can take. However, for consent to be valid it should be an active indication of the user’s wishes.”[1091]

The difference between direct marketing that’s based on the balancing provision (on an opt-out basis) and direct marketing that’s based on the legal basis consent (opt-in) isn’t merely theoretical. The balancing provision sometimes allows firms to process personal data for direct marketing on an opt-out basis, but in such cases the provision requires the firm to weigh the interests involved. By relying on fictitious opt-out consent, firms could try to escape the responsibility to balance its interests against those of the data subject.[1092]

A number of larger behavioural targeting firms, cooperating in the Interactive Advertising Bureau, offer people the chance to opt out of targeted advertising on a centralised website: youronlinechoices.com. But under this scheme, participating firms may continue to process information about people (phase 1 and 2 of behavioural targeting), as they merely promise to stop showing targeted advertising (phase 5) after people object.[1093] In short, the website offers the equivalent of Do Not Target, rather than Do Not Track.[1094] But even if the opt-out system did stop firms from tracking people, it’s hard to see how such an opt-out system could meet data protection law’s requirements for consent.[1095]

The Data Protection Directive says that consent must be “unambiguous.” This seems superfluous. As Kosta puts it, “the element that the consent has to be given unambiguously should be intrinsic in the concept of consent in order for it to qualify

---

[1091] Article 29 Working Party 2013, WP 208, p. 3.
[1092] The legal basis consent doesn’t legitimise disproportionate data processing. See section 5 of this chapter, and chapter 9, section 2.
[1093] The opt-out page of the Internet Advertising Bureau says: “Declining behavioral advertising only means that you will not receive more display advertising customised in this way” (Interactive Advertising Bureau Europe – Youronlinechoices).
[1094] See on the difference between Do Not Track (/Do Not Collect) and Do Not Target chapter 8, section 5.
[1095] Article 29 Working Party 2011, WP 188, p. 6.

as valid."[1096] The word "unambiguous" seems to have led to confusion.[1097] Some appear to believe that non-unambiguous consent – if there were such a thing – can be given by failing to object. Views along these lines were expressed in discussions about the e-Privacy Directive's consent requirement for tracking technologies (see section 4 of this chapter).

In sum, consent to personal data processing requires an "indication of wishes" to be valid. In some circumstances consent can be implied, but mere silence doesn't signify consent. The European Commission proposal for a Data Protection Regulation tightens the requirements for consent, and always requires consent to be explicit (see chapter 8).[1098] Just like in the 1990s, firms have reacted to the 2012 proposal by lobbying for a regime that accepts "implied consent."[1099]

***Specific and informed***

The Data Protection Directive also requires consent to be "specific" and "informed."[1100] Specific means that consent "must relate to a particular data processing operation concerning the data subject carried out by a particular controller and for particular purposes."[1101] For instance, consent to use personal data "for commercial purposes" would be too vague.[1102] The Working Party confirms that "blanket consent without specifying the exact purpose of the processing is not acceptable."[1103]

---

[1096] Kosta 2013a, p. 235.
[1097] See Traung 2012, p. 38.
[1098] Chapter 8, section 3, discusses the proposals regarding consent.
[1099] See on lobbying in the 1990s chapter 4, section 1. For examples of lobbying regarding consent and the 2012 proposals, see Facebook proposed amendments 2013, p. 23; Amazon proposed amendments (article 4(1)(8); International Chamber of Commerce 2013, p. 3; eBay proposed amendments 2012.
[1100] Kosta suggests that "specific" and "informed" are largely overlapping, and that the requirement of specificity may be superfluous (Kosta 2013a, p. 224).
[1101] European Commission amended proposal for a Data Protection Directive (1992), p. 12. See also Article 29 Working Party 2013, WP 202, p. 15.
[1102] See European Commission amended proposal for a Data Protection Directive (1992), p. 15. The European Commission gives "for commercial purposes" as an example of a processing purpose which isn't specified. But the same example can be applied to "specific" consent.
[1103] Article 29 Working Party, WP 187, p. 17.

Consent has to be informed. In a case on working hours (not regarding data protection law), the European Court of Justice required "full knowledge of all the facts" for consent to be valid.[1104] A firm can't establish whether somebody *is* informed when he or she consents. For instance, a firm can never guarantee that people read the text of a consent request. But as transparency is a precondition for valid consent, firms must provide information in accordance with the requirements of data protection law. If a consent request doesn't clearly explain how the firm wants to use the data, the consent can't be informed.

Obtaining consent of a data subject must be distinguished from the transparency requirement. The Data Protection Directive always requires data controllers to be transparent about data processing, whether they rely on consent or not.[1105] It's not possible to obtain consent by silently changing a privacy policy. If a data subject doesn't know about new terms and conditions, there can't be an expression of will.[1106] It would be absurd to argue that the person consented.

***Freely given***

Consent must be freely given, so consent given under pressure isn't valid. As Kosta puts it, "consent of the data subject is still freely given when positive pressure is exercised, while the exercise of any kind of negative pressure renders the consent invalid."[1107] An extreme example of negative pressure is holding a gun to somebody's head while asking whether he or she consents. The consent wouldn't be free. But to make consent involuntary, pressure doesn't have to be so great. For instance, if an employer asks an employee for consent, the consent might not be sufficiently free,

---

[1104] ECJ, C-397/01 en C-403/01, Pfeiffer and others, 5 October 2004, dictum (2) and par. 82.
[1105] See article 11 of the Data Protection Directive. This requires information "where the data have not been obtained from the data subject." In such cases there's no consent. See chapter 4, section 3.
[1106] As Radin puts it, there would be "sheer ignorance" on the side of the user (Radin 2013, p. 19-21).
[1107] Kosta 2013a, p. 256.

because of the imbalance of power.[1108] And the European Court of Justice says people applying for passports can't be deemed to have freely consented to have their fingerprints taken, because people need a passport.[1109]

But positive pressure is generally allowed. For instance, in most circumstances, data protection law probably allows firms to entice people to consent by offering something in return, such as a discount.[1110] In principle, a firm is allowed to say: you can use this service if you consent to being tracked. But it can be difficult to differentiate between positive and negative pressure, for instance if a data controller offers a take-it-or-leave-it choice. A service could be so important that people have no genuine choice not to use it. Bygrave suggests that the requirement of fair data processing implies that firms shouldn't pressure people too much into disclosing data, and that firms shouldn't abuse their market power.[1111] The European Data Protection Supervisor and national Data Protection Authorities have voiced similar opinions.[1112] The voluntariness of consent is discussed in more detail in the next section.[1113]

### 6.4 Consent for tracking technologies

European legal discussions on behavioural targeting often focus on the e-Privacy Directive's consent requirement for tracking technologies, rather than on the general data protection rules. The 2002 e-Privacy Directive was updated in 2009.[1114]

Article 5(3) of the e-Privacy Directive applies to anyone that wants to store or access information on a user's device, including if no personal data are involved.[1115] The

---

[1108] Kosta 2013a, p. 386; Article 29 Working Party, WP 187, p. 13-14. See also Naczelny Sąd Administracyjny [Supreme Administrative Court], 1 December 2009, I OSK 249/09 (Inspector General for Personal Data Protection), English translation: <www.giodo.gov.pl/417/id_art/649/j/en/> accessed 28 May 2014.
[1109] CJEU, C-291/12, Schwartz v. Stadt Bochum, 17 October 2013, par. 32.
[1110] See European Agency for Fundamental Rights 2014, p. 59.
[1111] Bygrave 2002, p. 58-59.
[1112] European Data Protection Supervisor 2011, p. 13-15.
[1113] See also section 3 and 4 of chapter 7, and section 3 and 5 of chapter 8.
[1114] The e-Privacy Directive 2002/58 was updated by Directive 2009/136. This study refers to the consolidated version from 2009.

preamble shows that article 5(3) aims to protect the device itself and its contents against unauthorised access. "Terminal equipment of users of electronic communications networks and any information stored on such equipment are part of the private sphere of the users requiring protection under the European Convention for the Protection of Human Rights and Fundamental Freedoms."[1116] The Working Party confirms that the provision applies, for instance, to apps that access information on a user's smartphone, such as location data or a user's contact list.[1117]

Another rationale for article 5(3) is protecting the user's device against parties that want to store information on a user's device, without the user's knowledge. The provision aims, for instance, to protect people against the secret installation of adware or spyware. Yet another rationale is protecting the user against surreptitious tracking, as explained in the preamble.[1118]

> So-called spyware, web bugs, hidden identifiers and other similar devices can enter the user's terminal without their knowledge in order to gain access to information, to store hidden information or to trace the activities of the user and may seriously intrude upon the privacy of these users.[1119]

Early proposals for the 2002 version of the e-Privacy Directive required firms to ask for consent before they placed certain kinds of cookies. After fierce lobbying by the marketing industry, the final version used ambiguous wording about a "right to refuse." The 2002 version of article 5(3) is usually interpreted as an opt-out

---

[1115] A user (article 2(a) of the e-Privacy Directive) isn't the same as a "subscriber" (article 2(k) of the Framework Directive 2002/21). We'll leave this complication aside for this study.
[1116] Recital 24 of the e-Privacy Directive.
[1117] Article 29 Working Party 2013, WP 202, p. 10.
[1118] See e.g. recital 24 and 25 of the e-Privacy Directive, and recital 65 and 66 of Directive 2009/136. See also Kierkegaard 2005; Kosta 2013.
[1119] Recital 24 of the e-Privacy Directive. Recital 25 adds that "so-called 'cookies', can be a legitimate and useful tool, for example, in analysing the effectiveness of website design and advertising."

system.[1120] Websites had an obligation to clearly inform people about the use of cookies, but few websites did.

***2009 revision***

Since 2009, article 5(3) of the revised e-Privacy Directive, sometimes called the Cookie Directive,[1121] requires firms to obtain the user's consent before using tracking technologies such as cookies. The general rule can be summarised as follows. Firms that want to store or access a cookie on a user's device must (i) give the user clear and complete information about the cookie's purpose, and (ii) obtain the user's consent. Certain functional cookies are exempted from the information and consent requirements. For example, no consent is needed for a cookie for a digital shopping cart or for a log-in procedure.[1122] For the definition of consent, the e-Privacy Directive refers to the definition in the Data Protection Directive: a free, informed, specific indication of will.[1123]

For ease of reading this study speaks of consent for "cookies" or for "tracking technologies", but article 5(3) applies to any information that can be stored on a user's device. Article 5(3) thus also applies to spyware and adware. Hence, if a firm wants to install adware, for instance coupled with a browser toolbar, it must give clear and comprehensive information to the user, and obtain the user's consent.[1124] It follows from the preamble of the amending directive that the provision also applies when spyware or similar files are distributed on USB sticks, music CDs etc.[1125]

---

[1120] Kierkegaard 2005. Some authors read the 2002 version as an opt-in system (see Traung 2010; Helberger et al. 2011).

[1121] See e.g. McStay 2012.

[1122] See in detail on the exempted cookies Article 29 Working Party 2012, WP 194.

[1123] Article 2(f) and recital 17 of the e-Privacy Directive.

[1124] The Dutch Telecommunications authority imposed a 1 million euro fine on a spyware distributor. On appeal, the fine was overturned (College van Beroep voor het bedrijfsleven [Trade and Industry Appeals Tribunal], 20 June 2013, ECLI:NL:CBB:2013:CA3716 (Dollarrevenue/Autoriteit Consument en Markt).

[1125] See recital 65 of Directive 2009/136. The provision would apply for instance to the CDs distributed by SONY in 2005, which installed spyware when people put the CD in their computer (Russinovich 2005). It has been argued that the provision also applies to accessing information in a digital TV decoder for behavioural targeting (Minister of Economic Affairs, Agriculture and Innovation of the Netherlands 2012).

Who has to comply? Article 5(3) states: “anyone” that wants to access information stored in a users’ device, or wants to store information in a user’s device. In principle, it’s the firm operating the cookie (such as an ad network) that must obtain consent. But from the beginning, the Working Party has said that a website publisher that allows third parties to place cookies shares the responsibility for information and consent.[1126]

The firm operating the cookie, or the website publisher, must at least explain the cookie’s purpose. The e-Privacy Directive says the information provided to users must be “clear and comprehensive” and must be in accordance with the Data Protection Directive. The latter requires more information if this is necessary to guarantee fairness.[1127] The Working Party gives several examples of how firms could ask for informed consent, including a pop-up window.[1128]

In short, article 5(3) requires informed consent for the use of most tracking technologies that are used for behavioural targeting. A problem with article 5(3) is that the provision is over inclusive. For instance, the provision also requires consent for many cookies that aren’t used for tracking people across the web. Chapter 8 returns to this topic.[1129]

***Browser settings***

A sentence from recital 66 of the 2009 directive that amended the e-Privacy Directive has caused much confusion and discussion. The recital says people can express consent with their browser under certain circumstances:

---

[1126] Article 29 Working Party 2010, WP 171, p. 24.
[1127] Article 5(3) of the e-Privacy Directive; article 10 and 11 of the Data Protection Directive.
[1128] Article 29 Working Party 2011, WP 188, p. 9-11.
[1129] Chapter 8, section 4.

> Where it is technically possible and effective, in accordance with the relevant provisions of [the Data Protection Directive], the user's consent to processing may be expressed by using the appropriate settings of a browser or other application.[1130]

Most browsers offer users the possibility to block first party cookies, third party cookies, or all cookies. Some conclude from recital 66 that default browser settings could be relied upon as an expression of consent for tracking cookies. For instance, the Interactive Advertising Bureau UK says: "We believe that default web browser settings can amount to 'consent' (…)".[1131] Perhaps the fact that the e-Privacy Directive doesn't speak of "unambiguous" consent has contributed to the confusion. In line with data protection law's requirement of an expression of will for valid consent, the Working Party has repeatedly rejected the idea that default settings of browsers could signify consent:[1132]

> Where the website operator can be confident that the user has been fully informed and actively configured their browser or other application then, in the right circumstances, such a configuration, would signify an active behaviour and therefore be respected by the website operator. (…) The process by which users could signify their consent for cookies would be through a positive action or other active behaviour, provided they have been fully informed of what that action represents.[1133]

---

[1130] Recital 66 of Directive 2009/136.
[1131] Interactive Advertising Bureau United Kingdom 2012 (emphasis original).
[1132] See e.g. Article 29 Working Party, WP 187, p. 32.
[1133] Article 29 Working Party 2013, WP 208, p. 4 (emphasis original).

Many commentators agree that default browser settings can't signify a specific and informed indication of wishes. It's unlikely that all people who that do *not* tweak their browser's default settings want to consent to all kinds of cookies. There wouldn't be an expression of wishes. And if a browser accepts a lot of cookies, including for the future, such "consent" can't be informed and specific.[1134] In addition, if browser settings could be relied upon for an expression of consent, this would imply that a party could assume that users consent to spyware or viruses if their browsers don't block such files.[1135]

There are more arguments against relying on default browser settings as a consent mechanism. For example, browser settings are merely mentioned in a recital.[1136] The informed consent requirement is laid down in article 5(3) of the e-Privacy Directive. Case law and literature suggest that if a recital and an article contradict each other, and both have a clear meaning, the article must prevail.[1137] Hence, a clear article such as article 5(3) should probably prevail over an ambiguous recital such as recital 66. Apart from that, recital 66 doesn't contradict article 5(3), but should be read as a reminder that consent can be given in any form.[1138]

Furthermore, European law suggests that a privacy-friendly interpretation of the e-Privacy Directive is called for. The e-Privacy Directive aims to protect the right to privacy and the right to data protection.[1139] These rights are included in the EU Charter of Fundamental Rights,[1140] and according to the European Court of Justice, the

---

[1134] See e.g. Traung 2012; McStay 2012; Kosta 2013. See also Article 29 Working Party, WP 171, p. 14.
[1135] Helberger et al. 2011, p. 63.
[1136] Recital 66 of Directive 2009/136.
[1137] Klimas & Vaiciukaite 2008. The European Court of Justice says "the preamble to a Community act has no binding legal force and cannot be relied on either as a ground for derogating from the actual provisions of the act in question or for interpreting those provisions in a manner clearly contrary to their wording" (ECJ, C-136/04, Deutsches Milch-Kontor GmbH, 24 November 2005, par. 32).
[1138] Traung 2010, p. 225.
[1139] See article 1 and article 5 of the e-Privacy Directive. See also the Data Protection Directive, which aims for a "high level of protection" of fundamental rights and in particular privacy (recital 10). Article 8(4)(c) of the Framework Directive 2002/21/EC (amended in 2009) requires national regulatory authorities to "contribut[e] to ensuring a high level of protection of personal data and privacy."
[1140] Article 7 and 8 of the EU Charter of Fundamental Rights.

e-Privacy Directive must be interpreted in line with fundamental rights.[1141] The e-Privacy Directive's preamble says that users' devices are part of the user's private sphere,[1142] and the European Court of Human Rights interprets the right to private life broadly.[1143] In addition, the Charter and other EU Treaties emphasise the importance of a high level of consumer protection.[1144]

Taking the requirements for consent into account, recital 66 should probably be read as follows. If browsers were developed with a function to express consent in line with the Data Protection Directive, such browsers could be used to consent to the use of cookies. However, for the moment most browsers aren't suitable to give informed consent for cookies. Chapter 8 discusses the Do Not Track standard, which could enable people to express their wishes with their browser.[1145]

The 2009 version of article 5(3) should have been implemented in national legislation in May 2011, but many member states missed this deadline.[1146] At the time of writing, enforcement of the consent requirement for tracking cookies is in its infancy, among other reasons because the national laws implementing the consent rule are rather new.[1147] Discussions about a Do Not Track standard may have delayed enforcement as well. It's unclear how national authorities will apply the implementation of article 5(3).[1148] The approaches seem to vary. For instance, the UK appears to accept a kind

---

[1141] ECJ, C-275/06, Promusicae, 29 January 2008, par. 67-68, and dictum. See also recital 62 of the Citizens' Rights Directive.
[1142] Recital 24 of the e-Privacy Directive; recital 65 of Directive 2009/136.
[1143] See chapter 3, section 2.
[1144] See article 38 and article 51(1) of the EU Charter of Fundamental Rights, and article 12, article 114(3) and article 169 of the Treaty on the Functioning of the EU (consolidated version 2012).
[1145] Chapter 8, section 5.
[1146] Article 4(1) of Directive 2009/136. According to the Working Party, all member states had implemented the amended e-Privacy Directive on 1 January 2013 (Article 29 Working Party 2013, WP 208, p. 2). It's not unusual that member states implement directives late.
[1147] Regulators have taken some action regarding the national implementation of article 5(3). For example, the Agencia Española de Protección de Datos (Spanish Data Protection Authority) issued a fine for non-compliance in January 2014 (Agencia Española de Protección de Datos 2014; see Pastor 2014). The Dutch Data Protection Authority has concluded in several investigations that article 5(3) was breached (see e.g. College bescherming persoonsgegevens 2013 (TP Vision); College bescherming persoonsgegevens 2014 (YD)). See regarding Google and article 5(3) chapter 8, section 1.
[1148] The Working Party has tried to align the implementation. In line with earlier Opinions, the Working Party says "an active indication of the user's wishes" is required for consent to cookies (Article 29 Working Party 2013, WP 208, p. 3).

of opt-out system,[1149] whereas the Netherlands requires, in short, opt-in consent for tracking cookies.[1150]

***Take-it-or-leave-it choices***

It's somewhat unclear what "free" consent means in the context of the e-Privacy Directive. The Dutch experience with the consent requirement for tracking cookies can serve as an illustration. In the Netherlands the consent requirement for tracking cookies came into effect in January 2013. The implementation law made clear that unambiguous (opt-in) consent was required for tracking cookies.[1151] Many websites reacted by denying entry to visitors that didn't accept third party tracking cookies, by installing "cookie walls" or "tracking walls" – barriers users could only pass if they allowed the website and its partners to track them.[1152] One could question whether consent is voluntary if a website installs a tracking wall.[1153] Among others, Kosta suggests that a tracking wall makes consent involuntary. "In such a case the user does not have a real choice, thus the consent is not freely given."[1154]

Indeed, in some cases consent may not be sufficiently "free" when a website uses a tracking wall. For example, the Dutch Data Protection Authority says that the national public broadcasting organisation isn't allowed to use a tracking wall.[1155] The Data Protection Authority says that the public broadcaster has a "situational monopoly", because the only way to access certain information online is through the broadcaster's website.[1156] This makes the consent involuntary. It remains to be seen whether Data

---

[1149] Information Commissioner's Office 2013a.
[1150] See below.
[1151] Article 11.7a of the Dutch Telecommunications Act (version applicable on 30 May 2014). The explanatory memorandum makes clear that opt-in consent is required for tracking cookies. See for a translation of the provision Zuiderveen Borgesius 2012, p. 5.
[1152] See Helberger 2013.
[1153] See Article 29 Working Party 2013, WP 208; Impact Assessment for the proposal for a Data Protection Regulation (2012), Annex 4, p. 76.
[1154] Kosta 2013, p. 17. See also Roosendaal 2013, p. 186.
[1155] Helberger 2013, p. 18.
[1156] College Bescherming Persoonsgegevens (Dutch DPA) 2013 (cookie letter).

Protection Authorities will use similar "situational monopoly" reasoning when commercial broadcasters and website publishers use tracking walls.

The Working Party is sceptical about tracking walls, but doesn't really prohibit them. It says people "should have an opportunity to freely choose between the option to accept some or all cookies or to decline all or some cookies."[1157]

> In some Member States access to certain websites can be made conditional on acceptance of cookies, however generally, the user should retain the possibility to continue browsing the website without receiving cookies or by only receiving some of them, those consented to that are needed in relation to the purpose of provision of the website service, and those that are exempt from consent requirement. It is thus recommended to refrain from the use of consent mechanisms that only provide an option for the user to consent, but do not offer any choice regarding all or some cookies.[1158]

Recital 25 of the e-Privacy Directive says "[a]ccess to specific website content may still be made conditional on the well-informed acceptance of a cookie or similar device, if it is used for a legitimate purpose." It is likely that the EU lawmaker didn't foresee that some websites would completely block visitors that don't accept third party tracking cookies. But the Working Party suggests that recital 25 isn't meant to allow firms to put the whole website behind a tracking wall: "[t]he emphasis on 'specific website content' clarifies that websites should not make conditional 'general access' to the site on acceptance of all cookies."[1159] The Working Party adds that

---

[1157] Article 29 Working Party 2013, WP 208, p. 5.
[1158] Article 29 Working Party 2013, WP 208, p. 5 (internal footnote omitted).
[1159] Article 29 Working Party 2013, WP 208, p. 5.

website publishers should "only limit certain content if the user does not consent to cookies."[1160]

The careful phrases suggest that the Working Party doesn't mean to say that all take-it-or-leave-it choices and tracking walls are prohibited.[1161] This seems to be the correct interpretation of current law. If there are alternative service providers, it is likely that data protection law will allow a firm to offer such a take-it-or-leave-it choice.[1162] When interpreting data protection law's consent rules, the general principle of freedom of contract can provide inspiration by analogy. True, contractual freedom isn't absolute.[1163] Nevertheless, the principle of contractual freedom would be hard to reconcile with reading a full prohibition of take-it-or-leave-it choices in current data protection law. That said, data protection law does require consent to be "free."

Several factors can be taken into account when assessing whether a firm is allowed to offer a take-it-or-leave-it choice, for instance with a tracking wall on its website. The following is a non-exhaustive list of circumstances in which the legality of tracking walls is particularly questionable. The firm has a monopoly position.[1164] There are no competitors that offer a similar, more privacy-friendly service.[1165] It's not a realistic option for people to go to a competitor, for instance because of a lock-in situation.[1166]

---

[1160] Article 29 Working Party 2013, WP 208, p. 5.
[1161] But see the English Information Commissioner's Office, which says: "Organisations should not coerce or unduly incentivise people to consent, or penalise anyone who refuses. Consent cannot be a condition of subscribing to a service or completing a transaction" (Information Commissioner's Office 2013b, p. 14).
[1162] See also European Agency for Fundamental Rights 2014, p. 59.
[1163] As Chang puts it, "[a]ll societies keep certain things off the market – human beings (slavery), human organs, child labour, firearms, public offices, health care, qualifications to practice medicine, human blood, educational certificates and so on" (Chang 2014, p. 395). See on inalienable rights, of which "transfer is not permitted between a willing buyer and a willing seller," also Calabresi & Melamed 1972 (p. 1092).
[1164] As Bygrave notes, "fairness (…) implies that a person is not unduly pressured into supplying data on him-/herself to a data controller or accepting that the data are used by the latter for particular purposes. From this, it arguably follows that fairness implies a certain protection from abuse by data controllers of their monopoly position" (Bygrave 2002, p. 58).
[1165] See section 28(3)(b) of the Federal Data Protection Act in Germany.
[1166] See on lock-in situations and transaction costs chapter 7, section 3. In some cases, the law aims to reduce the problem of lock-in. For instance, the Universal Services Directive (2002/22/EC) requires phone companies to offer number portability (article 30(1)). The European Commission proposal for a Data Protection Regulation (2012) introduces a right to data portability in article 18.

There are circumstances that make it difficult or burdensome to leave the service.[1167] (It makes little sense to join another social network if all of one's friends are on Facebook.) A service is aimed at, or often used by, children.[1168] Under the given circumstances, it's unfair to expose people to tracking.[1169] Lastly, if a tracking wall affects millions of people, it deserves more scrutiny than when it only affects a few people.[1170] In sum, to assess the voluntariness of consent, all circumstances have to be taken into account – as is usually the case when applying legal provisions.

***Confidentiality of communications***

Apart from article 5(3), article 5(1) of the e-Privacy Directive is also relevant for behavioural targeting. Article 5(1) concerns the confidentiality of communications and can be summarised as follows. Member states must ensure the confidentiality of communications and the related traffic data by means of publicly available electronic communications services. In particular, member states must prohibit tapping, storage or other types of communications surveillance, without the consent of the users. Hence, the provision emphasises member states' positive obligations regarding confidentiality of communications.[1171]

Certain forms of behavioural targeting are clearly covered by article 5(1). If an internet access provider employs deep packet inspection to analyse people's internet use, including email communication, article 5(1) applies.[1172] Email messages are a form of communication, and the e-Privacy Directive applies to telecommunications providers, such as internet access providers.[1173] But web browsing and using IPTV or

---

[1167] For instance, there could be transaction costs. See chapter 7, section 3.
[1168] The Article 29 Working Party says that tracking shouldn't be made a condition for the use of a social network service. Perhaps this remark is partly inspired by the fact that many children use such sites (Article 29 Working Party, WP 187, p. 18).
[1169] See chapter 4, section 4 on the interpretation of fairness. See also Bygrave 2002, p. 58.
[1170] See Radin 2013.
[1171] Steenbruggen 2009, p. 176; p. 356.
[1172] See for an example, Phorm, which was discussed in section 3 of this chapter, and in chapter 2, section 2.
[1173] See on the scope of the e-Privacy Directive chapter 5, section 6; chapter 9, section 5. An "electronic communications service" is, in short, a service that consists wholly or mainly in the conveyance of signals on

video-on-demand services also fall within the European legal definition of communication.[1174] Monitoring people's web browsing is thus only allowed upon obtaining their consent, as member states must prohibit "interception or surveillance of communications and the related traffic data by persons other than users, without the consent of the users concerned."[1175] It has been suggested, amongst others by the European Data Protection Supervisor, that article 5(1) doesn't only apply to telecommunications providers.[1176] This would imply that ad networks must also comply with the provision in many circumstances.[1177] Regardless of the debate surrounding the applicability of article 5(1), consent is required by article 5(3) for most tracking technologies.

### 6.5 A limited but important role for informed consent

Informed consent has an important but limited role in data protection law. Consent is important, because the data subject can allow, or choose not to accept, data processing that would otherwise be prohibited.

Consent could be seen as a legal basis for data processing activities for which there's no overriding interest. "If no consent is given," Gutwirth notes, "the other legitimate grounds in themselves seem to span the whole gamut of possibilities, unless one assumes that such consent legitimizes disproportionate and illegitimate processing – which is very questionable."[1178] In theory (and leaving aside the EU Charter of

---

electronic communications networks (article 2(c) of the Framework Directive 2002/21/EC (amended in 2009)). It's thus a transmission service.

[1174] The e-Privacy Directive defines communication in article 2(d): "any information exchanged or conveyed between a finite number of parties by means of a publicly available electronic communications service. This does not include any information conveyed as part of a broadcasting service to the public over an electronic communications network except to the extent that the information can be related to the identifiable subscriber or user receiving the information." See Steenbruggen 2009, p. 181; p. 354. Traung 2010, p. 227.

[1175] Article 5(1) of the e-Privacy Directive.

[1176] See Traung 2010, p. 227; González Fuster et al. 2010, p. 115; European Data Protection Supervisor 2008, par 33. See also Article 29 Working Party 2006, WP 118.

[1177] If somebody browses the web while using an electronic communications service that is not publicly available (perhaps a Wi-Fi network in a coffee shop), this might be different. A full discussion of the scope of article 5(1) would go beyond the scope of this study.

[1178] Gutwirth 2002, p. 100.

Fundamental Rights), a data protection regime without a consent provision could be envisaged.[1179] In such a regime, a firm that couldn't rely on a contract with the data subject would have to check whether it could rely on the balancing provision. But if the data subject's fundamental rights outweighed the firm's interests, the data processing couldn't legally take place. In the current regime, firms can ask consent for processing that isn't "necessary."[1180] But even after consent is obtained, firms have to comply with the other data protection provisions.[1181]

A strong believer in informational self-determination and data subject control might see consent as the primary condition for data processing, at least in the private sector.[1182] In this view, the other legal bases are exceptions for data processing that's "necessary" for overriding interests. If the other legal bases were seen as exceptions to the consent requirement, the balancing provision would be a peculiar provision, because of its vagueness.

In theory, a data protection regime without a balancing provision could also be imagined.[1183] But such a regime would require a lot of consent requests, including for relatively innocuous practices. The balancing provision protects people from too many consent requests for trivial matters. Some practices would be almost impossible to do legally if the balancing provision didn't exist. For instance, Data Protection Authorities allowed Google to rely on the balancing provision for the processing of personal data (pictures including people) for its Streetview service.[1184] It's difficult to

---

[1179] The EU Charter of Fundamental Rights mentions consent as a legal basis for personal data processing (article 8(1)).

[1180] A data protection regime without a consent provision isn't fully hypothetical. For instance, early data protection acts in Belgium and France didn't include a consent clause (see De Hert et al. 2013, p. 59).

[1181] See chapter 9, section 2.

[1182] See Purtova 2011, p. 235-237. In some countries, consent is seen as the primary legal basis for processing (Korff 2002, p. 71).

[1183] A data protection regime without a balancing provision isn't fully hypothetical. For example, Spain had a very narrow version of the balancing provision, which only applied to data that appeared in public sources. The European Court of Justice didn't accept this (CJEU, C-468/10 and C-469/10, ASNEF, 24 November 2011). And the 1992 Data Protection Act in Hungary (replaced in 2012) didn't have a clear balancing provision (Act LXIII of 1992 on the Protection of Personal Data and Public Access to Data of Public Interest).

[1184] See Van Der Sloot & Zuiderveen Borgesius 2012a.

see how Google could have obtained consent of all people whose images (personal data) were included on pictures.

In any case, in 1992 the European Commission suggested that there's no priority between the legal bases. "Consent is no longer the main criterion, subject to exceptions; it is now the first of several alternatives (new article 7(a))."[1185] In sum, the legal bases consent and the balancing provision both have a role to play in data protection law. Apart from that, it doesn't seem plausible that one of the legal bases would be abolished.[1186]

***e-Privacy Directive***

The e-Privacy Directive says its provisions "particularise and complement" the Data Protection Directive.[1187] Article 5(3) of the e-Privacy Directive complements the requirement in data protection law of a legal basis for personal data processing. If a firm uses a tracking cookie to process personal data, it needs a legal basis for the processing. Hence, usually an ad network would need to obtain "unambiguous consent" for personal data processing, even if it obtained consent for using the cookie. In practice it would make sense to merge the consent request for the cookie and the following personal data processing operation.[1188] If a firm could base personal data processing for behavioural targeting on the balancing provision, the firm would still have to obtain consent for the use of the tracking cookie. From the firm's perspective, it's thus hardly relevant on which legal basis it can rely upon for personal data processing for behavioural targeting.[1189]

Therefore, article 5(3) could be interpreted as blocking firms from relying on the balancing provision for behavioural targeting. Seen in this light, article 5(3) of the e-

---

[1185] European Commission amended proposal for a Data Protection Directive (1992), p. 16 (capitalisation adapted).

[1186] It would be difficult to abolish the legal basis consent, as it's included in the EU Charter of Fundamental Rights (article 8 (2)).

[1187] Article 1(2) of the e-Privacy Directive.

[1188] Article 29 Working Party 2013, WP 202, p. 14. Consent (article 7(a) of the Data Protection Directive) is usually the only available legal basis for behavioural targeting; see section 1-3 of this chapter,

[1189] See Article 29 Working Party 2014, WP 217, p. 46.

Privacy Directive codified an interpretation of the Data Protection Directive's legal basis requirement in the behavioural targeting context. As article 5(3) applies to the storing or accessing any information (personal data or not) on a user's device, article 5(3) implicitly sidesteps the discussion of whether tracking cookies and similar files are personal data.[1190]

Other rules in the e-Privacy Directive can also be seen as codifying an interpretation of the Data Protection Directive's legal basis requirement, for instance the rules on spam.[1191] In short, the e-Privacy Directive only allows sending marketing emails to non-customers after the receiver's prior consent is obtained (an opt-in system). "The use of (…) electronic mail for the purposes of direct marketing may be allowed only in respect of subscribers or users who have given their prior consent."[1192] Hence, firms can't rely on the balancing provision for sending commercial emails to non-customers. But within the context of an existing customer relationship, the e-Privacy Directive allows a firm to send marketing emails to offer similar products or services, if the email includes a clear opt-out possibility. There's thus an opt-out system for certain first party direct marketing emails, which resembles the regime of the balancing provision.[1193] The e-Privacy Directive has more rules that essentially block certain types of firms from relying on the balancing provision for processing for direct marketing. For instance, certain types of firms (in short: telecommunication providers) are required to obtain consent for processing traffic and location data, unless a specified exception applies.[1194]

In a nutshell, if direct marketing uses any other method than paper, human phone calls, or visits to people's houses, EU law requires the individual's consent – with the

---

[1190] However, the scope of article 5(3) is too broad, as it also requires consent for certain types of innocuous cookies. See chapter 8, section 4.
[1191] See on the right to object to direct marketing and the e-Privacy Directive Article 29 Working Party 2014, WP 217, p. 45-47.
[1192] Article 13 of the e-Privacy Directive. For direct marketing by automatic calling machines (robo calls) or by fax, consent is also required, subject to exceptions.
[1193] Article 13(2) of the e-Privacy Directive. See also recital 41.
[1194] See on the scope of the e-Privacy Directive chapter 5, section 6, chapter 8, section 4, chapter 9, section 5.

exception of some types of first party direct marketing. The opt-out regime for paper, human phone calls, and house visits can plausibly be explained by the fact that such marketing techniques are relatively costly. The higher costs of such practices reduce the chance of abusive practices. It's cheaper to send spam email to millions of people, than to hire workers to call millions of people.[1195]

***Default rules and mandatory rules***

Regarding direct marketing, the Data Protection Directive's consent provision and the balancing provision could be seen as mirror images. The legal bases consent (article 7(a)) and the balancing provision (article 7(f)) provide default positions that the data subject can alter.[1196]

Sometimes personal data processing for direct marketing is *only* allowed after consent. The default position is: data processing is not allowed. Without consent, a firm may not process personal data. But with consent the data subject can allow data processing that would otherwise be prohibited. In other words, the data subject can alter the default by giving consent to data processing. Sometimes data processing for direct marketing is allowed *without* consent. If a firm can rely on the balancing provision, the default is: data processing is allowed.[1197] But the data subject has the right to stop the data processing: to opt out. By opting out, the data subject can alter the default position to: data processing is not allowed.[1198]

In law and economics terms, the consent requirement lays down a "default" rule, also called a "non-mandatory" rule. Default rules "apply unless the parties make deviating arrangements."[1199] The data subject can make a deviating arrangement by giving

---

[1195] See recital 42 of the e-Privacy Directive.
[1196] Purtova 2014 makes a similar point, but refers to the default positions as "entitlements", a concept introduced by Calabresi & Melamed 1972.
[1197] Of course, firms need to comply with all data protection law's requirements, regardless of the legal basis for processing.
[1198] See article 14(b) of the Data Protection Directive. See section 2 of this chapter.
[1199] Hesselink 2005, p. 46. In a famous law and economics article on default rules, Ayres & Gertner speak of "rules that parties can contract around by prior agreement" (Ayres & Gertner 1989, p. 87). A rule that lays down a default

consent to data processing. Likewise, the regime for direct marketing that follows from the balancing provision could be seen as a default rule. The data subject can make a deviating arrangement by objecting to data processing (opting out).

The other data protection rules are "mandatory" (with arguably a few exceptions.[1200]) In law and economics terms, mandatory "rules cannot be contracted around; they govern even if the parties attempt to contract around them."[1201] People can't set data protection law's mandatory rules aside by contractual agreement, or with consent.[1202] For instance, the following declaration wouldn't be enforceable:

> I hereby consent to the use of my personal data for improving products and services (including more relevant advertising), and other business purposes.[1203] I hereby waive my rights to access, correction and erasure. I will not hold you liable in case of a data breach. The above applies not only to you, the data controller, but also to the selected parties that may obtain my personal data from you.[1204]

In sum, while consent plays an important role, that role is limited at the same time. The freedom to consent to data processing could be seen as an extremely limited version of contractual freedom.

---

is sometimes called "contractible." Mandatory rules can also be called ius cogens (versus default rules: ius dispositivum).

1200 First, in some cases (not regarding direct marketing) the data subject has a relative right to object (article 14(a). Second, with consent the data subject can allow data export to outside the EU (article 26(b)). Third, the data subject can allow the processing of special categories of data with "explicit consent" (article 8(2)(a)).

1201 Ayres & Gertner 1989, p. 87. The mandatory character of data protection law can also be framed differently. The right to protection of personal data (article 8 of the EU Charter of Fundamental Rights) can be seen as an inalienable right (see Calabresi & Melamed 1972).

1202 The Working Party says consent "is primarily a ground for lawfulness, and it does not waive the application of other principles" (Article 29 Working Party 2011, WP187, p. 7). See also chapter 9, section 2.

1203 The purpose ins't sufficiently "specified", and the consent isn't sufficiently "specific" and "informed" (article 6(1)(b) and article 2(h) of the Data Protection Directive).

1204 These rights are not waivable (see article 12 and 23 of the Data Protection Directive).

### 6.6 Data protection law unduly paternalistic?

Sometimes it's suggested that data protection law is too paternalistic, because it limits the data subject's contractual freedom. For example, Bergkamp says data protection law "is driven by paternalistic motives; individuals need to be protected and be given inalienable but vague fundamental rights, the scope of which government officials define *ex post* in specific cases."[1205] Even worse: data protection law "does not permit variation by contract."[1206]

This study does not find data protection law unduly paternalistic.[1207] There are at least three reasons why data protection law isn't unduly paternalistic. First, in line with positive law, this study takes the view that some paternalism can be justified. Second, pure paternalism is only present when a legal rule only aims at protecting a person against him- or herself. But there are other rationales for data protection law than protecting people against themselves. Third, data protection law leaves some important choices to the data subject.

There's a huge body of literature on paternalism from many disciplines.[1208] Cserne discusses paternalism in the context of contract law. His paternalism definition is apt for this study.

---

[1205] Bergkamp 2002, p. 37. See also Cuijpers 2007.
[1206] Bergkamp 2002, p. 38. It must be noted that Bergkamp's position seems rare.
[1207] Few authors argue explicitly that data protection law isn't too paternalistic, perhaps because data protection law is rarely accused of being too paternalistic. An implicit argument that data protection law isn't too paternalistic can be found in, for instance, De Hert & Gutwirth 2006; Blume 2012; Purtova 2011, p. 204.
[1208] See for good and easy to read introductions Cserne 2008; Dworkin 2010; Ogus 2010; Sunstein 2013. See on privacy law and paternalism, from a US perspective Solove 2013.

> There are three conditions for an act to be paternalistic. The paternalist
>
> (1) interferes with the subject's liberty,
>
> (2) acts primarily out of benevolence toward the subject (i.e., his goal is to protect or promote the interests, good or welfare of the subject),
>
> (3) acts without the consent of the subject.[1209]

Data protection law's mandatory rules comply with the definition's first element, because the data subject can't waive them. Such mandatory rules limit the data subject's choices, so they interfere with his or her liberty. (This study uses liberty in a narrow sense, roughly comparable with contractual freedom.[1210] A general discussion of the meaning of liberty and paternalism falls outside this study's scope.[1211]) Data protection law's mandatory rules also comply with the third element. The mandatory rules interfere with the data subject's liberty, without his or her consent.[1212]

The second element of the definition requires that the lawmaker "acts primarily out of benevolence toward the subject." This concerns the rationale for a rule. The legal system contains many prohibitions and mandatory rules that have nothing to do with paternalism. For instance, a rule can protect other parties by limiting a person's

---

[1209] Cserne 2008, p. 18. Outside the legal field, Dworkin 2010 gives a similar description. See on paternalism in the context of behavioural targeting Hoofnagle et al. 2012.

[1210] Liberty in the sense of contractual freedom is also called "party autonomy" in the context of contract law (see Grundmann 2002; Grundmann et al. 2001). See on party autonomy and rational choice theory chapter 7, section 2.

[1211] See for a general discussion of freedom, or liberty, in connection with the case law of the European Court of Human Rights: Marshall 2009.

[1212] It could be argued that the data subject gave some kind of broad consent to the democratically elected lawmaker. But we'll leave this line of argument aside. See critically on such arguments Cserne 2008, p 32-33.

freedom: thou shall not kill.[1213] Likewise, if a rule mainly aims to protect a public interest, it's not a purely paternalistic rule. Such rules aren't purely paternalistic, because the lawmaker doesn't act primarily out of benevolence toward the subject.

It's not always easy to establish the rationale for a rule. People might disagree about the rationale for a rule, even if they agree on the rule.[1214] For instance, an obligation to wear a motorcycle helmet could be defended on paternalistic grounds. But the helmet obligation could also be defended by pointing out the costs for society that would result from motorcyclists having accidents that lead to death or injury.[1215] Smoking bans could likewise be defended on both paternalistic and non-paternalistic grounds.[1216]

The Data Protection Directive aims to "protect the fundamental rights and freedoms of natural persons, and in particular their right to privacy."[1217] This could be seen as acting out of benevolence toward the data subject, and thus as paternalism. But protecting fundamental rights is also a public interest. Many scholars say that the right to data protection and the right to privacy are important for our society as a whole.[1218] The protection of privacy and the fair processing of personal data concern the question of what kind of society we want. This goes beyond individual interests.

The European Court of Human Rights suggests that respect for privacy is important for a democratic society.[1219] And the Court speaks of "[t]he interests of the data subjects *and the community as a whole* in protecting the personal data."[1220] Following

---

[1213] See Mill 2011 (1859). Protecting other parties can be seen as an answer to externalities (see chapter 7, section 3).
[1214] Sunstein 1995a.
[1215] Such costs for others could be seen as negative externalities. See chapter 7, section 3.
[1216] See Cserne 2008, p. 34-38.
[1217] Article 1(1) of the Data Protection Directive.
[1218] See e.g. Simitis 1987; Regan 1995; Schwartz 1999; Schwartz 2000; Westin 2003; Rouvroy & Poullet 2009; De Hert & Gutwirth 2006; Allen 2011; Van der Sloot 2012. See also chapter 3, section 1.
[1219] See ECtHR, Rotaru v. Romania, No. 28341/95, 4 May 2000, par. 59; ECtHR, Klass and others v. Germany, No. 5029/71, 6 September 1978, par. 49.
[1220] ECtHR, S. and Marper v. United Kingdom, No. 30562/04 and 30566/04. 4 December 2008, par. 104. (emphasis added).

that reasoning, data protection law isn't purely paternalistic.[1221] Literature on the right to confidentiality of communications contains similar reasoning. The right to confidentiality of communications protects the trust society has in a communication channel.[1222] Furthermore, chapter 7 shows that economic theory accepts several rationales for regulatory intervention that have nothing to do with paternalism.[1223] Some of these rationales can be invoked for data protection law.

That said, benevolence towards the data subject is undoubtedly among the rationales for data protection law. But rules that can be explained by paternalistic motives aren't necessarily *unduly* paternalistic. Looking at positive law in Europe, there are many rules that could plausibly be explained, at least in part, by paternalistic motives.[1224] The European legal system accepts, and perhaps even requires, some paternalism.[1225] Pursuant to the EU Charter of Fundamental Rights for instance, "Union policies shall ensure a high level of consumer protection."[1226] And the Treaty on the European Union says the Union aims for a "social market economy."[1227] Briefly stated, in Europe the question is not: "is legal paternalism acceptable?" The question is: "how much legal paternalism is acceptable?"

European consumer law, broadly defined, contains many rules that remind one of data protection law's transparency principle. The rules aim to empower consumers to make choices in their own best interests. For instance, rules that require firms to include information on packaging aim to empower consumers to make decisions in their own best interests.[1228] Such rules only mildly interfere with contractual freedom. But consumer protection law also contains rules that directly regulate the contents of contracts. As the European Commission puts it, "in some situations, providing a basis

---

[1221] See Sunstein, who says paternalism does not "include government efforts to promote certain familiar and widely held social goals; consider laws designed to protect privacy (…)" (Sunstein 2014, p. 80).
[1222] Asscher 2002, p. 18; p. 247; Steenbruggen, p. 44-49; p. 354.
[1223] See chapter 7, section 2 and 3.
[1224] Ogus 2010.
[1225] But see for another view Van Aaken 2013.
[1226] Article 38 of the EU Charter of Fundamental Rights.
[1227] Article 3(3) of the Treaty on EU (consolidated version 2012).
[1228] See Luth 2010.

for informed choice and legal redress has been regarded as insufficient, notably as regards protection of physical health and safety."[1229] For example, minimum safety standards could be seen as bans of products that don't comply with the requirements.[1230] Other products can't be legally bought at all. Many national consumer protection statutes contain a blacklist of contract terms that aren't enforceable.[1231] Such rules limit contractual freedom, and paternalistic motives are likely to be among the motives. On the other hand, many consumer protection rules can also be explained as a response to market failures, such as information asymmetries.[1232]

In the context of consumer law, Hesselink suggests that rules that aim to protect consumers must generally be mandatory to have any effect. Otherwise the firm, which is usually the one drafting the contract, can set the protective rules aside in the contract.

> Obviously, the main character of rules inspired by the policy of consumer protection is that they are protective. This means that the rules of contract law aim at the protection of the consumer against the other party to the contract (the professional). In order to make this protection effective such rules are typically mandatory, i.e. they cannot be waived.[1233]

---

[1229] European Commission 2002, p. 6.
[1230] See for instance the General Product Safety Directive. Food is heavily regulated as well (see Van Der Meulen & Van Der Velde 2004).
[1231] See Ebers 2007 (p. 344) on the implementation of the Unfair Contract Terms Directive.
[1232] See on information asymmetry and other market failures chapter 7, section 3.
[1233] Hesselink 2007, p. 339. The European Court of Justice uses similar reasoning in favour of mandatory rules (CJEU, ECJ, C-243/08, Pannon GSM, 4 June 2009, par. 22-25). See also recital 22 of the Consumer Sales Directive (1999/44/EC): "the parties may not, by common consent, restrict or waive the rights granted to consumers, since otherwise the legal protection afforded would be thwarted (…)"

Balancing protecting people and respecting their freedom of choice is common in the law.[1234] "Paradoxically", says Mak, "interference in the contractual relationship is sometimes required in order to guarantee that both contract parties can fully enjoy their freedom of self-determination."[1235] Similar reasoning applies to data protection law. Seen from this angle, data protection law aims to strike a balance between protecting and empowering people.

### 6.7 Conclusion

This chapter discussed the role of informed consent in the regulatory regime for privacy and behavioural targeting. Discussions about the regulation of behavioural targeting tend to focus on the consent requirement for tracking technologies in the e-Privacy Directive.

Since 2009, article 5(3) of the e-Privacy Directive requires any party that stores or accesses information on a user's device to obtain the user's informed consent. Article 5(3) applies to many tracking technologies such as tracking cookies. There are exceptions to the consent requirement, for example for cookies that are strictly necessary for a service requested by the user, and for cookies that are necessary for the transmission of communication.

For the definition of consent, the e-Privacy Directive refers to the Data Protection Directive, which states that valid consent requires a free, specific, informed indication of wishes. People can express their will in any form, but mere silence or inactivity isn't an expression of will. During the drafting of the Data Protection Directive in the early 1990s, many firms argued that they should be allowed to presume consent for processing, as long as people don't opt out. But the EU lawmaker rejected this idea.

---

[1234] See for instance Study Group on Social Justice in European Private Law 2004; Grundmann et al. 2001; Hesselink 2005.

[1235] Mak 2008, p. 26.

Nowadays, marketers often suggest that people who don't block tracking cookies in their browser give implied consent to tracking cookies. But this interpretation of the law seems incorrect. As the Article 29 Working Party notes, the mere fact that a person leaves the browser settings untouched doesn't mean that the person has expressed the will to be tracked. In sum, the e-Privacy Directive requires consent for the use of most tracking technologies. There's much debate on whether opt-out systems are sufficient to obtain the user's consent or not.

In line with the transparency principle, consent has to be specific and informed. Furthermore, only "free" consent can be valid. Nevertheless, in most circumstances, current data protection law will probably allow controllers to offer take-it-or-leave-it choices. Hence, in principle website publishers are allowed to install tracking walls that deny entry to visitors that do not consent to being tracked.

As far as personal data are being processed, the Data Protection Directive also applies to behavioural targeting. As we saw in the previous chapter, behavioural targeting does indeed entail personal data processing in most cases. The Data Protection Directive only allows personal data processing if it can be based on consent or another legal basis. For the private sector, the most relevant legal bases are: a contract, the balancing provision, and the data subject's consent.

As discussed in chapter 4, marketers feared that direct mail marketing would only be allowed with the data subject's prior consent when the European Commission presented a proposal for a Data Protection Directive in 1990. After lobbying by the direct marketing industry, the European Commission said in 1992 that personal data processing for certain types of direct mail marketing can be based on the balancing provision: on an opt-out basis.[1236] In brief, a firm can rely on the balancing provision when the processing is necessary for its legitimate business interests, and these interests are not overridden by the data subject's fundamental rights. The "necessary"

---

[1236] See chapter 4, section 1.

requirement sets a higher threshold than useful or profitable. If a firm relies on the balancing provision for direct marketing, data protection law grants the data subject the right to stop the processing: to opt out.

The Data Protection Directive doesn't state explicitly whether behavioural targeting (a type of direct marketing) can be based on the balancing provision. But the most convincing view is that behavioural targeting can't be based on the balancing provision, in particular if it involves tracking an internet user over multiple websites. In most cases the data subject's interests must prevail over the firm's interests, as behavioural targeting involves collecting and processing information about people's browsing behaviour, which many people regard as personal. Indeed, the Working Party says firms can almost never base personal data processing for behavioural targeting on the balancing provision.

A firm can also process personal data if the processing is necessary to perform a contract with the data subject. For instance, certain data have to be processed for a credit card payment, or for a newspaper subscription. Some internet companies suggest that a user enters a contract by using their services, and that it's necessary for this contract to track the user for behavioural targeting. As the Interactive Advertising Bureau US puts it, "visiting a web site is a commercial act, during which a value exchange occurs. Consumers receive content, and in exchange are delivered [targeted] advertising."[1237] But according to the Working Party, in general, firms can't rely on this legal basis for behavioural targeting. In any case, the practical problems with informed consent to behavioural targeting which are discussed in the next chapter would be largely the same if firms could base the processing for behavioural targeting on a contract with the data subject.

If firms want to process personal data, and can't base the processing on a legal basis such as a contract or on the balancing provision, they must ask the data subject for

---

[1237] Rothenberg (IAB US) 2013. See for a critical analysis of such claims: chapter 7, section 2.

consent. The Working Party says consent is generally the required legal basis for personal data processing for behavioural targeting. In sum, consent plays an important role in the EU legal regime for behavioural targeting. Data protection law is clearly influenced by the perspective of privacy as control over personal information.

While consent plays an important role in EU data protection law, that role is limited at the same time. The other provisions in the Data Protection Directive are mandatory (with a few exceptions). The data subject can't waive data protection law's safeguards, and can't contract around the rules. Therefore, data subjects don't enjoy full contractual freedom regarding personal data concerning them.

Nevertheless, this study takes the view that data protection law isn't unduly paternalistic. The European legal system accepts, and perhaps requires, a degree of paternalism. Furthermore, there are other rationales for data protection law than protecting people against themselves. The right to privacy and the right to data protection aim to contribute to a fair society, which goes beyond protecting individual interests. And from an economic perspective, regulatory intervention isn't paternalistic if it aims to reduce market failures, such as information asymmetries. The relevance of market failures for the regulation of behavioural targeting is elaborated in the next chapter.

* * *

# 7 Informed consent in practice

Considering the important role of informed consent in the current regulatory regime for behavioural targeting, this study can't ignore how people make privacy choices in practice. Is it feasible that people manage their privacy in the area of behavioural targeting through the legal instrument of informed consent?

For this chapter literature from the emerging field of the economics of privacy was analysed, as well as behavioural economics literature and social science studies on how people make privacy choices. The chapter could also be seen as a critical analysis of the privacy as control perspective, as the idea of informed consent is closely related to the control perspective.[1238]

Economics and behavioural economics provide useful tools to analyse certain problems with informed consent in practice. Even if one doesn't agree with economic rational choice theory (which is discussed in section 2), concepts such as information asymmetry, transaction costs and externalities can help to analyse different problems with the informed consent approach. While economists might use different phrases, the arguments derived from economics aren't necessarily new for legal scholars. To illustrate, if a lawyer says "[t]he opt-out options Google offers authenticated users are labour-intensive,"[1239] an economist might say that the transaction costs are too high.

---

[1238] See on the privacy as control perspective chapter 3, section 1, and chapter 4, section 5.
[1239] College bescherming persoonsgegevens (Dutch DPA) 2013 (Google) p. 31.

To apply economic theory, this chapter compares consenting to behavioural targeting with entering into a market transaction.[1240] This study does *not* argue that personal data should be seen as tradable goods on a market.[1241] Rather, the approach in this chapter is as follows. If one compares, for argument's sake, consenting to behavioural targeting with entering into a market transaction, economic theory suggests that there are market failures that justify more legal intervention.

Another reason to discuss economics in this study is that it's sometimes suggested that behaviourally targeted advertising is needed to fund the internet: "[w]hat powers the 'free' Internet are data collection and advertising."[1242] However, this chapter shows that such claims are too simple. For instance, in the long term behavioural targeting may decrease ad revenues for some website publishers. Furthermore, the chapter shows that it's an open question whether behavioural targeting is good or bad from an economic perspective.

Section 7.1 of this chapter discusses studies on people's attitudes towards behavioural targeting. Section 7.2 introduces the economic analysis of law, and the economic analysis of privacy. The section also discusses the limitations of the economic perspective on privacy. Section 7.3 analyses problems with informed consent through an economic lens. Section 7.4 turns to behavioural economics. The analysis in this chapter can help to explain the alleged privacy paradox (section 7.5): people say they care about privacy, but often fail to protect their information. Section 7.6 concludes.

---

[1240] For ease of reading, this chapter speaks of "consent to behavioural targeting." From a legal perspective, it would be more correct to speak of (i) unambiguous consent to personal data processing for behavioural targeting (in the sense of article 7(f) of the Data Protection Directive), and of (ii) consent to the use of tracking technologies (in the sense of article 5(3) of the e-Privacy Directive.

[1241] See on inalienability Calabresi & Melamed 1972.

[1242] Thierer 2010. See for similar claims e.g. Interactive Advertising Bureau Europe Youronlinechoices.

### 7.1 People's attitudes regarding behavioural targeting

Research suggests that, while some like the idea, most people don't want targeted advertising based on their online behaviour. People realise the possible benefits from targeted ads and content, but also find the underlying data processing creepy.

Turow et al. found in a nationally representative phone survey that 66% of adult Americans didn't want to receive advertisements that are tailored to their interests. The number was 55% for the age group between 18 and 24. When people were told that tailored advertisements would be based on their browsing behaviour, 87% didn't want targeted advertising. People were also asked whether they would allow marketers to "follow you online in an anonymous way in exchange for free content." 68% said they would definitely not allow it, and 19% probably wouldn't.[1243] The researchers conclude: "Contrary to what marketers say, Americans reject tailored advertising (…). Whatever the reasons, our findings suggest that if Americans could vote on behavioural targeting today, they would shut it down."[1244] The TRUSTe company found similar results: only 15% of the respondents would "definitely or "probably" consent to tracking for more relevant advertising.[1245]

In a survey by Cranor & McDonald, 18% of the respondents wanted behaviourally targeted advertising because it leads to more relevant advertising. 12% didn't mind being tracked. On the other hand, 46% found it "creepy" when advertisements are based on their browsing behaviour. 64% agreed with the statement "[s]omeone keeping track of my activities online is invasive."[1246] The researchers also questioned people about firms analysing the contents of email messages for targeted advertising. This is a common practice for so-called "free" email services such as Gmail and Yahoo. 4% liked their email being scanned because it could lead to more relevant

---

[1243] Turow et al. 2009, p. 16.
[1244] Turow et al. 2009, p. 4.
[1245] TRUSTe Research in partnership with Harris Interactive 2011.
[1246] Cranor & McDonald 2010, p. 23. See in detail about the demographics of the respondents p. 5-6. See also McDonald 2010.

advertising. About one in ten indicated "it's ok as long as the email service is free."[1247] But 62% found advertising based on email content creepy.[1248] A study among university students in Toronto found similar results.[1249]

Some studies find less negative attitudes to behavioural targeting. Hastak and Culnan found that 48% felt uncomfortable about their browsing behaviour being used for advertising. 23% were comfortable with it. That number grew to 40% if websites would give information about behavioural targeting and would offer an opt-out system.[1250] Some, but not all, industry-sponsored surveys find more positive attitudes towards behavioural targeting. For instance, one report says: "[m]ost consumers (84%) state they would *not* pay for access to online content that is free now, and instead, they would rather receive targeted advertising in exchange for free access to online content" (emphasis original). On the other hand, the report says: "Nearly all (93%) Internet users would use or already use the DNT button, however, only 22% of users are aware of this function."[1251] It should be noted that industry-sponsored studies aren't always clear on the methodology.[1252]

Ur et al. report on 48 in-depth interviews about online behavioural advertising. After being informed about behavioural targeting, people saw disadvantages and benefits. Almost half of the participants liked the idea of more relevant advertising. On the other hand, a majority mentioned privacy when asked whether there were downsides to behavioural targeting. "Participants commonly said they were scared about being tracked and monitored."[1253] People also complained about the lack of control.[1254] Most participants didn't like the idea of behavioural targeting. "However, this attitude seemed to be influenced in part by beliefs that more data is collected than actually

---

[1247] Cranor & McDonald 2010, p. 22.
[1248] Cranor & McDonald 2010, p. 21.
[1249] Foster et al. 2011.
[1250] Hastak & Culnan 2010.
[1251] Annalect 2012.
[1252] See for criticism on studies by Westin for instance Hoofnagle & Urban 2014.
[1253] Ur et al. 2012, p. 7.
[1254] Ur et al. 2012, p. 6.

is."[1255] The researchers conclude that people find behavioural targeting "smart, useful, scary, and creepy at the same time."[1256]

Results from European researchers are in line with the American results. A large study (26,574 people) in the European Union found that people were worried about privacy, and that they wanted more control over their information. "Nearly three-quarters of Europeans say their approval should be required in all cases before any kind of personal information is collected and processed."[1257] The study also found that seven out of ten people were concerned that firms might use data for new purposes such as targeted advertising without informing them.[1258] Only 22% indicated that they trusted search engines, social network sites, or email services to protect their information.[1259]

In interviews in the United Kingdom, Brown et al. found that people disliked third party data collection. "There was a strongly negative, almost emotional reaction in every group to the idea of third parties collecting data across a range of different devices and activities to develop an understanding of every aspect of consumers' lives."[1260] Interviews in the Netherlands suggest that few people were aware of behavioural targeting. People expressed privacy concerns after being told about it.[1261] A study by the Dutch Dialogue Marketing Association found that 70% of the respondents didn't want behavioural advertising.[1262] A 2012 representative study in the United Kingdom found that 8% of the respondents were comfortable with advertising based on their browsing history.[1263] 10% was conformable with Gmail scanning the contents of emails for targeted advertising. Around eight out of ten

---

[1255] Ur et al. 2012, p. 11.
[1256] Ur et al. 2012, p. 6.
[1257] European Commission 2011 (Eurobarometer), p. 172.
[1258] European Commission 2011 (Eurobarometer), p. 146.
[1259] European Commission 2011 (Eurobarometer), p. 138.
[1260] Brown et al. 2010, p. 83.
[1261] Helberger et al. 2012, p. 70.
[1262] Boogert 2011.
[1263] Bartlett 2012, p. 36-37.

people worried about firms using their data without consent and selling data to third parties.[1264]

Sometimes it's suggested that the younger generation doesn't care about privacy. "People have really gotten comfortable not only sharing more information and different kinds, but more openly and with more people. That social norm is just something that has evolved over time," said Mark Zuckerberg, Facebook's CEO.[1265] Such claims have some appeal at first glance. Some teenagers post "drunk" pictures or other information about private matters on Facebook. But research suggests that young people do care about privacy. Dana boyd concludes from her ethnographic research: "[m]uch to the surprise of many adults, teens actually care about privacy and take measures to make accessible content meaningless to outside viewers."[1266] An American study by the Pew Research Centre finds that young adults (18-29) are more likely than other older people to take steps like clearing cookies or browsing history.[1267] Other studies by the Pew Research Centre confirm that young people care about privacy.[1268] Given these outcomes, the claim that young people don't care about privacy seems incorrect. Furthermore, even if teens cared less about their privacy, this wouldn't prove that social norms have changed. Some teens drive too fast, drink too much, or take drugs recreationally. 10 or 20 years later, many have changed their habits.[1269]

Surveys and interviews give more reliable information than mere intuition, but they must be interpreted with caution. People often act differently in practice than might be expected from them based on survey results. This is the case for privacy choices as well. People say they care deeply about privacy, yet often divulge personal

---

1264 Bartlett 2012, 39.
1265 Zuckerberg, quoted in Kirkpatrick 2010. See generally on Zuckerburg on privacy The Zuckerberg files 2014.
1266 Boyd 2012, p. 16 (internal citations omitted). Ethnography is "a qualitative research methodology used by social scientists to understand and document cultural practices. Born out of anthropology – and embraced by many other disciplines – ethnographic work seeks to capture and explain the social meaning behind everyday activities" (boyd 2014, p. 23).
1267 Pew Research Center 2013, p. 10.
1268 Pew Research Center 2013a.
1269 See Richards 2014a, p. 17-18.

information in exchange for minimal benefits. Section 5 returns to this "privacy paradox."[1270] Furthermore, it's difficult to generalise findings from studies that use different methods. Many studies discussed above are from the US, and one should be careful when extrapolating the results to Europe. Another problem with surveys about privacy is that people who care a lot about their privacy may refuse to answer survey questions.[1271]

While caution is needed with interpreting the surveys and interviews, a couple of common themes emerge. People have mixed feelings. They see advantages in personalised advertising, but find it creepy at the same time. A small minority says it prefers behaviourally targeted advertising because it leads to more relevant ads. But a majority says it doesn't want behavioural targeting. Such survey results provide an argument in favour of legal intervention to improve privacy protection in the area of behavioural targeting.

## 7.2 Economics of privacy

In this chapter, economic theory is used to analyse problems with a legal construction: informed consent to data processing for behavioural targeting. This section gives a cursory introduction on the economic analysis of law.[1272] The section then introduces the emerging field of the economics of privacy. Finally, the limitations of the economic analysis of privacy are highlighted.

Law and economics is described by Posner as the "economic analysis of legal rules and institutions."[1273] Economics can be defined as "the science which studies human behaviour as a relationship between ends and scarce means which have alternative

[1270] See Acquisti 2010b, p. 6.
[1271] Heisenberg 2005, p. 39-40.
[1272] The introduction doesn't capture all the subtleties of economic theory and law and economics scholarship.
[1273] Posner 2011, p. xxi. This study uses the phrases "economic analysis of law" and "law and economics" interchangeably. See for an introduction to law and economics Kornhauser 2011.

uses."[1274] Like lawyers, economists look at the world in a particular way. Economics concerns the question of how parties make decisions when trying to maximise their preferences, with the limited means at their disposal.

In neoclassical economics (economics for short), it's usually assumed that parties want to maximise their own welfare, or their own utility.[1275] For example, a firm aims to maximise profit. But welfare doesn't merely concern money or things that are usually given a monetary value. An individual also aims to maximise welfare, which may include happiness, satisfaction, psychological well-being, or privacy.[1276]

Economists often use rational choice theory to predict human behaviour. Rational choice theory analyses behaviour assuming that people generally want to maximise their welfare, and that people can choose the best way to maximise their welfare. In short, it's assumed that people act "rationally" on average. Rational choice theory is a tool to predict human behaviour and doesn't aim to fully describe reality.[1277] "It is a *method* of analysis," says Becker, "not an assumption about particular motivations."[1278] Rational choice theory doesn't suggest that people always act rationally. But by assuming that people act rationally on average, the theory can still be used to predict human behaviour, and to reflect on how to regulate behaviour. For example, say a lawmaker raises the fines for speeding to deter people from driving too fast. The lawmaker assumes that people weigh the benefit of quick arrival against the potential cost of paying a fine. Even though some people might still drive too fast, on average, the measure could lead to less speeding.[1279]

---

[1274] Robbins 2007 (1934), p. 15.
[1275] This study speaks of economics for ease of reading, but it would be more correct to speak of "neoclassical economics." The neoclassical school of economics is merely one of a number of schools in economic thought, but neoclassical economics is presently the most influential school. Other schools include Austrian, Marxist, and Keynesian economics. See for an accessible overview of economic thought, distinguishing nine schools: Chang 2014, p. 109-169 (chapter 4) with further references (p. 165).
[1276] See Cooter & Ulen 2012, p. 12.
[1277] Posner 2011, p. 4.
[1278] Becker 1993, p. 385.
[1279] See Posner 1998, p. 1556-1557.

Law and economics literature often analyses which rule leads to the highest aggregate welfare for society (social welfare).[1280] In theory, there are situations in which social welfare increases: if somebody gains, and nobody incurs a loss. If nobody can increase their welfare without imposing costs on others, the situation is called "Pareto efficient."[1281] A different efficiency criterion is Kaldor Hicks efficiency, which refers to a situation where one person gains more than another person loses. According to this criterion, social welfare increases, if there's a change in which the gains of the winners are so great that they could compensate the losses of the losers. The Kaldor Hicks criterion doesn't require that winners actually compensate losers. In other words, the Kaldor Hicks criterion concerns the size of the pie and not how the pie is distributed. Any change that increases the pie is an improvement under the Kaldor-Hicks criterion.[1282] From this perspective, the question of how welfare is distributed within society is less relevant. In economics, tax is often seen as the best way to distribute wealth within society. Like this, for legal rules other than tax rules it makes sense to concentrate on how to enlarge the pie, rather than how to distribute the pie.[1283]

In economic theory, a (hypothetical) perfectly functioning free market leads to the highest social welfare – provided there are no market failures and setting aside how welfare is distributed within society. Private exchanges should lead to the highest social welfare, because people are assumed to enter contracts only when they expect to gain something from it, as they aim to maximise their expected welfare. Therefore, in theory unrestricted trade in a market without market failures leads to the highest aggregate welfare. This explains why economists are sometimes sceptical of laws that interfere with the free market, or that interfere with contractual freedom.[1284]

---

[1280] The economic analysis of law could thus be seen as a utilitarian approach.
[1281] See Kornhauser 2011.
[1282] See Kornhauser 2011.
[1283] See e.g. Kaplow & Shavell 1994. See also Hesselink 2011a, p. 298-301; Wagner 2010, p. 63.
[1284] Trebilcock 1997, p. 7; Hermalin et al. 2007, p. 24.

In reality, the ideal type of a perfectly functioning free market is exceedingly rare. From an economic perspective, there may be reason for the lawmaker to intervene when the market doesn't function as it ideally should. The law should aim at reducing market failures, such as information asymmetries, externalities, and market power. But legal intervention brings costs and economic distortions as well, and this has to be taken into account. From this perspective, legal intervention should thus be limited to situations where the costs of intervention are lower than the costs of the market failure.[1285]

Sometimes the law seems to be based implicitly on a kind of rational choice model. Put differently, sometimes the law appears to assume that people make choices in their own best interests, as long as they have enough information upon which to base their decisions.[1286] Contractual freedom, or party autonomy, is one of the primary principles of contract law – although it's never absolute.[1287] The notion of "informed consent" in data protection law, influenced by Westin's privacy as control perspective, also seems to be inspired by the idea that data subjects make "rational" choices. As Hoofnagle & Urban put it, "Westin's homo economicus (…) is expected to negotiate for privacy protection by reading privacy policies and selecting services consistent with her preferences."[1288]

***Economic analysis of privacy***

Economic theory can be used to analyse aspects of people's choices regarding privacy.[1289] One of the leading scholars in the economics of privacy is Acquisti. He

---

[1285] Market failure is "[a] general term describing situations in which market outcomes are not Pareto efficient" (Organisation for Economic Co-operation and Development (OECD) 1993, p. 55). See Hermalin et al. 2007, p. 30; Luth 2010, p. 15.

[1286] Ben-Shahar & Schneider 2011, p. 650; Sunstein & Thaler 2008, p. 6.

[1287] Grundmann summarises: "party autonomy dominates and the limits are seen as exceptions" (Grundmann 2002, p. 271). See also article II – 1:102 of the Draft Common Frame of Reference (Principles, Definitions and Model Rules of European Private Law), which contains the principle of contractual freedom: "Parties are free to make a contract or other juridical act and to determine its contents, subject to any applicable mandatory rules."

[1288] Hoofnagle & Urban 2014 (abstract).

[1289] See for an overview of the field of the economics of privacy Acquisti 2010a; Acquisti 2010b; Acquisti & Brandimarte 2012; Hui & Png 2006; Brown 2013.

explains: "the economics of privacy attempts to understand, and sometimes measure, the trade-offs associated with the protection or revelation of personal information."[1290]

An example of a trade-off is using a social network site. The user discloses personal data (a cost) to gain welfare: the use of a so-called "free" service. For instance, people don't pay with money for Facebook, which in turn analyses their behaviour for marketing purposes. Many email services offer a similar trade-off. They analyse the contents of messages for targeted advertising.[1291] As a US judge notes about Google: "in this model, the users are the real product."[1292] A website publisher that allows third party tracking on its website also offers a trade-off to visitors. Website visitors disclose personal information, and in exchange they can consult the website. Another example of a trade-off is joining a supermarket loyalty card programme. Customers disclose personal data, like their name and information about their shopping habits, in exchange for discounts.

Whether people realise that firms gather personal data is another matter. Acquisti notes that trade-offs can exist, even when people don't realise they disclose personal information: "the existence of such trade-offs does not imply that the economic agents are always aware of them as they take decisions that will impact their privacy."[1293] Hence, a "trade" could be analysed with economic theory, even when from a legal perspective there's no agreement to trade personal data for the use of a service. As noted, this study does not suggest that consenting to data processing *should* be seen as entering a contract from a legal perspective.[1294]

---

[1290] Acquisti 2010, p. 23.
[1291] See Acquisti & Brandimarte 2012, p. 548; European Data Protection Supervisor 2014, p. 10.
[1292] United States District Court, Northern District of California, San Jose division, Case C-12-01382-PSG, Order granting to dismiss (re: docket No. 53, 57, 59), 3 December 2013, In re Google, Inc, privacy policy litigation. See also Blue_beetle 2010: "If you are not paying for it, you're not the customer; you're the product being sold."
[1293] Acquisti 2010a, p. 4.
[1294] There is some debate on the question of whether consent to data processing should be seen as entering a (type of) contract. See chapter 6, section 1.

***Economic theory doesn't dictate the ideal level of privacy protection***

This study doesn't aim to answer the question of whether behavioural targeting leads to a net benefit or a net cost for society from an economic viewpoint. Like people who work in other disciplines, economists disagree on the ideal level of privacy protection. Neither economic theory nor empirical economic research has provided a definitive answer to the question of whether behavioural targeting – or a law that limits behavioural targeting – would lead to more or less social welfare. Some economists say that more legal protection of personal data is good, but others argue the opposite. "Economic theory", concludes Acquisti, "has brought forward arguments both supporting the view that privacy protection *increases* economic efficiency, and that it *decreases* it."[1295] Empirical economic research doesn't arrive at definitive conclusions either. "Considering the conflicting analyses", says Acquisti, "the only straightforward conclusion about the economics of privacy and personal data is that it would be futile to attempt comparing the aggregate values of personal data and privacy protection, in search of a 'final,' definitive, and all-encompassing economic assessment of whether we need more, or less, privacy protection."[1296] Other scholars agree that it's an open question whether more or less legal protection of privacy would be better from an economic perspective.[1297]

Why would it be "futile" to try to calculate the level of privacy protection that leads to the highest level of aggregate welfare? It's hard to agree on which costs and benefits to count, and many costs and benefits will only become clear after many years. Furthermore, many privacy-related costs are difficult, perhaps impossible, to quantify. Researchers have tried to measure the benefits of using of personal data and the benefits of legal limits on using personal data. They come to contradicting

[1295] Acquisti 2010a, p. 34 (emphasis original). But see Swire, who suggests "economists are largely privacy skeptics (Swire 2003, p. 24).
[1296] Acquisti 2010b, p. 19. See also Acquisti 2010a, p. 42.
[1297] See e.g. Irion & Luchetta 2013, p. 39; Strandburg 2013.

conclusions. Some say that legal privacy protection reduces social welfare, because it limits data flows.[1298]

For example, behavioural targeting has benefits, for firms and internet users. Behavioural targeting leads to profit for many firms. Internet users can benefit when revenue from targeted advertising is used to fund so-called "free" internet services. (However, in the end consumers pay for this advertising if firms pass on the advertising costs in product prices.) Behaviourally targeted advertising can bring products under the consumers' attention, which could save them searching costs. But it would be difficult to calculate the total benefits of behavioural targeting.[1299]

Likewise, aggregating all costs of behavioural targeting is difficult, or even impossible. Costs for firms include money spent on data processing systems. Furthermore, some estimate that billions of Euros are lost, because people would engage in more online consumption if they felt their privacy were better protected online.[1300] The European Commission says it would be good for the market if people worried less about their privacy. "Lack of trust makes consumers hesitate to buy online and adopt new services."[1301]

Not protecting personal data can incur costs for data subjects. Some privacy-related costs could be calculated, at least in theory. For example, when a firm experiences a data breach, the leaked data could lead to identity fraud. Such costs could materialise years after the data are collected. Or if a person's email address is disclosed too widely, this could lead to that person receiving spam. The time it takes to clean one's inbox is a cost.[1302] If people invest time in avoiding being tracked, this is costly as well.[1303] Other privacy-related costs are harder to quantify. Such costs include

---

[1298] Acquisti 2010b; Acquisti 2010a, p. 25-29.
[1299] Acquisti 2010b, p. 13; Acquisti 2010a, p. 42.
[1300] Acquisti 2010b, p. 13; Acquisti 2010a, p. 21.
[1301] European Commission proposal for a Data Protection Regulation (2012), p. 1. See also recital 5 of the e-Privacy Directive.
[1302] Acquisti & Brandimarte 2012.
[1303] Calo 2013, p. 30.

annoyance, a creepy feeling, and the long-term effects on society. In sum, while it could be attempted to quantify whether behavioural targeting leads to a net benefit or to a net loss for society, such an economic analysis would be riddled with imperfections. Moreover, as discussed below, there's more to life than economic analysis.

***Behavioural targeting and so-called "free" services***

Sometimes marketers suggest that behavioural targeting is needed to fund the so-called "free" internet, or that stricter rules would impose too much costs on businesses.[1304] Firms would lose income that they derive from personal data, and firms would spend money on compliance. But the observation that regulation imposes costs on firms doesn't conclude the economic analysis. In economics, the relevant question is whether society as a whole wins or loses. But as it's often claimed that behavioural targeting funds the "free" internet, this claim is unpacked a bit further here.

Advertising funds an astonishing amount of internet services. Without paying with money, people can use online translation tools, access many (although not all) quality newspapers, use email accounts, watch videos, listen to music, etc.[1305] It's also clear that a lot of money is at stake with behavioural targeting. For example, in 2007 Google paid 3.1 billion dollars for DoubleClick, which was a leading firm in the field of behavioural advertising.[1306] Facebook makes its money from advertising and many ads on its site are likely to behaviourally targeted.

Notwithstanding, there's reason for scepticism about the argument that the web wouldn't be "free" anymore without behavioural targeting. After reviewing the limited available data, Strandburg concludes that "apocalyptic predictions of this sort

---

[1304] See for instance Interactive Advertising Bureau United Kingdom 2012; Interactive Advertising Bureau Europe 2013.
[1305] People do usually pay for internet access at home or through a cellphone plan.
[1306] Google Investor Relations 2007.

should be taken with a large grain of salt."[1307] Even if behavioural targeting were completely banned, online advertising would remain possible. For instance, contextual advertising (such as advertising for wine on websites about wine) doesn't require monitoring people's behaviour. And for years Google didn't use behavioural targeting for its search ads.[1308] Moreover, it seems plausible that advertisers who couldn't use behavioural targeting anymore would spend some of the money saved on that type of advertising on other kinds of online advertising. Furthermore, there's little public information on how effective behavioural targeting is in improving the click-through rate on ads, when compared to contextual advertising.[1309] One industry-funded paper suggests that behaviourally targeted ads cost around 2.5 as much for advertisers than randomly presented ads.[1310] But scholars have criticised the paper for its methods and assumptions.[1311]

As a side note, behavioural targeting isn't limited to so-called "free" services. Many providers of paid services also engage in behavioural targeting. For instance, internet access providers have inspected their subscribers' internet use for behavioural targeting. Meanwhile they continued to charge their subscribers.[1312] Many paid smart phone applications also collect data for behavioural targeting.[1313]

There's little public information about the relative share of behavioural targeting income compared to other types of online advertising – let alone from independent sources.[1314] Industry organisations sometimes claim that many jobs are dependent on behavioural targeting.[1315] But other industry reports suggest that behavioural targeting

---

[1307] Strandburg 2013, p. 152. Similarly Mayer 2011.
[1308] See Hoofnagle 2009.
[1309] See Strandburg 2013, p. 10; Mayer & Mitchell 2012, p. 8.
[1310] Beales 2010.
[1311] Mayer & Mitchell 2012, p. 8.
[1312] See on deep packet inspection for behavioural targeting chapter 2, section 2.
[1313] Thurm & Iwatani Kane 2010. See chapter 2, section 3.
[1314] See Strandburg 2013, p. 10; Mayer & Mitchell 2012, p. 8.
[1315] See for high estimates Interactive Advertising Bureau Europe & McKinsey 2010. See also Direct Marketing Association (United States) 2013: "The DDME ["Data-Driven Marketing Economy"] added $156 billion in revenue to the U.S. economy and fueled more than 675,000 jobs in 2012 alone. (…) Regulation would impact all innovation, small businesses, jobs and economic growth."

isn't a major part of all online advertising income. The ValueClick firm estimated in 2008 that behavioural targeting makes up a 3.4% share of all online advertising income.[1316] A 2009 report for the Interactive Advertising Bureau US estimated the behavioural targeting share to be 18%.[1317] The Dutch Interactive Advertising Bureau concluded in 2011 that 2% of all online advertising income in the Netherlands is based on behavioural targeting.[1318] This is partly a question of definitions. For instance, Google's search ads were counted as non-behaviourally targeted in the report. Nowadays, Google's search ads are, or at least could be, behaviourally targeted.

As noted in chapter 2, in the long run, behavioural targeting may actually decrease ad revenues for some website publishers.[1319] Without behavioural targeting, advertisers that want to reach New York Times readers have to advertise on the New York Times website. Behavioural targeting enables advertisers to target people who received a cookie on the New York Times website. This implies that advertisers can reach New York Times readers without buying expensive advertising space on the New York Times website. In sum, while it can't be ruled out that some services would cease being offered for "free" if the law limited the possibilities for behavioural targeting, the long-term economic effects of legal intervention are uncertain.

The argument that behavioural targeting shouldn't be limited because it funds "free" services resembles a well-known economic argument to be cautious with consumer protection rules: consumers as a group pay the price for rules that aim to protect consumers. Firms that suffer costs from consumer protection rules are likely to pass on these to consumers by raising prices. For instance, it could be argued that legal minimum safety standards for a consumer product raise the price of that product. The higher price could mean that consumers who can only afford to buy low quality goods

---

[1316] Otlacan 2008.
[1317] Beales 2010, p. 13.
[1318] Interactive Advertising Bureau The Netherlands 2011; Interactive Advertising Bureau The Netherlands & Deloitte 2011.
[1319] See chapter 2, section 1 and 6.

can't buy the product at all.[1320] In practice, such arguments don't stop the lawmaker from requiring minimum safety standards or adopting consumer protection rules. This makes sense. As Wagner puts it, "[r]ational consumers will be prepared to pay extra in exchange for some protection from the delivery of defective products."[1321]

To conclude, it's contentious whether more legal protection of personal data would increase or decrease social welfare from an economic perspective. "In principle, there is an optimal level of data protection regulation, but, given the state of the art, it is not possible to locate it with any degree of precision," summarise Irion & Luchetta. "There is no indication whatsoever (…) whether more or less privacy would be beneficial."[1322] Acquisti adds that "it may not be possible to resolve this debate using purely economic tools."[1323]

***Limitations of economic analysis of privacy***

Economics and behavioural economics provide useful analytical tools to analyse certain practical problems with informed consent for behavioural targeting. But economic analysis has its limitations, especially when discussing fundamental rights. Policy questions can't be answered solely on economic grounds. As Posner notes in his law and economics handbook, "there is more to justice than economics."[1324]

> But there is more to notions of justice than a concern with efficiency. It is not obviously inefficient to allow suicide pacts; to allow private discrimination on racial, religious, or sexual grounds; to permit killing and eating the weakest passenger in the lifeboat in circumstances of genuine desperation, to force people to give self-incriminating

---

[1320] See Sunstein 2013a, p. 8; Luth 2010, p. 35, with further references.
[1321] Wagner 2010, p. 63.
[1322] Irion & Luchetta 2013, p. 39.
[1323] Acquisti 2010a, p. 34.
[1324] Posner 2011, p. 35.

> testimony; to flog prisoners; to allow babies to be sold for adoption; to permit torture to extract information; to allow the use of deadly force in defense of a pure property interest; to legalize blackmail; or to give convicted felons a choice between imprisonment and participation in dangerous medical experiments. Yet all these things offend the sense of justice of modern Americans, and all are to a greater or lesser (usually greater) extent illegal. An effort will be made in this book to explain some of these prohibitions in economic terms, but many cannot be. Evidently, there is more to justice than economics, and this is a point the reader should keep in mind in evaluating normative statements in this book.[1325]

Acquisti agrees that economic analysis isn't the end of the story: "the value of privacy eventually goes beyond the realms of economic reasoning and cost benefit analysis, and ends up relating to one's views on society and freedom."[1326] Certain privacy harms "not merely intangible, but in fact immeasurable."[1327] He warns against an "extremisation" of the debate.[1328] Too much attention to economics and trade-offs may take our attention away from privacy infringements that are harder to quantify. Indeed, sometimes it's suggested that there's no need to regulate behavioural targeting because the "harm" is difficult to quantify in monetary terms.[1329] In any case, European data protection law applies to personal data processing, whether there's

---

[1325] Posner 2011, p. 35. I don't suggest that Posner finds law and economics ill-equipped to discuss privacy. Posner suggests that the protection of personal information is bad from an economic perspective (Posner 1978).
[1326] Acquisti 2004, p. 27. See generally on the limitations of economic analysis of privacy Cohen 2012, chapter 6.
[1327] Acquisti 2010b, p. 3.
[1328] Acquisti 2011.
[1329] This line of argument seems to be more prevalent in the US than in Europe. See e.g. Lenard & Rubin 2010; Szoka & Thierer 2008.

(quantifiable) harm or not. The harm question is relevant where data protection law requires balancing different interests.[1330]

The problem that some types of costs and benefits are hard to quantify isn't unique for privacy. As Ramsay puts it, "[t]here is always the danger that the more measurable costs (e.g., compliance costs) to directly affected groups will be regarded as outweighing the intangible benefits to a large and diffuse consumer group."[1331] He adds that firms may be tempted to exaggerate the costs:

> If policy making is based on an economic cost-benefit analysis, then it will be in the interests of pressure groups (…) to demonstrate through their own analysis the benefits or costs of particular policies – to the extent that certain concentrated producer groups have greater access to information and expertise this may cause policy-making to be skewed in their interests, and there is always the danger therefore that cost-benefit analysis will simply become another technique to be abused to promote particular interests.[1332]

Fairness, fundamental rights, and privacy's value in a democratic society play a marginal role in the economic analysis of privacy.[1333] But such considerations are important. Irion & Luchetta note that data protection law isn't economic regulation, and that its success shouldn't be measured by looking at its economic impact.[1334] And in the European legal system, economic arguments don't trump other arguments – and they shouldn't. As Hesselink puts it, "the law should govern the market rather than the

---

[1330] The balancing provision (article 7(f) of the Data Protection Directive) is the main example, but applying open norms such as "excessive" also requires the balancing of interests.
[1331] Ramsay 1985, p. 358.
[1332] Ramsay 1985, p. 358. Baldwin et al. 2011 (p. 323) and Sunstein 2013a (p. 175) also warn for this effect.
[1333] See for an amusing text on the difficulties of combining the viewpoints of an economic approach and a EU data protection approach Kang & Buchner 2004.
[1334] Irion & Luchetta 2013, p. 23. Of course, examining the economic impact of regulation is useful.

other way round."[1335] With these caveats, let's see what economics and behavioural economics have to offer.

### 7.3 Informed consent and economics

The economic analysis of privacy decisions is largely based on the view of privacy as control over personal information. Through an economic lens, consent to behavioural targeting can be compared with entering into a market transaction with a firm. Under rational choice theory, there may be reason for the lawmaker to intervene in contractual freedom, for instance because of market failures such as information asymmetries, externalities, or market power.[1336]

***Information asymmetry***

Information asymmetry describes "a situation where one party possesses information about a certain product characteristic and the other party does not."[1337] Since the 1970s economists devote much attention to markets with asymmetric information, for example where consumers have difficulties evaluating the quality of products or services. Akerlof used the market for used cars as an example.[1338] Suppose sellers offer bad cars ("lemons") and good cars. Sellers know whether they have a bad or a good car for sale, but buyers can't detect hidden defects. A rational buyer will offer the price corresponding to the average quality of all used cars on the market. But this means that sellers of good cars are offered a price that is too low. Hence, owners of good cars won't offer their cars for sale. The result is that the average quality of used cars on the market decreases. Buyers will therefore offer lower prices, and fewer people will offer their cars for sale. The average quality of cars on the market will

---

[1335] Hesselink 2005, p. 179.
[1336] US legal scholars have applied insights from law and economics to consent to online data processing (e.g. Kang 1998; Schwartz 2003). In Europe, Brown 2013 gives an analysis of market failures in the area of online privacy.
[1337] Luth 2010, p. 23.
[1338] Akerlof 1970. He focuses on one problem resulting from information asymmetry: adverse selection. Another market failure that is related to information asymmetry falls outside the scope of this study: moral hazard.

drop. Sellers thus don't compete on quality in a market characterised by asymmetric information about quality, resulting in a race to the bottom. This can lead to products or services of low quality.

From an economic perspective, there may be reason for the lawmaker to intervene, because information asymmetries can lead to market failure. For instance, an economist might argue that one of the main rationales for consumer law is responding to information asymmetry.[1339] Seen from this angle, the main reason for responding to information asymmetry is protecting a well-functioning market, rather than paternalistic motives towards the consumer. If a lawyer said that consumer law aims to protect consumers because of their weaker bargaining position, an economist might add that the weaker bargaining position can be largely explained by information asymmetry.[1340]

***Information asymmetry and behavioural targeting***

The current state of affairs regarding behavioural targeting is characterised by large information asymmetries.[1341] Many firms track people for behavioural targeting without them even being aware. When one sees releasing personal data as "payment" for services, it's clear that there are information asymmetries. As Cranor & McDonald put it, "people understand ads support free content, but do not believe data are part of the deal."[1342] To make an informed choice, people must realise they are making a choice.

Research shows that most people are only vaguely aware that data are collected for behavioural targeting. For instance, Ur et al. found in interviews that participants were

---

1339 See e.g. Luth 2010, p. 15, p. 69; Howells 2005, p. 352; Grundmann 2002, p. 279.
1340 See Ramsay 1985, p. 369. The European Court of Justice combines the two views: "the [Unfair Contract Terms] Directive is based on the idea that the consumer is in a weak position vis-à-vis the seller or supplier, as regards both his bargaining power and his level of knowledge" (ECJ, C-243/08, Pannon GSM, 4 June 2009, par. 22). See on paternalism chapter 6, section 6.
1341 Acquisti & Grossklags 2007.
1342 Cranor & McDonald 2010, p. 21.

"surprised to learn that browsing history is currently used to tailor advertisements."[1343] In a survey, Cranor & McDonald found that 86% of respondents were aware that behavioural targeting takes place.[1344] But they also find that people know little about how data relating to their online behaviour is collected: "it seems people do not understand how cookies work and where data flows."[1345] Furthermore, only 40% of respondents thought that providers of email services scan the contents of messages for the purpose of targeted advertising. 29% thought this would never happen, either because the law prohibits it, or because the consumer backlash would be too great. Almost half of Gmail users didn't know about the practice,[1346] while Gmail has been scanning emails for advertising since 2004.[1347] Research in Europe also suggests that many people are unaware of behavioural targeting.[1348] Cranor & McDonald conclude that people generally lack the knowledge needed to make meaningful decisions about privacy in the area of behavioural targeting.[1349] In addition, people who have learned how to defend themselves against tracking must update their knowledge constantly.[1350] For example, many firms used flash cookies to re-install cookies that people deleted. Hoofnagle et al. summarise: "advertisers are making it impossible to avoid online tracking."[1351]

But if firms did ask for consent for behavioural targeting, information asymmetry would still be a problem, notes Acquisti.[1352] First, there are many firms involved in serving behaviourally targeted ads, and the underlying data flows are complicated. It's almost impossible for people to find out what happens to their data. Will their name

---

[1343] Ur et al. 2012, p. 4.
[1344] However, only 51% of the respondents though that this happens a lot at present (Cranor & McDonald 2010, p. 21).
[1345] Cranor & McDonald 2010, p. 16.
[1346] Cranor & McDonald 2010, p. 21.
[1347] Battelle 2005, chapter 8.
[1348] Helberger et al. 2012, p. 70.
[1349] Cranor & McDonald 2010.
[1350] Acquisti and Grossklags 2007 make a similar point, giving other examples.
[1351] Hoofnagle et al. 2012, p. 273. See on tracking technologies chapter 2, section 2.
[1352] Acquisti 2010b, p. 15-16; Acquisti 2010a, p. 38. Acquisti doesn't explicitly present these three categories.

be tied to the profile of their surfing behaviour? Will their data be shared with other firms? If a firm goes bankrupt, will its database be sold to the highest bidder?[1353]

Second, even if people knew what firms did with their data, it would be difficult to predict the consequences.[1354] If a firm shares data with another firm, will the data be used for price discrimination? Will visits to a website with medical information lead to higher health insurance costs? If there's a data breach at a firm, will this lead to identity fraud?

Third, it's difficult for people to attach a monetary value to information about their behaviour, so they don't know how much they "pay." For instance, people may not know how much profit a firm makes with their information, or what the costs are of a possible privacy infringement. The value of the so-called "payment", that is the piece of personal information, depends on the question of what the receiving parties do with the personal information.[1355] Put differently: the "price" paid by the website visitor only becomes clear when firms exploit the personal information. "To what, then," asks Acquisti, "is the subject supposed to anchor the valuation of her personal data and its protection?"[1356]

As Vila at al. note, if the privacy-friendliness of websites is seen as a product feature, the web has characteristics of a lemons market.[1357] It's hard for people to determine how much of their personal information is captured during a website visit and how the information will be used. And website publishers rarely use privacy, or the absence of tracking, as a competitive advantage. Virtually every popular website tracks the

---

[1353] See e.g. the Toysmart case in the US (In re Toysmart.com, LLC, Case no. 00-13995-CJK, in the United States Bankruptcy Court for the District of Massachusetts 2000), and the Broadcast Press case in the Netherlands (Voorzieningenrechter Rechtbank Amsterdam, 12 February 2004, ECLI:NL:RBAMS:2004:AO3649 (Broadcast Press)).
[1354] Acquisti & Grossklags 2007, p. 365.
[1355] See Schwartz 2000a, p. 775; Strandburg 2013, in particular p. 130-165.
[1356] Acquisti 2010a, p. 39.
[1357] Vila et al. 2004. A similar conclusion is drawn by Pasquale 2013; European Data Protection Supervisor 2014, p. 33.

behaviour of visitors for behavioural targeting, or allows third parties to track the visitors.[1358]

"This situation looks like the classic market for lemons problem", says Strandburg about behavioural targeting. "Consumers cannot recognize quality (here, absence of data collection for advertising) and hence will not pay for it. As a result, the market spirals downward."[1359] After interviewing people in the online marketing business, Turow concludes that competition pushes firms towards privacy invasive marketing practices, which seems to confirm the lemons situation.[1360] Furthermore, many website publishers don't have much power in negotiations with ad networks. There also seems to be a lemons problem in the market for smartphone applications and social network sites.[1361]

There are firms, such as a few search engine providers, that use privacy-friendliness as a selling point.[1362] But it's difficult for a firm to distinguish itself from others by offering privacy-friendly services. Virtually every privacy policy begins with phrases along the lines of: "the privacy of our users is and will continue to be a top priority for us."[1363] (In many cases, website publishers firms say later in the privacy policy that they allow third party tracking.) Therefore, it's difficult for a website publisher to use the fact that it doesn't allow third party tracking as an incentive for potential visitors to use its website. At first glance, its privacy policy wouldn't look much different than privacy policies of other websites that do allow third party tracking.[1364]

A hypothetical fully rational person would know how to deal with information asymmetry and uncertainty. For instance, the person could base his or her decision on what happens to people's personal data on average, and he or she wouldn't be

---

[1358] See chapter 2, section 3.
[1359] Strandburg 2013, p. 156.
[1360] Turow 2011, p 199.
[1361] See on social network sites and information asymmetry Bonneau & Preibusch 2010.
[1362] Two examples are: <www.duckduckgo.com> and <www.startpage.com>. See also Willis 2013a, p. 128-130.
[1363] This phrase is taken from the blog post in which Yahoo said it wouldn't honour Do Not Track signals (Yahoo 2014). See on Do Not Track chapter 8, section 5.
[1364] See Marti 2014.

optimistic about quality in a lemons situation. But people don't tend to deal with information asymmetry in a "rational" way (see section 4 of this chapter).

One caveat: most authors that apply law and economics to behavioural targeting discuss the American situation. In the US, there's no general data protection law; online privacy is mostly governed by self-regulation, the Federal Trade Trade's Commission norms on unfair business practices, and narrowly tailored sector-specific statutes.[1365] In theory, the information asymmetry problems should be less severe if all firms complied with European data protection law. For instance, if firms would always comply with the purpose limitation principle, unexpected data uses should be rare. In practice compliance with data protection law is not a given, partly because many popular services are from American origin.[1366]

***Transaction costs***

The obvious reaction to information asymmetries is requiring firms to provide information to data subjects. But this runs into problems as well, because of transaction costs among other reasons. "Transaction costs are any costs connected with the creation of transactions themselves, apart from the price of the good that is the object of the transaction."[1367] Examples are the time a consumer spends on reading contracts, or searching for a product. Transaction costs aren't a market failure, but they can help to explain why the information asymmetry problem is difficult to solve.[1368]

***Transaction costs and behavioural targeting***

In the behavioural targeting area, the time it would take people to inform themselves is a transaction cost. Hence, because of transaction costs the information asymmetry

---

[1365] See Schwartz & Solove 2009.
[1366] See on the purpose limitation principle chapter 4, section 3. See on the (lack) of compliance chapter 8 section 1, and chapter 9, section 1.
[1367] Luth 2010, p. 20 (emphasis omitted). The classic article on transaction costs is Coase 1960.
[1368] See Dahlman 1979.

problem is likely to persist. Law and economics literature on consumer law suggests that consumers don't read standard contracts, partly because of the transaction costs. As consumers don't read standard contracts, there's information asymmetry, and firms don't compete on the quality of standard contracts. This can lead to a lemons situation, with contracts that are unfavourable to consumers.[1369] The situation is similar for behavioural targeting.

As noted, the transparency requirements in European data protection law should be distinguished from the obligation to obtain consent for data processing, or for using tracking technologies.[1370] In practice, many firms seek consent in their terms and conditions, or in their privacy policies. But hardly anyone reads privacy policies or consent requests. To illustrate, an English computer game store obtained the soul of 7500 people. According to the website's terms and conditions, customers granted "a non transferable option to claim, for now and for ever more, your immortal soul," unless they opted out. By opting out, people could save their soul and could receive a five pound voucher. But few people opted out. The firm later said it wouldn't exercise its rights.[1371]

Marotta-Wurgler researched the readership of end user license agreements (EULAs) of software products. She analysed the click streams of almost 50,000 households, and found an "average rate of readership of EULAs (…) on the order of 0.1 percent to 1 percent." On average, those readers didn't look long enough at EULA to read them.[1372] "The general conclusion is clear: no matter how prominently EULAs are disclosed, they are almost always ignored."[1373] There's little reason to assume the readership of privacy policies is much higher.

---

[1369] See e.g. Faure & Luth 2011, p. 342; Wagner 2010, p. 61-62; Schäfer & Leyens 2010, p. 105, p. 108.
[1370] See chapter 4, section 3.
[1371] Fox News 2010.
[1372] Marotta-Wurgler 2011, p 168.
[1373] Marotta-Wurgler 2011, p. 182.

There are several reasons why people don't read privacy policies. First, life is too short. Cranor & McDonald calculate that it would cost the average American 244 hours per year to read the privacy policies of the websites she visits. This would be about 40 minutes a day, or about half of the time that the average American spent online every day (in 2006). Expressed in money, this cost would be around 781 billion dollars, in lost productivity and lost value of leisure time, if people actually were to read privacy policies.[1374] The costs for individuals to inform themselves exceeded the revenues from the ad industry they might try to protect themselves from. All online advertising income in the US was estimated to be 21 billion dollar in 2007.[1375] Moreover, people have better things to do than reading privacy policies. In daily life, people encounter information everywhere. For instance, many services and products come with terms and conditions. And the law often requires firms to disclose information to people. For example, European consumer law also relies heavily on information requirements.[1376]

Privacy policies are often long and difficult to read. In one study, more than half of the examined privacy policies were too difficult for a majority of American internet users.[1377] A quarter of Europeans say privacy policies are too difficult.[1378] And privacy policies are often vague, and fail to make data processing transparent.[1379] (The author of this study often has trouble deducing from a privacy policy what a firm plans to do with personal data.)

And if people understood a privacy policy, it's questionable whether they'd realise the consequences of the combination and analysis of their data. A user might only release scattered pieces of personal data here and there, but firms could still construct detailed

---

[1374] It would be more correct to speak of the "opportunity costs" for the individual.
[1375] Cranor & McDonald 2008. The study only looked at the time to read first party notices, with no time estimates for third party privacy policies.
[1376] Luth 2010.
[1377] Jensen & Potts 2004.
[1378] European Commission 2011 (Eurobarometer), p 112-114.
[1379] Verhelst 2012, p. 221.

profiles by combining data from different sources.[1380] In addition, even if somebody manages to decipher a privacy policy, his or her quest might not be over. A website's privacy policy often refers to the privacy policies of ad networks or other firms. Hence, people might have to consult dozens of privacy policies to learn about data collection on one website. Some firms change their privacy policies without notice, so people would have to check a privacy policy regularly. All these transaction costs hinder meaningful decisions regarding behavioural targeting.

The accepting without reading problem isn't unique to the privacy field. Most consumers don't read (other) contracts either.[1381] Some have argued that an "informed minority" of consumers disciplines the market by reading contracts. The idea is that firms adapt their contracts to the few people who read contracts.[1382] But many authors are sceptical about the informed minority argument. If an informed minority is too small, it won't discipline the market.[1383] It seems there aren't enough people who read privacy policies to discipline the market in the behavioural targeting area. True, a change in a firm's privacy policy could lead to media attention, and sometimes firms react to that.[1384] But such cases are rare.

If somebody read and understood a privacy policy, transaction costs could still be a problem. Moving to another service often involves transaction costs for the user. For instance, transferring emails and contacts to another email provider costs time. Furthermore, "when the costs of switching from one brand of technology to another are substantial, users face *lock-in*."[1385] If iTunes changes its privacy policy, many people might just accept. And when all one's friends are on Facebook, it makes little

---

1380 Barocas & Nissenbaum 2009, p. 6.
1381 EU consumer law makes certain contract terms invalid, which makes it, to some extent, safe for consumers not to read contracts (see for instance the Unfair Contract Terms Directive).
1382 Schwartz & Wilde 1978, p. 638.
1383 See e.g. Luth 2010, p. 149; Bakos et al. 2009.
1384 For instance, after attention in the press, Facebook offered people a way to opt out of their "Beacon" service (Debatin et al. 2009).
1385 Shapiro & Varian 1999, p. 104.

sense to join another social network site.[1386] In addition, there might not be any privacy friendly competitors, especially since there's information asymmetry in the market. As noted, most popular websites allow third parties to track their visitors for behavioural targeting. To illustrate, it's hard to find a tracking-free news website.

Some firms use transaction costs strategically. Firms can discourage people from opting out, by requiring multiple mouse clicks for an opt-out. For instance, people face transaction costs when they want to opt out of receiving behaviourally targeted ads on the website Youronlinechoices, managed by the Interactive Advertising Bureau. It takes three clicks and a waiting period to opt out of receiving behaviourally targeted ads.[1387] Opting out of Google's advertising cookies takes five mouse clicks from its search page.[1388]

Lastly, reading privacy policies doesn't guarantee that somebody knows what will happen with his or her data. For instance, some firms don't act according to their privacy policy. Google said on a website that people who used the Safari browser on certain devices were effectively opted out of tracking, because Safari blocks third party cookies. But Google bypassed Safari's settings.[1389] It would take people too much time to keep track of whether firms actually comply with their privacy policies. Furthermore, things can go wrong. A firm could experience a data breach for example.

Because of transaction costs, there may be an economic argument for having policymakers set standards. As Baldwin et al. note, "if information disclosure rules were employed instead of [other] regulation in relation to food safety, a visit to the

---

[1386] The European Commission proposal introduces a right to data portability to mitigate the problem of lock-in. See article 18 and recital 55 of the European Commission proposal for a Data Protection Regulation (2012).

[1387] In a non-scientific test, I had to wait forty-five seconds. First I had to choose a country (click 1), then I had to click on "your ad choices" (click 2). Next I had to wait until the website contacted the participating ad networks. Then I could opt out of receiving targeted advertising (click 3). For several ad networks the website gave an error message. (See Interactive Advertising Bureau Europe - Youronlinechoices.) See Leon et al. 2012 for a more academic discussion of the (non) user friendliness of industry opt-out systems.

[1388] College bescherming persoonsgegevens (Dutch DPA) 2013 (Google), p. 82.

[1389] Felten 2012; Mayer 2012. See chapter 2, section 2.

supermarket would involve a very lengthy process of scrutinizing labels."[1390] Therefore, there could be an economic rationale for having regulators ensuring a reasonable level of food safety. "It might, in many circumstances, be far more efficient for consumers to rely on the expertise and protection of public regulators and inspectorates, rather than depend on their own individual assessments of risks.[1391] A similar argument can be made in the area of behavioural targeting.

Outside data protection law, rules that require firms to disclose information to people are ubiquitous as well. Lawmakers often choose this regulatory technique in the hope people will make decisions in their own best interests. In European consumer law, for instance, this is the predominant approach.[1392] But there's little evidence that providing information helps to steer people towards decisions in their own best interests. Many scholars are sceptical.[1393] Ben-Shahar & Schneider summarise: "[n]ot only does the empirical evidence show that mandated disclosure regularly fails in practice, but its failure is inevitable."[1394]

Privacy policies fail to inform people who use computers, and it's even more difficult to inform people who use mobile devices with smaller screens. Soon it may become even harder to make data processing transparent, if more objects will be connected to the internet. Common phrases in this context are the Internet of Things, ubiquitous computing, and ambient intelligence.[1395] It's hard to give people effective information about behavioural targeting when they use computers and smart phones, but transparency would be even harder to achieve if firms use objects without a screen for

[1390] Baldwin et al. 2011, p. 120.
[1391] Baldwin et al. 2011, p. 120. See also Helberger 2013a, p. 37.
[1392] Grundmann et al. 2001; Luth 2010, p. 228. See on the US: Ben-Shahar & Schneider 2011.
[1393] See for an overview, with references Luth 2010.
[1394] Ben-Shahar & Schneider 2011, p. 651.
[1395] See chapter 2, section 2.

data collection. And it's not straightforward how informed consent could work in such an environment.[1396]

The foregoing doesn't imply that data protection law's transparency principle is useless. The transparency requirements can help to make behavioural targeting controllable for Data Protection Authorities and lawmakers. Without data protection law more problems might remain hidden. If problems are brought to light, the lawmaker could intervene.[1397] Hence, data protection law's transparency requirements could serve an important purpose, even if they fail to empower the individual.

***Externalities***

From an economic viewpoint, one reason for legal intervention in markets is when an activity has negative effects on people other than the contract parties. Economists refer to costs or damage suffered by third parties as a result of economic activity as negative externalities. Externalities occur because contract parties that aim to maximise their own welfare don't let costs for others influence their decisions.

An example of an externality is environmental pollution. Suppose a firm produces aluminium, and sells it to another party. If producing aluminium causes pollution, it imposes costs on others. Rational producers and buyers ignore these costs. When the costs of pollution for others are taken into account, too much aluminium is produced from a social welfare perspective. Global warming could be seen as an enormous externality problem. Externalities can also be positive. If somebody hires a gardener to craft a beautiful garden in front of her house, other people in the street might enjoy

[1396] See Article 29 Working Party 2014, WP 223. There's research on how to enable informed consent in a ubiquitous computing environment. See e.g. Le Métayer & Monteleone 2009.

[1397] This is one of the rationales for the obligation for data controllers to notify Data Protection Authorities of processing operations (article 18-21 of the Data Protection Directive). The 2012 proposals abolish this requirement.

the sight. These neighbours gain welfare from the garden without paying for it; they enjoy a positive externality.[1398]

Many legal rules, such as the rules in environmental law, can be explained as a response to an externalities problem. Even a rule that makes a contract to commit a murder void could be seen in this light. The rule protects a third party, namely the intended victim. Similarly, a prohibition of falsely yelling “fire” in a crowded theatre could be seen as a response to an externality problem.[1399] Legal responses to externalities often limit an individual’s freedom. Generally speaking, if the lawmaker wants to reduce negative externalities resulting from contracting practices, the rules have to be mandatory. If the lawmaker would use non-mandatory default rules, the contract parties would set the rules aside.[1400] After all, the externality is caused by the fact that the contract parties don’t take the interests of non-contract parties into account.[1401] Legal responses to externalities have nothing to do with paternalism, as the rules don’t aim to protect people against themselves.

***Externalities and behavioural targeting***

Are externalities relevant for consent to behavioural targeting? If somebody consents to sharing his or her data with a firm, there are no negative externalities at first glance. The person merely gives up an individual interest. But people’s consent to behavioural targeting may lead to the application of knowledge to others. This could be seen as an externality imposed on others.[1402] For instance, say a supermarket can track the shopping behaviour of thousands of customers that joined a loyalty programme and consented to having their data analysed. The supermarket constructs the following predictive model: 90% of the women who buy certain products will give birth within two months. Out of privacy considerations, Alice didn’t join the

---

[1398] See on externalities Coase 1960; Dahlman 1979; Trebilcock 1997, chapter 3; Luth 2010, p. 22.
[1399] See US Supreme Court, Schenck v. United States - 249 U.S. 47 (1919).
[1400] See on the difference between mandatory rules and default rules chapter 6, section 5.
[1401] Wagner 2010, p. 53.
[1402] See MacCarthy 2011; Brown 2013; Hildebrandt et al. 2008. See also Hirsch 2006, who compares negative externalities in the context of environmental law and privacy law.

loyalty programme. But when she buys certain products, the shop can predict with reasonable accuracy that she's pregnant.[1403] This could be seen as an externality imposed on Alice, which is a result from the fact that people consented to having their personal information processed. Hence, firms can also learn information about people who do not agree to data collection. This topic is completely separate from the issue of people tending to click "I agree" to many requests.[1404]

Moreover, if almost everybody consents to being tracked, *not* consenting could make somebody conspicuous. Does he or she have something to hide? Sometimes not divulging information, or not participating, can raise suspicion.[1405] Osama Bin Laden was found, partly because it was suspicious that his large compound didn't have internet access.[1406] And some intelligence services find it suspicious if internet users use privacy enhancing technologies.[1407]

There may be positive externalities when people consent to behavioural targeting. For instance, firms might use behavioural targeting data that are collected with consent for innovative products that other people can use. It could be seen as a positive externality if innovative products benefit other parties than the firm and the person that consented.[1408] An oft-cited example of a positive externality resulting from commercial data collection is Google Flu trends. In short, Google uses people's search behaviour to deduce information about the spread of flu.[1409] However, the usefulness of the service has been questioned.[1410]

---

[1403] The example is based on a news report on the US supermarket Target, which reportedly found that a woman was pregnant, based on the products she bought (see chapter 2, section 5).
[1404] See Barocas 2014, p. 159.
[1405] See Posner 2011, p. 25. Peppet 2011.
[1406] Ambinder 2011.
[1407] See for instance Greenwald & Ball 2013.
[1408] New uses of personal data may breach the purpose limitation principle, but we'll leave that topic aside for now (see chapter 4, section 3). Some might argue that so-called "free" websites are a positive externality, enjoyed by web users, of contracts between website publishers and advertisers (see Strandburg 2013, p. 108, who is critical of that claim).
[1409] Ginsberg et al. 2009.
[1410] Ohm 2013, p. 342. Furthermore, research suggests that Flu Trends isn't very accurate (Hodson 2014; Lazer et al. 2014).

The phrase "big data" has become a buzzword. There's no commonly accepted definition, but "big data" roughly refers to the analysing large data sets. Some have high hopes for "big data", and speak of "a revolution that will transform how we live, work and think."[1411] Others are sceptical.[1412] According to Arnbak for instance, "the concept of 'big data' [is] a carefully constructed frame by proponents of systematic surveillance for commercial purposes."[1413] As an aside, legal limits on the use of personal data don't imply that all advantages of large-scale data analysis are lost. Many positive externalities could also be generated by using aggregated data, rather than personal data. And not all large-scale data analysis ("big data") relies on data about individuals.

In this chapter, the focus is on externalities resulting from an individual consenting to a firm processing his or her personal data. Another example of a negative privacy externality is a firm that sells Alice's contact information to other firms, thereby increasing the chance that Alice is subjected to invasive marketing, such as spam.[1414] And privacy invasive tracking that results from a contract between an ad network and a website publisher could be seen as an externality imposed on website visitors.

In conclusion, it would be difficult to assess whether the positive externalities of behavioural targeting outweigh the negative externalities or vice versa. But consent to behavioural targeting does have negative externalities. If lawmakers want to respond to negative externalities, they generally need to use mandatory rules rather than default rules.

***Market power***

Market power, like a monopoly situation, may be a reason for legal intervention from an economic viewpoint. In a perfectly competitive market, many firms must compete

---

1411 Mayer-Schönberger & Cukier 2013. See also Manyika et al. 2011; Tene & Polonetsky 2012a; Tene & Polenetsky 2013; Moerel 2014; World Economic Forum 2014.

1412 See for instance boyd & Crawford 2013; Ohm 2013; Morozov 2013.

1413 Arnbak 2013. See on behavioural targeting as surveillance chapter 3, section 3.

1414 Varian 2009, p. 103.

for consumers and firms have no market power. Without problems such as information asymmetries, competition should lead to products that consumers want, for prices close to the productions costs. Competition should thus lead to the highest social welfare, and to consumer-friendly services. This is the rationale for laws that aim to mitigate market power, such as competition law. The opposite of a perfectly competitive market is a monopoly situation. A monopolist has market power and can raise prices without fearing the reaction of competitors.[1415]

***Market power and behavioural targeting***

Privacy scholars often complain that people lack real choice if firms offer take-it-or-leave-it-choices.[1416] This is a valid concern. As noted, from a data protection law perspective, sometimes the position of a firm asking consent is such that consent wouldn't be sufficiently "free."[1417] However, data protection law and economics use different frameworks. From an economic perspective the question of whether there's too much market power depends on the specifics of that particular market. The conclusion would be different for search engines, social networks sites, online newspapers, or games for phones.

Many take-it-or-leave-it choices regarding behavioural targeting may not be an abuse of market power from the viewpoint of competition law or economics.[1418] For instance, there could be a situation of monopolistic competition, where many firms compete by differentiating similar products. This often occurs in markets for magazines or newspapers. For online services, such as websites and smart phone apps, monopolistic competition is common as well. Monopolistic competition is usually not regarded as a market power problem from an economic viewpoint. If a

---

[1415] See Bar-Gill 2012, p. 16.
[1416] See Solove 2013, p. 1898; Blume 2012, p. 29; Rouvroy & Poullet 2009, p. 50; p. 70-74; Bygrave 2002, p. 58-59.
[1417] See chapter 6, section 3 and 4, and chapter 8, section 3 and 5.
[1418] See on the interplay between competition law and data protection law European Data Protection Supervisor 2014.

user said a website doesn't give a real choice whether to allow tracking or not, an economist might counter that the user could visit another website.[1419]

Even in a perfectly competitive market, many problems described in this chapter could remain. For example, information asymmetries can lead to a lemons situation with services that offer low privacy levels, even if a market is perfectly competitive.[1420] Therefore, market power may not be the main problem for consent to behavioural targeting.

Nevertheless, market power may be relevant for consent to behavioural targeting.[1421] As noted in chapter 2, the online marketing industry is becoming increasingly centralised.[1422] If in ten years a couple of firms are responsible for all behavioural targeting in the world, this calls for different regulatory answers than if thousands of firms engage in behavioural targeting.

In conclusion, people face severe difficulties when deciding whether to consent to behavioural targeting. One of the main problems is asymmetric information. Transaction costs make this information asymmetry difficult to overcome. From an economic perspective, information asymmetry can lead to market failure, which justifies regulatory intervention, provided that legal intervention doesn't bring too many costs or economic distortions. The next section shows that there are also "behavioural market failures" in the area of behavioural targeting.[1423]

### 7.4 Informed consent and behavioural economics

Behavioural economics highlights more problems with informed consent to behavioural targeting. Behavioural economics aims to improve the predictive power

---

[1419] In practice, there's a good chance that the same ad networks would track people on other websites. Chapter 8, section 3 and 5, and chapter 9, section 5 and 7, return to the topic of take-it-or-leave-it-choices.
[1420] Bar-Gill 2012, p. 16.
[1421] See on privacy and market power Brown 2013; European Data Protection Supervisor 2014.
[1422] Chapter 2, section 2.
[1423] The phrase "behavioral market failure" comes from Bar-Gill 2012.

of economic rational choice theory by including findings from psychology and behavioural studies. Research shows that people structurally act differently than rational choice theory predicts.[1424]

If many people made decisions that didn't conform to rational choice theory, but did so in different ways, on average their decisions might still conform to rational choice theory. Random deviations from rational choice theory would not influence the theory's predictive power in the aggregate.[1425] But people tend to make decisions that are *systematically* different from what rational choice theory predicts. Sunstein summarises: "[p]eople are not always 'rational' in the sense that economists suppose. But it does not follow that people's behaviour is unpredictable, systematically irrational, random, rule-free or elusive to scientists. On the contrary, the qualifications can be described, used, and sometimes even modeled."[1426]

One difference between people who conform to rational choice theory and people in the real world is that people in the real world have bounded rationality. Human attention is scarce. Simon explains: "[t]he term 'bounded rationality' is used to designate rational choice that takes into account the cognitive limits of the decision maker – limitations of both knowledge and computational capacity."[1427] The human mind has limited capacity for decisions that require taking many factors into account. People tend to be bad at calculating risks and at statistics in general.

Because of their bounded rationality, people often rely on rules of thumb, or heuristics. Kahneman defines a heuristic as "a simple procedure that helps find adequate, though often imperfect, answers to difficult questions."[1428] Most of the time such mental shortcuts work fine. "Do as the others do" is often a useful heuristic, for

---

[1424] There are heated debates among economists on the question of whether behavioural economics really adds something to neoclassical economics (see e.g. Posner 1998). This study doesn't take sides in this debate. Some might argue that certain biases discussed in this section could partly be explained under neoclassical economic theory (see e.g. Cofone 2014).
[1425] Posner 1998.
[1426] Sunstein 2000, p. 1.
[1427] Simon 1997 (1987).
[1428] Kahneman 2011, p. 98.

instance. When you are in a department store and everybody starts to flee for the exit, leaving the building too might be a good idea. But sometimes, heuristics lead to decisions that people later regret. “Humans predictably err.”[1429] Such systematic deviations from rational choice theory, or common mistakes, are called biases.

Biases are studied and used in marketing and advertising.[1430] As Bar-Gill explains, “competition forces sellers to exploit the biases and misperceptions of their customers.”[1431] Apart from questions of fairness, this can lead to “behavioural market failures”, and thus decrease social welfare.

> The basic claim is that market forces demand that sellers be attentive to consumer psychology. Sellers who ignore consumer biases and misperceptions will lose business and forfeit revenue and profits. Over time, the sellers who remain in the market, profitably, will be the ones who have adapted their contracts and prices to respond, in the most optimal way, to the psychology of their customers.[1432]

Privacy scholars have started to take behavioural economics insights into account.[1433] Important behavioural research on how people make privacy choices is done by scholars such as Acquisti, Cranor and McDonald, who all work, or worked, at the Carnegie Mellon University in Pittsburgh. Acquisti & Brandimarte note that even fully informed people often have difficulties making privacy choices in their own interests.

---

1429 Sunstein & Thaler 2008, p. 7.
1430 Howells 2005, p. 361-362; Bar-Gill 2012.
1431 Bar-Gill 2012, p. 2.
1432 Bar-Gill 2012, p. 8. Luth 2010 reaches a similar conclusion (p. 81, p. 107-108, p. 288). See also Sunstein 2013a, p. 90; Sunstein 2013.
1433 An influential paper is Acquisti & Grossklags 2007.

> As a matter of fact, the information available to individuals when making decisions regarding privacy is often incomplete (…). Moreover, due to bounded rationality, the individual cannot obtain and retain all information necessary to make a perfectly rational decision. Even if she could access all that information, and even if she had unlimited capability of information storage and processing, her choices would nonetheless be influenced by several psychological biases and heuristics (…) All these factors influence the individual's privacy decision-making processes in such a way that even if she was willing, in theory, to protect her privacy, in practice she may not do so.[1434]

Somebody who wants to make a rational choice to consent to behavioural targeting would have to take a number of factors into account. Making "rational" choices about complex matters such as privacy is difficult, and people often rely on heuristics for such choices. Relying on heuristics for privacy decisions can lead to biases, such as the status quo bias and present bias.

***Status quo bias***

The status quo bias, or inertia, refers to the power of the default.[1435] Most people don't change the default option. This means that the default setting will have a big impact on the dynamics between the firm and its users. A famous example of the status quo bias concerns the percentage of organ donors. Countries that use an opt-out system (people donate their organs unless they express that they don't want to donate) have many donors, while countries that use an opt-in system have few donors.[1436] The status quo bias is surprising from a rational choice perspective. Rational choice theory

---

[1434] Acquisti & Brandimarte 2012, p. 564.
[1435] See Samuelson & Zeckhauser 1988.
[1436] Johnson & Goldstein 2003.

would predict that people choose according to their preferences, regardless of the default option – assuming there are no transaction costs to changing the default.[1437]

Marketers can leverage the status quo bias. Free trial periods of newspapers can lead to subscriptions for years, because – in line with the status quo bias – people don't get around to cancelling. "Buy this pack of shampoo, and get a 2 euro refund", relies on transaction costs and the status quo bias. With such mail-in rebates, many people fail to send in the coupon. As an aside, sending in the coupon would also disclose one's name and bank account number to the firm.

The status quo bias is relevant for behavioural targeting. As Sunstein puts it, "true, we might opt out of a website policy that authorizes a lot of tracking (perhaps with a simple click) – but because of the power of inertia, many of us are not likely to do so."[1438] Few people tweak the settings of their browser or their social network site accounts.[1439] The effect of the status quo bias is aggravated if switching to another service also entails transaction costs.[1440]

Insights into the status quo bias help to understand the decades-old discussion about opt-in versus opt-out systems for direct marketing and behavioural targeting. This is basically a discussion on who profits from the status quo bias. Firms often prefer to collect personal data, unless people object. This illustrates that marketers understand the power of the default.[1441] Privacy advocates tend to prefer opt-in systems for privacy-intrusive practices.[1442] As noted, a purely dogmatic analysis of the law also leads to the conclusion that an expression of will is required for valid consent.[1443]

---

[1437] Of course, that assumption rarely holds in practice.
[1438] Sunstein 2013, p. 1893. See along similar lines Sunstein 2013a, p. 102.
[1439] On the settings of social media accounts Acquisti & Gross 2006.
[1440] See on transaction costs section 3 of this chapter.
[1441] As the DoubleClick ad network puts it, a default browser setting that doesn't allow third party cookies "is basically equivalent to not allowing them at all, because 99% of the population will see no reason to change the default." (Kristol p. 188.)
[1442] See Willis 2013a, especially p. 81.
[1443] See chapter 6, section 3 and 4.

*Myopia and other biases*

More biases are relevant for consent to behavioural targeting, such as myopia. Literally myopia means limited sight, or short sightedness. In behavioural economics, myopia refers to the effect that people tend to focus more on the present than on the future. People often pursue immediate gratification, thereby ignoring future costs.[1444] For example: "I can finish these footnotes on Sunday." People who are planning to lose weight might still eat a piece of cake, because it looks so good now, thereby forgetting they were planning to eat less sugar. Myopia also helps to explain why many people find it difficult to save money for their retirement.[1445]

People might choose immediate access to a service, even if this means they have to consent to behavioural targeting, contrary to earlier plans. Say Alice reads about behavioural targeting and decides not to accept any more tracking cookies. That night, she wants to read an online newspaper, and wants to watch the news online. Both websites deny entry to visitors that don't accept third party tracking cookies.[1446] Contrary to her earlier plans, Alice clicks "yes" on both websites. Hence, people don't always stick with default options. Sometimes this can be explained by myopia, or present bias.[1447]

Overconfidence and optimism biases are related to myopia. People tend to underestimate the risk of accidents and diseases, and overestimate the chances of a long and healthy life, or winning the lottery. Most drivers think they drive better than the average driver, and most newlywed couples think there's an almost 100% chance that they will stay together, even when they know that roughly one in two marriages

---

[1444] Luth 2010, p. 53.
[1445] Sunstein & Thaler 2008, chapter 6.
[1446] Early 2013 this was the case in the Netherlands. The National Public Broadcasting Organisation and one of the larger newspapers (Volkskrant) both installed a cookie wall (<www.publiekeomroep.nl> and <www.volkskrant.nl> accessed 15 February 2013). See chapter 6, section 4, and chapter 8, section 3 and 5.
[1447] In one Dutch survey, 30% doesn't want tracking cookies at all, and 41% only wants tracking cookies from some sites. However, 50% usually clicks "OK" to consent requests for cookies (Consumentenbond (Dutch Consumer Organisation) 2014).

ends in divorce.[1448] The success of "buy now, pay later" deals can be partly explained by myopia and optimism bias.[1449] Research suggests people also tend to underestimate the risks of identity fraud and of re-identification of anonymised data.[1450]

The way information is presented can also influence decisions. This is known as the framing effect.[1451] For example, many people see a link to a privacy policy as a quality seal. 41% of Europeans don't read privacy policies, because they think it's enough to check whether a website has one.[1452] In a California survey, the majority thought that the mere fact that a website had a privacy policy meant that their privacy was protected by law.[1453] Turow at al. argue that the phrase "privacy policy" is misleading.[1454] Facebook speaks of a "data use policy", which seems a more apt name.[1455]

Research suggests that privacy policies with vague language give people the impression that a service is more privacy-friendly than privacy policies that give more details.[1456] Another study suggests that "any official-looking graphic" can lead people to believe that a website is trustworthy.[1457] Böhme and Köpsell find that people are more likely to consent if a pop-up looks more like an end user license agreement (EULA). The researchers varied the design of consent dialog boxes and tested the effect by analysing the clicks of more than 80,000 people. They conclude that people are conditioned to click "agree" to a consent request if it resembles a EULA.

---

1448 Sunstein & Thaler 2008, p. 31-33.
1449 Sunstein & Thaler 2008, p. 35.
1450 Acquisti & Grossklags 2005.
1451 For example, Kahneman found that even among doctors, "[t]he statement that 'the odds of survival one month after surgery are 90%' is more reassuring than the equivalent statement that 'mortality within one month of surgery is 10%'" (Kahneman 2011, p. 88).
1452 European Commission 2011 (Eurobarometer), p. 118-120.
1453 Hoofnagle & King 2008; Turow 2003; Turow et al. 2005.
1454 Turow et al. 2007.
1455 Facebook, 'Data Use Policy'.
1456 Good et al. 2006.
1457 Moores 2005.

> [U]biquitous EULAs have trained even privacy-concerned users to click on "accept" whenever they face an interception that reminds them of a EULA. This behaviour thwarts the very intention of informed consent. So we are facing the dilemma that the long-term effect of well-meant measures goes in the opposite direction: rather than attention and choice, users exhibit ignorance.[1458]

Furthermore, Acquisti et al. discuss a "control paradox." People share more information if they *feel* they have more control over how they share personal information. The researchers conclude that control over personal information is a normative privacy definition: control *should* ensure privacy. But in practice, "'more' control can sometimes lead to 'less' privacy in the sense of higher objective risks associated with the disclosure of personal information."[1459]

Several authors conclude that there's a behavioural market failure regarding online privacy. Firms wouldn't stay in business if they didn't exploit people's biases. As Strandburg puts it, "[t]he behavioral advertising business model seems almost designed to take advantage of (…) bounded rationality."[1460] Firms often have larger data sets than scientists to discover biases. For instance, some internet firms can analyse the behaviour of hundreds of millions of people to test various designs and opt-out systems. Calo warns against "the mass production of bias."[1461]

### 7.5 Privacy paradox

There seems to be a privacy paradox. In surveys, people say they care about privacy. But people often divulge personal data in exchange for minimal benefits or

---

[1458] Böhme & Köpsell 2010.
[1459] Acquisti et al. 2012, p. 6.
[1460] Strandburg 2013, p. 149. See along similar lines Calo 2013; Acquisti 2010a, p. 6.
[1461] Calo 2013, p. 12. See for an example of a large-scale experiment by Facebook chapter 3, section 3.

convenience, and relatively few people use technical tools to protect their privacy online. Declared preferences (what people say in surveys) are often less reliable than revealed preferences (how people act). Sometimes it's suggested that people only care about privacy when they don't have to deal with other interests. "Consumers may tell survey takers they fear for their privacy, but their behaviour belies it. People don't read privacy policies, for example."[1462]

Scholars from various disciplines counter that people do care about privacy, but have difficulties acting according to their privacy preferences.[1463] Similarly, people who care about the environment might not study the label of every supermarket product to establish if it was produced in an environmentally friendly way.[1464] Another similarity with privacy policies is that merely studying the ingredients on a package may not be enough to assess how environmentally friendly a product is.

Regarding privacy decisions, it's doubtful whether revealed privacy preferences can be used to estimate how much people value their privacy in monetary terms. It's easy to manipulate the value people attach to their personal data.[1465] For instance, in a study by Cranor & McDonald, most participants believe they wouldn't pay one dollar a month to keep a website from using behavioural targeting. At first glance, this might suggest that few value protecting their information more than one dollar a month. But 69% would *not* accept a one dollar discount in exchange for having their data collected for behavioural targeting. This suggests that most people think their personal data is worth more than one dollar a month. In short, people's willingness to pay for privacy is different to their willingness to accept (a discount) to forego privacy.[1466] If it were assumed that people make "rational" choices to maximise their own welfare, in this case their privacy, the results would be surprising.

---

[1462] Goldman 2002.
[1463] See for instance Trepte et al 2014; Acquisti & Grossklags 2007; Solove 2013; Cranor & McDonald 2010. See also Cofone 2014. Moreover, fundamental rights also apply if people don't care about fundamental rights.
[1464] Thanks to Lauren Willis, who pointed this out at the Privacy Law Scholars Conference in Berkeley (2013).
[1465] See Acquisti et al. 2013a.
[1466] Cranor & McDonald 2010, p. 25. The effect that people value things more when they own them is called the endowment effect. See on that effect in the privacy context Acquisti et al. 2013a.

In follow-up interviews and a survey, Cranor & McDonald "found people generally unwilling to pay for privacy, not because they do not value it, but because they believe it is wrong to pay."[1467] 69% of the respondents agreed with the statement "Privacy is a right and it is wrong to be asked to pay to keep firms from invading my privacy."[1468] 61% agreed it would be "extortion" if a firm would ask them to pay for not collecting data. The researchers suppose "that one reason people will not pay for privacy is because they feel they should not have to: that privacy should be theirs by right."[1469] This suggests that the EU legal regime comes closer to the expectations of the US respondents in this research than a free market model regarding privacy.

Self-help tools exist to protect privacy in the area of behavioural targeting. For instance, people can install browser plug-ins that blocks ads and limit tracking, and millions of people do so.[1470] But many people find technical privacy protection tools too complicated.[1471] The time it would take people to learn to use the tools is a transaction cost. And even if a tool is easy, people might refrain from using it because they think it's difficult.[1472] In any case, so far most people seem to be losing the technological arms race. Some firms seem to be on a quest for more effective and opaque tracking technologies. For instance, it would be very difficult to detect or to protect oneself against device fingerprinting. If technology alone determined the level of online privacy, behavioural targeting firms would be likely to emerge as winners, and data subjects as losers.[1473]

This study doesn't suggest that all privacy problems can be attributed to behavioural biases. Even if people wouldn't have difficulties making decisions in accordance with

---

[1467] Cranor & McDonald 2010, p. 28.
[1468] Cranor & McDonald 2010, p. 26.
[1469] Cranor & McDonald 2010, p. 26.
[1470] The ad blocking software Adblok Plus was reportedly downloaded 200 million times (Adblock Plus 2014). Some estimate that between 9 and 23% of internet users use ad blocking software (Hill 2013). And in April 2014 there were about 2.5 million people connected users to the anonymity service Tor at any given moment (Tor 2014).
[1471] Leon et al. 2012. See for an amusing account of trying to use self-help tools Angwin 2014.
[1472] Willis 2013, p. 1164.
[1473] See chapter 2, section 2.

their declared interests, they still wouldn't be able to fully protect their privacy. For instance, it's very hard to defend oneself against group profiling.[1474] A firm that has a predictive model may need only a few data points to predict other information about somebody. Nevertheless, behavioural economics insights can help to explain the alleged privacy paradox.

Because privacy choices are context-dependent, caution is needed when drawing conclusions about the effect of biases. One bias might influence a privacy decision in one direction, while another bias might influence the same decision in another direction.[1475] Still, it would be naive to ignore behavioural economics when making laws that rely, in part, on the decisions of people whose privacy the law aims to protect.[1476]

### 7.6 Conclusion

This chapter analysed practical problems with informed consent, and thus with the privacy as control perspective. The chapter also discussed the economics of privacy and behavioural targeting.

As noted previously, this study offers suggestions to improve privacy protection, without being unduly prescriptive.[1477] If rules impose unreasonable costs on society, this study considers them unduly prescriptive. From an economic perspective, it's unclear whether behavioural targeting leads to a net benefit or a net loss for society. On the one hand, using personal data can increase social welfare. For instance, firms such as ad networks and website publishers profit from behavioural targeting. Income from online advertising could be used to fund so-called "free" web services. On the other hand, using personal data can decrease social welfare. For instance, if

[1474] Gürses 2010, p. 51. See section 3 of this chapter on externalities.
[1475] Acquisti & Grossklags 2007, p. 371. Luth 2010 arrives at a similar conclusion regarding consumer protection (p. 279-283).
[1476] Acquisti & Grossklags 2007, 374. In the context of EU consumer law Gomez reaches a similar conclusion (Gomez 2010, p. 110).
[1477] See chapter 1, section 1.

somebody's information ends up in the wrong hands, this could lead to receiving spam or to identity fraud. Other privacy related costs are harder to quantify, such as annoyance, a creepy feeling, and chilling effects. As it's unclear whether more or less privacy protection would be better from an economic perspective, more legal limits on behavioural targeting wouldn't necessarily be too costly.

From an economic perspective, consenting to personal data processing for behavioural targeting, or consenting to the use of a tracking cookie, can be seen as entering into a market transaction with a firm. But this "transaction" is plagued by information asymmetries. Many people don't know their behaviour is tracked, so their "choice" to disclose data in exchange for the use of a service isn't informed. But if firms sought consent for behavioural targeting, information asymmetry would remain a problem. People rarely know what a firm does with their personal data. And it's hard for people to predict the consequences of future data use. From an economic perspective, information asymmetry can lead to market failure, which can justify regulatory intervention. If people can't assess the quality of products or services, sellers won't compete on quality. This can lead to low quality products or services: a "lemons" market. Indeed, websites rarely compete on privacy. Virtually every popular website allows third parties to track its visitors.

Through an economic lens, data protection law's requirements for firms to be transparent about their data processing practices can be seen as an attempt to mitigate the information asymmetry. Website publishers can comply with the transparency requirements by disclosing the information in a privacy policy. But the information asymmetry problem is difficult to solve because of transaction costs. Reading privacy policies would cost too much time, as they are often long, difficult to read, and vague. "Only in some fantasy world do users actually read these notices and understand their implications before clicking to indicate their consent."[1478]

[1478] White House (Podesta J et al.) 2014, p. xi; see also p. 38.

Behavioural economics insights highlight more practical problems with informed consent. For instance, the status quo bias describes people's tendency to stick with default options. If people are assumed to consent if they fail to object, most people will "consent." With an opt-in system that requires an affirmative action for valid consent, people are less likely to consent.

Present bias, or myopia, suggests that people often choose immediate gratification and don't pay attention to future costs or disadvantages. If a website has a tracking wall, and people can only use the site if they agree to being tracked, they're likely to consent, ignoring the costs of future privacy infringements. The following chapters return to the topic of take-it-or-leave-it-choices.[1479]

In sum, behavioural economics can help to understand the alleged privacy paradox. People who say they care about their privacy often disclose information in exchange for small benefits. Part of this is conditioning: many people click "yes" to any statement that is presented to them. Exaggerating slightly: people don't read privacy policies; if they were to read, they wouldn't understand; if they understood, they wouldn't act.[1480]

* * *

[1479] See in particular chapter 8, section 3 and 5, and chapter 9, section 5 and 7.
[1480] Ben-Shahar and Schneider arrive at a similar conclusion on the regulatory technique of mandated disclosure of information in general: people "often do not read disclosed information, do not understand it when they read it, and do not use it even if they understand it" (Ben-Shahar & Schneider 2011, p. 665).

# 8 Improving empowerment

To defend privacy in the area of behavioural targeting, this study argues for a combined approach of protecting and empowering people. This chapter discusses how the law could improve individual *empowerment*. The following chapter focuses on *protection* of the individual.[1481] The behavioural economics analysis in the previous chapter suggests that fostering individual control over personal data won't suffice to protect privacy in the behavioural targeting area.

Why still aim for empowerment? In theory, it might be possible to have a legal regime that strictly defines all data processing practices that are prohibited, or those practices that are allowed. In such a hypothetical regime, there would be no need to give choices to the data subject with an informed consent provision or opt-out possibilities. This study doesn't explore such a hypothetical regime, for several reasons.[1482]

First, it's not feasible that the EU would abolish data protection law and would start from scratch to develop a new privacy regime. And a data protection regime without a consent provision is unlikely, if only because the EU Charter of Fundamental Rights lists consent as a legal basis for processing.[1483] Second, it would be almost impossible to define all beneficial and all harmful data processing activities in advance.[1484] Third, people's tastes differ. Some people would approve of a certain practice, while others wouldn't. As noted, the privacy-as-control perspective, and regulation with a consent

---

[1481] As noted, this study distinguishes protection and empowerment rules to structure the discussion, but it's not suggested that there's a formal legal distinction (see chapter 4, section 5).
[1482] I'm not aware of any serious proposals for a legal privacy regime without any role for consent or opt-out procedures.
[1483] Article 8 of the EU Charter of Fundamental Rights.
[1484] See Solove 2013, p. 1895. In theory, a regime without consent might be possible. See chapter 6, section 5.

provision, has the advantage of respecting people's individual preferences.[1485] Taking away *all* privacy choices from the individual would probably make the legal regime unduly paternalistic.[1486] Indeed, several scholars that are extremely sceptical of informed consent as a privacy protection measure still say that a legal privacy regime without any role for informed consent is neither feasible nor desirable.[1487] The foregoing doesn't mean that the lawmaker should stay away from mandatory rules that limit people's choices. On the contrary, such mandatory rules are needed, and are discussed in the next chapter.

In sum, it's likely that there will always be many circumstances where relying on informed consent, in combination with data protection law's safeguards, is the appropriate legal approach. For those cases, transparency and consent should be taken seriously. And compared with the current situation of very limited individual control over personal information in the behavioural targeting area, some improvement must be possible.[1488]

This chapter is structured as follows. Section 8.1 discusses enforcement. Section 8.2 and 8.3 discuss measures to improve transparency and to make consent more meaningful. Section 8.4 gives suggestions to improve the consent requirement for the use of tracking technologies. Section 8.5 discusses the Do Not Track standard. Section 8.6 concludes.

## 8.1 Enforcement

It's difficult to quantify the effect of data protection law. "With data protection," notes Bennett, "it is not clear how one could measure or even observe success. Impact has to be evaluated according to complex changes in the treatment of a very

---

[1485] See chapter, 3, section 1.
[1486] See Solove 2013, p. 1894.
[1487] See e.g. Barocas & Nissenbaum 2009; Nissenbaum 2011; Solove 2013, p. 1899; Barocas & Nissenbaum 2014.
[1488] Data protection is only relevant as far as it applies to behavioural targeting. As noted, this study argues data protection law should generally apply to behavioural targeting (see chapter 5).

intangible, elusive, and ephemeral commodity – personal information."[1489] Even so, there's wide agreement that there's a compliance deficit with data protection law.[1490] In the area of behavioural targeting, non-compliance seems especially rampant. For instance, transparency regarding behavioural targeting often leaves something to be desired, and many firms fail to ask prior consent for using tracking technologies in compliance with the law. Hence, stricter enforcement of the law is needed to improve data subject control in the area of behavioural targeting.

Stricter enforcement is easier said than done. Data Protection Authorities are understaffed, and lack resources.[1491] Data protection law applies to the private and the public sector, and supervising the law for the private sector alone is an immense task.[1492] Enforcement is more difficult because many firms using behavioural targeting are based outside the EU. Even if the law applies, international investigations are costly. And until recently, behavioural targeting took place largely below the radar.[1493] Furthermore, many Data Protection Authorities lack effective enforcement powers.[1494] Some authorities can only impose low fines – in one member state the maximum fine is 290 Euro.[1495] In some countries, Data Protection Authorities can't impose firms penalties for many types of violations. Additionally, there are Data Protection Authorities that appear to prefer a light touch approach.[1496] For instance, the Irish Data

---

[1489] Bennett 1992, p. 238. See also Irion & Luchetta 2013, p. 23, p. 28.

[1490] See for instance Bennett 2011a, p. 493; Irion & Luchetta 2013, p. 50; Borghi et al. 2013. Empirical research seems to confirm a lack of compliance with data protection law (see e.g. Burghardt et al. 2010; Birnhack & Elkin-Koren 2010). In some member states, it's not the Data Protection Authority but another regulator that oversees compliance with article 5(3) of the e-Privacy Directive. For ease of reading, this study speaks of Data Protection Authorities.

[1491] Irion & Luchetta 2013, p. 28; European Agency for Fundamental Rights 2010, p. 8; European Agency for Fundamental Rights 2014a, p. 46-47.

[1492] Some parts of the public sector are outside the scope of the 1995 Data Protection Directive (see chapter 4, section 2).

[1493] Behavioural targeting hasn't been ignored earlier. For instance, the Article 29 Working Party discussed tracking and profiling since 1997 (see Article 29 Working Party 1997, WP 6; 1999, WP 17; WP 37, p. 16). In the US, the Federal Trade Commission has discussed online privacy since 1996 (see Federal Trade Commission 2012, appendix A).

[1494] Impact Assessment for the proposal for a Data Protection Regulation (2012), p. 17-18; annex 1, p. 36-38, annex 2, p. 41-44; European Agency for Fundamental Rights 2010, p. 8.

[1495] In Lithuania the maximum administrative fine is 290 euro (Impact Assessment for the proposal for a Data Protection Regulation (2012), annex 1, p. 37). See also European Data Protection Supervisor 2014, p. 16; European Agency for Fundamental Rights 2014a, p. 46-49.

[1496] Irion & Luchetta 2013, p. 29.

Protection Commissioner is criticised for not enforcing the law against Facebook.[1497] On the other hand, some Data Protection Authorities receive criticism for being too aggressive.[1498]

Another problem that relates to the enforcement deficit is that data protection law contains many general rules with rather open norms. For example, there's still discussion on the question of whether data protection law applies when firms don't tie a name to data they process for behavioural targeting.[1499] Some Data Protection Authorities may be hesitant to impose sanctions in cases that are likely to lead to discussion about the material scope of the law. And for data subjects it may be unclear what they can expect. The next chapter returns to the topic of data protection law's open norms.[1500]

Causal relationships are hard to prove, but data protection law does seem to have effect. For instance, while many European websites don't ask consent for using tracking cookies in compliance with the e-Privacy Directive, they do offer some information about cookies. The consent requirement for tracking technologies from the 2009 e-Privacy Directive has led many European website publishers to behave in a manner that complies with the 2002 e-Privacy Directive, which required transparency and an opt-out option for cookies.[1501] And the fact that many firms lobbied in Brussels to influence the proposals for a Data Protection Regulation suggests that they don't think data protection law can be ignored.[1502]

Sometimes Data Protection Authorities take action in the area of behavioural targeting. For instance, the Dutch Authority has investigated the use of tracking

---

[1497] Max Schrems from Austria is one of the most vocal critics of the Irish Data Protection Authority (see Europe versus Facebook 2014).

[1498] Bamberger & Mulligan 2013 report on criticism on the aggressive approach of the Spanish DPA (p. 1593-1616).

[1499] See chapter 5.

[1500] Chapter 9, section 1.

[1501] See chapter 6, section 4.

[1502] See on lobbying chapter 5, section 5, chapter 6, section 3, chapter 8, section 3, and chapter 9, section 6.

cookies on smart TV sets, and the use of cookies by a behavioural targeting firm.[1503] And Data Protection Authorities have examined Google's data processing practices. In 2012, Google consolidated most of its more than 60 privacy policies into one overarching policy that governs the majority of its services. The new policy allows Google to combine user data over its various services. Google embarked on a large-scale information campaign that alerted people to the changes, with banners on its search page and on other Google websites. The Working Party had asked Google to postpone introducing the new policy, so Data Protection Authorities could gather more information. Google refused.[1504]

The Working Party sent Google long questionnaires about the privacy policy changes, but Google didn't answer all the questions in detail. The Working Party summarised its preliminary findings in a letter to Google.[1505] Among other things, the Working Party complains that Google doesn't offer enough transparency and fails to properly aks for consent for combining the data.[1506] Furthermore, Google doesn't ask consent for cookies in accordance with the e-Privacy Directive.[1507] Several privacy authorities from outside Europe jointly wrote an open letter to express their support to the Working Party's conclusions.[1508] Data Protection Authorities in six member states continued the investigation. At the time of writing, Data Protection Authorities in Spain and France have imposed fines of 900,000 and 150,000 Euros.[1509]

***Enforcement strategies***

An important avenue for further research is how compliance with the data protection rules could be improved. While this isn't a study on enforcement, some preliminary

---

[1503] College bescherming persoonsgegevens 2013 (TP Vision); College bescherming persoonsgegevens 2014 (YD).
[1504] See for a summary of the events College bescherming persoonsgegevens (Dutch DPA) 2013 (Google), p. 7-11.
[1505] Along with the French CNIL, the DPAs from the following countries continued the investigation: Germany, Italy, the Netherlands, Spain and the United Kingdom. See the website of CNIL, with further references (CNIL 2012 (Google)).
[1506] Formally it's a letter signed by 28 national Data Protection Authorities.
[1507] See Article 29 Working Party 2013 (Google letter), appendix, p. 5. See also CNIL 2014 (Google), p. 17-20.
[1508] The signatories of the letter include authorities from Mexico, Hong Kong, and Australia (Asian Pacific Privacy Authorities 2012, Google letter).
[1509] See Agencia Española de Protección de Datos (Spanish Data Protection Authority) 2013; CNIL 2014 (Google).

remarks are made on the topic. In the field of regulation studies, much has been written on the best way to make firms comply with the law, for instance with environmental law.[1510] Adapting a categorisation by Baldwin et al., firms can be categorised by looking at their intentions and their know-how. Grossly simplifying, a firm could be well-intentioned or ill-intentioned, and could be informed or ignorant.[1511] This way, four types of firms can be distinguished. The categories are simplifications. In reality, a firm will have characteristics of several categories. The classification is meant to illustrate that for some firms hard enforcement is needed. For other firms, raising awareness of the legal requirements may be the most effective tool to make them comply with data protection law.

The first category of firms is informed and well-intentioned.[1512] An example might be a large firm with skilled technologists and data protection lawyers. The firm understands the law, wants to comply, and can comply. The lawyers know every detail of the law and can translate the data protection principles into practical guidelines for the technologists to implement. Generally speaking, large-scale privacy violations are not to be expected from firms in the first category. The firms in this category are aware of the legal requirements. Hence, raising awareness of data protection law isn't needed for such firms. And threatening with sanctions isn't needed, as these firms are well-intentioned and want to comply with the law.

Second, a firm can be ignorant and well-intentioned. Such firms want to comply with the law, but might break the law by accident. For instance, a website publisher might use social media buttons or a web analytics programme on its website, without realising these expose visitors to privacy-invasive tracking. Or a developer of smart phone apps might use an ad network's services to include ads in its app. The

[1510] Regulation studies can be described as follows: "a multi-disciplinary field, with substantial contributions to regulatory debates being made by political scientists, lawyers, sociologists, anthropologists, and others. Writings on regulation are well-represented across scholarly publication outlets and there has also been the inevitable arrival of a journal with the word regulation in its title, *Regulation and Governance*" (Baldwin et al. 2010).
[1511] Baldwin et al. 2010 speak of "ill-disposed" and "well-disposed" firms, and of "highly capable" firms and "low capacity" firms (p. 304-306).
[1512] Baldwin et al. 2010, p. 304.

developer might consciously include a snippet of code from the ad network in the app. An app developer might also unwittingly enable third party tracking, when using "libraries"; these are blocks of ready-made code. A library might include code that enables an ad network to track the activities of the app's users.[1513] And a firm that doesn't tie a name to the data it processes might not realise it processes personal data.[1514]

Unwillingness isn't the main problem for this second category of firms. The problem is ignorance. For well-intentioned but ignorant firms, awareness raising is likely to be the most effective way of ensuring that they comply with the law. If Data Protection Authorities wanted to do more to raise awareness regarding the law, there would be various ways to do so. For instance, the Working Party's opinions, although sometimes hard to read for non-specialists, also receive attention in the press, which could bring the legal requirements to the attention of firms. And Data Protection Authorities might speak at conferences and other events. But another approach is also possible. Strict enforcement with respect to ill-intentioned firms may raise awareness regarding the law, and incentivise firms to educate themselves. To illustrate, the Dutch Data Protection Authority decided in 2007 that it "will concentrate on carrying out investigations and enforcement actions – the core task of any independent supervisory authority – to ensure a more effective promotion of the awareness of standards, and a stronger, more efficient enforcement of the compliance with legislation."[1515]

Third, a firm can be informed and ill-intentioned. The firm is an "amoral calculator", aims for maximum profit, and sees the risk of fines as a business risk.[1516] This type of firm could also be described as fully rational in the economic sense.[1517] The firm will

---

[1513] See Article 29 Working Party 2013, WP 202. See also the firm Flurry, which was discussed in chapter 2, section 2 (Yahoo 2014 (Flurry)).
[1514] See chapter 5, section 2.
[1515] College bescherming persoonsgegevens, Annual report 2007, p. 69-70.
[1516] Baldwin et al. 2010, p. 305. See also Becker 1993.
[1517] See chapter 7, section 2.

choose to bend or break the rules, as long as the expected profit from breaking the rules is higher than the chance of being fined, multiplied with the expected fine. As Black notes, “when compliance becomes a matter of risk management, non-compliance becomes an option.”[1518] For a firm with billions of profit, a fine of one million euro isn’t a dissuasive threat. In the context of environmental law, Faure observes: “fining a polluter with a too low fine can have a perverse learning effect.”[1519]

But high penalties alone aren’t enough. To incentivise a firm to comply with the law, the firm must believe there’s a considerable chance that it will get caught and will have to pay the penalty.[1520] Suppose the expected fine is one million euro, and there’s a 1% probability that such a fine is imposed. The expected loss is thus 1% of one million euro = 10,000 euro. To ensure a credible chance of enforcement, Data Protection Authorities should receive sufficient funding.

There may be other reasons for firms to comply with the law than avoiding fines.[1521] For instance, some firms offer consumer services, and may fear that people will switch to another service. Fear of consumer backlash is mainly relevant for firms that also offer consumer services, such as a search engine, a social network site, or computer software.[1522] For such firms, naming and shaming by the press or by Data Protection Authorities may be a worse punishment than a fine. Some Data Protection Authorities already use the shaming approach. For instance, the French Data Protection Authority obliged Google to publish on its search homepage that it had violated French law.[1523] The lawmaker could consider introducing the possibility for data Protection Authorities to publish the names of firms that breach data protection

---

[1518] Black 2008, p. 454.
[1519] Faure 2010, p. 263.
[1520] Faure 2010. See for a similar conclusion Schneier 2012, chapter 9; chapter 13; p. 241.
[1521] Like individuals, firms are not fully “rational” in the economic sense. See Chang 2014, p. 176 and further.
[1522] However, switching to another service may be difficult for a consumer, for instance because of transaction costs or network effects. And there might not be any competitors with better privacy policies. See chapter 7, section 3 and 4.
[1523] See e.g. CNIL 2014 (Google). See generally on reputational sanctions Van Erp 2007.

law. For some firms naming and shaming is less worrisome. For example, it’s hard for people to boycott an ad network, if they don’t know which websites work with the ad network.[1524] In sum, for the third category, well-informed but ill-intentioned firms, dissuasive penalties and a credible threat of enforcement are needed. Raising awareness regarding the law won’t help to make these firms in comply with the law.

This study doesn’t suggest that some firms enjoy breaking the law, although the phrase “ill-intentioned” was used above. As noted in the last chapter, market forces may push firms towards exploiting information asymmetries and people’s biases, and towards more privacy invasive tracking.[1525] If the trend towards centralisation in the online marketing industry continues, at some point perhaps a handful of well-informed large firms are responsible for the majority of behavioural targeting. It can’t be ruled out that some of these firms would be ill-intentioned.

The fourth category of firms is ill-intentioned and ignorant. They’re not aware of the law, but wouldn’t mind breaking it anyway. For example, it would be difficult to make criminals operating spyware comply with European data protection law, especially if they’re based in a far-away country. But sometimes the law could be enforced to other players. For example, a European website publisher could be held responsible if it allows third parties to distribute spyware.[1526]

In sum, the best methods of ensuring that firms comply depend on the intentions and the legal and technical know-how of the firm. For some firms dissuasive penalties and a credible threat of enforcement are needed. For others raising awareness of the law may be the best approach to foster compliance. Faure arrives at a similar conclusion about environmental law.

---

[1524] See Schneier 2012, p. 183. There might be an indirect effect: website publishers might be hesitant to work with an ad network that receives criticism from the public.
[1525] Chapter 7, section 3 and 4.
[1526] See Article 29 Working Party 2010, WP 171: publishers and ad networks are often joint controllers. See also Castelluccia & Narayanan 2012, p. 22.

> Deterrence may be the primary goal in case of intentionally violating perpetrators (…) (who could only be brought to compliance by threatening them with high penalties) whereas a softer compliance strategy (providing information leading towards following the law) may be the more appropriate strategy with firms that merely breach because of lacking information.[1527]

The European Commission has realised that Data Protection Authorities have insufficient powers. Therefore, the proposal for a Data Protection Regulation aims to strengthen their enforcement powers. For instance, the proposal would enable Data Protection Authorities, in some circumstances, to impose sanctions of up to 2% of a firm's annual worldwide turnover. The European Parliament has proposed to increase the maximum to 5%.[1528] The proposal also calls for adequate resources for Data Protection Authorities.[1529]

***Enforcement by data subjects***

In principle, enforcement could also come from data subjects. But people rarely go to court when their data protection rights are breached. Litigation is expensive, and people aren't likely to go to court if litigation costs outweigh the damages that can be won.[1530] This problem isn't unique for data protection law. For example, if a consumer buys a product for ten euro that doesn't function as promised, it's not worth suing the producer.[1531] But if millions of consumers lose ten euro, the aggregate costs for society can be enormous. Similarly, privacy violations can concern millions of

---

[1527] Faure 2010, p. 263.
[1528] Article 79 of the European Commission proposal for a Data Protection Regulation (2012); article 70(2a)(c) of the LIBE Compromise, proposal for a Data Protection Regulation (2013).
[1529] Article 47(5) of the European Commission proposal for a Data Protection Regulation (2012); article 47(5) the LIBE Compromise, proposal for a Data Protection Regulation (2013).
[1530] See Impact Assessment for the proposal for a Data Protection Regulation (2012), p. 38; European Agency for Fundamental Rights 2014a, p. 39-44.
[1531] Baldwin et al. 2011, p. 126-127.

individuals that each bear small costs, such as annoyance. Solove compares privacy violations to bee stings. One isn't a problem, but many together would be.[1532] The problem of mass harm situations provides an argument for enforcement by regulatory authorities, such as consumer protection agencies or Data Protection Authorities.

An option that could be explored is introducing collective action procedures in the area of data protection law.[1533] Collective action procedures should make it possible for people to sue a firm collectively. Like this, a firm can be held accountable when it imposes small costs to many people that amount to large costs in the aggregate. The Commission proposal for a Data Protection Regulation would allow organisations to take firms to court for breaching people's data protection rights.[1534]

The Commission has published a recommendation on collective redress, which could also have an impact on data protection practice.[1535] The recommendation aims to "facilitate access to justice, stop illegal practices and enable injured parties to obtain compensation in mass harm situations caused by violations of rights granted under Union law, while ensuring appropriate procedural safeguards to avoid abusive litigation."[1536] The preamble states that data protection law is an area where collective action could be important.[1537] The recommendation encourages, but doesn't require, member states to adapt their laws. It could take years before a legally binding instrument is adopted.[1538] Another problem with enforcement by data subjects is that winning compensation for non-monetary damages can be difficult. Hence, it would be

---

[1532] Solove 2013, p. 1891. See also Haggert & Ericson 2000, who speak of a "surveillant assemblage."
[1533] The Article 29 Working Party has also suggested the introduction of class action suits in data protection law (Article 29 Working Party 2010, WP 168, p. 16). See also European Agency for Fundamental Rights 2014a, p. 32; p. 53.
[1534] See article 73(2), 74, 75 and 76(1) of the European Commission proposal for a Data Protection Regulation (2012).
[1535] European Commission 2013 (Collective Redress Recommendation).
[1536] Article 1(1) of European Commission 2013 (Collective Redress Recommendation).
[1537] Recital 7 of European Commission 2013 (Collective Redress Recommendation).
[1538] Hodges 2013 argues that it would be very difficult to develop a Europe-wide collective redress system, because of the different national legal systems. See also par. 41 of the European Commission 2013 (Collective Redress Recommendation).

worthwhile to examine whether the law should enable people to claim compensation for non-monetary damages that result from data protection law violations.

### 8.2 Transparency

The last chapter showed that information asymmetry is a problem in the area of behavioural targeting. For some information asymmetry problems data protection law already suggests an answer, but for others it doesn't. Information asymmetry is a problem from an economic perspective and from the perspective of privacy as control.[1539] But information asymmetry is also a problem under current law.

The main problem is many people don't know that their activities are monitored for behavioural targeting. At first glance, the answer is straightforward. The Data Protection Directive requires a firm to tell data subjects its identity and the processing purpose, and all other information that's necessary to guarantee fair processing.[1540] The Directive doesn't explicitly require firms to publish an easily accessible privacy policy, but it's general practice. The European Commission proposal for a Data Protection Regulation codifies this practice.[1541] And, as discussed in the next section, asking prior consent does more to alert people to tracking than offering an opt-out possibility.

A second category of information asymmetry is that people have scant idea about what firms do with their personal data. Again, at least the beginning of the answer is straightforward. Data protection law requires firms to disclose their processing purposes. And firms must clearly describe a specified purpose that isn't too vague or too general, and must not use personal data for unrelated purposes that the data

---

[1539] See on economics chapter 7, section 3. See on the privacy as control perspective chapter 3, section 1. Information asymmetry is also a problem from the privacy as identity construction perspective.

[1540] Article 10 and 11 of the Data Protection Directive. See on the information that should be provided when firms apply profiling techniques also: article 4 of the Council of Europe, Committee of Ministers, Recommendation (2010)13 to member states on the protection of individuals with regard to automatic processing of personal data in the context of profiling, 23 November 2010.

[1541] Article 11 of the European Commission proposal for a Data Protection Regulation (2012).

subject doesn't expect.[1542] Data Protection Authorities summarise that firms must aim for "surprise minimisation."[1543] As discussed in chapter 4, the purpose limitation principle isn't as strict as it might seem.[1544] Nevertheless, the principle could help to protect people against unexpected uses of their data. Transparency about data processing can only be meaningful if the purpose limitation principle is complied with.[1545]

The information asymmetry is partly caused by transaction costs, such as the time it would take people to inform themselves.[1546] Reading privacy policies would take too much time. They're often long and difficult to read and sometimes refer the reader to policies from other firms. According to the Article 29 Working Party, long privacy policies full of legalese aren't acceptable. "Internet companies should not develop privacy notices that are too complex, law-oriented or excessively long." [1547] Furthermore, privacy policies that obfuscate relevant information by pointing to other privacy policies are unlikely to comply with data protection law's transparency principle.

In its Google investigation, the Working Party complains that Google's privacy policy is too vague. "Google has not indicated what data is combined between which services."[1548] Furthermore, "Google gives incomplete or approximate information about the purposes and the categories of data collected. The privacy policy is a mix of particularly wide statements and of examples that mitigate these statements and mislead users on the exact extent of Google's actual practices."[1549] Indeed, while

---

1542 Article 6(1)(b) of the data Protection Directive; Article 29 Working Party 2013, WP 203. See chapter 4, section 3.
1543 Kohnstamm & Wiewiórowski 2013.
1544 See chapter 7, section 3 and 4.
1545 But see Moerel 2014, who suggests the purpose limitation principle should be deleted from the Data Protection Regulation (p. 55).
1546 See chapter 7, section 3.
1547 Article 29 Working Party 2013 (Google letter), p. 2. See also Article 29 Working Party 2004, WP 100, p. 5.
1548 Article 29 Working Party 2013 (Google letter), appendix, p. 3.
1549 Article 29 Working Party 2013 (Google letter), appendix, p. 2 (capitalisation adapted). For similar conclusions about earlier versions of Google's privacy policy, see Van Hoboken 2012, p. 329-333; Van Der Sloot & Zuiderveen Borgesius 2012, p. 102-108.

Google’s privacy policy deserves praise for staying away from legalese, it uses confusing terms that leave the reader guessing which personal data are processed for which purposes. To illustrate, it’s unclear which types of data Google sees as personal data.

The European Commission proposal for a Data Protection Regulation gives more detailed transparency rules.[1550] For instance, it requires firms to have “easily accessible policies (…) in an intelligible form, using clear and plain language, adapted to the data subject.”[1551] The clear language requirement is in line with European consumer law, which requires firms to disclose “information in a clear and comprehensible manner.”[1552] The preamble stresses the importance of clear information in the area of online advertising.[1553] Codifying the clear language requirement could discourage firms from using unreadable policies. And the requirement would make it easier for Data Protection Authorities to intervene when firms use vague policies or consent requests. The rule wouldn’t be enough to ensure actual transparency, but it could help to lower the costs of reading privacy policies.

An important aspect of effectively informing people is not overwhelming them with information.[1554] Less is more. Therefore, making privacy policies simpler seems like a good idea. But privacy isn’t simple.[1555] Describing complicated data processing practices accurately leads to a long text. If the text is too concise, it doesn’t provide enough information. Reducing transaction costs by making privacy policies simpler is

---

[1550] Unlike the Data Protection Directive’s article 11(c), the European Commission proposal’s article 14(1)(h) doesn’t mention “the categories of data” as an example of the information that firms must give to guarantee fair processing. See critically Korff 2012, p. 33.

[1551] Article 11 of the European Commission proposal for a Data Protection Regulation (2012). See generally chapter III, section 1 of the proposal, “Transparency and modalities.”

[1552] For instance, the Consumer Rights Directive requires firms to disclose “information in a clear and comprehensible manner (article 6(1)), and in “plain and intelligible language” (article 7(1); article 8(1)). The preamble discusses traders that supply digital content, such as apps or software. Such firms must inform consumers in particular about “the ways in which digital content can be used, for instance for the tracking of consumer behaviour (recital 19).”

[1553] Recital 46 of the European Commission proposal for a Data Protection Regulation (2012). See also Impact Assessment for the proposal for a Data Protection Regulation (2012), annex 2, p. 31.

[1554] Helberger 2013a, p. 34.

[1555] Daniel Solove used a similar phrase during the Symposium 2012: Privacy & Technology, 9 November 2012, Harvard University, Boston (<www.harvardlawreview.org/privacy-symposium.php> accessed 15 August 2013).

hard to reconcile with reducing information asymmetry.[1556] And reading privacy policies, even short ones, takes time. Many short notices together still add up to a lot of information. And each day, people have to deal with more information than only privacy policies. For instance, consumer law requires firms to disclose information on many products.[1557]

Some improvement must be possible over the current situation, as now privacy policies are often long, unreadable texts.[1558] The Working Party suggests using layered privacy policies. A firm should explain in a few sentences what it wants to do with personal data. People should be given the chance to click through to more detailed information.[1559] However, research shows it's questionable whether people would ever read the second and third layer. In any case, we shouldn't hope for too much when aiming to make people read privacy statements, simplified or not. Research suggests that "even the most readable policies are too difficult for most people to understand and even the best policies are confusing."[1560]

Maybe icons could be useful to communicate the data processing practices of firms. The Working Party and the European Commission encourage the use of icons,[1561] and the European Parliament has proposed to require firms to use icons to inform people about data processing practices.[1562] There are self-regulatory bodies that give seals,

---

[1556] See Nissenbaum 2011, p. 36; Solove 2013, p. 1885; Bar-Gill 2012, p. 37.
[1557] See about the cumulative effect of legal transparency requirements Ben-Shahar & Schneider 2011.
[1558] See for an overview of research on the comprehensibility of texts: Lentz et al, Knowledge Base Comprehensible Text. Some lawmakers adopted detailed rules regarding the readability of information. For example, in Brazil the law requires a minimum font of at least size 12 in standard terms for consumer contracts (article 54(3) of the Federal law n. 8.078, of September 11th, 1990). In Florida, the law has strict requirements regarding the presentation of insurance policies. "Every policy shall be readable as required by this section. (…) An insurance "policy is deemed readable if (…) [t]he text achieves a minimum score of 45 on the Flesch reading ease test (…) or an equivalent score on any other test comparable in result and approved by the office" (Florida Statutes: Insurance Rates and Contracts, Title XXXVII, chapter 627, Insurance Rates and Contracts, article 627.4145, par. 1(a).)
[1559] Article 29 Working Party 2004, WP 100.
[1560] Mcdonald et al. 2009, p. 50.
[1561] European Commission 2007 (PETs), par. 4.3.2.
[1562] See article 13(a), and the annex, of the LIBE Compromise, proposal for a Data Protection Regulation (2013). I have to admit that to me, the proposed six icons don't seem very clear. But it's possible that after a while, people would start to recognise the icons.

but such seals don't always imply that a website has high standards.[1563] Some providers have awarded seals to any firm, without a prior check. One paper found that websites with a seal from a particular organisation were generally less trustworthy than websites without that seal.[1564]

In the field of consumer law, scholars have suggested the introduction of intermediaries that help people to benefit from information.[1565] Regulators could audit intermediaries to ensure honesty. A similar approach could be considered for personal data processing practices. For instance, firms could be required to disclose their data processing practices to intermediaries that give ratings or seals. An organisation could make "white lists" or "block lists" for cookies that people can install in their browsers. Researchers at Stanford University are working on such a project.[1566] The European Parliament's LIBE Compromise enables firms to request a Data Protection Authority, for a reasonable fee, to certify that the personal data processing is performed in compliance with the Regulation.[1567]

In view of the limited effect that privacy policies have in informing people, more research is needed on alternative ways of presenting information. The current "failure of mandated disclosure" doesn't prove that legal transparency requirements will always fail.[1568] Calo argues that we shouldn't forget about transparency and informed consent, before better ways of presenting information have been tried.[1569] There's

---

[1563] See Rodrigues et al. 2013, p. 52-54; Tschofenig et al. 2013, p. 7-8. See also Schneier 2012, p. 183. Under the Unfair Commercial Practices Directive, one of the practices that's always unfair is: "Displaying a trust mark, quality mark or equivalent without having obtained the necessary authorisation" (Annex I (2)).

[1564] Edelman 2011. In 2014, the organisation in question, TRUSTe, agreed to settle Federal Trade Commission charges that it deceived consumers about its recertification program (Federal Trade Commission 2014a). See generally on trust marks and European law: Balboni 2008.

[1565] For instance, an intermediary could offer a website where people can easily compare cell phone contracts, adapted to their own usage. See for ideas along these lines Bar-Gill 2010, p. 41-42; Luth 2010, p. 243-247.

[1566] Cookie Clearinghouse 2014.

[1567] Article 39 of the LIBE Compromise, proposal for a Data Protection Regulation (2013). The Working Party is critical about the idea as it is phrased in the LIBE Compromise (Article 29 Working Party 2013 (draft LIBE comments, p. 4-5).

[1568] Calo 2011a. The phrase "failure of mandated disclosure" is taken from Ben-Shahar & Schneider 2011.

[1569] Calo 2011a.

research on better ways of presenting privacy policies.[1570] Cooperation between disciplines is needed, such as technology design, computer interface design, and psychology.[1571] There are firms that experiment with novel ways of presenting information about privacy.[1572] Some smart phone apps show that it's possible to communicate basic information in an intuitive way on small screens. But it appears firms put more effort in communicating the functions of an app than communicating their privacy policies.[1573]

But even if effective ways to present privacy policies could be developed, it might be difficult to make firms use them, because incentives are lacking. A firm that wants to distract people from information has many ways to do so, for instance by giving more information than needed, by using ambiguous language, or by framing information.[1574] "Click here for more relevant advertising" doesn't have the same ring to it as "Click here for continuous surveillance." But as long as information isn't misleading, the Data Protection Directive doesn't seem to have an answer to framing. In some cases, consumer law could be applied by analogy to framing. For example, it's unfair to present rights given to consumers in law as a distinctive feature of the trader's offer.[1575] In this light, a privacy policy raises questions if it presents people's data protection rights, such as the right to access, as a favour. Perhaps standardised privacy policies could help.[1576] The European Commission proposal for a Data Protection Regulation would make it possible to require firms to use a standard form to communicate their privacy policies.[1577]

---

[1570] See in the privacy field for instance Calo & Vroom 2012. Calo argues that the difference between effective information and nudges is a matter of degree rather than kind (Calo 2013a).
[1571] See in this context the work of the interdisciplinary research projects SPION (Security and privacy in online social networks), <www.spion.me/publications>, and USEMP (User Empowerment for Enhanced Online Management), <www.usemp-project.eu> accessed 28 May 2014.
[1572] For instance, Google publishes videos about cookies (Google (How Google uses cookies)).
[1573] See Helberger 2013a.
[1574] See Ben-Shahar & Schneider 2011; Willis 2013.
[1575] Annex 1 (10) of the Unfair Commercial Practices Directive. See on fairness in consumer law and data protection law chapter 4, section 4.
[1576] Verhelst 2012, p. 222-225; Kelley et al. 2010; Helberger 2013a, p. 30.
[1577] Article 14(8) of the European Commission proposal for a Data Protection Regulation (2012).

For some types of information asymmetry, current data protection law simply doesn't have an answer. It's impossible for people to predict the possible consequences of future uses of personal data. Education about privacy risks seems to be the appropriate answer. In some other contexts, the law requires information about risks, such as on cigarette warnings. Thus, perhaps firms could be required to disclose information about privacy risks.[1578]

Furthermore, it's hard to make an informed decision whether to disclose personal data in exchange for the use of a "free" service, because people don't know the value of their data. Data protection law doesn't have an answer here either. But the transparency principle could provide inspiration. It has been suggested in literature that firms should be required to tell the data subject how much profit they'll make with his or her personal data.[1579] Consumer law prohibits firms from advertising a product as "free" if there are hidden costs.[1580] By analogy, this makes some privacy policies suspicious if the firm captures personal data by way of "payment." In this light, Facebook's claim that "it's free and always will be" deserves scepticism.[1581]

***Risk of manipulation***

Some fear that personalised ads and other content could surreptitiously steer people's behaviour. In short, behavioural targeting could be used to manipulate people. As noted, it's an open question how serious the threat is at present. But in some contexts, such as political advertising, undue influence would be more worrying than in

---

[1578] Such information could include, for instance, the number of data breaches that have occurred the year before. Thanks to Oren Bar-Gill for this suggestion.

[1579] Traung 2012, p. 42.

[1580] Annex 1 (20) of the Unfair Commercial Practices Directive. "Commercial practices which are in all circumstances considered unfair (…) [include:] Describing a product as 'gratis', 'free', 'without charge' or similar if the consumer has to pay anything other than the unavoidable cost of responding to the commercial practice and collecting or paying for delivery of the item."

[1581] "It's free and always will be", says Facebook on the page where people can register for an account (<www.facebook.com> accessed 28 May 2014). See on framing chapter 7, section 4.

others.[1582] As in some cases personalisation could become a problem, scholars and policymakers should keep a close eye on the developments.

Data protection law can help to keep track of developments and perhaps to lessen some risks. The transparency principle also applies if a firm processes personal data to personalise ads or services. The law requires firms to tell data subjects the processing purpose and to give all information that's necessary to guarantee fair data processing.[1583] This suggests a firm must say so if the processing purpose is personalising content. For example, the firm could explain it uses people's browsing behaviour to personalise content.[1584]

If the lawmaker wanted to preclude problems related to surreptitious personalisation, the law could require an icon to accompany personalised content.[1585] A requirement to distinguish certain content wouldn't be a novelty. EU law requires advertising to be clearly labelled as such.[1586] Furthermore, data protection law can be interpreted as generally requiring an option to opt out of personalisation. If personal data processing for personalisation is based on the legal basis consent, people can withdraw their consent. If the processing is based on the balancing provision or on a contract, people have the right to object on compelling legitimate grounds. If the processing concerns personalised advertising, people have an absolute right to object: the right to stop the

---

1582 See chapter 2, section 7, and chapter 3, section 3.

1583 Article 10 and 11 of the Data Protection Directive. When a firm applies a predictive model to an individual (phase 5 of the behavioural targeting process), the firm processes personal data, and data protection law applies (see chapter 5, section 2). Therefore, the firm has to inform the data subject about the processing purpose.

1584 See also Bozdag & Timmersmans 2011, who call for transparency to mitigate the risk of filter bubbles.

1585 See Helberger 2011; Koops 2008, p. 336; Oostveen 2012. An icon to accompany personalised content wouldn't be a complete novelty. When Google started to personalise search results in 2009, for a while it included a link that could alert people that the results were personalised (Horling 2009).

1586 Article 9(1)(a) and 19 of the Audiovisual Media Services Directive; Article 6 of the E-Commerce Directive, Unfair Commercial Practice Directive, Annex I (11). See Helberger 2013, p. 8. The effectiveness of icons is an open question. Whether an icon alerts people to personalisation would have to be assessed in behavioural studies.

processing.[1587] The lawmaker could consider explicitly codifying a requirement for firms to offer people the possibility to stop or pause personalisation.[1588]

Data protection law is silent on the lawfulness of price discrimination and personalised prices.[1589] But if an online shop personalises prices, for instance, on the basis of a cookie representing a customer, it singles out a person and processes personal data. Data protection law requires the data controller to disclose the processing purposes to the data subject.[1590] Therefore, a firm is also obliged to disclose the purpose if the purpose is personalising prices.[1591] Apart from that, data protection law has a specific provision for certain automated decisions, which may be relevant for personalised pricing as well. This provision is discussed in the next chapter.[1592]

Regarding the transparency principle, there's a potential loophole in the Data Protection Directive. Article 11 states which information firms must disclose "where the data have not been obtained from the data subject." This provision applies, for instance, when a data controller obtains data without the individual's consent. But the second paragraph could be interpreted as softening the transparency requirement in case of predictive modelling. "Paragraph 1 shall not apply where, in particular for processing for statistical purposes or for the purposes of historical or scientific research, the provision of such information proves impossible or would involve a disproportionate effort (…)."[1593] Firms could use statistical data to construct predictive models. A firm could try to argue that informing people about its plans to build a predictive model on the basis of their personal data would take "disproportionate

---

[1587] Article 14(a) and 14(b) of the Data Protection Directive. See on opting out chapter 6, section 2; on withdrawing consent chapter 6, section 3.
[1588] A requirement to offer people the chance to pause processing wouldn't be a novelty. Article 9(2) of the e-Privacy Directive requires firms to offer people the possibility to temporarily refuse the processing of location data. Turow proposes an alternative: firms should be required to offer people the chance to see which ads somebody with another cookie profile would see (Turow 2011, p. 198-199).
[1589] See on personalised pricing chapter 2, section 7 and the references there.
[1590] Article 10 and 11 of the Data Protection Directive.
[1591] See on price discrimination chapter 2, section 7 and the references there. See also chapter 9, section 7.
[1592] Article 15 of the Data Protection Directive. See Chapter 9, section 6.
[1593] Article 11(2) of the Data Protection Directive. See also recital 38-40.

effort."[1594] Following that reasoning, the firm wouldn't have to inform the people whose data it uses for building the predictive model. Therefore, the lawmaker could consider stating in a recital that this provision doesn't legitimise building predictive models without transparency for the people from whom the input data were collected. On the other hand, such a rule could hamper scientific or medical research. This suggests the lawmaker should consider drafting separate rules for behavioural targeting or for electronic direct marketing. (The next chapter returns tot this idea.[1595])

***Access rights***

To foster transparency, data protection law requires more from firms than privacy policies and consent requests. For instance, people have the right to access data concerning them.[1596] Again, this calls for enforcement of existing rules and for the development of user-friendly solutions. There's work in this area. For example, Google lets a person see the interest categories that Google tied to the cookie that represents the person. A person can rectify the categories Google has associated with the cookie.[1597] However, Google doesn't show people all information it has on them, and Google doesn't explain how it inferred the interest categories.[1598] Notwithstanding, the interest manager shows that creative solutions to enable access rights are possible.

Access rights to cookie-based profiles could have drawbacks. An ad network could design a system where a user could inspect all data that an ad network has attached to his or her cookie, such as his or her browsing history. But such a system would also

---

[1594] Aggregating personal data to construct a predictive model could be seen as the destruction of personal data, if the personal data are deleted. The destruction of personal data is included in the definition of processing. Hence, in principle a data controller should be transparent about this purpose. See Article 29 Working Party 2014, WP 216, p. 7.

[1595] Chapter 9, section 2 and section 7.

[1596] Article 12 of the Data Protection Directive; article 8(2) of the EU Charter of Fundamental Rights.

[1597] The "Ads Preferences Manager (…) lets you view, delete, or add interest categories associated with your browser so that you can receive ads that are more interesting to you" (Google 2009). See <www.google.com/settings/ads>. See also Van Der Sloot & Zuiderveen Borgesius 2012, p. 102-108.

[1598] "To some extent," notes Van Hoboken, "the control and transparency is merely a façade, behind which a (for the end-user) opaque sophisticated data processing architecture is doing the real work" (Van Hoboken 2009).

create a privacy risk. If Eve found Alice's device, he could see all the websites she visited by inspecting her cookie-profile. If this problem is indeed unsolvable, it could be argued that cookie-based profiling by ad networks is unlawful, as ad networks can't comply with data protection law's access rights. On the other hand, if Eve found Alice's device, it's likely he could also access other information on the device. So perhaps the fact that Eve can inspect her browsing history isn't Alice's main problem.

The European Commission approaches the problem of access rights to pseudonymous data differently in its proposal for a Data Protection Regulation. "If the data processed by a controller do not permit the controller to identify a natural person, the controller shall not be obliged to acquire additional information in order to identify the data subject for the sole purpose of complying with any provision of this Regulation."[1599] This provision could have unfortunate effects. A firm could invoke the provision to deny a data subject access to the browsing history in a cookie-based profile, if the firm can't establish whether the access request comes from the person whose browsing history is stored. If this rule were combined with a provision that allows behavioural targeting on an opt-out basis, people could be tracked and profiled without consent, and wouldn't even be able to exercise their access rights.[1600] Transparency and data subject control would be almost completely absent. Furthermore, not enabling data subject access to personal data seems hard to reconcile with the EU Charter of Fundamental Rights, which states: "[e]veryone has the right of access to data which has been collected concerning him or her."[1601]

***Caveat and conclusion***

As previously mentioned, one policy instrument to reduce information asymmetry is educating the public. Many people lack basic knowledge of internet technology and of

---

[1599] Article 10 of the European Commission proposal for a Data Protection Regulation (2012). The LIBE Compromise confirms this approach (article 10(1) of the LIBE Compromise, proposal for a Data Protection Regulation (2013)). See also article 15(2) of the LIBE Compromise.
[1600] See chapter 6, section 2.
[1601] Article 8(2) of the EU Charter of Fundamental Rights

security and privacy risks. As Cranor & McDonald put it, "consumers cannot protect themselves from risks they do not understand."[1602] However, learning takes time. It seems people are only vaguely aware of behavioural targeting, although it has been happening since the mid 1990s.[1603] And it's questionable whether education could keep up with the pace of the developments in the online marketing industry. Nevertheless, some knowledge is better than none. But the law shouldn't put unreasonable burdens on people's shoulders. In the European legal system, the state has positive obligations to protect people's privacy.[1604] Hence, empowerment shouldn't turn into responsibilisation.[1605] This term describes "the process whereby subjects are rendered individually responsible for a task which previously would have been the duty of another – usually a state agency – or would not have been recognized as a responsibility at all."[1606] While this caveat should be borne in mind, education could help.

In conclusion, stricter enforcement of data protection law, at least how it's interpreted by the Working Party, could help to reduce the information asymmetry. But there's room for refinement of the current legal framework. More transparency could give people a bit more control over information concerning them. Interdisciplinary research is needed to develop better ways to communicate privacy policies. But without a credible threat of enforcement and dissuasive sanctions, firms may lack incentives to make behavioural targeting transparent.

### 8.3 Consent for personal data processing processing processing

EvenEven though the last chapter showed that expectations of informed consent as a privacy protection measure shouldn't be too high, some improvement over the current

[1602] Cranor & McDonald 2010, p. 27. Castelluccia & Narayanan 2012 also call for education (p. 18-19).
[1603] As noted in chapter 2, section 2, cookies have been used for tracking since at least 1996.
[1604] See for instance ECtHR, Z v. Finland, No. 22009/93, 25 February 1997, par. 36. See chapter 3, section 2.
[1605] See Gürses 2010, p. 97. See also Acquisti et al. 2013, p. 2.
[1606] Wakefiel & Flemicg 2009, p. 276. See on responsibilisation in the privacy field the research project SPION, Security and Privacy for Online Social Networks, <www.spion.me> accessed 26 May 2014. Thanks to Seda Gürses for pointing out this concept to me.

situation must be possible. As noted, unambiguous consent is generally the only available legal basis for personal data processing for behavioural targeting, and the e-Privacy Directive requires consent for most tracking technologies.[1607]

It's sometimes suggested that firms can obtain the data subject's consent for personal data processing through their terms and conditions. But the Working Party doesn't accept this. "Consent must be specific. (…) Rather than inserting the information in the general conditions of the contract, this calls for the use of specific consent clauses, separated from the general terms and conditions"[1608] Case law of the European Court of Justice also suggests a consent request shouldn't be hidden in terms and conditions.[1609] Furthermore, obtaining consent by quietly changing a privacy policy isn't possible under data protection law, as there wouldn't be an expression of will by the data subject.[1610] A data subject thus shouldn't have to keep checking a privacy policy to see whether he or she accidentally consents to a new practice by continuing to use a service.

In its Google investigation, the Working Party says that "passive users" weren't informed, and weren't asked for consent. In brief, passive users are people who are tracked by Google on non-Google websites, for instance through its DoubleClick ad network.[1611] Such "users are generally not informed that Google is processing personal data, such as IP addresses and cookies."[1612] The Working Party adds that Google doesn't ask consent for using tracking cookies, as the e-Privacy Directive requires.[1613]

The European Commission proposal for a Data Protection Regulation reaffirms that mere inactivity doesn't signal consent. The proposal requires consent to be "explicit."

---

[1607] Chapter 6.
[1608] Article 29 Working Party, WP 187, p. 33-35. "The information must be provided directly to individuals. It is not enough for it to be merely available somewhere" (p. 35).
[1609] CJEU, C-92/09 and C-93/09, 9 November 2010, Volker und Markus Schecke and Eifert.
[1610] See chapter 6, section 3.
[1611] Article 29 Working Party 2013 (Google letter), appendix, p. 2, footnote 2. Passive users are "users who does not directly request a Google service but from whom data is still collected, typically through third party ad platforms, analytics or +1 buttons."
[1612] Article 29 Working Party 2013 (Google letter), appendix, p. 3.
[1613] Article 29 Working Party 2013 (Google letter), appendix, p. 5.

Consent requires a "statement" or "a clear affirmative action."[1614] "Silence or inactivity should (…) not constitute consent," adds the preamble.[1615] Furthermore, the proposal prohibits hiding a consent request in terms and conditions. "If the data subject's consent is to be given in the context of a written declaration which also concerns another matter, the requirement to give consent must be presented distinguishable in its appearance from this other matter."[1616] Just like in the early 1990s, when the Commission presented its proposal for a Data Protection Directive, many firms reacted to the 2012 proposal by lobbying to soften the requirements for consent.[1617]

***Nudging and take-it-or-leave-it choices***

The status quo bias suggests that requiring opt-in consent could lead to people disclosing fewer data. Requiring opt-in consent could be seen as a kind of "nudging", a phrase coined by Thaler & Sunstein.[1618] A lawmaker nudges when it uses insights from behavioural economics to gently push people's behaviour in a certain direction, without actually limiting their freedom of choice.[1619] Setting defaults is a classic example of nudging. Furthermore, a regime that requires affirmative action for consent (in line with legal doctrine) does more to alert people to data processing than a regime that accepts mere silence as "implied" or "opt-out" consent.

---

[1614] Article 4(8) of the European Commission proposal for a Data Protection Regulation (2012).

[1615] Recital 25 of the European Commission proposal for a Data Protection Regulation (2012). Facebook doesn't agree: "We (…) propose that the reference that consent must be given 'explicitly' and 'silence and inactivity should not constitute consent' should be deleted from Recital 25" (Facebook proposed amendments 2013).

[1616] Article 7(2) European Commission proposal for a Data Protection Regulation (2012).

[1617] See Facebook proposed amendments 2013, p. 23; Amazon proposed amendments (article 4(1)(8); International Chamber of Commerce 2013, p. 3; eBay proposed amendments 2012. See on the 1990s chapter 6, section 3.

[1618] Sunstein gives an opt-in requirement for tracking as an example of a nudge (Sunstein 2013a, p. 38; Sunstein 2013b, p. 13). See on nudging also chapter 9, section 2.

[1619] They describe nudging as follows: "A nudge, as we will use the term, is any aspect of the choice architecture that alters people's behavior in a predictable way without forbidding any options or significantly changing their economic incentives. To count as a mere nudge, the intervention must be easy and cheap to avoid. Nudges are not mandates. Putting the fruit at eye level counts as a nudge. Banning junk food does not" (Sunstein & Thaler 2008, p. 6). If the lawmaker aims to use default settings to keep people in the default setting, some speak of "policy defaults" (Ayres & Gertner 1989, Willis 2013a).

However, Willis warns that it's hard for a lawmaker to make firms use nudges, if those firms don't want to nudge people in the same direction as the lawmaker. Firms have many ways to entice people to opt in.[1620] As Sunstein puts it, "if regulated institutions are strongly opposed to a default rule and have easy access to their customers, they may well be able to use a variety of strategies, including behavioral ones, to encourage people to move in the direction the institutions prefer."[1621] For instance, firms can offer take-it-or-leave-it choices, such as tracking walls on websites. Hence, even if firms offered transparency and asked prior consent for behavioural targeting, people might still feel they have to consent.[1622]

The European Commission proposal for a Data Protection Regulation retains the requirement that consent must be free. The preamble adds: "consent does not provide a valid legal ground where the individual has no genuine and free choice and is subsequently not able to refuse or withdraw consent without detriment."[1623] This recital could be applied to tracking walls, but it doesn't give much more guidance than the existing requirement that consent must be "free."

The LIBE Compromise contains a provision that can be read as a prohibition of tracking walls under certain circumstances: "[t]he execution of a contract or the provision of a service shall not be made conditional on the consent to the processing of data that is not necessary for the execution of the contract or the provision of the service pursuant to article 6(1), point (b)."[1624] That latter provision concerns the legal basis that applies when the processing is necessary to perform a contract with the data subject. However, the LIBE Compromise would also allow firms to rely on the balancing provision for some behavioural targeting practices with pseudonymous

---

1620 Willis 2013; Willis 2013a.

1621 Sunstein 2013a, p. 119. See also Solove 2013, p. 1898. See in detail about the strategies firms can use to make people agree to tracking Willis 2013a, especially p. 111 and further.

1622 See European Commission 2011 (Eurobarometer), p. 27.

1623 Recital 33 of the European Commission proposal for a Data Protection Regulation (2012). Facebook has proposed an amendment that says: "a data controller may legitimately make consent to the processing a condition of access to a service, particularly when the service is free of charge to the data subject" (Facebook proposed amendments 2013, p. 27, amendment to recital 34).

1624 Article 7(4) of the LIBE Compromise (capitalisation adapted).

data. Hence, for many behavioural targeting practices the practical effect of this prohibition of tracking walls would seem to be limited.[1625]

Should the law do anything about take-it-or-leave-it choices regarding the enjoyment of privacy when using websites and other internet services? This is a hard question that invokes discussions on how much legal paternalism is justified. Some authors suggest tracking walls should be prohibited.[1626] (A few suggest tracking walls are already prohibited under the Data Protection Directive.[1627]) A blanket prohibition of take-it-or-leave-it choices would prohibit people from disclosing their personal information in exchange for using a service. As far as protecting the data subject is the main rationale for the ban, a ban on tracking walls would fall within the paternalism definition used in this study.[1628] It doesn't follow that banning tracking walls would be *unduly* paternalistic. That said, some take-it-or-leave-it choices might concern relatively innocuous data processing practices, and it isn't evident that such choices should be prohibited.

The principle of contractual freedom can be applied by analogy to consent to tracking, but contractual freedom isn't absolute. And while insights from contract law can be applied by analogy to consent in data protection law, the two legal fields are different. Furthermore, if a ban on tracking walls would protect the data subject's interests and societal interests at the same time, it wouldn't be purely paternalistic. The next chapter discusses whether there are circumstances in which tracking walls should be prohibited, apart from the general rule that consent must be "free" to be valid.[1629]

---

1625 See article 2(a), article 6(f), and recitals 38 and 58a of the LIBE Compromise, proposal for a Data Protection Regulation (2013). See chapter 6, section 2.

1626 See for instance Irion & Luchetta 2013, p. 78; Brussels declaration 2011 (I am one of the signatories). At least one country prohibits take-it-or-leave-it choices. Article 16(2) of the Personal Information Protection Act of South Korea says: "The personal information processor shall not deny the provision of goods or services to the data subjects on ground that they would not consent to the collection of personal information exceeding minimum requirement." See also Strandburg 2013, p. 88.

1627 Roosendaal 2013, p. 186. In contrast, I think current data protection law often allows take-it-or-leave-it choices (see chapter 6, section 3 and 4).

1628 See chapter 6, section 6.

1629 Chapter 9, section 5 and 7.

Some have suggested the law could require firms to offer a tracking-free version of their service, which has to be paid for with money.[1630] Such a rule would enable people to compare the prices of websites. Now the "price" of a website is usually hidden because people don't know what information about them is captured, nor how it will be used.[1631] Some commentators suggest the price of a tracking-free version shouldn't be left to the market alone.[1632] There are precedents for legal intervention in the prices of media. For instance, EU law limits the amount of advertising that can be shown on television.[1633] As Helberger notes, such an advertising maximum could be seen as a price cap, as the time people spend watching advertising on TV could be seen as payment for content.[1634]

A requirement for firms to offer a tracking-free but paid-for version of their service would be less protective of privacy than a ban on tracking walls. Myopia might lead most people to choose the free version, because they focus on the short-term loss of paying for a service, even if this means they have to consent to behavioural targeting, contrary to earlier plans.[1635] Furthermore, many say it's "extortion" if they have to pay for privacy.[1636]

In conclusion, behavioural economics insights are in line with the formal legal conclusion. Firms aren't allowed to infer consent from mere silence, and shouldn't be allowed to do so. But even if firms offered transparency and asked for opt-in consent for tracking in compliance with the law, the problem of take-it-or-leave-it choices and tracking walls would remain. As long as the law allows take-it-or-leave-it choices, opt-in systems won't be effective privacy nudges.

---

[1630] Traung 2012, p. 42; Irion & Luchetta 2013, p. 38; Calo 2013, p. 50.
[1631] Helberger 2013, p. 19. See also Strandburg 2013, p. 90-91, and chapter 7, section 3 and 4.
[1632] Irion & Luchetta 2013, p. 38.
[1633] Article 23(1) of the Audio Visual Media Services Directive says: "The proportion of television advertising spots and teleshopping spots within a given clock hour shall not exceed 20 %."
[1634] Helberger 2013, p. 18. See also Smythe 1977.
[1635] See myopia chapter 7, section 4, and on the attraction of "free" offers Ariely 2008 (chapter 3); Hoofnagle & Whittington 2013.
[1636] See Cranor & McDonald 2010, p. 27.

### 8.4 Consent for tracking technologies

This section discusses how the e-Privacy Directive's consent requirement for the use of tracking technologies could be improved. Human attention is scarce and requiring consent too often overwhelms people. Requiring consent too often also imposes too much transaction costs on people. There's little reason to require consent for truly innocuous practices. In the Data Protection Directive, the balancing provision is an appropriate legal basis for such practices.[1637] Article 5(3) of the e-Privacy Directive already has exceptions for, in short, cookies that are necessary for establishing communication, and cookies that are necessary for a service that's requested by the user.[1638] More exceptions to the cookie consent requirement could be introduced.

The Working Party suggests, in short, that an exception should be introduced for innocuous analytics cookies.[1639] Some analytics cookies could be relatively innocent, for instance if they can only be used to count website visitors and for some basic analysis of which pages are most popular. In such cases, the processing could probably be based on the balancing provision in many circumstances – if it weren't for the e-Privacy Directive. A right to opt out might suffice under general data protection law, assuming the firm complies with all other data protection principles.[1640] As an aside: it's questionable whether the popular analytics software Google analytics would fall within the exception suggested by the Working Party. Google could use the system to track people across the web.[1641]

It might be better if the lawmaker phrased the consent requirement for tracking in a more technology neutral way. Such a rule could be included in the general data

---

[1637] See chapter 6, section 2, on the balancing provision (article 7(f) of the Data Protection Directive).
[1638] See chapter 6, section 4.
[1639] Article 29 Working Party 2012, WP 194, p. 10-11. A similar exception for innocuous analytics cookies is proposed in the Netherlands (Proposal to amend the Telecommunicatiewet (Telecommunications Act): Eerste Kamer, vergaderjaar 2014–2015, 33 902, A <www.eerstekamer.nl/wetsvoorstel/33902_wijziging_artikel_11_7a> accessed 17 November 2014).
[1640] It's also conceivable that no personal data are processed, depending on how the analytics software works.
[1641] It's unclear whether Google uses Google Analytics to track people from website to website. Google says on one of its web pages: "The Google Analytics Tracking Code also reads the double-click [advertising] cookie (…)" (Google Developers 2014). See on DoubleClick: chapter 2, section 2.

protection regime, rather than in the e-Privacy Directive. The law could require consent for collecting and further processing of personal data, including pseudonymous data, for behavioural targeting and similar purposes – regardless of the tracking technology.[1642] As outlined in chapter 6, one of the aims of article 5(3) is to protect people against surreptitious tracking.[1643] It doesn't make sense if the law only protects people against surreptitious tracking if it involves storing or accessing information on a user's device.[1644]

Phrasing the consent requirement for behavioural targeting in a more technology neutral way could also mitigate another problem. In some ways the scope of article 5(3) seems too narrow. For instance, it's unclear whether the provision applies if firms use passive device fingerprinting for behavioural targeting. Passive device fingerprinting relies on looking at information that a device discloses, such as the type of browser, installed fonts, and other settings. The device could send such information as a part of standard network traffic.[1645] It could be argued that passive device fingerprinting doesn't involve "access to information already stored" on a device.

In theory the lawmaker could try to ensure, for instance in a recital, that article 5(3) also applies to information that is emitted by devices. But this might make the scope of article 5(3) too wide. Take the following hypothetical. A train company estimates how many people there are in each carriage, by capturing the signal from their phones. The company immediately deletes all unique identifiers and aggregates the data, thereby anonymising the data.[1646] The company only knows that there are 50

---

[1642] Perhaps the profiling definition (article 4(3)(a)) of the LIBE Compromise, proposal for a Data Protection Regulation (2013) could serve as a starting point for a legal definition of behavioural targeting. The Dutch lawmaker has tried to capture behavioural targeting in legal language in the Telecommunications Act (for a translation see Zuiderveen Borgesius 2012, p. 5).
[1643] Article 5(3) also has other aims; see chapter 6, section 4.
[1644] If article 5(3) were revised, it should be remembered that the current provision also aims to protect people against unauthorised access to information on their devices. See chapter 6, section 4.
[1645] See chapter 2, section 2. The Working Party said in December 2013 that it was planning to release guidance on device fingerprinting, but at the time of writing this isn't published yet (Article 29 Working Party (Work programme 2014-2015)).
[1646] For this example, we will assume anonymisation is possible. See chapter 5, section 3 for the difficulties of anonymisation.

people in car A, 3 people in car B, and so on. The company uses this information to display on electronic signs which cars still have seating. The processing is limited to counting people and deleting the personal data. Assuming the company offers a clear and easy way to opt out and complies with all data protection principles, it could be argued that the processing can be based on the balancing provision. However, if article 5(3) would apply to capturing any signals emitted by user devices, the company would have to ask consent. Such a consent requirement might annoy travellers and hamper the introduction of a useful service. Following this line of thinking, it would be best not to apply article 5(3) to all information that is disclosed by devices. True, it could also be argued that the risks involved in the hypothetical service are too high and that, therefore, an opt-in system should be required. In any case, general data protection law allows for a more nuanced assessment than the hard consent requirement of article 5(3) of the e-Privacy Directive.

Even if people realise that they are being tracked through device fingerprinting or through a built-in device identifier, it's difficult to defend themselves. It's hard for users to hide their device's fingerprint, or to change the device identifier. The Working Party says "[u]nique, often unchangeable, device identifiers should not be used for the purpose of interest based advertising and/or analytics, due to the inability of users to revoke their consent."[1647] Perhaps the law could explicitly prohibit behavioural targeting that relies on identifiers that are difficult to delete or change. Or the law could prohibit firms from using tracking technologies that are likely to be unknown for the average user, unless firms take measures to make the tracking transparent and controllable.[1648] Such a requirement could already be read in the current transparency principle.

Firms can behave in a manner that might formally comply with the e-Privacy Directive's consent requirement, while breaching the spirit of the law.[1649] For instance,

---

[1647] Article 29 Working Party 2013, WP 202, p. 17.
[1648] See 35th International Conference of Data Protection and Privacy Commissioners 2013.
[1649] See on such "creative compliance" chapter 8, section 1.

website publishers can ask repeated consent for every website visit, or show people an avalanche of pop-up windows. It could be argued that such behaviour doesn't comply with the preamble of the 2009 directive, which amended the e-Privacy Directive. "The methods of providing information and offering the right to refuse should be as user-friendly as possible."[1650] But that doesn't give much guidance. It's hard to preclude firms from breaching the spirit of the law. This is a general problem with laws that require firms to implement opt-in systems to nudge people in a certain direction – if the firm wants to nudge people in the opposite direction.[1651]

### 8.5 Do Not Track

To foster data subject control, user-friendly systems should be developed to enable people to express their choices. This section discusses an example of such a system: the Do Not Track standard. European Data Protection Authorities have asked browser vendors since 1999 not to allow third party cookies by default.[1652] However, Data Protection Authorities have little legal power to regulate browser vendors.[1653] Data protection law imposes obligations on data controllers. But with behavioural targeting the browser vendor is rarely the data controller. The ad network and the website publisher are joint controllers if they determine the purposes and means of the processing.[1654] At the time of writing most browser vendors allow third party cookies by default. This can probably be partly explained by the fact that the major browser

[1650] Recital 66 of Directive 2009/136/EC.
[1651] See section 3 of this chapter.
[1652] Article 29 Working Party 1999, WP 17. "Cookies should, by default, not be sent or stored" (p. 3). See similarly Article 29 Working Party 2010, WP 171.
[1653] More generally, Data Protection Authorities have little legal power to regulate the technical architecture that enables and shapes data processing. An important question is whether there are ways to ensure democratic input and societal debate on the development of such technologies. This research avenue falls outside the scope of this thesis.
[1654] Article 29 Working Party 2010, WP 171, p. 10-12. See on "controllers": chapter 4, section 2.

vendors are connected to firms that use behavioural targeting. The browser users aren't paying customers.[1655]

In the US, the Federal Trade Commission (FTC) has called upon the online advertising industry to adopt a Do Not Track system since 2010. The FTC didn't have a particular system in mind, but did explain what such a system should offer. Among other things, the system should be user-friendly and should stop firms from collecting information if people express a choice not to be tracked.[1656]

The 2009 directive that amended the e-Privacy Directive hints at a user-friendly system for users to give or withhold consent. "Where it is technically possible and effective, in accordance with the relevant provisions of [the Data Protection Directive], the user's consent to processing may be expressed by using the appropriate settings of a browser or other application."[1657] In 2011, EU Commissioner Kroes suggested that a Do Not Track system could enable firms to comply with the e-Privacy Directive's consent requirement.[1658] The Working Party later confirmed that, under certain conditions, a Do Not Track standard could enable firms to comply with the e-Privacy Directive's consent requirement.[1659]

***World Wide Web Consortium's DNT Group***

Since September 2011, a Tracking Protection Working Group of the World Wide Web Consortium ("DNT Group") has been engaged in a discussion about a Do Not Track standard.[1660] The World Wide Web Consortium (W3C) is an international organisation where member organisations cooperate to develop technical web

---

[1655] See Kristol 2001, p. 169-170; Soghoian 2010; Soghoian 2010a; Wingfield 2010. Mozilla (of the Firefox browser) is an exception. Mozilla receives funding from Google, but doesn't seem to have other connections to behavioural targeting. Apple does have an ad network, but its Safari browser blocks third party cookies. Google (of the Chrome browser) and Microsoft (of the Internet Explorer browser) both use behavioural targeting,

[1656] Federal Trade Commission 2010, p. 63-69. The FTC repeated its call in Federal Trade Commission 2012, p. 53. See also Department of Commerce United States 2010, p. 51; p. 72. See on the early history of Do Not Track Soghoian 2011.

[1657] Recital 66 of Directive 2009/136/EC.

[1658] Kroes 2011.

[1659] Article 29 Working Party 2011, WP 188, p. 10; Kohnstamm (chairman of the Article 29 Working Party) 2012.

[1660] W3C Tracking Protection Working Group (website).

standards.[1661] The W3C standards aren't legally binding; the success of a W3C standard is measured by its rate of adoption.[1662] The DNT Group mainly consists of representatives from firms. But several non-governmental organisations and academics also participate in the discussion, as does a representative of the Article 29 Working Party.[1663] The DNT Group could thus be seen as a multi-stake-holder negotiation.[1664]

The Do Not Track standard should enable people to use their browser to signal to websites that they don't want to be tracked. A website publisher or another firm that receives a "Do not track me" signal could reply to the browser: "OK, I won't track you."[1665] Hence, the Do Not Track standard doesn't actually block third party cookies or other tracking technologies. But if the firm continued to track a person after it replied to that person "OK, I won't track you", the law could come into play. In principle, general contract law could be applied. In contract law an indication of wishes can be expressed in any form, and also implicitly. An automatic "I won't track you" reply to a browser request could be seen as an expression of will to enter an agreement, in which the firm promises it won't monitor browsing behaviour.[1666]

A Do Not Track system could dramatically reduce the transaction costs of opting out of each behavioural targeting firm separately.[1667] In that way, the Do Not Track standard is somewhat comparable with a centralised Do Not Call registry where

---

1661 See <www.w3.org>.
1662 See Doty & Mulligan 2013.
1663 Rob van Eijk (of the Dutch Data Protection Authority) participates for the Working Party. I presented a paper at a workshop that was organised by the DNT Group (Zuiderveen Borgesius 2012), and I have given a presentation on the Dutch Telecommunication Act during a conference call in January 2013.
1664 See Doty & Mulligan 2013. See generally on self-regulation in the internet context: Bonnici 2008, on technical standards p. 115-150.
1665 The above is a simplification. The DNT Group foresees more possible answers from firms (W3C, DNT Last Call Working Draft 24 April 2014, section 6.2).
1666 See on the legal requirements for an expression of will chapter 6, section 1, 3 and 4. See for a US perspective on applying contract law to Do Not Track Fairfield 2012.
1667 And, unlike the cookie-based opt-out systems offered by the industry, such as the Youronline choices website that is discussed below, Do Not Track doesn't rely on cookies. Therefore, people don't lose their Do Not Track setting if they clear their cookies.

people can opt out of telemarketing. Similarly, some countries have "Robinson lists": databases with names of people who don't want to receive direct marketing mail.[1668]

It's not immediately apparent how Do Not Track – an opt-out system – could help firms to comply with the e-Privacy Directive. But an arrangement along the following lines could be envisioned. Firms should refrain from tracking internet users in Europe that haven't set a Do Not Track preference. Only if a person signals to a specific firm "Yes, you can track me" after receiving sufficient information, that firm may place a cookie to track that user. Hence, in Europe not setting a preference would have the same legal effect as setting a preference for "Do not track me." In Europe, Do Not Track would thus be a system to opt in to tracking.[1669] In countries without a legal requirement to obtain consent for tracking, firms might be allowed to track people who don't set a Do Not Track preference. Do Not Track would thus be a system to opt out of tracking in the US. Since 1 January 2014, a Californian law requires, in short, website publishers to disclose how they respond to Do Not Track signals.[1670]

At the time of writing, after almost three years of discussion, the DNT Group still hasn't reached consensus regarding certain major topics. The most contentious topic is what firms should do when they receive a "Do not track me" signal from somebody. Research shows that most people expect that activating Do Not Track will result in firms not collecting data, in phase 1 of the behavioural targeting process.[1671] In short, people expect Do Not Track really to mean Do Not Collect. Like the Federal

---

[1668] See on Robinson lists Tempest 2007.
[1669] In Europe Do Not Track would be a system to opt in to tracking, as data processing for behavioural targeting is only allowed after consent, and the e-Privacy Directive requires consent for most tracking technologies (see chapter 6). The territorial scope of the e-Privacy Directive and the Data protection Directive is complicated. A full discussion of the territorial scope falls outside this study's scope. See on the territorial scope of EU data protection law the references in chapter 4, section 1, and chapter 1, section 4.
[1670] Business and Professions Code, section 22575-22579.
[1671] McDonald & Peha 2011; Hoofnagle et al. 2012a.

Trade Commission, European Data Protection Authorities say firms should stop collecting data if somebody signals "Do not track me."[1672]

But many firms prefer Do Not Target. They want to continue collecting data when they receive a "Do not track me" signal. The firms merely want to stop showing targeted ads (phase 5). Members of the Digital Advertising Alliance, a large marketing trade group, don't even want to offer Do Not Target. The Digital Advertising Alliance has proposed a system in which firms can continue collecting data, and can continue targeting ads to people who signal "Do not track me." The firms say they'll keep a profile with inferred interests of somebody who signals "Do not track me", but will delete that person's browsing history.[1673] The DNT Group rejected the proposal of the Digital Advertising Alliance.[1674] At the time of writing, there's no agreement in the DNT Group about which data uses should still be allowed when people signal "Do not track me."

Another point of discussion is whether a signal from a browser, or other user agent, with a default setting of "Do not track me" should be respected.[1675] In 2012, Microsoft announced that the next version of its Internet Explorer browser would be set on "Do not track me" by default.[1676] Many marketers responded angrily. Some firms say that default Do Not Track signals don't express a user's choice, and can thus be ignored. Yahoo for instance, one of the largest behavioural targeting firms, said it would ignore the DNT signals from Microsoft Internet Explorer.[1677] There's some irony in this, as currently the behaviour of hundreds of millions of people is monitored while they were never given a choice. And as noted, the Interactive Advertising Bureau UK

---

[1672] See for instance Kohnstamm (chairman of the Article 29 Working Party) 2012: "According to European laws Do Not Track should be 'do not collect'."
[1673] The Digital Advertising Alliance thus proposes to delete some data in phase (2) of the behavioural targeting process.
[1674] W3C, DNT Last Call Working Draft 24 April 2014, par. 4.
[1675] In theory, this shouldn't be an issue in Europe. As noted, in Europe Do Not Track would be a system to opt in to tracking.
[1676] Lynch 2012.
[1677] Yahoo Public Policy Blog 2012. The Digital Advertising Alliance, a marketing trade group, also said companies don't have to honour the Do Not Track signals from Microsoft's browser (Mastria 2012).

suggests that people can give consent to tracking cookies by leaving the default settings of their browser untouched.[1678]

At the time of writing, the question of how to treat browsers that signal "Do not track me" by default is still subject to debate. In brief, the DNT Group's current view is that browser vendors should not make their browsers signal "Do not track me" by default. This might be different if a browser is explicitly marketed as a privacy-preserving browser, for instance with a brand name like "SuperDoNotTrack."[1679]

Meanwhile, major browser vendors have already technically implemented a system that enables people to signal Do Not Track preferences. Many people have selected the "Do not track me" setting. Some estimate that "Do Not Track is already set in about 20% of browser requests to European websites."[1680] However, most behavioural targeting firms ignore Do Not Track signals, saying they don't know what "Do not track me" means.[1681] For instance, the Chief Privacy Officer of Yahoo reportedly said in 2011: "[r]ight now, when a consumer puts Do Not Track in the header, we don't know what they mean."[1682] Google has reportedly expressed similar opinions.[1683]

From the start, proposals for a Do Not Track standard have excluded tracking within one website.[1684] In brief, there's agreement within the DNT Group that tracking within one website shouldn't be affected by "Do not track me" signals. This would imply that firms such as Amazon or Facebook are allowed to analyse people's behaviour within their own website, regardless of whether people signal "Do not track me." In contrast, the e-Privacy Directive's consent rule also applies to first party tracking

---

[1678] See chapter 6, section 4.
[1679] W3C, DNT Last Call Working Draft 24 April 2014, par. 4.
[1680] Baycloud Systems 2014. The US Interactive Advertising Bureau has claimed: "My members [are] seeing 20-25% of user base sending flag. (…) We expect DNT:1 signals to approach 50% in short-term" (Zaneis 2013).
[1681] Some firms, such as Twitter, say they stop collecting data when they receive a "Do not track me" signal (Twitter 2012). Mayer & Narayanan (Donotrack.us website) give a list of firms that are taking steps to honour Do Not Track signals.
[1682] Quoted in Mullin 2011.
[1683] Mullin 2011.
[1684] Schunter & Swire 2013, p. 12. Some complain that Do Not Track helps larger firms such as Google and Facebook and hurts ad networks that don't offer consumer services (see Chapell 2014).

cookies.[1685] Therefore, it's hard to see how a Do Not Track standard that doesn't apply to first party tracking could help firms to comply with the e-Privacy Directive.

In April 2014 the DNT Group published a "last call working draft" of the Tracking Preference Expression document, with the *technical* requirements for a Do Not Track standard. A last call is an invitation for people inside and outside W3C to comment on the technical soundness of a proposed standard. But many major issues remain undecided, and must be set out in another document (the Tracking Compliance and Scope specification). For instance, the DNT Group still has to decide which types of data can be processed according to the standard when people signal "Do not track me."

> Of note, this document does not define site behavior for complying with a user's expressed tracking preference (…). The Tracking Compliance and Scope (TCS) specification which standardizes how sites should respond to Do Not Track requests, including what information may be collected for limited permitted uses despite a Do Not Track signal, is under discussion.[1686]

A few days after the DNT Group published the last call working draft, Yahoo announced it wouldn't honour Do Not Track signals.[1687] Hence, it seems questionable whether the standard will be widely respected by firms. And meanwhile, the Do Not Target versus Do Not Collect debate continues.

To enable websites to comply with EU law, the Do Not Track standard should at least comply with the following two conditions. First, firms must not collect data for

---

[1685] See chapter 6, section 4..
[1686] W3C, DNT Last Call Working Draft 24 April 2014, introduction. See section 6.2.1 of the document for the proposed definition of tracking.
[1687] Yahoo Public Policy Blog 2014.

behavioural targeting about people in the EU who don't set a preference. Silence is not consent after all.[1688] Second, if a person visits a website and signals "Do not track me", the website and its partners shouldn't follow that person's activities. No tracking should generally mean no data collection.[1689] Some minor exceptions may be needed for this rule. For instance, in some cases it may be necessary for website publishers to store the IP address of certain visitors for a short period, for security reasons.[1690]

***Tracking walls and take-it-or-leave-it choices***

From the beginning of the discussions, the Do Not Track standard would allow a website to ask a visitor who signals "Do not track me" for an exception, along the following lines. "We see your Do Not Track signal. But do you make an exception for me and my ad network partners so we can to track you?"[1691] Hence, if a standard were developed that complied with EU law, many websites would probably respond by installing tracking walls. This would be comparable with the situation that would result from strictly implementing article 5(3) of the e-Privacy Directive.[1692]

The possibility of tracking walls and take-it-or-leave-it choices isn't a flaw of the Do Not Track system, but a logical consequence of the general principle of contractual freedom, and of the consent rules in the Data Protection Directive.[1693] If a "Do not track me" setting leads to being confronted with tracking walls on many websites, people might change their setting to forego that extra click.[1694] And people might just click "yes" to requests for exceptions.[1695] In sum, a hypothetical Do Not Track standard that complied with EU law would probably bring us back to the problem of tracking walls.

---

[1688] See chapter 6, section 3.
[1689] See Kohnstamm (chairman of the Article 29 Working Party) 2012.
[1690] See on that topic Soghoian 2011a.
[1691] See for instance W3C, DNT Last Call Working Draft 24 April 2014, section 7.
[1692] See section 3 of this chapter, and chapter 6, section 3 and 4.
[1693] See on tracking walls and take-it-or-leave-it choices chapter 6, section 3 and 4, and chapter 8, section 3.
[1694] See Strandburg 2013, p. 169-170.
[1695] See chapter 7, section 3 and 4.

*Other possibilities for user-friendly consent mechanisms*

Do Not Track could be seen as a system that aims to make consent more meaningful. There would be other possibilities to enable people to express their choices. For instance, a centralised system could be developed where people can choose to be tracked.[1696] The Interactive Advertising Bureau (IAB) shows such as system would be possible. The IAB runs a website where people can opt out of receiving targeted ads: youronlinechoices.com. There are, however, serious problems with the website. For instance, the website merely offers the equivalent of Do Not Target. Firms may continue to track people who have opted out.[1697] The website's FAQ explains: "[d]eclining behavioral advertising only means that you will not receive more display advertising customised in this way."[1698] But it seems plausible that people expect the website to offer Do Not Collect.[1699]

Additionally, the site works with opt-out cookies. Hence, if a person clears his or her cookies – a measure that is often suggested to limit tracking – the opt-outs are lost.[1700] Furthermore, in 2011 the Working Party noted that the Youronlinechoices website included code that enables user tracking, while users weren't informed about this.[1701] Nevertheless, the website does show that a centralised system for firms to obtain consent for tracking would be possible.

In sum, if a Do Not Track standard were developed that complied with EU law, many websites would probably respond by installing tracking walls. Even if firms provided

[1696] See Article 29 Working Party 2011, WP 188, p. 6.
[1697] Article 29 Working Party 2011, WP 188, p. 7. As an aside, suggesting to people that they can opt out of tracking while they can only opt out of receiving behaviourally targeted ads is hard to reconcile with article 7 of the Unfair Commercial Practices Directive on "misleading omissions". See on consumer law chapter 4, section 4.
[1698] Interactive Advertising Bureau Europe – Youronlinechoices.
[1699] In the US there's a similar website. Research suggests that many people expect it to offer Do Not Collect rather than Do Not Target (Cranor & McDonald 2010, p. 18).
[1700] In reaction to the Federal Trade Commission's call for a Do Not Track system, Google has released an extension for its Chrome browser in 2011: "Keep My Opt-Outs". This extension "enables you to opt out permanently from ad tracking cookies." See Google Public Policy Blog 2011.
[1701] Article 29 Working Party 2011, WP 188, p.7.

clear information, even if people understood the information, and even if firms asked prior consent, people might still feel they have to consent to behavioural targeting.

### 8.6 Conclusion

This chapter discussed how the law could improve individual empowerment in the behavioural targeting area. Strictly enforcing the data protection principles would be a good start. The law also needs amendments.

Of course, the Data Protection Directive is only relevant if the practice of behavioural targeting is found to come within the directive's scope. This will be the case if behavioural targeting is seen as processing personal data. Hence, from a normative perspective, data protection law should apply to behavioural targeting, including when firms use pseudonymous data. Apart from that, as discussed in chapter 5, a sensible interpretation of data protection law implies that data that are used to single out a person should be seen as personal data.

To reduce the information asymmetry in the area of behavioural targeting, the transparency principle should be enforced. In line with European consumer law, the lawmaker should require firms to phrase privacy policies and consent requests in a clear and comprehensible manner. Codifying the clear language requirement could discourage firms from using legalese in privacy policies. The rule wouldn't be enough to ensure actual transparency, but it could help to lower the costs of reading privacy policies. Furthermore, interdisciplinary research is needed to develop tools to make data processing transparent in a meaningful way.

Regarding consent, the existing rules must be enforced. Even though website publishers have started to inform visitors about cookies, many fail to ask consent for behavioural targeting, or don't even offer an option to opt out of tracking. Firms shouldn't be allowed to infer consent from mere silence. This follows from legal doctrine. Furthermore, behavioural economics insights suggest that requiring opt-in consent could nudge people towards disclosing fewer data. The European Commission proposal reaffirms that consent requires a clear expression of will.

Human attention is scarce and too many consent requests can overwhelm people. One problem with the consent requirement for tracking technologies in article 5(3) of the e-Privacy Directive is that the scope of article 5(3) has proven to be too broad. Article 5(3) also applies to some cookies that pose little privacy risks and that aren't used to collect detailed information about individuals, such as certain types of cookies that are used for website analytics. But there's little reason to ask consent for truly innocuous practices. It would probably be better if the lawmaker phrased the consent requirement for tracking in a more technology neutral way. The law could require consent for the collection and further processing of personal data, including pseudonymous data, for behavioural targeting and similar purposes – regardless of the technology that's used. An option that could be explored is whether a separate legal instrument is needed for behavioural targeting (see section 7 of the next chapter).

Furthermore, a user-friendly system should be developed to make it easier for people to give or refuse consent. Work is being done in this area. The Tracking Protection Working Group of the World Wide Web Consortium (DNT Group) is in the process of trying to develop a Do Not Track standard. The Do Not Track standard should enable people to signal with their browser that they don't want to be tracked. But even a hypothetical Do Not Track system that would comply with European law would probably lead to tracking walls. The next chapter examines whether specific rules regarding such take-it-or-leave-it choices are needed in some circumstances.[1702]

How should the suggestions in this chapter be assessed in the light of the central question of this thesis: how could European law improve privacy protection in the area of behavioural targeting, *without being unduly prescriptive*? In this study, the

---

[1702] Chapter 9, section 5 and 7.

"not unduly prescriptive" requirement means that measures shouldn't be unreasonably costly for society, or unreasonably paternalistic.

Enforcing and tightening data protection law's transparency requirements wouldn't be unduly paternalistic, if at all. Requiring firms to be transparent about behavioural targeting doesn't interfere with the data subject's liberty.[1703] Furthermore, from an economic perspective, markets don't function well when there's information asymmetry. Protecting a well-functioning market has nothing to do with paternalism. Requiring firms to use an opt-in system for valid consent (rather than an opt-out system) could be seen as a measure to nudge people towards disclosing less personal information. As the data subject can still allow tracking, by giving consent, such a rule hardly interferes with the data subject's liberty. This implies that an opt-in requirement isn't very paternalistic. Apart from the fact that a nudge hardly interferes with liberty, there are other rationales for an opt-in requirement than protecting the data subject against him or herself.[1704] Again this implies that opt-in requirements aren't unduly paternalistic.

Drafting readable privacy policies costs time and money. The costs of relatively simple measures, such as avoiding legalese in consent requests and privacy policies, may be manageable. While not too costly, the effectiveness of such measures remains to be seen; they must be tested in practice. However, making data processing transparent in a meaningful way may require serious investments, for instance in design and research.[1705] In some cases other measures, such as mandatory rules or prohibitions, may be cheaper.[1706] In sum, the costs of empowering the individual shouldn't be underestimated, and in some cases they can be considerable. But in general it can't be said that the costs are unreasonable.

---

[1703] See the paternalism definition in chapter 6, section 6.

[1704] In US literature, nudges are sometimes called "libertarian paternalism" (Sunstein & Thaler 2008, introduction). Some see nudges as (too) paternalistic; see e.g. Mitchell 2004. This depends largely on the paternalism definition one uses.

[1705] See on transparency enhancing tools (TETs): chapter 9, section 6.

[1706] See Helberger 2013a, p. 28.

In conclusion, aiming for data subject control isn't a panacea, but compared to the current situation, where hundreds of millions of people are tracked without being aware, some improvement must be possible. Enforcing and tightening the data protection principles could help to empower the data subject. However, aiming for individual empowerment alone won't suffice to defend privacy in the area of behavioural targeting. Even if firms provided clear information, even if people understood the information, and even if firms asked prior consent, many people might still feel they must consent to behavioural targeting when encountering take-it-or-leave-it choices. Hence, protection of the individual is needed as well. This approach is discussed in the next chapter.

* * *

# 9 Improving protection

How could the law protect, rather than empower, the individual? The protective data protection principles should be enforced more strictly. But this won't be enough to improve privacy protection in the area of behavioural targeting. In addition to data protection law, more specific rules regarding behavioural targeting are needed. If society is better off if certain behavioural targeting practices don't take place, the lawmaker should consider banning them.

Section 9.1 discusses the strengths and weaknesses of data protection law's general rules with open norms, compared to more specific rules. Section 9.2 argues that more attention to protecting the individual wouldn't necessarily make the law unduly paternalistic. Section 9.3 discusses the data minimisation principle. Section 9.4 shows that the transparency principle can be read as a prohibition of surreptitious data processing. Section 9.5 concerns sensitive data and chilling effects. Section 9.6 discusses data protection law's provision on automated decisions. A conclusion is provided in section 9.7.

## 9.1 General and specific rules

If fully complied with, the data protection principles could give reasonable privacy protection in the area of behavioural targeting.[1707] But there are at least two problems with data protection provisions that aim to protect the data subject. First, as discussed

---

[1707] As discussed in chapter 4, section 5 and chapter 6, section 5, data protection law contains many protective rules. See also Bygrave 2002, who discusses the implication of the data protection principles for profiling and behavioural targeting (p. 334-362).

in the last chapter, compliance and enforcement are lacking.[1708] Second, a common complaint is that the Data Protection Directive uses too many general rules with open norms. The open norms can help to explain the lack of compliance, as discussed below.

Because the Data Protection Directive lays down an omnibus regime and aims to cover many different situations, it contains many general rules with rather open norms.[1709] The strength of this regulatory strategy is that the law doesn't leave many gaps. Open norms can be applied to unforeseen situations, for instance, when new technologies are developed. Open norms also allow firms to decide how to achieve compliance. For example, firms can choose the best technical solution to comply with data protection law's security principle.[1710]

But open norms also have weaknesses. Opens norms can make the law hard to apply for firms, hard to understand for data subjects, and hard to enforce for Data Protection Authorities. Phrases such as "fairly", "necessary", and "not excessive" leave ample room for interpretation.[1711] Basic definitions of data protection law are subject to significant discussion.[1712] It has been said about data protection law that "the unclear definitions of legal terms are a major problem, potentially the greatest problem."[1713]

The distinction between specific rules and general rules with open norms is a matter of degree rather than kind. Lawyers can find ambiguity in the most detailed and specific rules. Hence, a rule is always relatively general or relatively specific.[1714] Besides, the complicated nature of data protection law shouldn't be exaggerated. Data protection law gives a relatively objective checklist for firms. Data protection law can

---

[1708] See chapter 8 section 1.
[1709] See chapter 4, section 2. See also ECJ, C-101/01, Lindqvist, 6 November 2003, par. 83; CJEU, C-468/10 and C-469/10, ASNEF, 24 November 2011, par. 35.
[1710] Article 17 of the Data Protection Directive. See on the security principle chapter 4, section 2 and the references there.
[1711] Article 6(1)(a), 6(1)(c), and 6(1)(c) of the Data Protection Directive.
[1712] See chapter 5: even the scope of "personal data", the key term of data protection law, is hotly debated. See also chapter 4, section 2.
[1713] Zwenne 2013, p. 37. See also Zwenne at al. 2007; Winter et al. 2008p. 161-162.
[1714] Hesselink 2011, p. 639. See also Sunstein 1995.

be applied without engaging in discussions about the scope and meaning of privacy.[1715] Notwithstanding, many data protection provisions are rather general.

Using a phrase from regulation studies, parts of data protection law can be characterised as principles-based regulation.[1716] "In principles-based regulation," explain Baldwin et al., "principles are used to outline regulatory objectives and values, and regulatees are left free to devise their own systems for serving such principles."[1717] Principles-based regulation "is a method of encouraging regulatees to think for themselves and assume responsible approaches."[1718] This approach works best for trustworthy firms. "Central to the success of PBR [principles-based regulation] is, accordingly, trust in the competence and responsibility of the regulatees."[1719]

Firms and regulators come from different backgrounds, and have different ideas. Therefore, firms may have genuinely different interpretations of what is meant by an open norm, according to Baldwin et al. "Firms and regulators are liable to interpret regulatory requirements in divergent ways because they see the world differently – even if the regulatees are well-disposed and highly capable."[1720] For example, if a firm saw incorrectly targeted ads as a problem, it might disagree with regulators when data processing is "excessive."[1721] Cultural differences between countries can also play a role when interpreting open norms.[1722] Furthermore, firms may see an open norm as an invitation for discussion, instead of as a rule they must follow, say Baldwin et al.

---

[1715] See chapter 4, section 2, and De Hert & Gutwirth 2006, p. 94.
[1716] See Busch 2010, p. 9. See on regulation studies chapter 8, section 1.
[1717] Baldwin et al. 2011, p. 302.
[1718] Baldwin et al. 2011, p. 303.
[1719] Baldwin et al. 2011, p. 303.
[1720] Baldwin et al. 2011, p. 306. This study calls such firms well-intentioned and (well-)informed. See on the appropriate enforcement strategies for different types of firms chapter 8, section 1.
[1721] See section 3 of this chapter.
[1722] For instance, US firms might not see privacy and data protection rights as fundamental rights.

> Even if there is general agreement on the governing principles for a regime, the relevant group of regulatory actors may treat those principles not as a statement of objectives but as starting points for debates on substantive aims – debates that they engage in with different conceptions of the game being participated in and different understandings regarding key aspects of that game (such as what constitutes "compliance" or a "reasonable practice").[1723]

Indeed, in the behavioural targeting area, some firms appear to see data protection rules as a starting point for discussion, rather than as rules they have to comply with.[1724] To illustrate, the Interactive Advertising Bureau UK (IAB) says the e-Privacy Directive's consent requirement for tracking technologies should be implemented in a way "that leaves space for innovative new business models to develop."[1725] The IAB suggests that it can be assumed that people consent to tracking cookies if they don't change their browsers' default settings. It appears the IAB sees the requirements for valid consent as open norms.

Specific rules are easier to follow and to enforce than general principles. To borrow an example from Sunstein, the rule "don't drive faster than 120", gives more guidance than "don't drive unreasonably fast", or "don't endanger other road users."[1726] Specific rules also provide more predictability regarding enforcement. Moreover, the *lex certa* principle requires the law to clearly describe which activities can lead to penalties.[1727] This would be especially relevant if Data Protection Authorities were

---

[1723] Baldwin et al. 2011, p. 304.
[1724] Some firms might simply not care about data protection law, for example because they don't expect it will be enforced (see chapter 8, section 1).
[1725] Interactive Advertising Bureau United Kingdom 2012, p. 2. See also Stringer 2013, on the Interactive Advertising Bureau arguing for a lighter regime for pseudonymous data.
[1726] The first two examples are taken from Sunstein, and slightly rephrased (Sunstein 1995, p. 959).
[1727] See on the foreseeability of rules ECtHR, Sunday Times v. The United Kingdom, No. 6538/74, 26 April 1979, par. 49. See also Zwenne 2013, p. 35.

given the authority to impose large penalties.[1728] In sum, there are good reasons for using more specific rules.

The main weakness of specific rules is that they're less flexible than more general norms. For instance, sometimes driving 140 mph is perfectly safe, and sometimes 60 mph is too fast. A maximum speed of 100 mph doesn't reflect such nuances. Another downside of specific rules is the possibility of "creative compliance."[1729] A firm could comply with the letter of the law, while breaching the spirit of the law. Creative compliance sometimes occurs in the field of tax law for example.[1730] Baldwin et al. suggest the lawmaker can mitigate the risk of creative compliance by ensuring that general principles apply in the background.[1731] To stay with the traffic law example, the law can generally prohibit endangering other road users, in addition to specific rules such as maximum speeds.[1732]

As far back as 1994 Simitis argued that the Data Protection Directive should be supplemented with sector specific rules. "Omnibus regulations of data processing are merely a first step. The more specific the processing issues are, the less general rules help. Although they may indicate the direction to be followed, they do not specify solutions appropriate for particular processing contexts."[1733] Simitis concludes the European Union "must complete the Directive with a series of regulations focusing on particular processing issues", for instance for "research and statistics, marketing, and credit agencies."[1734] The Data Protection Directive's preamble says its principles "may be supplemented or clarified, in particular as far as certain sectors are concerned, by

---

[1728] See chapter 8, section 1.
[1729] Baldwin et al. 2011, p. 306.
[1730] Baldwin et al. 2011, p. 232.
[1731] Baldwin et al. 2011, p 305-306. Arguably such a relationship exists between the e-Privacy Directive and the general Data Protection Directive.
[1732] See for instance article 5.1 of the Dutch Road Traffic Act: "It is an offence for any road user to act in such manner as to cause a hazard (or a potential hazard) on the public highway or to obstruct other road users in any way." And as noted in chapter 4, section 4, the good faith requirement in contract law can be used if more specific contract law provisions leave a gap.
[1733] Simitis 1994 p. 466. See also De Hert & Gutwirth 2006, p. 102.
[1734] Simitis 1994, p. 467. See also Blume 2012 (p. 32-34) who discusses whether the public and the private sector should be subject to different data protection regimes.

specific rules based on those principles."[1735] But with the e-Privacy Directive as the major exception, there hasn't been much activity on this front.[1736] That said, there are many norms, legal and non-legal, that protect privacy in addition to data protection law. For instance, the medical profession has its own norms, while some countries have specific rules for CCTV.[1737]

In conclusion, the Data Protection Directive open norms are flexible, but this flexibility comes at a cost for legal certainty and clarity. If specific rules were adopted for behavioural targeting, the general data protection principles should continue to apply as well, to ensure that the law doesn't leave any gaps.

## 9.2 Mandatory rules and paternalism

This section discusses factors that the lawmaker can take into account when deciding whether to use more protective rules in addition to data protection law. The section also considers, and rejects, the idea that using mandatory protective rules would make the law unduly paternalistic.

The behavioural economics analysis in previous chapters shows that more protective rules are needed to improve privacy protection in the area of behavioural targeting. Several scholars have hinted at the need for prohibitions in privacy law, because they lost faith in informed consent.[1738] But when should the lawmaker use prohibitions? De Hert & Gutwirth discuss five factors that the lawmaker can take into account when choosing between general data protection law and stricter "opacity tools." As discussed in chapter 4, De Hert & Gutwirth distinguish data protection law, a "transparency tool", from more prohibitive "opacity tools", such as the legal right to

[1735] Recital 68 of the Data Protection Directive.
[1736] See on the e-Privacy Directive chapter 5, section 6, chapter 6, section 4, and chapter 8, section 4.
[1737] See on CCTV Hempel & Töpfer 2004.
[1738] See e.g. Barocas & Nissenbaum 2009; Solove 2013; Radin 2013; Sloan & Warner 2013; Tene & Polonetsky 2012. See generally about mandatory rules regarding privacy Allen 2011. It must be noted that US scholars are critiquing the US "notice and consent" regime, which, unlike data protection law, doesn't include many mandatory rules.

privacy in the European Convention on Human Rights. Opacity tools aim "to guarantee non-interference in individual matters."[1739] Some of the suggestions for stricter rules and prohibitions that are given below in this chapter can be defended on the grounds suggested by de Hert & Gutwirth.[1740]

Opacity tools are appropriate in the following circumstances, according De Hert & Gutwirth.[1741] First, the sanctity of the home, not only in a literal sense, should be protected. "People need places where they can rest and come to terms with themselves in a sphere of trust and security (…)."[1742] Second, opacity tools are "required when other firmly rooted (in tradition or in law) human rights are at stake, such as the right to have correspondence and the content of communication protected."[1743] These first two reasons to choose opacity tools thus are reminiscent of the perspective of privacy as limited access, or as confidentiality.[1744]

Third, De Hert & Gutwirth note that the Data Protection Directive contains some opacity tools, rules of a more prohibitive nature. An example given by the authors is data protection law's stricter regime for "special categories" of data, such as data regarding health or political opinions.[1745] A second example is data protection law's in-principle prohibition of certain automated decisions with far-reaching effects for the individual (see section 6 of this chapter). The authors suggest that the stricter rules regarding automated decisions and special categories of data can be explained by the risk of unfair social sorting, or "discriminatory effects."[1746]

---

[1739] De Hert & Gutwirth 2006, p. 66. See chapter 4, section 3.
[1740] See section 3 and 5 of this chapter.
[1741] De Hert & Gutwirth 2006, p. 101.
[1742] De Hert & Gutwirth 2006, p. 101.
[1743] De Hert & Gutwirth 2006, p. 101.
[1744] Web browsing is a form of "communication" according to the legal definitions in the EU telecommunications framework (see chapter 6, section 4).
[1745] Article 8 of the Data Protection Directive.
[1746] De Hert & Gutwirth 2006, p. 102. See also Bennett 2011a, p. 490-491.

Fourth, "a need for opacity can be drawn from the function of human rights in promoting and encouraging citizenship."[1747] The lawmaker should use opacity tools if data processing threatens the "formation of the free and equal citizen." This rationale could be extended: if data processing threatens values that are important for a democratic society, rules of a more prohibitive nature are needed. In general, De Hert & Gutwirth seem especially inclined to argue for opacity tools when, apart from individual interests, societal interests are at stake as well.[1748] Lastly, similar to Simitis, De Hert & Gutwirth call for opacity tools if data protection regulation leaves too much room for different interpretations.[1749] This mainly seems to be an argument for clear and specific rules, rather than for prohibitive rules.

If the data subject can override a rule by giving consent, this study doesn't see it as a prohibition. In the terminology of chapter 6, prohibitions are "mandatory", and rules that can be overridden with consent are "default rules."[1750] De Hert & Gutwirth don't limit their category of opacity tools to mandatory rules. For instance, the authors see the e-Privacy Directive's opt-in requirement for commercial email as an opacity tool, "which inherently implies the prohibition of unsolicited marketing mail unless the user makes an explicit request to receive it."[1751] Similarly, they see the data protection regime for special categories of data an opacity tool, even though in many member states the prohibition of processing can be overridden with explicit consent.[1752] This study classifies such opt-in requirements as default rules.

***Paternalism***

The previous chapter discussed ways to make consent more meaningful. If firms want to process personal data, and can't base the processing on the balancing provision or

---

1747 De Hert & Gutwirth 2006, p. 102.
1748 This line of reasoning is related to the economic concept of externalities (see chapter 7, section 3).
1749 De Hert & Gutwirth 2006, p. 102.
1750 See chapter 6, section 5.
1751 De Hert & Gutwirth 2006, p. 95.
1752 De Hert & Gutwirth 2006, p. 77. They note that the prohibition of processing special categories of data isn't absolute.

another legal basis, they must ask the data subject for consent. Hence, by default, certain data processing activities aren't allowed, but the data subject can change this default situation by consenting to processing.[1753] Such a default rule leaves the choice to the data subject. In contrast, mandatory rules can't be overridden with consent, and limit the data subject's contractual freedom. As discussed in chapter 6, paternalism involves, in short, limiting somebody's contractual freedom in order to protect that person.[1754] Therefore, unlike default rules, mandatory rules could be unduly paternalistic in some cases.[1755]

But using more mandatory rules that protect the data subject wouldn't necessarily make the law unduly paternalistic. A rule is purely paternalistic if it only aims at protecting people against themselves. But there are other rationales for legal privacy protection than protecting people against themselves. The right to privacy and the right to data protection aim to contribute to a fair society, which goes beyond individual interests.

Additionally, an economic argument can be made in favour of adopting mandatory rules in the area of behavioural targeting.[1756] As discussed, an economic analysis of informed consent to behavioural targeting suggests there are market failures, such as information asymmetry. It may be impossible to reduce the information asymmetry problem to manageable proportions.[1757] Reducing market failures has nothing to do with paternalism. Furthermore, using protective mandatory rules could be more efficient than giving people the choice to give or refuse consent. It would take people

---

[1753] Article 7(a) and 8(2)(a) of the Data Protection Directive. Data processing practices that aren't allowed without consent are, in short, those practices that can't be based on article 7(b)-7(f) of the Data Protection Directive.
[1754] See chapter 6, section 6.
[1755] Some scholars see default rules as mildly paternalistic (see for instance Sunstein & Thaler 2008).
[1756] See chapter 7, section 3 (on transaction costs).
[1757] See chapter 7 and 8.

several weeks a year to read all online privacy policies they encounter. The aggregate costs for society would be enormous.[1758]

Furthermore, the European Court of Human Rights requires protection of the right to private life that is "effective, not theoretical and illusory."[1759] Because behavioural research shows that data protection law's informed consent requirement is problematic in practice, more protective measures are needed to provide effective privacy protection.[1760] If informed consent requirements don't succeed in protecting privacy in the area of behavioural targeting, it's likely to affect millions of people.[1761] In addition, the current situation is that hundreds of millions of people are being tracked and profiled without being aware. As Hoofnagle et al. note, tracking millions of people without their consent could be seen as a unilateral intervention imposed by the marketing industry, without prior debate.[1762]

Moreover, bothering people dozens of times per day with choices that they don't understand doesn't empower them in any real sense. The time somebody spends on such choices can't be spent on pursuing other goals. "Time is limited," notes Sunstein, "and some issues are complex, boring, or both."[1763] In daily life, there are many decisions people don't have to worry about: "how best to clean tap water, or how to fly an airplane, or what safety equipment should be on trains."[1764] "If we did not benefit from an explicit or implicit delegation of choice-making authority, we would be far worse off, and in an important sense less autonomous, because we would have less time to chart our own course."[1765]

---

[1758] Expressed in money, in 2007 the cost of reading privacy policies would be around 781 billion dollars, while all online advertising income in the US was estimated to be 21 billion dollar. (Cranor & McDonald 2008).

[1759] ECtHR, Christine Goodwin v. the United Kingdom, No. 28957/95, July 11, 2002, par 74.

[1760] See chapter 7, section 3 - 6.

[1761] Radin suggests that the amount of people affected should be taken into account when regulating standard contract terms (Radin 2013, chapter 9).

[1762] Hoofnagle et al. 2012.

[1763] Sunstein 2013, p. 1884.

[1764] Sunstein 2013, p. 1884.

[1765] Sunstein 2013, p. 1884. See also Wagner 2010, p. 68.

Solove makes a similar point. "With the food we eat and the cars we drive, we have much choice in the products we buy, and we trust that these products will fall within certain reasonable parameters of safety. We do not have to become experts on cars or milk, and people do not necessarily want to become experts on privacy either."[1766] He adds: "many people do not want to micromanage their privacy. They want to know that someone is looking out for their privacy and that they will be protected from harmful uses."[1767]

It doesn't follow that we should outsource all our choices to the state.[1768] But the foregoing does suggest that, sometimes, prohibitions can give people more time to lead their lives. In sum, somewhat paradoxically, sometimes taking choices away from the individual with mandatory rules can foster real individual empowerment.[1769] It is, of course, necessary to arrange democratic legitimacy and sufficient checks and balances regarding the entity that sets the rules.[1770]

***Nudging and using transaction costs strategically***

Formally a mandatory rule can be distinguished from a non-mandatory default rule. But in practice the distinction isn't as hard as it may seem. Default rules can be "sticky", because many people stick with default options.[1771] As noted in the previous chapter, requiring opt-in consent for tracking could be seen as nudging.[1772] The lawmaker could also use an option in between mandatory and default rules: the strategic use of transaction costs.[1773]

---

[1766] Solove 2013, p. 1901.
[1767] Solove 2013, p. 1901.
[1768] To avoid misunderstandings: Solove and Sunstein don't suggest we outsource all our (privacy) decisions to the state. In fact, they seem more worried about legal paternalism than many European scholars (including me).
[1769] See along similar lines, in the context of contract law Mak 2008, p. 26.
[1770] As noted in chapter 1, section 4, a discussion of the democratic deficit of the EU falls outside the scope of this study.
[1771] Ayres 2012. See on the stickiness of defaults in the context of tracking Tene & Polonetsky 2012, p. 335.
[1772] Chapter 8, section 3.
[1773] Thanks for Oren Bar-Gill for suggesting this idea to me. See on the strategic use of transaction costs in the context of privacy and tracking Willis 2013a, especially p. 82-84, p. 122-128. See also Guibault, who suggests that

For example, the lawmaker could strengthen a nudge by adding transaction costs.[1774] Perhaps the lawmaker could require one mouse click for valid consent, if the consent concerns relatively innocuous types of tracking. The lawmaker could require three mouse clicks for more worrying practices. "Sticky defaults", says Ayres, "should be thought of as an intermediate category falling between ordinary defaults and traditional mandatory rules."[1775] Transaction costs could come in different shades, to introduce different degrees of stickiness for the default. In theory the law could require a thirty second waiting period, a phone call, or a letter by registered mail to opt in to certain practices.[1776]

The law does add friction to some decisions. For instance, formalities in contract law add transaction costs. Sometimes the law requires the involvement of a notary for a valid contract, for example when buying a house. And under Italian law, certain types of onerous contract clauses in standard contract terms must be signed separately.[1777] Data protection law requires "explicit" consent for the processing of special categories of personal data. About half of the member states require such explicit consent to be in writing.[1778] Some legally imposed transaction costs can be explained, at least in part, by the wish to reduce the chance of careless decisions. As noted, marketers are aware of the importance of transaction costs – and sometimes use them strategically. Opting out of behavioural targeting and other types of direct marketing

---

individually negotiated contracts regarding copyright involve more transaction costs than standard contracts, but could be regulated less strictly than standard contracts (Guibault 2002, p. 303).

[1774] If a nudge is made stronger by using transaction costs strategically, it might not count as a "nudge" anymore, since it's not "easy and cheap to avoid" (Sunstein & Thaler 2008, p. 6).

[1775] Ayres 2012, p. 2087 (including a helpful schedule).

[1776] See Ayres 2012, p. 2103. Ayres also gives more exotic examples. People could be required to answer a question before they can alter a default. "Will other corporations have the opportunity to purchase your mailing address and shopping information?" (p. 2077). See for a critique on the strategic use of transaction costs in the area of behavioural targeting Willis 2013a, p. 122-128.

[1777] Article 1341 of the Italian Civil Code (from 1942). See for a translation Gorla 1962, p. 2.

[1778] Article 8(2)(a) of the Data Protection Directive; Impact Assessment for the proposal for a Data Protection Regulation (2012), annex 2, p. 29. There are exceptions to the explicit consent requirement; see article 8(2)(b) - 8(5).

often takes more effort than opting in.[1779] In principle, the lawmaker could do something similar.

But caution is needed if the lawmaker considers adding friction to consent procedures in the area of behavioural targeting. A legal regime that adds transaction costs and allows firms to offer take-it-or-leave-it choices could lead to an unpleasant situation. Website publishers could use tracking walls, including if the lawmaker required three mouse clicks for consent.[1780] People would not enjoy clicking three times “I agree” if they want to visit a website, and accept they have to agree to tracking. With that caveat, the conclusion still stands: the distinction between mandatory rules and opt-in systems (default rules) isn’t a black and white issue. In principle the lawmaker has a range of options.

To conclude, there are good reasons to supplement the general data protection regime with specific rules, or with prohibitions, in the area of behavioural targeting. Taking into account the limited potential of informed consent as a privacy protection measure, using mandatory rules doesn’t imply undue paternalism.

### 9.3 Data minimisation

Many data protection provisions always apply, regardless of whether the data subject has consented to the processing. For instance, the data minimisation principle is mandatory. Several Data Protection Directive provisions express the data minimisation principle. For example, the amount of personal data must be “not excessive” in relation to the processing purposes. And firms must not keep data

---

[1779] See chapter 7, section 3.

[1780] Some website publishers impose transaction costs on visitors to improve advertising income. For instance, many websites cut articles in parts, so the reader has to click to reach the next part. Each click causes the website to refresh, which enables the website to display new ads.

"longer than is necessary" for the processing purposes.[1781] The European Data Protection Supervisor describes the data minimisation principle as follows.

> The principle of "data minimization" means that a data controller should limit the collection of personal information to what is directly relevant and necessary to accomplish a specified purpose. They should also retain the data only for as long as is necessary to fulfil that purpose. In other words, data controllers should collect only the personal data they really need, and should keep it only for as long as they need it.[1782]

Limiting the amount of data stored and shortening retention periods could mitigate some risks that are inherent to personal data processing. The vast scale of data processing for behavioural targeting aggravates chilling effects and the lack of individual control over personal information. And data storage brings risks, such as data breaches.[1783] Compliance with the data minimisation principle could mitigate such privacy problems.

Enforcing the data minimisation principle could also limit the amount of data that's available to construct predictive models.[1784] Predictive models based on the personal data of one group of people can be used to infer confidential information about people who weren't part of that group.[1785] Respect for the data minimisation principle limits the amount of information that firms can use for such practices. On the other hand, a lack of data can lead to incorrect predictive models, which in turn may cause unfair

---

[1781] Article 6(d) en article 6(e) of the Data Protection Directive. See for an overview of the data protection principles chapter 4, section 2.
[1782] European Data Protection Supervisor (Glossary). The Parliamentary Assembly of the Council of Europe stresses the importance of data minimisation in article 18(8) of its Resolution 1843 (2011) The protection of privacy and personal data on the Internet and online media, 7 October 2011.
[1783] See chapter 3, section 3.
[1784] See Hildebrandt et al. 2008, p. 245; Calo 2013, p. 44.
[1785] See chapter 2, section 5; chapter 7, section 3.

outcomes.[1786] For instance, an incorrect predictive model could say that a person is likely to default on credit, while more data might help to preclude such errors. This line of thought could lead to the conclusion that enough data should be available to create correct predictive models.[1787] But with behavioural targeting, the risks resulting from collecting too many personal data seem greater than the risks resulting from not having enough data to construct accurate predictive models. Besides, predictive models for behavioural targeting are rarely accurate. Accuracy in individual cases isn't the goal of behavioural targeting. A model can be useful for behavioural targeting if it correctly predicts that 0.5 % of the people who see an ad will click on it, if the click-through rate for untargeted ads is lower.[1788]

It follows from the structure of the Data Protection Directive that the data minimisation requirements from article 6 always apply, regardless of the legal basis for personal data processing in article 7 (such as consent or the balancing provision).[1789] In the words of the European Court of Justice: "all processing of personal data must comply, first, with the principles relating to data quality set out in article 6 of the directive and, secondly, with one of the criteria for making data processing legitimate listed in article 7 of the directive."[1790] A couple of national courts have ruled that data processing can be unlawful because it's disproportionate, even though the data subject has consented.[1791] As the Working Party puts it, "consent (…) is not a license for unfair and unlawful processing. If the purpose of the data processing is excessive and/or disproportionate, even if the user has consented, the [data controller] will not have a valid legal ground and would likely be in violation of

---

[1786] See Barocas 2014, chapter V.
[1787] See Schermer 2013, p. 147; Van Der Sloot 2013.
[1788] See chapter 2, section 5.
[1789] See on the relation between consent and the other data protection provisions chapter 6, section 5 and 6.
[1790] CJEU, C-131/12, Google Spain, 13 May 2014, par. 71 (capitalisation adapted). This could be different when an exception on the basis of article 13 of the Directive applies. See similarly ECJ, C-465/00, C-138/01 and C-139/01, Österreichischer Rundfunk, 20 May 2003, par, 65; CJEU, C-468/10 and C-469/10, ASNEF, 24 November 2011, par. 26.
[1791] Hoge Raad [Dutch Supreme Court], 9 September 2011, ECLI:NL:HR:2011:BQ8097 (Santander), English summary in Valgaeren & Gijrath 2011; Naczelny Sąd Administracyjny [Polish Supreme Administrative Court], 1 December 2009, I OSK 249/09 (Inspector General for Personal Data Protection), English translation: <www.giodo.gov.pl/417/id_art/649/j/en/> accessed 28 May 2014.

the Data Protection Directive."[1792] Scholars concur that consent can't legitimise disproportionate data processing.[1793]

A firm could try to argue it needs all the information it can get its hands on, because its processing purpose is targeting ads as precisely as possible. Or a firm could argue that collecting large amounts of data is "necessary" to build accurate predictive models. But it seems unlikely that judges or Data Protection Authorities would agree with such reasoning. Kuner has analysed how Data Protection Authorities apply the proportionality principle.[1794] He concludes that for data controllers, "the risk of legal problems caused by application of the proportionality principle can be particularly high" for some data processing practices. As an example he gives "the large-scale collection of data over the internet."[1795]

In its investigation of Google's 2012 privacy policy changes, the Working Party says that Google doesn't respect the data minimisation principle. "Google empowers itself to collect vast amounts of personal data about internet users, but Google has not demonstrated that this collection was proportionate to the purposes for which they are processed."[1796] The Working Party adds that "the Privacy policy suggests the absence of any limit concerning the scope of the collection and the potential uses of the personal data."[1797]

Few would probably argue that "excessive" personal data processing should be allowed. But when is data processing excessive? Acquisti argues for a strict interpretation of the data minimisation principle, although, being an economist rather than a data protection lawyer, he doesn't use the phrase data minimisation. According to Acquisti, firms should explain why they need personal data and why they can't

---

[1792] Article 29 Working Party 2013, WP 202, p. 16. See also Article 29 Working Party 2014, WP 217, p. 33.
[1793] Bygrave & Schartum 2009, p. 164; p. 166; Rouvroy & Poullet 2009, p. 73; Kosta 2013, p. 27; Gellert & Gutwirth 2013, 527; Dinant & Poullet 2006.
[1794] The principles of data minimization and proportionality are related. Kuner says "Proportionality has also led to creation of the concept of 'data minimisation' (Kuner 2008, p. 1618)
[1795] Kuner 2008, p. 1620 (capitalisation adapted).
[1796] Article 29 Working Party 2013 (Google letter), appendix, p. 7. See similarly CNIL 2014 (Google), p. 20-22.
[1797] Article 29 Working Party 2013 (Google letter), p. 1. See also the appendix, especially p. 4 and 7.

reach the same goal processing fewer data, for instance, by using privacy-preserving technologies. Like this, "the burden of proof for deciding whom and how should protect consumers privacy would go from *prove that the consumer is bearing a cost* when her privacy is not respected, to *prove that the firm cannot provide the same product*, in manners that are more protective of individual privacy."[1798] Acquisti concludes regulation may be needed to change the incentives for firms, to push them towards more privacy friendly practices.[1799]

A very strict interpretation of the data minimisation principle would imply that most data collection for behavioural targeting is prohibited. In principle, behavioural targeting would be possible without large-scale data collection, because behavioural targeting systems exist that don't involve sharing one's browsing behaviour with a firm. For example, a browser plug-in called Adnostic builds a profile based on the user's browsing behaviour, and uses that profile to target ads. Minimal information leaves the user's device, as the behavioural targeting happens in the browser. "The ad network remains agnostic to the user's interests."[1800] Mozilla is conducting research on a similar system for the Firefox browser.[1801] As behavioural targeting would be possible without large-scale data collection, it could be seen as "excessive" if firms collect large amounts of personal data for behavioural targeting. At present, the data minimisation principle is rarely interpreted as requiring such privacy-friendly behavioural targeting systems.

The lawmaker should consider making it more explicit, for instance in a recital, that consent can't legitimise disproportionate data processing. Such a recital could remind firms that the data subject's consent doesn't legitimise collecting personal data at will,

---

[1798] Acquisti 2010a, p. 43. See along similar lines Acquisti 2010b, p. 19-20.
[1799] Acquisti 2010b, p. 19-20; Acquisti 2010a, p. 43. Mayer & Narayanan 2013 arrive at a similar conclusion (p. 95-96).
[1800] Barocas et al. 2010. See also Castelluccia & Narayanan 2012, p. 16. See on privacy preserving analytics and click-fraud prevention Mayer & Narayanan (Donottrack.us website). See on click-fraud prevention also Soghoian 2011a.
[1801] Scott 2013.

and that Data Protection Authorities can intervene in the case of excessive data processing.

The European Commission proposal for a Data Protection Regulation makes the data minimisation principle more explicit. Personal data must be "limited to the minimum necessary in relation to the purposes for which they are processed; they shall only be processed if, and as long as, the purposes could not be fulfilled by processing information that does not involve personal data."[1802] This formulation allows for a stricter interpretation of the data minimisation principle. A proposal to modernise the Council of Europe's Data Protection Convention also provides inspiration. The proposal suggests adding the proportionality principle to the main principles of the Data Protection Convention, as follows: "[d]ata processing shall be proportionate in relation to the legitimate purpose pursued and reflect at all stages of the processing a fair balance between all interests concerned, be they public or private interests, and the rights and freedoms at stake."[1803]

Perhaps the law could prohibit storing data for behavioural targeting longer than a set period of, to give an example, two days. Such a hard and fast rule provides more legal certainty than general principles. Compared to estimating when the data minimisation principle requires deletion, complying with a maximum retention period of two days is easy. As noted, De Hert & Gutwirth call for specific rules if data protection regulation leaves too much room for different interpretations.[1804] However, limiting retention periods (phase 2) won't do much for people who think the tracking itself (phase 1) is the main problem, for instance, because of chilling effects. The most

---

[1802] Article 5(c) of the European Commission proposal for a Data Protection Regulation. Article 5(e) adds that data may not be kept longer than necessary. The LIBE Compromise text speaks of "data minimisation" and "storage minimisation" (article 5(c) and 5(e)). Article 6(1) the e-Privacy Directive is an example of a strict data minimisation requirement. Traffic data "must be erased or made anonymous when it is no longer needed for the purpose of the transmission of a communication" (subject to exceptions).

[1803] Council of Europe, The Consultative Committee Of the Convention for the Protection of Individuals with Regard to Automatic Processing of Personal Data [ETS No. 108] 2012a, article 5(1). See on the proportionality principle chapter 4, section 2, and chapter 6, section 1 and 2.

[1804] See section 1 of this chapter.

effective way to reduce chilling effects is not collecting data (phase 1).[1805] As an aside, it's unclear whether storing tracking data for longer than a few days helps much to improve the click-through rate on ads.[1806]

### 9.4 Transparency

The transparency principle can be read as a prohibition of surreptitious data processing. Hence, while the last chapter discussed the transparency principle as a means to empower the individual, the principle could also be seen as more prohibitive. As the European Commission put it in 1992, the fair and lawful principle "excludes the use for example of concealed devices which allow data to be collected surreptitiously and without the knowledge of the data subject."[1807]

Data processing is only allowed if it's done in compliance with the transparency principle. Of course, firms are allowed to use sophisticated technology that's difficult to explain to people. A different interpretation of the transparency principle might make the whole internet illegal. But the Data Protection Directive requires the data controller to inform data subjects about its identity and about the processing purposes, and to give all other information that's necessary to guarantee fairness.[1808]

With some behavioural targeting practices, it would be hard for a website publisher to comply with the law's transparency requirements, even if it were to try its best. For example, some ad networks allow other ad networks to buy access to individuals (identified through cookies or other identifiers) by bidding on an automated auction.[1809] In such situations, the website publisher doesn't know in advance which ad networks will display ads on its site, and which ad networks will track its website

---

[1805] See along similar lines Diaz & Gürses 2012, p. 2-3.
[1806] See Strandburg 2013, p. 104-105.
[1807] European Commission amended proposal for a Data Protection Directive (1992), p. 15. See also European Agency for Fundamental Rights 2014, p. 76-78.
[1808] Article 10 and 11 of the Data Protection Directive (see chapter 4, section 3). Moreover, article 5(3) of the e-Privacy Directive requires "clear and comprehensive information" for the use of most tracking technologies (see chapter 6, section 4).
[1809] See on ad exchanges, real time bidding, and cookie synching chapter 2 section 6.

visitors. In data protection parlance: the publisher doesn't know who the joint data controllers are.[1810] Neither does the publisher know for which purposes the ad networks will use the data they collect.[1811] As noted, the Data Protection Directive obliges data controllers to provide a data subject information about their identity, the processing purpose, and all other information that's necessary to guarantee fair processing.[1812] Therefore, it's hard to see how the publisher could comply with the law's transparency requirements.

Some websites use phrases along the following lines in their privacy policies. "We or other companies may use cookies to suggest and deliver content which we believe may interest you."[1813] The Working Party doesn't accept such vague information: "[s]tatements such as 'advertisers and other third parties may also use their own cookies or action tags' are clearly not sufficient."[1814] Furthermore, a user's consent can't be specific and informed if a website can't explain to visitors for which ad networks it asks consent.[1815]

If it's indeed impossible for firms to comply with data protection law's transparency requirements, only one conclusion seems possible: the processing isn't allowed. As Blume notes, "it must be considered whether a lack of transparency should have consequences and maybe imply that data processing cannot take place."[1816] The lawmaker should consider making it more explicit that processing is prohibited, unless firms can comply with the transparency principle.

---

[1810] The Working Party says ad networks and website publishers are often joint data controllers, as they jointly determine the purposes and means of the processing. See Article 29 Working Party 2010, WP 171, p 11.

[1811] In principle, it's the firm operating the cookie (such as an ad network) that must obtain consent. But the Working Party says that a website publisher that allows third parties to place cookies shares the responsibility for information and consent. See chapter 6, section 4.

[1812] Art 10 and 11 of the Data Protection Directive (see chapter 4, section 3, and chapter 8, section 2). See also article 5(3) of the e-Privacy Directive.

[1813] This phrase is taken from the privacy policy of the Guardian (Guardian, privacy policy).

[1814] Article 29 Working Party 2010, WP 171, p. 18.

[1815] See on the requirements for valid consent chapter 6, section 3 and 4, and chapter 8, section 3 and 4.

[1816] Blume 2012, p. 32.

The transparency principle could also limit what firms can lawfully do with personal data. As noted, transparency about data processing can only be meaningful if the purpose limitation principle is complied with. The purpose limitation principle prohibits firms from using data for goals that the data subject can't expect, unless an exception applies.[1817] Some online marketing practices, such as selling copies of data to other firms, seem hard to reconcile with the purpose limitation principle. It would be difficult for the seller to ensure that the buyer doesn't use the data for unexpected purposes.

Transparency isn't only important to make personal data processing controllable for the individual. Transparency can also help to make data processing controllable for Data Protection Authorities and the lawmaker. Data protection law's transparency requirements can help to uncover problems that might call for regulatory intervention.[1818] Hence, also in cases when hard prohibitions are a better approach than data protection law, data protection law could still be useful in bringing problems to light that need the attention of policymakers.

## 9.5 Sensitive data

The mere collection of data about people's behaviour can have a chilling effect. For example, if people fear surveillance, they might refrain from looking for medical information on the web.[1819] Research confirms that people don't like it when information regarding their health is used for behavioural targeting.[1820] Many marketers seem to realise people's uneasiness with such practices, as some self-regulatory codes for behavioural targeting have stricter rules for data regarding health.[1821]

---

[1817] See chapter 4, section 3; chapter 8, section 2.
[1818] See chapter 4, section 3. See also Bennett 2011a, p. 491.
[1819] Behavioural targeting can be seen as a type of surveillance; see chapter 3, section 3.
[1820] See chapter 7, section 1, and Leon et al 2013. See on chilling effects chapter 3 section 3.
[1821] See for instance Direct Marketing Association (United States) 2014.

There's a long tradition of protecting personal data regarding health, as illustrated by the Hippocratic oath that requires doctors to keep patient information confidential. Medical secrecy protects individual privacy interests of patients, and a public interest: the trust in medical services.[1822] The European Court of Justice confirms that the right to privacy "includes in particular a person's right to keep his state of health secret"[1823] The Court adds that protecting health data is important for the individual, and for society's trust in health services.[1824] The European Court of Human Rights uses similar reasoning:

> The protection of personal data, in particular medical data, is of fundamental importance to a person's enjoyment of his or her right to respect for private and family life as guaranteed by Article 8 of the Convention. Respecting the confidentiality of health data is a vital principle in the legal systems of all the Contracting Parties to the Convention. It is crucial not only to respect the sense of privacy of a patient but also to preserve his or her confidence in the medical profession and in the health services in general.[1825]

Furthermore, processing special categories of data can lead to unfair treatment. If a cookie representing somebody says that person is in the "lesbian, gay, bisexual, and transgender" category,[1826] or the "handicapped" category,[1827] the person could be

---

[1822] Ploem 2004, p. 129-133.
[1823] European Union Civil Service Tribunal, Civil Service Tribunal Decision F-46/095, V & EDPS v. European Parliament, 5 July 2011, par. 163.
[1824] European Union Civil Service Tribunal, Civil Service Tribunal Decision F-46/095, V & EDPS v. European Parliament, 5 July 2011, par 123.
[1825] I. v. Finland, App. No. 25011/03, 17 Jul. 2008, par. 38. See along similar lines Z v. Finland (9/1996/627/811) 25 February 1997, par. 95.
[1826] Flurry (audiences). Flurry is firm offering analytics and advertising for mobile devices. Among the demographic data that advertisers can select, Flurry lists "race" (Flurry, factual).
[1827] Rocket Fuel, Health Related Segments 2014.

discriminated against on this basis, even if no name is tied the cookie.[1828] Likewise, a cookie profile could be used for unfair discriminatory practices, if the profile says somebody is poor, rich, or from a certain neighbourhood, and decisions are based on that profile.[1829] And if a name is tied to the information, it could lead to embarrassment or worse if the information leaks.[1830]

Does data protection law's regime for health-related personal data (a "special category of data") apply to behavioural targeting?[1831] As discussed in chapter 5, many firms operate in a grey area. Much depends on the type of behaviour that firms track, and how they use that information.[1832] An ad network that tracks daily visits to website with kosher recipes could conclude that somebody is Jewish. Ad networks don't have an interest in harming people on the basis of sensitive information. Ad networks aim to increase the click-through rate on ads. For an ad network the topic of the website that somebody visits is of little relevance, as long as a correlation can be found between a visit to that website and clicking on certain ads.[1833] On the other hand, there are ad networks that enable advertisers to advertise to people based on special categories of data.[1834] Some ad networks use interest categories such as "arthritis", or "cardiovascular general health."[1835]

Case law suggests that the phrase "special categories of data" must be given a wide interpretation.[1836] Hence, tracking on websites with medical information should probably be seen as the processing of "data concerning health or sex life." Such

---

[1828] It appears an US data broker also sold addresses of people in the "rape victims" category (Hill 2013a).
[1829] Non-discrimination law might apply to some discriminatory practices. See section 6 below.
[1830] See on data breaches regarding health related data I. v. Finland, App. No. 25011/03, 17 Jul. 2008, par. 38.
[1831] See article 8 of the Data protection Directive.
[1832] See chapter 5, section 6.
[1833] See Van Hoboken 2012, who arrives at a similar conclusion regarding search engines (p. 332).
[1834] Assuming that behavioural targeting entails personal data processing.
[1835] Yahoo! Privacy. See also the interest category "lesbian, gay, bisexual, and transgender" highlighted previously in this section.
[1836] ECJ, C-101/01, Lindqvist, 6 November 2003, par. 50: "the expression 'data concerning health' (…) must be given a wide interpretation." This suggests that special categories of data generally must be interpreted generously (see Bygrave 2014, p. 167). The Office of the Privacy Commissioner of Canada, applying PIPEDA, the Canadian equivalent of data protection law, concluded that "Google is delivering tailored ads in respect of a sensitive category, in this case, health" (Office of the Privacy Commissioner of Canada (Google) 2014).

tracking is thus prohibited, or only allowed after obtaining the data subject's explicit consent.[1837] The privacy risks involved in using health data for behavioural targeting outweigh the possible societal benefits in allowing such practices. Therefore, the EU lawmaker should consider prohibiting the use of any data regarding health for behavioural targeting, whether the data subject gives consent or not.[1838]

Data protection law's regime for special categories of data can be criticised for being too data-centred. As Nissenbaum notes, sensitivity often depends on the context, rather than on the type of data.[1839] Say a website offers information about diseases. The website publisher allows an ad network to track the website visitors. In theory, the ad network could only record that a person (or the cookie with ID *xyz*) visited a website in the Netherlands, and disregard it's a website about health problems. But even if the ad network doesn't collect or infer special categories of data, a chilling effect could occur if people expect that visits to health websites are tracked.

Therefore, the lawmaker should consider whether prohibitions are needed in certain contexts. Such prohibitions have been suggested. For instance, the European Consumers' Organisation says tracking on health related websites should be prohibited.[1840] A difficult question would be how to phrase such prohibitions in a way that doesn't make them over or under inclusive. How to define "health related websites"? Is it enough if the website presents itself as a health related website, for instance by including a picture of a doctor in a white coat? And would a prohibition of using any "health data" for behavioural targeting also cover tracking of daily visits to a website with gluten free recipes? And which rules should cover smart health apps? Furthermore, legal limits on the use of health related data shouldn't unnecessarily hamper socially beneficial processing practices. For instance, rules shouldn't unduly

---

[1837] See article 8 of the Data Protection Directive.
[1838] See for a similar idea Turow 2011, p. 200. As noted, some member states have chosen not to allow people to override the prohibition of processing special categories of data with explicit consent. See chapter 5, section 6.
[1839] Nissenbaum 2010. See in detail on sensitive data (from a US perspective) Ohm 2014.
[1840] European Consumer Organisation BEUC 2013, p. 8. See also Willis 2013a, p. 87.

hinder medical practice or scientific research. In sum, drafting and agreeing on prohibitions would be hard. But that shouldn't be a reason to ignore this legal tool.

***Politics***

A second example of chilling effects that can result from behavioural targeting concerns reading about politics online.[1841] People might refrain from reading about certain political opinions or topics if they fear surveillance. People may have an individual interest in keeping their political views confidential, and in not having others drawing the wrong conclusions about their political opinions. Somebody might visit a website about communism or extreme right wing ideas out of curiosity, or because of strong disagreement. There's also a societal interest in respecting the confidentiality of political opinions. In a democratic society people are expected to vote, and arguably they should be able to inform themselves without fear of surveillance. It's widely accepted that information about people's political opinions deserves protection.[1842]

The freedom to receive and impart information protects individual interests and the common good. Article 10 of the European Convention on Human Rights says the right to freedom of expression includes the freedom to receive information and ideas without interference by public authority. The Court emphasises the role of freedom of expression for a democratic society. "Freedom of expression constitutes one of the essential foundations of such a society, one of the basic conditions for its progress and for the development of every man."[1843] Furthermore, "the public has a right to receive information of general interest"[1844] and "the internet plays an important role in enhancing the public's access to news and facilitating the sharing and dissemination

---

[1841] See chapter 2, section 5, and chapter 3, section 1 and 3.
[1842] See on the processing of personal data regarding one's political opinion ECtHR, Rotaru v. Romania, No. 28341/95, 4 May 2000. There's also a tradition of secret voting. See Jacobs 2011.
[1843] ECtHR, Handyside v. the United Kingdom, No. 5493/72, 7 December 1976, par. 49. See on the function of freedom of expression also Van Hoboken 2012, chapter 4.
[1844] ECtHR, Társaság a Szabadságjogokért v. Hungary, No. 37374/05, 14 April 2009, par. 26. See for an overview of case law on the right to receive information (with a different focus than this study) Herr 2011; Hins & Voorhoof 2007.

of information generally (…).”[1845] Article 10 doesn’t merely require states to refrain from interfering with the right to freedom of expression. States may have to take action: “the genuine and effective exercise of freedom of expression under Article 10 may require positive measures of protection, even in the sphere of relations between individuals.”[1846] The International Covenant on Civil and Political Rights is phrased in stronger terms than the European Convention on Human Rights, and also protects the right to “seek” information.[1847]

***News services***

Neither article 10 of the Convention, nor the related case law, grants a right not to have one’s browsing behaviour monitored. But arguably the values underlying freedom of expression imply that people should be able to read the news without fear of undue surveillance.[1848] Helberger emphasises the value of news services for a democratic society, and questions whether tracking walls on such services are acceptable.

> [T]here might be situations in which policymakers might decide that the acceptance of profiling and targeting is not an acceptable price at all, comparable e.g. to the existing prohibition on the sponsoring on news or religious programs. Taking e.g. into account the particular importance that news content has for political participation and democratic life, and argument could be made that in order to avoid chilling effects

---

[1845] ECtHR, Fredrik Neij and Peter Sunde Kolmisoppi (The Pirate Bay) v. Sweden, No. 40397/12, 19 February 2013 (inadmissible), capitalisation adapted.
[1846] ECtHR, Khurshid Mustafa v. Sweden, No. 23883/06 16 March 2009, par. 32.
[1847] Article 19(2) of the International Covenant on Civil and Political Rights.
[1848] Richards makes a similar point in the US context (Richards 2008, p. 428).

> people should never been required to accept tracking of their news consumption.[1849]

Furthermore, if somebody wants to read one source (website A), another source (website B) may not be a valid alternative for that person. As Helberger puts it, "media is speech, and when consuming media content it does matter who the speaker is. Accordingly, turning away and/or listening to another speaker is not necessarily an option."[1850]

The Audiovisual Media Services Directive prohibits sponsoring for news programmes.[1851] That prohibition only applies to television broadcasting and comparable moving images, so it doesn't apply to news websites with only text and pictures.[1852] Nevertheless, the rule shows that specific regulation for marketing in the context of news services wouldn't be a novelty. As De Hert & Gutwirth suggest, prohibitive rules are an appropriate response when data processing threatens important values for a democratic society.[1853] A chilling effect is hard to prove empirically.[1854] But if a chilling effect occurred in relation to reading about politics, it would threaten the democratic society.[1855] Furthermore, processing information about people's medical conditions or political opinions brings the risk of discrimination.

A full prohibition of any third party tracking on news services may be too blunt an instrument. For instance, it would be hard to define the scope of the ban. Would the ban apply to political blogs and to online newspapers that only gossip about

---

[1849] Helberger 2013, p. 18.
[1850] Helberger 2013, p. 12.
[1851] Article 10(4) of the Audio Visual Media Services Directive. "News and current affairs programmes shall not be sponsored." See along similar lines article 18(3) of the European Convention on Transfrontier Television. See for commentary Kabel 2008, p. 640.
[1852] Recital 28 of the Audio Visual Media Services Directive.
[1853] De Hert & Gutwirth 2006, p. 101-102. See section 2 of this chapter.
[1854] A survey by Cranor & McDonald 2010 suggests behavioural targeting has a chilling effect, but the research concerns declared (not revealed) preferences. A survey among 520 writers in the US finds many writers self-censor their work because they fear surveillance by intelligence agencies (PEN America 2013). Another study analysed Google search results, and suggests people "were less likely to search using search terms that they believed might get them in trouble with the U. S. government" (Marthews & Tucker 2014).
[1855] See chapter 3, section 3.

celebrities? And such a ban could lower the advertising income for news websites, at least in the short term. It wouldn't make sense if a rule that aims to ensure that people feel free to read about politics causes news services to go bankrupt. However, as discussed, in the long run behavioural targeting might decrease ad revenues for some website publishers.[1856]

The lawmaker should consider separate rules for tracking on websites of public service media, such as public broadcasters. The Council of Europe says public service media should promote democratic values, and should offer "universal access."[1857] In many European countries public service broadcasters receive public funding.[1858] Some public service broadcasters expose website visitors to third party tracking. For example, people could only access the website of the Dutch public broadcaster if they "consented" to tracking by various third parties. According to the Dutch Data Protection Authority, the universal access requirement implies that the broadcaster shouldn't make website visitors "pay" again with their personal data.[1859] Helberger concurs:

> It is (…) at least questionable whether in a situation in which access to the website is made conditional upon the acceptance of cookies, the website is still accessible for everyone. Very much will depend on whether users will find this too high are price, taking also into account that these contents have already been financed with public money.[1860]

---

[1856] See chapter 2, section 1 and 6, and chapter 7, section 2.
[1857] Recommendation CM/Rec(2007)3 of the Committee of Ministers to member states on the remit of public service media in the information society, 31 January 2007. See on public service media McGonagle 2011, chapter 4.
[1858] See on this topic European Commission 2009 (State Aid).
[1859] College bescherming persoonsgegevens (Dutch DPA) 2013 (NPO). See also chapter 6, section 3 and 4, and chapter 8, section 3 and 5. See on "paying" with personal data chapter 7, section 2.
[1860] Helberger 2013, p. 20 (internal reference omitted).

This study agrees with this line of reasoning. The lawmaker should prohibit public service broadcasters to make the use of their services dependent on consent to third party tracking. Such a prohibition shouldn't be limited to websites. For instance, certain types of digital television also enable tracking for behavioural targeting.[1861] The lawmaker should consider banning all personal data collection for behavioural targeting and similar purposes on public service media – at least when third parties collect the data.[1862]

***Public sector websites***

More generally, people should be able to visit important government websites without exposing themselves to tracking by third parties. As noted, under current law tracking walls and similar take-it-or-leave-it choices are prohibited if people must use a website, because the consent wouldn't be "free."[1863] For instance, say people are required to file their taxes online. If the tax website had a tracking wall that imposes third party tracking, people's consent to tracking wouldn't be voluntary. The EU lawmaker should consider make it explicit that public sector websites shouldn't offer visitors take-it-or-leave-it choices regarding commercial tracking.[1864]

Apart from the question of whether people are required to use a website, it's questionable whether it's appropriate for public sector bodies to allow third party tracking for commercial purposes on their websites – even if people consent. If a website is funded by the state, people paid for that website through taxes. It's hard to see why the state should facilitate tracking for commercial purposes on public sector websites. In practice, public sector websites might use third party widgets such as

---

[1861] See College bescherming persoonsgegevens (Dutch DPA) 2013 (TP Vision); Hessische Datenschutzbeauftragte (Data Protection Authority Hesse, Germany) 2014.

[1862] Data use by third parties tends to be riskier and less transparent than data use by the website publisher.

[1863] See chapter 6, section 3 and 4, and chapter 8, section 3 and 5.

[1864] For instance, the EU lawmaker could state that in a recital regarding consent in data protection law, or regarding (the successor of) article 5(3) of the e-Privacy Directive. Data analytics for fraud prevention may be necessary for some public sector websites.

social media buttons.[1865] The website publisher might not realise that the inclusion of such code exposes visitors to privacy-invasive tracking.[1866] The lawmaker could consider banning any third party tracking for commercial purposes on public sector websites. The exact scope of such a ban would require further debate.[1867] At the time of writing, in the Netherlands a bill to amend the implementation law of the e-Privacy Directive is being discussed, that contains, in short, a prohibition of tracking walls on public sector websites.[1868]

***Traffic and location data***

For traffic data and for location data, the e-Privacy Directive has specific rules, which resemble the rules for special categories of data in the Data Protection Directive.[1869] But the rules on traffic and location data only apply to providers of publicly available electronic communications services, such as internet access providers or phone operators – telecommunication providers for short.[1870] Telecommunication providers may only process traffic and location data with the user's consent, unless a specified exception applies.[1871] Hence, telecommunication providers can't rely on the balancing provision for processing such data.[1872] But many firms, such as ad networks and providers of smart phone apps, process more traffic and location data than telecommunication providers. The scope of the regime for traffic and location data

---

[1865] To illustrate: Van Der Velden found third party tracking on 60% of a set of Dutch governmental websites she examined (Van Der Velden 2014).
[1866] Using the taxonomy of chapter 8, section 1, the website publisher may be well-intentioned but ignorant. As an aside: websites can include "greyed out" buttons, which don't track people unless people click on the button to activate the button, for instance to "like" a page (see Schmidt 2011; Schneier 2013b).
[1867] For instance, what to do about organisations that are partly funded by the state? And some (first party) tracking could be necessary for website security purposes.
[1868] Proposal to amend the Telecommunicatiewet (Telecommunications Act): Eerste Kamer, vergaderjaar 2014–2015, 33 902, A <www.eerstekamer.nl/wetsvoorstel/33902_wijziging_artikel_11_7a> accessed 17 November 2014.
[1869] See for traffic data article 5 and article 6, and for location data article 9 of the e-Privacy Directive. See chapter 5, section 6.
[1870] An "electronic communications service" is, in short, a service that consists wholly or mainly in the conveyance of signals on electronic communications networks (article 2(c) of the Framework Directive 2002/21/EC (amended in 2009)). It's thus a transmission service.
[1871] See article 5(1) and 6 (traffic data) and article 9 (location data) of the e-privacy Directive. The e-Privacy Directive distinguishes users from subscribers. This distinction isn't further explored in this study.
[1872] See on article 7(f) of the Data Protection Directive, the balancing provision chapter 6, section 2.

must probably be broadened. Traffic and location data are sensitive, and deserve extra protection – also when they are processed by firms other than telecommunication providers.

The lawmaker should consider introducing specific rules for using traffic and location data for behavioural targeting. Hence, such rules would focus more on the processing purpose than on the type of firm.[1873] Some scholars suggest that using traffic and location data shouldn't be allowed at all in some situations: "location-based services should not even offer the option (to minors) to share their location with third parties and/or use it for behavioural tracking purposes."[1874] As these authors note, specific rules regarding tracking children may be needed.[1875]

In conclusion, strictly enforcing the existing rules on special categories of data could reduce privacy problems such as chilling effects. For instance, if they fear surveillance people might be hesitant to look for medical information on the web, or to read about politics on the web. As chilling effects can result from the collection context, the lawmaker should consider additional rules that focus on the context, rather than on the data type. For example, the lawmaker should consider banning third party tracking for behavioural targeting on public service media.

### 9.6 Automated decisions

"Data processing may provide an aid to decision-making, but it cannot be the end of the matter; human judgment must have its place," said the European Commission in 1992.[1876] This is the rationale for article 15 of the Data Protection Directive, the provision on automated decisions. Article 15 is based on the French Data Protection Act from 1978, which prohibits automated court decisions. The French Act also

---

[1873] It could be called a functional approach if the lawmaker focuses on the purpose of behavioural targeting, rather than on certain types of firms (see, in a different context Arnbak 2013a).
[1874] Ausloos et al. 2012, p. 25. See also Turow 2011, p. 200.
[1875] As noted in chapter 1, section 4, the question of whether special privacy rules are needed for children falls outside this study's scope. See on such issues Van Der Hof et al. 2014.
[1876] European Commission amended proposal for a Data Protection Directive (1992), p. 26.

prohibits other automated decisions with legal effect for the individual, unless a specified exception applies.[1877] Article 15 of the Data Protection Directive, sometimes called the Kafka provision, could be seen as an in-principle prohibition of certain fully automated decisions with far-reaching effects. The analysis below mainly relies on literature, because the provision hasn't been applied much in practice.[1878]

The Directive's provision on automated decisions applies to data processing by firms, and by the state. Within the private sector, the provision applies to a wide range of activities, such as credit scoring.[1879] The provision doesn't concern the legal basis for collecting data. Hence, in principle firms that gather personal data to use for automated decisions, could base such collection on various legal bases, including consent and the balancing provision.[1880] The main rule of the Directive's provision on automated decisions is as follows:

> Member States shall grant the right to every person not to be subject to a decision which produces legal effects concerning him or significantly affects him and which is based solely on automated processing of data intended to evaluate certain personal aspects relating to him, such as his performance at work, creditworthiness, reliability, conduct, etc.[1881]

In short, people may not be subjected to certain automated decisions with far-reaching effects. The Directive says a person has "the right not to be subject to" certain

---

[1877] Article 10, Loi Informatique Et Libertes [Act on Information Technology, Data Files and Civil Liberties] (Act N°78-17 Of 6 January 1978), last amended 17 March 2014: "No court decision involving the assessment of an individual's behaviour may be based on an automatic processing of personal data intended to assess some aspects of his personality." See Korff 2010b, p. 24-27; Kabel 1999, p. 281-282.
[1878] See Korff 2012, p. 26.
[1879] See European Agency for Fundamental Rights 2014, p. 117. Certain public sector activities are outside the Directive's scope. See chapter 4, section 2.
[1880] See on the legal basis requirement for data processing chapter 6.
[1881] Article 15(1) of the Data Protection Directive.

decisions.[1882] But literature suggests this implies an in-principle prohibition of such decisions.[1883] Several countries emphasise the prohibitive character of the provision in their implementation laws. For instance, the Austrian act says that "nobody shall be subjected to" such decisions.[1884] Other countries phrased it less strictly.[1885]

Does article 15 apply to behavioural targeting? Four conditions must be met for the provision to apply, says Bygrave. Slightly rephrased, the conditions are as follows: (i) There must be a decision, (ii) that decision is based solely on automated processing of data, (iii) the data used for the decision are intended to evaluate certain personal aspects of the person concerned, and (iv) the decision must have legal or other significant effects for the person.[1886] With behavioural targeting, an algorithm decides to show the right ad at the right time to the right person, based on analysing that person's behaviour. Data processed for behavioural targeting are "intended to evaluate certain personal aspects" about a person. Therefore, the first three conditions are met.[1887] The fourth condition requires the decision to have "legal effects", or to "significantly" affect the person.

An automated court decision would be an example of a decision with legal effect. The Belgian Data Protection Authority suggests that a targeted ad that includes "a reduction and therefore a price offer" has legal effect as well.[1888] Presumably, the Authority sees a price offer as an invitation to enter an agreement, which could indeed be seen as having a legal effect. This interpretation would make article 15 applicable

---

[1882] Article 15 uses the phrase "every person", rather than "data subject." Some suggest article 15 also applies if a firm can't identify a person about whom it makes an automated decision. See Konarski et al. 2012, p. 34.
[1883] Korff 2012 (p. 26) and De Hert & Gutwirth 2008 (p. 283) see article 15 as an in-principle prohibition. But see Bygrave 2001, who suggests that the provision might allow the automated decisions if the data subject doesn't object. See also Hildebrandt 2012, p. 50.
[1884] Article 49(1) of the Datenschutzgesetz of Austria. See also article 17(1) of the Personal Data Protection Act in Estonia, and article 12bis of the Data Protection Act in Belgium.
[1885] See for instance the Data Protection Act in Portugal (article 13), in Spain (article 13), and in Norway (article 22 and 25).
[1886] Bygrave 2002, p. 320. See for a similar analysis with three conditions: European Commission amended proposal for a Data Protection Directive (1992), p. 26.
[1887] Bygrave 2002, p. 320. See also International Working Group on Data Protection in Telecommunications (Berlin Group) 2013, p. 6.
[1888] Commission for the Protection of Privacy Belgium 2012, par. 80. See also Vermeulen 2013, p. 12.

to certain types of price discrimination.[1889] From here on, this chapter focuses on decisions that "significantly" affect people, rather than on decisions with "legal effects."

The Data Protection Directive doesn't explain when a decision "significantly" affects a person. But it seems questionable whether one targeted ad falls within the scope of an automated decision that "significantly affects" a person within the meaning of article 15. However, Bygrave argues that in some cases behavioural targeting – "cybermarketing" as he referred to it in 2002 – can have significant effects, for example if a firms charges higher prices to somebody, or denies somebody access to a service:[1890]

> For instance, a cybermarketing process could be plausibly said to have a significant (significantly adverse) effect on the persons concerned if it involves unfair discrimination in one or other form of "weblining" (e.g., the person visiting the website is offered products or services at a higher price than other, assumedly more valuable customers have to pay, or the person is denied an opportunity of purchasing products/services that are made available to others).[1891]

It's dubious whether one targeted ad should generally be seen as an automated decision with significant effects in the sense of article 15. Somebody might not even notice the ad. In many cases, receiving one single targeted ad probably doesn't lead to

---

[1889] See on price discrimination chapter 2, section 7, chapter 8, section 2, and chapter 9, section 7.

[1890] Bygrave 2002, p. 323-324. Church & Millard note: "[t]here is no further definition of a "significant effect", though it is very unlikely that this would be limited to decisions having a pecuniary effect" (Church & Millard 2010, p. 84). See on the vagueness of "significant effect" also Article 29 Working Party 2012, WP 191, p. 14.

[1891] Bygrave 2002, p. 323-324 (punctuation adapted, internal footnote omitted). The word "weblining" refers to "redlining", where city areas are used as a proxy to discriminate people based on race (Stepanek 2000). See critically on that phrase Zarsky 2002, p. 35.

an effect for the individual that should be regarded as "significant" in the sense of article 15.

Behaviourally targeting as a *practice* does have significant effects for society and for individuals. For instance, large-scale data collection can lead to chilling effects, and people lack control over what happens to their data. Furthermore, the very point of advertising is to change views, attitudes, actions, and behaviours over time.[1892] Thus, in aggregate, behavioural targeting may well significantly affect someone.

It could also be argued that one targeted ad should generally be seen as a decision that "significantly affects" somebody in the sense of article 15. One could focus less on the effects of one automated decision on the individual, and more on the effects of automated decisions generally on individuals and society. Following that line of reasoning, article 15 can be triggered *because* behavioural targeting as a practice has significant effects.[1893]

In sum, the text of article 15 seems to suggest that the provision only applies if one specific automated decision significantly affects an individual. The correct interpretation of the provision must come from the courts. But, as noted, so far the provision hasn't been applied much in practice.

***Exceptions to the in-principle prohibition***

For this study, there are two relevant exceptions to the in principle prohibition of automated decisions with significant effects, which are summarised now.[1894] First, an automated decision is allowed when it "is taken in the course of the entering into or performance of a contract, provided the request for the entering into or the

---

[1892] See on the privacy implications of behavioural targeting chapter 3.
[1893] See generally Hildebrandt 2012.
[1894] A third exception isn't discussed here, because of its limited relevance for behavioural targeting. An automated decision is allowed if it's authorised by a law that includes measures to safeguard the data subject's legitimate interests (article 15(2)(b) of the Data Protection Directive).

performance of the contract, lodged by the data subject, has been satisfied."[1895] For instance, an insurance firm might use software to decide whether or not it will offer people an insurance contract. The provision allows such an automated decision if it leads to offering somebody a contract, because the person's request to enter a contract has been met.[1896]

Second, firms are allowed to automatically refuse to enter a contract with somebody, if "there are suitable measures to safeguard his legitimate interests, such as arrangements allowing him to put his point of view."[1897] Hence, a firm that uses software to automatically deny somebody an insurance contract could ensure that the person can ask a human to reconsider the decision. This makes it trivial for a firm, such as an insurance company, to comply with the provision.[1898] It might be enough if the insurance company included a phone number on the website, where people can ask a human to reconsider the automated decision to deny the insurance contract.

For many types of unfair social sorting the provision offers little help.[1899] Suppose an ad network refrains from showing certain ads to people who visited a price comparison website, or to people whose IP address suggests that they are from a poor neighbourhood. Those people may not realise the ad network excludes them from the campaign. Therefore, it's difficult to challenge the decision. Likewise, the provision doesn't help much to reduce the risk of filter bubbles and manipulation.[1900] One automated decision to personalise a website might not "significantly" affect a person within the meaning of article 15; and therefore the decision may remain outside the provision's scope.[1901] However, as noted in the previous chapter, data protection law could help to make personalisation more transparent – including if article 15 doesn't

---

[1895] Article 15(2)(a) of the Data Protection Directive.
[1896] In any case, somebody probably wouldn't object to an automated decision, if the decision were in the person's favour. See Kabel 1996, p. 281. But see Bygrave 2002, p. 327.
[1897] Article 15(2)(a) of the Data Protection Directive.
[1898] Article 15 doesn't require the firm to amend the criteria for the decision. Bygrave 2002, p. 324; Rubinstein 2013, p. 6.
[1899] See on social sorting chapter 3, section 3.
[1900] See on filter bubbles chapter 3, section 3.
[1901] Article 15(1) of the Data Protection Directive.

apply. After all, firms are required to disclose the processing purpose, and this requirement also applies when the purpose is personalising content.[1902]

The Data Protection Directive grants the data subject the right to learn the underlying logic of an automated decision with significant effects.[1903] Hence, an insurance firm that denies somebody a contract based on an automated decision must explain the logic behind that decision, if the person who was denied the contract requests so. For instance, the firm could explain why the software denied the insurance contract, and which factors were taken into account. In some cases the right to ask for the decision's logic could help the data subject, but there are several reasons not to expect too much from this right.

First, the provision on automated decisions is hardly ever applied in practice. Second, the person has to ask for the information. Hence, if somebody isn't aware of an automated decision, the provision is of little help. For instance, if an ad network only shows an offer to certain people, a person who doesn't receive the offer is probably unaware of being excluded. Third, the Directive's recital 41 limits the right to learn the logic behind the automated decision. The right "must not adversely affect trade secrets or intellectual property."[1904] A firm might claim it can't fully explain an automated decision, because that would disclose too much about the software it uses. However, the recital doesn't allow the firm to refuse all information: "these considerations must not (…) result in the data subject being refused all information."[1905] The issue isn't merely theoretical. Facebook has invoked the recital to limit information it gives to people who exercised their right to access.[1906]

---

[1902] Article 10 and 11 of the Data Protection Directive. See chapter 4, section 3, and chapter 8, section 2.
[1903] Article 12(a) of the Data protection Directive. The right to ask the logic behind an automated decision can be characterised as a rule that aims to empower the data subject by granting her a right. But the rule is discussed in this chapter, because of its relevance for the automated decisions provision.
[1904] Recital 41 of the Data Protection Directive. See about the legal effect of recitals chapter 6, section 4.
[1905] Recital 41 of the Data Protection Directive.
[1906] Facebook invoked article 4(12) of the Irish Data Protection Act, which is based on recital 41. See Hildebrandt 2011, p. 3-4; Europe versus Facebook 2014.

***Data Protection Regulation proposals***

The European Commission proposal for a Data Protection regulation amends the provision on automated decisions. Article 20 of the proposal is called "measures based on profiling."[1907] The main rule is similar to the one in the Data Protection Directive: in principle a person should not be subjected to measures based on profiling that significantly affect him or her:

> Every natural person shall have the right not to be subject to a measure which produces legal effects concerning this natural person or significantly affects this natural person, and which is based solely on automated processing intended to evaluate certain personal aspects relating to this natural person or to analyse or predict in particular the natural person's performance at work, economic situation, location, health, personal preferences, reliability or behaviour.[1908]

The provision's second paragraph says profiling measures with significant effects are only allowed if an exception applies. The exceptions are similar to those in the Data Protection Directive. But a new exception is introduced. A profiling measure with significant effects is allowed if people give their consent, and if there are suitable safeguards.[1909] The proposal thus introduces yet another default rule that can be overridden with consent.[1910]

It's unclear to what extent the proposed provision applies to behavioural targeting. As the Belgian Data Protection Authority notes, it's "not easy to determine whether

---

[1907] Article 20 of the European Commission proposal for a Data Protection Regulation.
[1908] Article 20(1) of the European Commission proposal for a Data Protection Regulation (2012).
[1909] Article 20(2)(c) of the European Commission proposal for a Data Protection Regulation (2012). The text of article 20(2)(c) isn't very clear. The European Commission might mean that *consent* is subject to suitable safeguards. But presumably the Commission means suitable safeguards to protect the data subject's interests.
[1910] See on the distinction between default rules and mandatory rules chapter 6, section 5.

profiling for direct marketing purposes in the form of specific advertising messages is part of the scope of this article."[1911] The Authority adds that the provision ought to apply: "this kind of profiling should be subject to the specific conditions set out in article 20."[1912]

The European Commission proposal prohibits profiling measures that are based only on special categories of data. "Automated processing of personal data intended to evaluate certain personal aspects relating to a natural person shall not be based solely on the special categories of personal data."[1913] This prohibition also applies to profiling measures that don't "significantly affect" a person.[1914] But the prohibition has a narrow scope, as it concerns measures that are based "solely" on special categories of data. Some firms use special categories of data for behavioural targeting.[1915] However, the prohibition doesn't apply as long as a firm also uses non-special personal data. This is generally the case, so the rule seems to be a dead letter. Regardless of whether the profiling provision applies, a firm needs the data subject's explicit consent for processing special categories of data for behavioural targeting.[1916]

Presumably the aim of preventing unfair discrimination is one of the rationales for the prohibition of profiling measures based solely on special categories of data. But the prohibition fails to take into account that measures based on profiling could also lead to unfair discrimination if no special categories of data are used. For instance, non-special personal data could be used to generate special categories of data. Or non-special personal data could end up being used as a proxy for special categories of

---

[1911] Commission for the Protection of Privacy Belgium 2012, par. 80. See also Information Commissioner's Office 2013; Federation of European Direct and Interactive Marketing (FEDMA) 2013, p. 3.
[1912] Commission for the Protection of Privacy Belgium 2012, par. 80.
[1913] Article 20(3) of the European Commission proposal for a Data Protection Regulation (2012).
[1914] Article 20(3) doesn't mention "significant effects", and doesn't refer to another article that mentions "significant effects."
[1915] See section 5 of this chapter and chapter 5, section 6.
[1916] Like the 1995 Directive, the European Commission proposal for a Data Protection Regulation (2012) allows member states to decide that special categories of data can't be processed on the basis of explicit consent (article 9(1) and 9(2)(a)).

data. As Korff notes, automated decisions could "reinforce societal inequality" and have discriminatory effects, even if only prima facie innocuous data are used.[1917]

> Crucially, this discrimination-by-computer does not rest on the use of overtly discriminatory criteria, such as race, ethnicity or gender. Rather, discrimination of members of racial, ethnic, national or religious minorities, or of women, creeps into the algorithms in much more insidious ways, generally unintentionally and even unbeknown to the programmers. But it is no less discriminatory for all that.[1918]

For example, a bank could use software to deny credit to people who live in a particular neighbourhood, because many people in that neighbourhood don't repay their debts. If primarily immigrants live in that neighbourhood, such profiling measures might discriminate against immigrants, by accident or on purpose. But such practices wouldn't be covered by the prohibition of profiling measures based solely on special categories of data. Similarly, the software could deny credit to somebody who lives in a poor neighbourhood, even though that person always repays his or her debts.

Following a suggestion by Korff and several civil rights organisations, the European Parliament proposes to prohibit profiling measures that have the effect of discriminating on the basis of special categories of data, intentional or not.[1919] "Profiling that has the effect of discriminating against individuals on the basis of race or ethnic origin, political opinions, religion or beliefs, trade union membership, sexual

---

[1917] Korff 2012, p. 18.

[1918] Korff 2012, p. 22-23 (emphasis original, enter omitted). See also White House (Podesta J et al.) 2014, p. 45-47; p. 51-53; Article 29 Working Party 2013, WP 203, p. 47; p. 59; p. 61; Siegel 2013, p. 62-65; Calders & Žliobaitė 2013. See in detail on data mining (and profiling) and discrimination Barocas 2014.

[1919] Korff 2012, p. 37; Bits of Freedom 2012; EDRi (European Digital Rights) 2014 (less explicitly).

orientation or gender identity, or that results in measures which have such effect, shall be prohibited. (…)"[1920]

A topic for further research is where non-discrimination law and data protection law overlap, and where the two fields could usefully supplement each other.[1921] One important difference is that data protection law applies as soon as personal data are collected or otherwise processed. In contrast, non-discrimination law becomes relevant in later phases: when there's a difference in treatment of a person or a group.[1922] Furthermore, many non-discrimination rules only apply to certain protected grounds, such as sex, sexual orientation, disability, age, race, ethnic origin, national origin, and religion or belief.[1923] Hence, non-discrimination law may be less effective to combat discrimination against, for instance, people who live in poor neighbourhoods.[1924]

The profiling provision in the European Commission proposal adds a new transparency requirement to data protection law's general transparency requirements. In short, a firm must tell the person concerned that it takes a profiling measure with significant effect, and must inform the person about the measure's envisaged effects.[1925] This rule obliges a firm to inform the data subject about profiling measures, including if the person hasn't asked for it.[1926] This is an improvement in comparison with the Data Protection Directive's provision, which only requires firms

---

[1920] Article 20(3) of the LIBE Compromise, proposal for a Data Protection Regulation (2013). In his draft report, Rapporteur Albrecht had proposed to prohibit all profiling that includes or generates special categories of data (Draft Albrecht report, amendment 162, article 20(3)).

[1921] As noted, non-discrimination law falls outside the scope of this study. There are still many open questions regarding the interplay of data protection law and non-discrimination law. See on this topic Hildebrandt et al. 2008; De Vries et al. 2013. See generally on discriminatory effects of profiling Zarsky et al. 2013; Barocas 2014. See generally on non-discrimination law in Europe: European Agency for Fundamental Rights 2010a.

[1922] Using the five phases of data processing that were distinguished in chapter 2, non-discrimination law would apply to phase (5), but not to earlier phases.

[1923] Hildebrandt et al. 2008; De Vries et al. 2013.

[1924] See chapter 2, section 5. It's possible, for instance, to infer whether people are likely to default on credit based on their shopping behaviour.

[1925] Article 20(4) of the European Commission proposal for a Data Protection Regulation (2012).

[1926] Article 20(4) refers to article 14, and article 14 suggests a requirement of proactive transparency. Nevertheless, there's some debate on the question of whether a firm must proactively provide this information, or whether it only has to provide information upon request (Hildebrandt 2012, p. 51; Rubinstein 2013, p. 7).

to give information about automated decisions upon request.[1927] But it's only a minor improvement. The main problem is that the new transparency requirement only applies to profiling measures that "significantly affect" a person.[1928] Hence, the new transparency requirement probably wouldn't apply to most targeted ads, or to personalised websites.[1929] Furthermore, the new transparency obligation doesn't require firms to provide information about the logic involved in the profiling measure.[1930] And as discussed, merely ensuring that firms offer transparency isn't enough to empower people in any real sense.

Like the Data Protection Directive, the European Commission proposal contains a general right of access. Data subjects have the right to obtain information about the processing of their data from a firm.[1931] In the proposal such information must include "the significance and envisaged consequences of such processing, at least in the case of measures [based on profiling] referred to in article 20."[1932] Unlike the Directive, the European Commission proposal doesn't grant people the right to ask for the logic involved in the profiling measure.[1933]

The new transparency requirement should be amended, to improve privacy protection. A firm should inform people about profiling measures, also when no legal or

---

[1927] Article 12(a) of the Data Protection Directive. The Directive does have a general transparency requirement in article 10 and 11.

[1928] Article 20(4), containing the transparency requirement, refers to "the cases referred to in paragraph 2." Paragraph 2 refers to "measures of the kind referred to in paragraph 1." Paragraph 1 speaks of "a measure which produces legal effects concerning this natural person or significantly affects this natural person (…)." Korff is critical of the new transparency provision (Korff 2012, p. 33).

[1929] But see the discussion of article 15 of the Data Protection Directive above in this section.

[1930] See article 12(a) of the Data Protection Directive.

[1931] Article 15 of the European Commission proposal for a Data Protection Regulation (2012).

[1932] Article 15(1) (h) of the European Commission proposal for a Data Protection Regulation (2012). It's unclear what the European Commission means by the "significance" of profiling measures. (Korff 2012, p. 33.)

[1933] Recital 51 of the European Commission proposal for a Data Protection Regulation (2012) suggests that data subjects have the right to learn the logic behind profiling measures, but the recital is oddly phrased, as it speaks of a right to know "the logic of the data that are undergoing the processing" (see Korff 2012, p. 33). There are more references to profiling in the proposal. For instance, firms must carry out a data protection impact assessment if profiling is "systematic and extensive" (article 33(2)(a)). The preamble suggests that children shouldn't be subjected to measures based on profiling (recital 58). The Commission can adopt delegated acts regarding the suitable measures to safeguard the data subject's interests (article 20(5)).

significant effects are foreseen.[1934] The provided information should include an explanation of the logic behind profiling measures. The requirement could be coupled with a reasonable, and not too broadly phrased, exception for trade secrets etc.[1935] Such transparency requirements could reduce the risk of filter bubbles or manipulative practices enabled by behavioural targeting.[1936] A firm that personalises ads or other content should be transparent about the personalisation. For instance, a website could include a button that leads to an explanation of why and how a website is personalised. While transparency requirements are not a panacea to protect privacy and fairness, such requirements could be helpful.[1937]

Scholars call for the development of TETs, transparency-enhancing technologies, to enable meaningful transparency regarding profiling.[1938] Such technologies should "aim at making information flows more transparent through feedback and awareness thus enabling individuals as well as collectives to better understand how information is collected, aggregated, analyzed, and used for decision-making."[1939]

The European Commission proposal's profiling provision "was not warmly welcomed by representatives from the direct marketing and the online advertising industry."[1940] The American Chamber of Commerce, a business lobbying organisation, says it would be best to get rid of the rules on profiling.[1941] "At minimum, the Regulation should make clear that the restrictions on profiling do not extend to beneficial activities such as fraud prevention, service improvement, and marketing/content customization."[1942] The Interactive Advertising Bureau UK and the Federation of

---

1934 This is required in article 14(1)(ga) of the LIBE Compromise, proposal for a Data Protection Regulation (2013).
1935 See Korff 2012, p. 33-34; Bits of Freedom 2012.
1936 See on the risk of manipulation resulting from behavioural targeting chapter 3, section 3.
1937 Apart from helping data subjects, legal transparency requirements can help regulators and policymakers to assess industry practices. See chapter 4, section 3.
1938 See for instance Hildebrandt & Gutwirth (eds.) 2008; Hildebrandt 2012. See also Bozdag & Timmersmans 2011.
1939 Diaz & Gürses 2012, p. 3-4.
1940 Vermeulen 2013, p. 12.
1941 International Chamber of Commerce 2013, p. 2.
1942 International Chamber of Commerce 2013, p. 2.

European Direct and Interactive Marketing (FEDMA) say profiling measures should be allowed on an opt-out basis, similar to the regime of the balancing provision.[1943]

While data protection rules regarding profiling could protect people against some forms of unfair social sorting, other social sorting practices remain outside the ambit of data protection law.[1944] For example, advertising that isn't targeted at individuals can have an effect that resembles social sorting through behavioural targeting. Predatory lending schemes or junk food could be advertised on a website that's visited primarily by poor people. If ads are adapted to the website rather than to individuals, it concerns a form of contextual advertising rather than behavioural targeting.[1945] Data protection law doesn't apply if people aren't singled out or otherwise identified. If social sorting through contextual ads is – or becomes – a problem, the lawmaker will have to seek a solution outside data protection law.

## 9.7 Conclusion

This chapter discussed how the law could improve protection of the individual, rather than empowerment. To start with, better enforcement of the current rules is needed. Many data protection provisions are mandatory; they always apply, regardless of whether the data subject has consented to the processing. If the data protection principles were fully complied with, they could give reasonable privacy protection in the area of behavioural targeting.

For example, it follows from the Data Protection Directive that excessive data processing isn't allowed, not even after the data subject's consent. Other data protection principles can defend privacy interests as well, also after somebody

---

[1943] Interactive Advertising Bureau United Kingdom 2012a; Federation of European Direct and Interactive Marketing (FEDMA) 2013, p. 4-5. As noted in chapter 6, section 2, the LIBE Compromise, proposal for a Data Protection Regulation (2013), allows, under certain circumstances, profiling based on the balancing provision. But the LIBE Compromise requires consent for profiling that has legal effects or significantly affects a person, unless a specified exception applies.
[1944] See Gürses 2010, p. 49; p. 55.
[1945] See chapter 2, section 1 and 3.

consents to processing. For instance, the purpose limitation principle and the security principle always apply. However, many provisions of the Data Protection Directive are rather general, and leave ample room for discussion. It's thus useful that the European Commission proposal for a Data Protection Regulation phrases some data protection principles more explicitly. But enforcing and tightening the data protection principles won't suffice to protect privacy in the behavioural targeting area. Additional rules are needed.

Article 5(3) of the e-Privacy Directive could be seen as a sector-specific rule for behavioural targeting, which supplements the general data protection regime.[1946] But article 5(3) requires too much, and at the same time, doesn't require much. On the one hand, article 5(3) is too blunt. The provision is over inclusive, as it also requires consent for certain innocuous types of cookies that pose few privacy threats. On the other hand, article 5(3) isn't very strict, as it merely obliges firms to obtain the individual's informed consent for the use of tracking cookies and similar technologies. Article 5(3) doesn't say much about the processing that takes place after a firm obtained consent for storing or accessing information on a user's device. Hence, the provision is mainly relevant for phase 1 of the behavioural targeting process (data collection). But, as far as personal data are processed, data protection law does regulate the processing after consent for the use of tracking technologies.

As noted in the previous chapter, it might be better if the lawmaker phrased the consent requirement for behavioural targeting in a more technology neutral way than article 5(3) of the e-Privacy Directive. The law could require consent for processing personal data, including pseudonymous data, for behavioural targeting and similar purposes – regardless of the technology that's used.

One option that could be explored is whether a separate legal instrument is needed for behavioural targeting. This study doesn't aim to propose a detailed sector-specific

---

[1946] See on article 5(3) chapter 6, section 4; chapter 8, section 4.

regime for behavioural targeting. Rather, some starting points for the discussion are given. In principle, specific rules could address different behavioural targeting phases: (1) data collection, (2) data storage, (3) data analysis, (4) data disclosure, and (5) the use of data for targeted advertising.[1947]

The most effective way to reduce chilling effects is not collecting data (phase 1).[1948] This could be partially achieved by applying and enforcing data protection law's regime for special categories of data, such as data regarding medical conditions, or political opinions. In a few EU member states, using special categories of personal data for direct marketing is prohibited; in many member states it is only allowed with the data subject's explicit consent. Because the privacy risks involved in using health data for behavioural targeting outweigh the possible societal benefits of allowing such practices, the EU lawmaker should consider prohibiting the use of any data regarding health for behavioural targeting, whether the data subject gives consent or not.

In many cases, sensitivity depends on the context, rather than on the types of data. Therefore, it should be considered whether data collection for behavioural targeting should be restricted or prohibited in certain contexts. For each situation where the lawmaker could consider banning certain practices, it could also opt for a lighter measure: banning tracking walls and similar take-it-or-leave-it choices.

To illustrate the possibility of regulating the collection context rather than a data type: for health related websites and services, the lawmaker should consider a ban on third party tracking for behavioural targeting. Specific rules should be considered as well for public sector services and websites. For instance, because of the special task of public service media, a chilling effect should be prevented. The lawmaker should consider banning all personal data collection for behavioural targeting and similar purposes on public service media – at least when third parties collect the data.

---

[1947] See on the five phases of behavioural targeting chapter 2.
[1948] See Diaz & Gürses 2012, p. 2-3.

As noted, under current law, a tracking wall could make consent involuntary if people must use a website. For many public sector websites, it could be the case that people are required to use them. Hence, if such public sector websites allow third party tracking, people should be able to use the website without consenting to such tracking. The lawmaker should consider making more explicit, for instance in a recital, that tracking walls and comparative take-it-or-leave-it choices are generally prohibited for public sector websites.

More generally, it doesn't seem appropriate for public sector websites to allow third party tracking for commercial purposes. Even if website visitors consent to tracking, it's far from evident why the state should facilitate firms to track people's behaviour for commercial purposes. Therefore, the lawmaker should consider a ban on third party tracking for commercial purposes on public sector websites.[1949]

Rules could also focus on phase 2 of the behavioural targeting process: data storage. For example, the data minimisation principle could be supplemented with more specific rules, in the form of maximum retention periods. The vast scale of data processing for behavioural targeting aggravates the chilling effects and the lack of individual control over personal information. Many risks would be reduced if fewer data were stored. With shorter retention periods, there would simply be fewer data that could be used for unexpected purposes. Shortening retention periods could mitigate some of the chilling effects.[1950] And restricting data collection in phase 1, or limiting retention periods in phase 2, would reduce the amount of information that's available to construct predictive models in phase 3.

Strict data minimisation requirements wouldn't be a novelty. The e-Privacy Directive says, in short, that traffic and location data must be erased when they're no longer

---

[1949] If such a ban were considered, many details, such as the scope of the ban, need further attention. For instance, what to do about organisations that are partly funded by the state?

[1950] See on empirical research on whether retention periods matter to users Leon et al. 2013: "participants who were told that data would be retained only for one day were significantly more willing to disclose browsing information" (p. 6).

required for conveying a communication or for billing, unless the user has given consent for another use. However, the e-Privacy Directive's rules for traffic and location data only apply to a narrow category of firms: providers of publicly available electronic communications services, such as internet access providers or phone operators – telecommunication providers for short.[1951] But many firms, such as ad networks and providers of smart phone apps, process more information of a more sensitive nature than telecommunication providers. This asymmetric situation calls for reconsideration.

Phase 3 concerns data analysis. Predictive models are outside the scope of data protection law.[1952] But as long as the data in phase 3 are (still) personal data, data protection law applies. Data protection law's transparency requirements can help to make personal data processing controllable for policymakers, as transparency can help to bring problems to light that might call for regulatory intervention.

Regulation could also focus on phase 4, data disclosure. In phase 4, firms make data available to advertisers or other firms. For example, an ad network can sell copies of data to other firms, or can enable advertisers to target specific persons with ads. This phase illustrates the importance of the purpose limitation principle. Maybe, in addition to the purpose limitation principle, data trade should be banned or restricted in certain contexts. It's not evident, for instance, that insurance companies should be allowed to obtain behavioural targeting data for the purpose of conducting risk calculations. And arguably, because of their special position, banks shouldn't be allowed to monetise their client's payment history through behavioural targeting.[1953]

The e-Privacy Directive prescribes an opt-in regime for using traffic and location data for direct marketing, but these rules only apply to telecommunications providers. The lawmaker should consider specific rules for traffic and location data for behavioural

---

[1951] It could be argued that the rules on traffic data (as far as they are included in article 5(1)) also apply to other types of firms. See chapter 6, section 4, and chapter 5, section 6.
[1952] See chapter 5, section 2 (and on predictive models chapter 2, section 5).
[1953] See Van Eijk 2014.

targeting. Such rules shouldn't only apply to telecommunications providers, but also to firms such as ad networks and providers of smart phone apps. In some contexts, collecting or using traffic and location data may have to be restricted or prohibited.

Sometimes, website publishers don't know in advance who will display ads on their websites, and who will track their website visitors. But if a publisher can't give data subjects the information that's required by the Data Protection Directive, the processing isn't allowed – and shouldn't be allowed. The transparency principle could thus limit what firms can lawfully do in phase 4. The lawmaker should consider making it more explicit that processing is prohibited, unless firms can comply with the transparency principle.

Phase 5 concerns the use of data for personalised advertising (or other purposes), and rules could focus on this phase as well. As far as the Data Protection Directive's provision on automated decisions applies at all to behavioural targeting, it applies to this phase. The provision could protect people against some forms of unfair social sorting. It follows from the provision that somebody may not be subjected to certain fully automated decisions that "significantly" affect her, unless a specified exception applies.[1954] But for behavioural targeting the relevance of the provision seems limited, because it's unclear whether one targeted ad "significantly" affects somebody in the sense of the provision. Furthermore, if an ad network only shows an offer to some people, somebody who doesn't receive the offer is probably unaware of being excluded.

The successor of the automated decisions provision in the European Commission proposal is called "measures based on profiling."[1955] The new provision obliges a firm to tell the person concerned that a profiling measure with significant effect is taken, and to inform the person about the measure's envisaged effects. The lawmaker should amend this provision, and prohibit profiling measures that have the effect of

---

[1954] Article 15 of the Data Protection Directive.
[1955] Article 20 of the European Commission proposal for a Data Protection Regulation.

discriminating on the basis of special categories of data, intentional or not. Such a prohibition would also apply if a firm used non-special data as a proxy for special categories of data. And firms should inform people about profiling measures and their underlying logic, and not only about profiling measures with significant effects. Interdisciplinary research is needed to develop tools to make profiling transparent in a meaningful way.

Usually non-discrimination law doesn't apply to the earlier phases of personal data processing, but it could apply to phase 5. Other rules that focus on phase 5 could also be envisaged. For example, it appears that a substantial part of the population would advocate a ban on personalised pricing, or a ban on personalised pricing in certain contexts.[1956] In any case, as noted in the previous chapter, data protection law requires transparency regarding personalised pricing. The data controller must disclose the processing purpose; this also applies if the purpose is personalising prices.[1957]

This study strongly argues against only focusing on data use (in phase 5) and leaving collection unregulated.[1958] Many privacy problems occur prior to phase 5. Apart from that, a regime that leaves collection unregulated would be difficult to reconcile with fundamental rights case law in Europe, and with the European Union Charter of Fundamental Rights.[1959]

As noted, a specific legal instrument for behavioural targeting, or for electronic direct marketing, would be one option to consider. In such a sector-specific regime, it would be easier to draft relatively specific rules that don't impose unreasonable burdens on

---

[1956] For instance, in a nationally representative survey in the US, Turow et al. 2005 "found that they [US adults] overwhelmingly object to most forms of behavioral targeting and all forms of price discrimination as ethically wrong" (p. 4). Whether personalised pricing is a good thing or not, and under which circumstances, is a complicated topic, which falls outside the scope of this study. See on personalised pricing chapter 2, section 8 and the references there.
[1957] Article 10 and 11 of the Data protection Directive. See chapter 8, section 2.
[1958] See for suggestions to regulators to focus (mainly or only) on use White House (Holdren JP et al.) 2014; Mayer-Schönberger & Cukier 2013. See for an argument against only regulating use (in phase 5): Hoofnagle 2014.
[1959] The European Court of Human Rights says the mere storage of data can interfere with privacy (see chapter 3, section 2). Furthermore, article 8 of the EU Charter of Fundamental Rights concerns personal data "processing", and processing includes collection (article 2(b) of the Data Protection Directive). See Irion & Luchetta 2013, p. 58.

other sectors. To illustrate, the legal regime for health related data shouldn't unduly hamper socially beneficial processing practices, such as research in the medical field or other scientific research.

Another option would be to include specific rules in other legal instruments. For example, rules for tracking on public service media could be included in media law. Other rules could be included in consumer law. Perhaps a black list could be drawn up of prohibited behavioural targeting practices.[1960] And the lawmaker could consider drawing up a list of circumstances to take into account in order to assess the voluntariness of consent.[1961]

In conclusion, enforcing and tightening the data protection principles could help to protect privacy in the area of behavioural targeting, even if people agree to consent requests. But additional rules are needed. The lawmaker shouldn't be afraid of prohibitions in the area of behavioural targeting. Taking into account the practical problems with informed consent to behavioural targeting, protecting the data subject with specific prohibitions or other mandatory rules wouldn't imply undue paternalism. True, it would be difficult to define prohibitions in such a way that they're not over or under inclusive. And banning certain practices implies that the lawmaker must make difficult normative choices. In an informed consent regime, such choices largely fall on the shoulders of the individual. Agreeing on prohibitions would be difficult, but that shouldn't be a reason to ignore the possibility.

* * *

[1960] See for instance the Unfair Commercial Practices Directive. The black list could be supplemented with a grey list, with practices that are presumed to be unfair. Hence, the "grey" practices are considered unfair, unless a firm can prove that the practice isn't unfair. (See on grey lists Centre for the Study of European Contract Law (CSECL) & Institute for Information Law (IViR) 2011, p. 228). The lists may have to be updated regularly.

[1961] See for a list of circumstances that could serve as a starting point for discussions chapter 6, section 4.

# 10 Summary and conclusion

This chapter summarises the study's main findings, draws conclusions, and answers the research question: how could European law improve privacy protection in the area of behavioural targeting, without being unduly prescriptive?

To protect privacy in the area of behavioural targeting, the EU lawmaker mainly relies on the consent requirement for the use of tracking technologies in the e-Privacy Directive, and on general data protection law. With informed consent requirements, the law aims to empower people to make choices in their best interests. But behavioural studies cast doubt on the effectiveness of the empowerment approach as a privacy protection measure. Many people click "I agree" to any statement that is presented to them. Therefore, to mitigate privacy problems such as chilling effects and the lack of individual control over personal information, this study argues for a combined approach of protecting and empowering the individual. Compared to the current approach, the lawmaker should focus more on protecting people.

The chapter is structured as follows. Section 10.1 gives an overview of how behavioural targeting works, and section 10.2 outlines privacy problems in the area of behavioural targeting. Section 10.3 discusses current data protection law. Section 10.4 discusses practical problems with informed consent to behavioural targeting, through the lens of behavioural economics. Section 10.5 gives suggestions to improve empowerment of the individual, and section 10.6 to improve protection of the individual. Section 10.7 concludes.[1962]

### 10.1 Behavioural targeting

In a common arrangement for online advertising, advertisers only pay if somebody clicks on an ad. Click-through rates are low: in the order of 0.1 % to 0.5 %. In other words, when an ad is shown to a thousand people, on average between one and five people click on it. Behavioural targeting was developed to increase the click-through rate on ads, and involves monitoring people's online behaviour to target ads to specific individuals.

In a simplified example, behavioural targeting involves three parties: an internet user, a website publisher, and an advertising network. Advertising networks are firms that serve ads on thousands of websites, and can recognise users when they browse the web. An ad network might infer that a person who often visits websites about fishing is a fishing enthusiast. If that person visits a news website, the ad network might display advertising for fishing rods. When simultaneously visiting that same website, another person who visits a lot of websites about cooking might see ads for pans.

This study analyses the behavioural targeting process by distinguishing five phases: (1) data collection, (2) data storage, (3) data analysis, (4) data disclosure, and (5) the use of data for targeted advertising. In phase 1 firms collect information about people's online activities. People's behaviour is monitored, or tracked. Information captured for behavioural targeting can concern many online activities: what people read, which videos they watch, what they search for, etc. Individual profiles can be enriched with up-to-date location data of users of mobile devices, and other data that are gathered on and off line.

---

[1962] Roughly, section 10.1 summarises chapter 2. Section 10.2 summarises chapter 3 and chapter 7, section 1. Section 10.3 summarises chapter 4 to 6. Section 10.4 summarises chapter 7. Section 10.5 and 10.6 summarise chapter 8 and 9.

A commonly used technology for behavioural targeting involves cookies. A cookie is a small text file that a website publisher stores on a user's computer to recognise that device during subsequent visits. Many websites use cookies, for example to remember the contents of a virtual shopping cart (first party cookies). Ad networks can place and read cookies as well (third party cookies). As a result, an ad network can follow an internet user across all websites on which it serves ads. Third party tracking cookies are placed through virtually every popular website. A visit to one website often leads to receiving third party cookies from dozens of ad networks.

In addition to cookies, firms can use many other technologies for data collection, such as flash cookies and other "super cookies", which are usually harder to delete than conventional cookies. Other tracking methods don't rely on storing an identifier on a device. For example, passive device fingerprinting involves recognising a device by analysing the information it transmits.

In phase 2, firms store the information about individuals, usually tied to identifiers contained within cookies, or via similar technology. Some firms have profiles on hundreds of millions of people. Many behavioural targeting firms can tie a name or an email address to the data they have on individuals.

In phase 3 the data are analysed. A firm could construct a predictive model, for instance along the following lines: if a person visits website A, B, C and D, there's a 0.5 % chance the person clicks on ads for product E. For behavioural targeting to be useful, a predictive model doesn't have to be accurate when applied to an individual. If a behaviourally targeted ad has a click-through rate of 0.5 %, this is a major improvement compared to a 0.1 % click-through rate of non-targeted ads.

With behavioural targeting and other types of profiling, a predictive model based on information about a group of people can be applied to somebody who isn't part of that group. Suppose an online shop obtains the consent of thousands of people to analyse their shopping habits over time. Based on the information it collected, the shop constructs a predictive model that says that 95% of the women who buy certain products will give birth within two months. Alice is a customer, but wasn't among the people who consented to data collection that formed the basis of the predictive model. When Alice buys certain products, the shop can infer with reasonable accuracy that she's pregnant. Hence, the shop can predict something about Alice, based on other people's information.

In phase 4, data disclosure, firms make data available to advertisers or other firms. For example, a social network site can enable advertisers to target specific persons with ads based on their behavioural profiles. Or a firm can sell copies of data to other firms. Firms can combine information from different sources to enrich profiles. Many types of firms are involved in behavioural targeting, and the resulting data flows are complicated. For example, an ad network that displays ads on a website can allow other ad networks to bid in an automated auction for the possibility to show ads to individuals. Data about individuals are auctioned off within milliseconds, and billions of such auctions take place every day. Such practices are referred to as real time bidding, or audience buying. A website publisher often doesn't know in advance who will serve ads on its website, and may not have a direct business relationship with the advertiser.

In phase 5 firms show targeted ads to specific individuals. Firms can personalise ads and other website content for each visitor. A firm might also refrain from showing an ad to certain people, based on their profiles. Behavioural targeting enables advertisers to reach a user, wherever he or she is on the web.

A website publisher can increase its income by allowing ad networks to track its visitors and to display behaviourally targeted ads. But in the long term behavioural targeting may decrease ad revenues for some website publishers. For example, an ad network doesn't have to buy expensive ad space on a large professional news website to advertise to a reader of that website. The ad network can show an ad to that person when he or she visits a random website, where advertising space is cheaper. One marketer summarises: "advertisers are buying audiences with data, rather than using content as a proxy to reach the people they want to reach."[1963]

### 10.2 Privacy and behavioural targeting

Surveys show that most people don't want behaviourally targeted advertising, because they find it creepy or privacy-invasive. A small minority says it doesn't mind the data collection and prefers behaviourally targeted advertising because it can lead to more relevant ads.

Privacy is notoriously difficult to define. Borrowing from Gürses, three privacy perspectives were distinguished in this study: privacy as limited access, privacy as control over personal information, and privacy as the freedom from unreasonable constraints on identity construction. The three perspectives partly overlap, and highlight different aspects of privacy.

The privacy as limited access perspective concerns a personal sphere, where people can be free from interference. The limited access perspective is similar to approaches of privacy as confidentiality, seclusion, or a right to be let alone. This perspective implies that too much access to a person interferes with privacy. For instance, if somebody wants to keep a website visit confidential, there's a privacy interference if others learn about the visit. A second privacy perspective focuses on the control people should have over information concerning them. Seeing privacy as control is common since the 1960s, when state bodies and other large organisations started to amass increasing amounts of information about people, often using computers. The control perspective has deeply influenced data protection law. Privacy as control is interfered with, for example, if personal information is collected surreptitiously. Third, privacy can be seen as the freedom from unreasonable constraints on identity construction. The privacy as identity construction perspective largely includes the

---

[1963] Collective 2014.

other two perspectives, but also highlights other concerns regarding modern data processing practices in the digital environment, such as profiling and behavioural targeting. There could be an interference with privacy if somebody is manipulated by the environment, which can include technology.

This study focuses on three main privacy problems of behavioural targeting: chilling effects, a lack of control over personal information, and the risk of unfair social sorting and manipulation. First, chilling effects can occur because of the massive collection of information about people's online activities. People may adapt their behaviour if they know their activities are monitored. For instance, somebody who fears surveillance might hesitate to look for medical information on the web, or to read about certain political topics.

Second, people lack control over data concerning them. The reality of current behavioural targeting practices is far removed from the ideal of privacy as control. People don't know which information about them is collected, how it's used, and with whom it's shared. The feeling of lost control is a privacy problem. And large-scale personal data storage brings risks. For instance, a data breach could occur, or data could be used for unexpected purposes, such as identity fraud.

Third, behavioural targeting enables social sorting. There's a risk of unfair discriminatory practices: firms can sort people into "targets" and "waste", and treat them accordingly.[1964] And some fear that behavioural targeting could be used to manipulate people. Personalised advertising could become so effective that advertisers have an unfair advantage over consumers. There could also be a risk of "filter bubbles" or "information cocoons", especially when behavioural targeting is used to personalise not only ads, but also other content and services.[1965] Briefly stated, the idea is that personalised advertising and other content could surreptitiously steer

---

[1964] Turow 2011.
[1965] The phrases are from Pariser 2011 and Sunstein 2006.

people's choices. In sum, from each of the three privacy perspectives, behavioural targeting is problematic.

### 10.3 Data protection law

The right to respect for private life, the right to privacy for short, is a fundamental right in the European legal system, and is included in the European Convention on Human Rights (1950). The European Court of Human Rights interprets the Convention's privacy right generously, and refuses to define the right's scope of protection. This way, the Court can apply the right to privacy in unforeseen situations and to new developments. For instance, the Court says information derived from monitoring somebody's internet usage is protected under the right to privacy.

To protect privacy in the area of behavioural targeting, the main legal instrument in Europe is the Data Protection Directive, coupled with the e-Privacy Directive's consent requirement for tracking technologies. Data protection law is a legal tool, which aims to ensure that the processing of personal data happens fairly and transparently. Data protection law grants rights to people whose data are being processed (data subjects), and imposes obligations on parties that process personal data (data controllers, limited to and referred to as firms in this study). Since its inception in the early 1970s, data protection law has evolved into a complicated field of law. Borrowing from Bygrave, the core of data protection law can be summarised in nine principles: the fair and lawful processing principle, the transparency principle, the data subject participation and control principle, the purpose limitation principle, the data minimisation principle, the proportionality principle, the data quality principle, the security principle, and the sensitivity principle.

The right to data protection and the right to privacy aren't the same. The EU Charter of Fundamental Rights (2000) includes a right to privacy, and a separate right to the protection of personal data. This study agrees with De Hert & Gutwirth, who characterise the right to privacy as an "opacity tool", and data protection law as a "transparency tool."[1966] The right to privacy in the European Convention on Human Rights prohibits intrusions into the private sphere. The right to privacy aims to give the individual the chance to remain shielded, or to remain opaque. This prohibition isn't absolute; privacy must often be balanced against other interests, such as the rights of others. Data protection law takes a different approach than the legal right to privacy, say De Hert & Gutwirth. In principle data protection law allows data processing, if the data controller complies with a number of requirements. Data protection law aims to ensure fairness, and one of the means to foster fairness is requiring firms to be transparent about personal data processing. Hence: a transparency tool.

In January 2012 the European Commission presented a proposal for a Data Protection Regulation, which should replace the 1995 Data Protection Directive. At the time of writing, it's unclear whether the proposal will be adopted. The most optimistic view seems to be that the Regulation could be adopted in 2015.[1967] While based on the same principles as the Directive, the proposal would bring significant changes. For instance, unlike a directive, a regulation has direct effect and doesn't have to be implemented in the national laws of the member states, so it should lead to a more harmonised regime in the EU. The proposal introduces new requirements for data controllers, such as the obligation to implement measures to ensure and demonstrate compliance. The proposal also aims to make it easier for people to delete their data from the web, and to transfer their personal data from one service provider to another. The proposal's preamble emphasises the ideal of data subject control. "Individuals should have control of their own personal data."[1968]

---

[1966] De Hert & Gutwirth 2006.
[1967] See European Council, 2014, p. 2.
[1968] Recital 6.

***Material scope of data protection law***

Whether data protection law applies to behavioural targeting is hotly debated. Data protection law only applies when "personal data" are processed: data that relate to an identifiable person. For behavioural targeting, firms often process individual but nameless profiles. Many behavioural targeting firms claim they only process "anonymous" data, and that data protection law thus doesn't apply. While the European Court of Justice, the highest authority on the interpretation of EU law, hasn't ruled on behavioural targeting yet, its case law is relevant. The discussion about nameless behavioural targeting profiles resembles the one about IP addresses. In a decision about IP addresses in the hands of an internet access provider, the Court said that those IP addresses were personal data.[1969] Furthermore, European Data Protection Authorities, cooperating in the Article 29 Working Party, say behavioural targeting generally entails personal data processing, even if a firm can't tie a name to the data it has on an individual. If a firm aims to use data to "single out" a person, or to distinguish a person within a group, these data are personal data, according to the Working Party.[1970] Although not legally binding, the Working Party's opinions are influential. National Data Protection Authorities often follow its interpretation.

The 2012 proposal for a Data Protection Regulation stirred up the debate about the material scope of data protection law. There has been much lobbying to make the proposal less burdensome for businesses. Many firms say that pseudonymous data, such as nameless behavioural targeting profiles, should be outside the scope of data protection law, or should be subject to a lighter regime. In March 2014, the European Parliament adopted a compromise text, which the Parliament's LIBE Committee prepared on the basis of the 3999 amendments by the members of parliament. This LIBE Compromise introduces a new category of personal data, pseudonymous data, and the rules are less strict for such data. Under certain conditions, the LIBE

---

[1969] ECJ, Sabam/Scarlet (C-70/10)

[1970] See e.g. Article 29 Working Party 2010, WP 171, p. 9.

Compromise allows firms to use behavioural targeting with pseudonymous data without the data subject's consent.

This study argues that data protection law should apply to behavioural targeting, and argues against a lighter regime for pseudonymous data. First, many risks remain, regardless of whether firms tie a name to the information they hold about a person. For instance, surveillance can cause a chilling effect, including if firms collect pseudonymous data. And a cookie-based profile that says a person is handicapped or from a poor neighbourhood could be used for unfair social sorting. Second, a name is merely one of the identifiers that can be tied to data about a person, and is not even the most practical identifier for behavioural targeting. For an ad network that wants to track somebody's browsing behaviour, or wants to target somebody with online advertising, a cookie works better than a name. Third, the behavioural targeting industry processes large amounts of information about people, and this brings risks. If data protection law didn't apply, this industry could operate largely unregulated. For these reasons, data that are used to single out a person should be considered personal data. In addition, it's often fairly easy for firms to tie a name to pseudonymous data.

***Informed consent***

Informed consent plays a central role in the current regulatory framework for behavioural targeting. Therefore, this study examined the role of informed consent in data protection law, and its value for regulating privacy in the area of behavioural targeting. The Data Protection Directive only allows firms to process personal data if they can base the processing on consent or on one of five other legal bases. The European Commission proposal for a Regulation duplicates the same legal bases without major revisions. For the private sector, the most relevant legal bases are: a contract, the balancing provision, and the data subject's consent.[1971]

---

[1971] The legal bases are listed in article 7 of the Data Protection Directive, and in article 6 of the European Commission proposal for a Data Protection Regulation.

A firm can process personal data if the processing is necessary for the performance of a contract with the data subject. For instance, certain data have to be processed for a credit card payment, or for a newspaper subscription. The "necessary" requirement sets a higher threshold than useful or profitable. Some internet companies suggest a user enters a contract by using their services, and that it's necessary for this contract to track the user for behavioural targeting. This interpretation seems incorrect. According to the Working Party, a firm can only rely on the legal basis contract if the processing is genuinely necessary for providing the service. The Working Party's view implies that, in general, firms can't rely on this legal basis for behavioural targeting. In any case, the practical problems with informed consent to behavioural targeting which are discussed below would be largely the same if firms could base the processing for behavioural targeting on a contract.

The balancing provision allows data processing when it's necessary for the firm's legitimate interests, except where such interests are overridden by the data subject's interests or fundamental rights. When weighing the interests of the firm and the data subject, all circumstances have to be taken into account, such as the sensitivity of the data and the data subject's reasonable expectations. The balancing provision is the appropriate legal basis for innocuous standard business practices. For example, a firm can generally rely on the balancing provision for postal direct marketing for its own products to current or past customers. If a firm relies on the balancing provision for direct marketing, data protection law grants the data subject the right to stop the processing: to opt out. The Data Protection Directive doesn't say explicitly whether behavioural targeting can be based on the balancing provision. But the most convincing view is that behavioural targeting can't be based on this provision, in particular when it involves tracking a person over multiple websites. In most cases the data subject's interests must prevail over the firm's interests, as behavioural targeting involves collecting and processing information about personal matters such as people's browsing behaviour. Indeed, the Working Party says firms can almost never rely on the balancing provision to process personal data for behavioural targeting.

If firms want to process personal data, and can't base the processing on the balancing provision or another legal basis, they must ask the data subject for consent. With consent, the data subject can allow data processing that would otherwise be prohibited. The Working Party says consent is generally the required legal basis for personal data processing for behavioural targeting. It follows from the Data Protection Directive's consent definition that consent requires a free, specific, informed indication of wishes. People can express their will in any form, but mere silence or inactivity isn't an expression of will. This is also the predominant view in general contract law. During the drafting of the Data Protection Directive in the early 1990s, firms have argued that opt-out systems should be sufficient to obtain "implied" consent for direct marketing. But the EU lawmaker rejected this idea.

A number of larger behavioural targeting firms offer people the chance to opt out of targeted advertising on a centralised website: youronlinechoices.com. However, participating firms merely promise to stop showing targeted ads, so they may continue to track people who have opted out. In short, the website offers the equivalent of Do Not Target, rather than Do Not Collect. But even if the firms stopped collecting data after somebody opts out, they couldn't use the website's opt-out system to obtain valid consent. Valid consent requires an expression of will, which generally calls for an opt-in procedure.

In line with the transparency principle, consent has to be specific and informed. Consent can't be valid if a consent request doesn't include a specified processing purpose and other information that's necessary to guarantee fair processing. Furthermore, consent must be "free." Negative pressure would make consent invalid, but positive pressure is generally allowed. In most circumstances, current data protection law allows firms to offer take-it-or-leave-it choices.

Hence, in principle website publishers are allowed to install "tracking walls" that deny entry to visitors that don't consent to being tracked for behavioural targeting. But a tracking wall could make consent involuntary if people must use a website. For instance, say people are required to file their taxes online. If the tax website had a tracking wall that imposed third party tracking, people's consent to tracking wouldn't be voluntary. Similarly, if students must use a university website, a tracking wall would make consent involuntary. According to the Dutch Data Protection Authority, the national public broadcasting organisation isn't allowed to use a tracking wall, because the only way to access certain information online is through the broadcaster's website. The Working Party emphasises that consent should be free, but doesn't say that current data protection law prohibits tracking walls in all circumstances.

Since 2009, article 5(3) of the e-Privacy Directive requires any party that stores or accesses information on a user's device to obtain the user's informed consent. Article 5(3) applies regardless of whether personal data are processed, and applies to many tracking technologies such as tracking cookies. There are exceptions to the consent requirement, for example for cookies that are strictly necessary for a service requested by the user, and for cookies that are necessary for transmitting communication. Hence, no prior consent is needed for cookies that are used for a digital shopping cart, or for log-in procedures.

Recital 66 of the 2009 directive that amended the e-Privacy Directive has caused much discussion: "in accordance with the relevant provisions of [the Data Protection Directive], the user's consent to processing may be expressed by using the appropriate settings of a browser or other application."[1972] Many marketers suggest that people who don't block tracking cookies in their browser give implied consent to behavioural targeting. For instance, the Interactive Advertising Bureau UK, a trade organisation, says "default web browser settings can amount to 'consent'."[1973] But this doesn't seem plausible. As the Working Party notes, the mere fact that a person leaves his or her browser's default settings untouched doesn't mean that the person expresses his or her will to be tracked.

---

[1972] Directive 2009/136, recital 66.
[1973] Interactive Advertising Bureau United Kingdom 2012, p. 2..

In sum, firms are required to obtain consent for most tracking technologies that are used for behavioural targeting. Therefore, firms must usually obtain the data subject's consent for behavioural targeting, regardless of the legal basis of ensuing personal data processing. Hence, even if, under rare circumstances, a firm could rely on the balancing provision to process personal data for behavioural targeting, the firm would generally need consent for using the tracking technology. Article 5(3) isn't widely enforced yet, among other reasons because the national implementation laws are rather new. Many member states missed the 2011 implementation deadline. The approaches in the member states vary. For example, the Netherlands requires, in short, opt-in consent for tracking cookies. In contrast, the UK appears to allow firms to use opt-out systems to obtain "implied" consent. However, the Working Party insists that the data subject's inactivity doesn't signify consent.

***A limited but important role for consent***

While consent plays an important role in data protection law, its role is limited at the same time. Consent can provide a legal basis for personal data processing. But if a firm has a legal basis for processing, the other data protection provisions still apply. Those provisions are mandatory. The data subject can't waive the safeguards or deviate from the rules by contractual agreement. For example, the security principle requires an appropriate level of security for personal data processing. And it follows from the purpose limitation principle that personal data must be collected for specified purposes, and should not be used for incompatible purposes. Hence, a contract between a firm and a data subject wouldn't be enforceable if it stipulated that the firm doesn't have to secure the personal data, or can use the data for new purposes at will. Data protection law thus limits the data subject's contractual freedom. On the other hand, data protection law leaves some important choices to the data subject. For instance, the data subject can give or withhold consent, and has the right to stop data processing for direct marketing which is based on the balancing provision. In sum, data protection law embodies an inherent tension between protecting and empowering the data subject.

## 10.4 Informed consent and behavioural economics insights

For this study the choice was made to incorporate insights from other disciplines than law. Literature from the emerging field of the economics of privacy was analysed, as well as behavioural economics literature and social science studies on how people make privacy choices in practice. The analysis shows that there are reasons for more regulatory intervention. Informed consent largely fails as a privacy protection measure.

***Economics***

From an economic perspective, it's unclear whether behavioural targeting leads to a net benefit or a net loss for society. The benefits include profit for ad networks and other firms. And income from online advertising could be used to fund so-called "free" web services. People gain utility from using a search engine or reading an online newspaper. As an aside, it's unclear whether behavioural targeting is needed to fund "free" websites. Advertising that doesn't require monitoring people's behaviour is also possible, such as contextual advertising: ads for cars on websites about cars.

Behavioural targeting can also decrease welfare. For instance, it can be costly for people if their information ends up in the wrong hands. People could receive invasive marketing such as spam, or they could fall victim to identity fraud. Personalised ads could be used to exploit people's weaknesses or to charge people higher prices. And it's costly if people invest time in evading tracking. Furthermore, it may hamper electronic commerce if people don't trust that their personal information is adequately protected when they buy online, or when they use internet services. Other privacy related costs are harder to quantify, such as annoyance, chilling effects, and the long term effects on society. In sum, it seems unlikely that economics could offer a definitive answer to the question of whether more or less legal privacy protection would be better in the behavioural targeting area. Apart from that, the European legal system doesn't give precedence to economic arguments. Nevertheless, economics provides a useful tool to analyse practical problems with consent to behavioural targeting.

Economists often use rational choice theory to predict human behaviour. Rational choice theory analyses behaviour assuming that people generally want to maximise their welfare, and that people are generally able to choose the best way to maximise their welfare. In economics, a (hypothetical) perfectly functioning free market leads to the highest social welfare – provided there are no market failures, and setting aside how welfare is distributed within society. But there may be reason for the lawmaker to intervene when the market doesn't function as it ideally should. From an economic perspective, the law should aim at reducing market failures, such as information asymmetries, externalities, and market power. However, legal intervention brings costs and economic distortions as well, which must be taken into account.

Through an economic lens, consenting to behavioural targeting can be seen as entering into a market transaction with a firm. But this "transaction" is plagued by information asymmetries. Research shows that many people don't know to what extent their behaviour is tracked, so their "choice" to disclose data in exchange for using a service can't be informed. Even if firms sought consent for behavioural targeting, information asymmetry would remain a problem. People rarely know what a firm does with their personal data, and it's difficult to predict the consequences of future data usage. Information asymmetry is a form of market failure. Firms won't compete on quality if people can't assess the quality of products. This can lead to low quality products. Websites rarely compete on privacy, as illustrated by the fact that people are tracked for behavioural targeting on virtually every popular website. There seems to be a comparable situation on the market for smart phone apps.

Data protection law aims to reduce the information asymmetry by requiring firms to disclose certain information to data subjects. The law obliges firms to provide data subjects with information about their identity and the processing purpose, and all other information that's necessary to guarantee fair processing. Website publishers can use a privacy policy to comply with data protection law's transparency requirements. These requirements also apply if a firm doesn't seek the data subject's consent, but relies on another legal basis for data processing.

However, the information asymmetry problem is hard to solve because of transaction costs for data subjects, and again, information asymmetries regarding the meaning of privacy policies. Reading privacy policies would cost too much time, as they're often long, difficult to read, and vague. It would take people several weeks per year if they read the privacy policy of every website they visit. The language in privacy policies is too difficult for many. It's thus not surprising that almost nobody reads privacy policies. In practice, data protection law thus doesn't solve the information asymmetry problem.

Externalities are another example of market failure. Economists refer to costs or damage suffered by third parties as a result of economic activity as negative externalities. Externalities occur because parties that aim to maximise their own welfare don't let costs for others influence their decisions. An example of a negative externality is environmental pollution from traffic or industry. Many legal rules, such as those in environmental law, can be seen as responses to an externalities problem. If the lawmaker wants to reduce negative externalities resulting from a contract, it generally needs to use mandatory rules. If the lawmaker used non-mandatory default rules, the contract parties would set the rules aside. After all, the externality is a result of the fact that contract parties don't take the interests of non-contract parties into account.

At first glance there are no negative externalities if somebody consents to sharing his or her data with a behavioural targeting firm. The person merely gives up an individual interest. But people's consent to behavioural targeting may lead to the application of knowledge to others. This can be illustrated with the example of a shop that uses a predictive model to predict the pregnancy of Alice, while the model is based on other people's data. This could be seen as an externality imposed on Alice, which is a result of the fact that people consented to having their personal information processed.

Market power, such as a monopoly situation, is a third example of market failure. Whether a firm has too much market power depends on the specifics of a particular market. The conclusion would be different for search engines, social network sites, online newspapers, or games for phones. Many take-it-or-leave-it choices regarding behavioural targeting may not be an abuse of market power from the viewpoint of competition law or economics. In any event, even in a market without market power problems, the practical problems with consent resulting from information asymmetries could persist.

***Behavioural economics***

Behavioural economics aims to improve the predictive power of economic theory, by including insights from psychology and behavioural studies. Behavioural economics suggests that people act structurally different than rational choice theory predicts. Because of their bounded rationality, people often rely on rules of thumb, or heuristics. Usually such mental shortcuts work fine, but they can also lead to behaviour that is not in people's self-interest. Systematic deviations from rational choice theory are called biases. Several biases influence privacy choices, such as the status quo bias and the present bias.

The status quo bias, or default bias, describes people's tendency to stick with default options. People are less likely to consent under an opt-in regime that requires an affirmative action for valid consent, than under an opt-out regime where people are assumed to consent if they don't object. In this light, the continuous opt-in/opt-out discussion about behavioural targeting and other types of direct marketing concerns the question of who benefits from the status quo bias, the firm or the data subject.

Present bias, or myopia, suggests that people often choose for immediate gratification and disregard future costs or disadvantages. For example, many find it hard to stick with a diet, or to save money for later. If a website has a tracking wall, and people can only use the site if they agree to behavioural targeting, they're likely to consent, thereby ignoring the costs of future privacy infringements. Behavioural economics can thus help to explain the alleged privacy paradox. People who say they care about their privacy, often disclose information in exchange for small benefits. Part of this is conditioning; many people click "yes" to any statement that is presented to them. It's only a slight exaggeration to say: people don't read privacy policies; if they were to read, they wouldn't understand; if they understood, they wouldn't act.

In conclusion, an economic analysis doesn't dictate the ideal level of legal privacy protection. It's not straightforward whether more or less legal privacy protection in the area of behavioural targeting would be better from an economic perspective. Therefore, it remains unclear whether legal limits on behavioural targeting would be too costly for society. In any case, the lawmaker shouldn't act too bluntly. Just like environmental law doesn't aim to undo the industrial revolution (and is unlikely to do so), legal privacy protection shouldn't undo the advantages of information technology (and is unlikely to do so).

The economic analysis does show that if consenting to behavioural targeting were compared to entering into market transaction, this transaction would take place in a market plagued by market failures. There also seems to be a behavioural market failure in the behavioural targeting area. If all competitors exploit people's biases, a firm has to do the same to stay in business. In sum, insights from economics and behavioural economics suggest more regulatory intervention is needed in the area of behavioural targeting.

### 10.5 Improving empowerment

Considering the limited potential of informed consent as a privacy protection measure, this study argues for a combined approach of empowering and protecting the individual. The study concludes that certain practices simply shouldn't be allowed (see below). But it doesn't seem feasible to define all beneficial or all harmful data processing activities in advance. Apart from that, the EU Charter of Fundamental Rights lists consent as a legal basis for personal data processing. Relying on informed consent, in combination with data protection law's other safeguards, will probably remain the appropriate approach in many circumstances. For those cases, transparency and consent should be taken seriously. While fostering individual control over personal information won't suffice to protect privacy in the area of behavioural targeting, some improvement must be possible, compared to the current situation of almost complete lack of control by individuals over their own data.

To improve privacy protection in the area of behavioural targeting, data protection law should be more strictly enforced, and needs amendments. The European Commission has realised that compliance with data protection law is lacking, and aims for better enforcement. For instance, under the proposal for a Data Protection Regulation, Data Protection Authorities could impose high penalties, and organisations could take a firm to court on behalf of data subjects if the firm breaches data protection law. An important avenue for further research is how compliance with the rules could be improved. One option that should be examined is the introduction of collective action procedures that enable groups of people to sue a firm if it breaches privacy or data protection rights. Another topic for further research is enforcement of European data protection law against firms that are based outside Europe, a topic that was outside this study's scope.

How could the law improve *empowerment* of the individual? To reduce the information asymmetry in the area of behavioural targeting, the transparency principle should be enforced. In line with European consumer law, the lawmaker should require firms to phrase privacy policies and consent requests in a clear and comprehensible manner. The European Commission proposal for a Data Protection Regulation requires firms to have easily accessible privacy policies "in an intelligible form, using clear and plain language."[1974] Codifying the clear language requirement could discourage firms from using legalese in privacy policies. And the requirement would make it easier for Data Protection Authorities to intervene when a firm uses a privacy policy or a consent request that is too vague. The rule wouldn't be enough to ensure actual transparency, but it could help to lower the costs of reading privacy policies. Also, interdisciplinary research is needed to develop tools to make data processing transparent in a meaningful way.

Regarding consent, the existing rules should be enforced. Requiring informed consent for tracking wouldn't guarantee transparency, but at least a consent request would alert people to the tracking, unlike an opt-out system. And because of the default bias, requiring opt-in consent for tracking could nudge people towards disclosing fewer data. The European Commission proposal reaffirms that consent must be expressed "either by a statement or by a clear affirmative action."[1975] The proposal also codifies the Working Party's view that a consent request may not be hidden in a privacy policy or in terms and conditions.

Human attention is scarce and too many consent requests can overwhelm people. Therefore, the scope of article 5(3) of the e-Privacy Directive is too broad. Article 5(3) requires consent for storing or accessing information on a user's device. This means consent is also required for some cookies that pose few privacy risks and that aren't used to collect detailed information about individuals, such as certain types of cookies that are used for website analytics. But there's little reason to seek consent for truly innocuous practices. The Data Protection Directive contains the balancing provision for such innocuous practices.

It would probably be better if the lawmaker phrased the consent requirement for tracking in a more technology neutral way. The law could require consent for the collection and further processing of personal data, including pseudonymous data, for

[1974] Article 11 of the European Commission proposal for a Data Protection Regulation (2012).
[1975] Article 4(8) of the European Commission proposal for a Data Protection Regulation (2012).

behavioural targeting and similar purposes – regardless of the technology that's used. Phrasing the rule in a more technology neutral way could also mitigate another problem. In some ways the scope of article 5(3) is too narrow. For instance, it's unclear to what extent article 5(3) applies if firms use device fingerprinting for behavioural targeting.

A user-friendly system should be developed to make it easier for people to give or refuse consent. Work is being done in this area, among others by the World Wide Web Consortium, an organisation that works on the standardisation of web technologies. The Consortium's Tracking Protection Working Group (DNT Group) is trying to develop a Do Not Track standard, which should enable people to signal with their browser that they don't want to be tracked. This way, people could opt out of tracking with a few mouse clicks. The system could thus lower the transaction costs of opting out of data collection by hundreds of firms.

It's not immediately apparent how Do Not Track – an opt-out system – could help firms to comply with the e-Privacy Directive's consent requirement for tracking technologies. But an arrangement along the following lines could be envisaged. Firms should refrain from tracking European internet users that haven't set a Do Not Track preference. If somebody signals to a firm "Yes, you can track me" after receiving sufficient information, that company may track that user. Hence, in Europe not setting a preference would have the same legal effect as setting a preference for "Do not track me." In Europe, Do Not Track would thus be a system to opt in to tracking.

At the time of writing, after almost three years of discussion, the DNT Group still hasn't reached consensus in relation to some major issues. The most contentious topic is what firms should do when they receive a "Do not track me" signal. Many firms that participate in the DNT Group want to continue to collect data from people who signal they don't want to be tracked. In brief, the firms want to offer Do Not Target, rather than Do Not Collect. Some firms even want to continue targeting ads to people who signal "Do not track me." The firms offer to delete people's browsing history, while retaining the inferred interest categories tied to people's profiles. There's no agreement in the DNT Group about which data uses should still be allowed when people signal "Do not track me."

From the start, the DNT Group agreed that the Do Not Track standard should allow a website to ask somebody who signals "Do not track me" for an exception, roughly as follows. "We see your Do Not Track signal. But do you make an exception for me and my ad network partners so we can track you?" As noted, data protection law allows take-it-or-leave-it choices in many circumstances. Hence, if a Do Not Track standard were developed that complied with European law, many websites would probably respond by installing tracking walls. Therefore, even if firms provided clear information, even if people understood the information, and even if firms asked for prior consent, many people might still feel that they're forced to consent to behavioural targeting. Even if Do Not Track emerges as a W3C standard, it seems unlikely that without additional legislative support it will solve the privacy problems posed by behavioural targeting.

To conclude, a lack of individual control over personal information aptly describes many privacy problems. But this doesn't mean that aiming for data subject control is the best regulatory tactic. Enforcing and tightening the data protection principles could improve data subject control. However, aiming for individual empowerment alone won't suffice in protecting privacy in the area of behavioural targeting.

## 10.6 Improving protection

A second legal approach to improve privacy protection in the area of behavioural targeting involves *protecting,* rather than empowering, people. If fully complied with, the data protection principles could give reasonable privacy protection in the behavioural targeting area, even if people agreed to consent requests. But additional regulation is needed as well.

The study offers suggestions on how the law could improve privacy protection, without being unduly prescriptive. In this study, rules are considered unduly prescriptive if they impose unreasonable costs on society, or if they're unduly paternalistic. As noted, from an economic perspective it's unclear whether more or less legal privacy protection in the area of behavioural targeting would be better. Therefore, stricter rules wouldn't necessarily be too costly for society. Additionally, the existence of market failures in the area of behavioural targeting suggests a need for regulatory intervention.

A greater focus on protecting the data subject wouldn't make the law unduly paternalistic either. Paternalism involves limiting a person's contractual freedom, predominantly to protect that person. The law in Europe accepts a degree of paternalism, and this study agrees with that approach. Many rules, such as consumer protection rules and minimum safety standards for products, could plausibly be explained, at least in part, by paternalistic motives, although such rules could also be seen as a response to market failures.

Pure paternalism is only present when a legal rule only aims at protecting somebody against him- or herself. But there are other rationales for legal privacy protection than protecting people against themselves. The right to privacy and the right to data protection aim to contribute to a fair society, which goes beyond individual interests. And responding to market failures has nothing to do with paternalism. Moreover, behavioural economics insights suggest that more protective rules are needed. After all, the European Court of Human Rights requires privacy protection that's "practical and effective, not theoretical and illusory."[1976]

The data minimisation principle, if effectively enforced, is an example of a data protection principle that could protect people's privacy, even after people consent to behavioural targeting. The vast scale of data processing for behavioural targeting

---

[1976] ECtHR, Christine Goodwin v. the United Kingdom, No. 28957/95, July 11, 2002, par 74.

aggravates the chilling effects, and the lack of individual control over personal information. And large-scale data storage brings risks, such as data breaches. Compliance with the data minimisation principle could mitigate such privacy problems. Furthermore, setting limits to data collection would reduce the amount of information that's available to construct predictive models. The Data Protection Directive states that data processing must be "not excessive" in relation to the processing purpose.[1977] It follows from the Directive's structure that this requirement also applies if the processing is based on the data subject's consent. The data minimisation principle should be phrased more clearly, which the European Commission proposal for a Data Protection Regulation does. "Personal data must be (…) limited to the minimum necessary in relation to the purposes for which they are processed."[1978] The lawmaker should explicitly codify that the data subject's consent doesn't legitimise disproportionate data processing. Such a rule could remind firms that consent doesn't give them carte blanche to collect personal information at will, and that a Data Protection Authority could intervene if they did.

The transparency principle can be interpreted as a prohibition of surreptitious data processing. With some behavioural targeting practices, it would be difficult for a website publisher to comply with data protection law's transparency requirements, even if it tried its best. For example, some ad networks allow other ad networks to buy access to individuals by bidding on an automated auction. In such situations, the website publisher doesn't know in advance who will display ads on its site, and who will track its website visitors. Therefore, it's hard to see how the publisher could comply with the law's transparency requirements. If a publisher can't give data subjects the information that's required by the Data Protection Directive, the processing isn't allowed – and shouldn't be allowed. The lawmaker should make it more explicit that processing is prohibited, unless firms can comply with the transparency principle.

---

[1977] Article 6(1)(c) of the Data Protection Directive.
[1978] Article 5(c) of the European Commission proposal for a Data Protection Regulation (2012).

Data protection law has a stricter regime for "special categories of data", such as data revealing race, political opinions, health, or sex life.[1979] Using special categories of data for behavioural targeting and other types of direct marketing is only allowed after the data subject's explicit consent is obtained, and in some member states prohibited. Strictly enforcing the existing rules on special categories of data could reduce privacy problems such as chilling effects. For instance, people might be hesitant to look for medical information on the web if they fear leaking information about their medical conditions. Because the privacy risks involved in using health data for behavioural targeting outweigh the possible societal benefits from allowing such practices, the EU lawmaker should consider prohibiting the use of any health related data for behavioural targeting, whether the data subject gives explicit consent or not. The rules on special categories of data could be interpreted in such a way that the collection context is taken into account. For example, tracking people's visits to websites with medical information should arguably be seen as processing "special categories of data", as the firm could infer data regarding health from such tracking information.

For providers of publicly available electronic communications services, such as internet access providers or phone operators, the e-Privacy Directive contains stricter rules for certain data types. For example, such providers may only process location data and traffic data with consent, unless a specified exception applies. But many firms, such as ad networks and providers of smart phone apps, process more data of a more sensitive nature than providers of publicly available electronic communications services. This asymmetric situation calls for reconsideration.

An option that should be explored is whether a separate legal instrument is needed to protect privacy in the behavioural targeting area. The current sector-specific rules in the e-Privacy Directive have major shortcomings. With a separate legal instrument for privacy protection in the area of behavioural targeting, the lawmaker could adopt appropriate rules for behavioural targeting, without imposing unnecessary burdens on

---

[1979] Article 8 of the Data Protection Directive.

other sectors. These specific rules could address the different behavioural targeting phases: (1) data collection, (2) data storage, (3) data analysis, (4) data disclosure, and (5) the use of data for targeted advertising. But specific rules could also be included in other legal instruments. For instance, rules regarding tracking and public service media could be included in media law. Other rules could be included in consumer law.

What should the lawmaker do about take-it-or-leave-it choices such as tracking walls? The law could prohibit take-it-or-leave-it choices in certain circumstances or contexts. For instance, public service broadcasters often receive public funding, and they have a special role in informing people. But if people fear surveillance, they might forego using public service media. Therefore, the lawmaker should prohibit public service broadcasters from installing tracking walls on their websites. The lawmaker could also go one step further, and prohibit all third party tracking for behavioural targeting on public service media.

More generally it's questionable whether it's appropriate for websites of state bodies to allow third party tracking for behavioural targeting – even when people consent. It's not evident why the public sector should facilitate tracking people's behaviour for commercial purposes. Therefore, the lawmaker should consider prohibiting all tracking for behavioural targeting on public sector websites.

The Data Protection Directive's provision on automated decisions could protect people against certain forms of unfair social sorting and discrimination. The provision says that a person may not be subjected to certain fully automated decisions that "significantly affect" him or her.[1980] But there are exceptions. For example, the law allows a firm to automatically refuse to enter into a contract with an individual, if there are safeguards in place for that person, which may include a possibility to ask for human intervention. By way of illustration, an insurance company that lets

[1980] Article 15 of the Data Protection Directive.

software automatically deny a website visitor an insurance contract could ensure that the person can ask a human to reconsider the decision. But for behavioural targeting the relevance of the automated decisions provision seems limited, as it's unclear whether one targeted ad qualifies as an automated decision that "significantly affects" somebody in the sense of the provision. However, in aggregate, behavioural targeting may well significantly affect a person. Indeed, the very point of advertising is to change views, attitudes, actions, and behaviours over time.

The successor of the automated decisions provision in the European Commission proposal is entitled "measures based on profiling."[1981] The provision introduces a new transparency requirement, which obliges a firm to tell the person concerned that a profiling measure with significant effect is taken, and to inform the person about the measure's envisaged effects. The provision should be amended. First, to improve transparency, firms should inform people about profiling measures and their underlying logic, even if no significant effects of the measure are foreseen. Also, interdisciplinary research is needed to develop tools to provide people with meaningful transparency regarding data processing and profiling. Second, profiling measures that have the effect of discriminating on the basis of special categories of data, intentional or not, should be prohibited, as proposed by the European Parliament. Such a prohibition would also apply if a firm used non-special data as a proxy for special categories of data.

### 10.7 Conclusion

In summary, the law could improve privacy protection in the area of behavioural targeting, by combining the empowerment and the protection approach, along with better enforcement of the existing rules. Collecting and storing fewer data, and not collecting data without meaningful consent, could reduce chilling effects. But the most effective way of preventing chilling effects is by not collecting data. Therefore,

[1981] Article 20 of the European Commission proposal for a Data Protection Regulation (2012).

data collection for behavioural targeting may have to be further restricted or banned in certain contexts.

To improve individual control over personal information, strictly enforcing the data protection principles would be a good start. Covert data collection is a problem from the normative perspective of privacy as control. But if behavioural targeting happens surreptitiously, this usually implies a breach of existing laws as well. The study provides suggestions on how to apply and enforce the data protection principles.

To mitigate the risk of unfair social sorting, data protection law can help as well. As long as the data aren't applied to an individual (phase 5), the sorting doesn't happen. But analysing vast amounts of data (phase 3) is a crucial step. Hence, limiting the amount of data that is available could mitigate the risks. Requiring firms to be transparent about personalisation could also mitigate the risk of manipulation. And the data protection principles can be interpreted as generally requiring firms to offer an option to opt out of personalisation. Furthermore, the legal transparency requirements can help to make data processing controllable for policymakers, as transparency can help to uncover problems that might call for regulatory intervention.

In sum, enforcing and tightening the data protection principles could help to protect privacy in the area of behavioural targeting. But this may not be enough. If society is better off if certain behavioural targeting practices don't take place, the lawmaker should consider banning them. The study shows that there are no reasons *never* to use prohibitions in the area of behavioural targeting. But it would be difficult to define prohibitions in such a way that they're not over or under inclusive. Here lies a challenge for further research. Hard questions are ahead for researchers and policymakers. The legal protection of privacy will remain a learning process. If new rules were adopted, their practical effect would have to be evaluated. The problems with the current informed consent requirements demonstrate that regulation that looks good on paper may not effectively protect privacy in practice. The way the online marketing industry evolves also has implications for the best regulatory approach. If in ten years a couple of firms are responsible for all the behavioural targeting in the world, this calls for different regulatory answers than if thousands of firms engage in behavioural targeting.

Behavioural targeting illustrates the difficulties that privacy protection faces in the twenty-first century. Gradually more objects are being connected to the internet. Transparency and individual control over personal data are difficult to achieve when people use computers and smart phones, but will be even harder to achieve when objects without a screen are used to collect data.

In conclusion, there's no silver bullet to improve privacy protection in the area of behavioural targeting. While current regulation emphasises empowerment, without much reflection on practical issues, this study argues for a combined approach of protecting and empowering people. To improve privacy protection, the data protection principles should be more strictly enforced. But the limited potential of informed consent as a privacy protection measure should be taken into account. Therefore, the lawmaker should give more attention to rules that protect, rather than empower, people.

* * *

***Legal texts***

***International***

Universal Declaration of Human Rights.

International Covenant on Civil and Political Rights.

Organisation for Economic Co-operation and Development, Guidelines governing the protection of privacy and transborder flows of personal data (1980, amended in 2013).

UN General Assembly, Guidelines for the Regulation of Computerized Personal Data Files, 14 December 1990.

Vienna Convention on International Sale of Goods.

***Council of Europe***

***Organisation for Economic Co-operation and Development***

***European Union, Charter and Treaties***

***European Union, regulations and directives***

***European Union, documents regarding the data protection reform***

ECtHR, Ciubotaru V. Moldova, No. 27138/04, 27 April 2010.
ECtHR, Uzun v. Germany, No. 35623/05, 2 September 2010.
ECtHR, Köpke v. Germany, No. 420/07 (inadmissible), 5 October 2010.
ECtHR, Mosley v. United Kingdom, 48009/08, 10 May 2011.
ECtHR, Von Hannover v. Germany (II), Nos. 40660/08 and 60641/08, 7 February 2012.
ECtHR, Romet v. Netherlands, No. 7094/06, 14 February 2012.
ECtHR, M.M. v. United Kingdom, No. 24029/07, 13 November 2012.
ECtHR, Fredrik Neij and Peter Sunde Kolmisoppi (The Pirate Bay) v. Sweden, No. 40397/12, 19 February 2013 (inadmissible)

***Court of Justice of the European Union***

ECJ, C-414/99 to C-416/99, 20 November 2001, Zino Davidoff.
ECJ, C-465/00, C-138/01 and C-139/01, Österreichischer Rundfunk, 20 May 2003.
ECJ, C-101/01, Lindqvist, 6 November 2003.
ECJ, C-397/01 and C-403/01, Pfeiffer et al., 5 October 2004.
ECJ, C-136/04, Deutsches Milch-Kontor GmbH, 24 November 2005.
ECJ, C-275/06, Promusicae, 29 January 2008.
ECJ, C-524/06, Huber, 16 December 2008.
ECJ, C-73/07, Satamedia, 16 December 2008.
ECJ, C-243/08, Pannon GSM, 4 June 2009.
ECJ, C-28/08 and T-194/04, Bavarian Lager, 29 June 2010.
CJEU, C-577/08, 29 June 2010, Brouwer.
CJEU, C-400/10, J. McB. v L. E., 5 October 2010.
CJEU, C-92/09 and C-93/09, 9 November 2010, Volker und Markus Schecke and Eifert.
CJEU, C-543/09, 5 May 2011, Deutsche Telekom.
European Union Civil Service Tribunal, Civil Service Tribunal Decision F-46/095, V & EDPS v. European Parliament, 5 July 2011.
CJEU, C-482/09, 22 September 2011, Budějovický Budvar.
CJEU, C-70/10, Scarlet v Sabam, 24 November 2011.
CJEU, C-468/10 and C-469/10, ASNEF, 24 November 2011.
CJEU, C-428/11, 18 October 2012, Purely Creative.
CJEU, C-291/12, Schwartz v. Stadt Bochum, 17 October 2013.
CJEU, Case C-473/12, 7 November 2013, Institut professionnel des agents immobiliers.
CJEU, C-293/12 and C-594/12, Digital Rights Ireland Ltd, 8 April 2014.
CJEU, C-131/12, Google Spain, 13 May 2014.

***Germany***

Bundesverfassungsgericht 25 March 1982, BGBl.I 369 (1982), (Volks-, Berufs-, Wohnungsund Arbeitsstättenzählung (Volkszählungsgesetz)) [Census case], translation by Riedel, E.H., Human Rights Law Journal 1984, vol. 5, no 1, p. 94.
Bundesverfassungsgericht 27 February 2008, decisions, vol. 120, p. 274-350 (Online Durchsuchung) [Online Search].
Bundesverfassungsgericht 2 March 2010, BvR 256/08 vom 2.3.2010, Absatz-Nr. (1 345), (Vorratsdatenspeicherung) [Data Retention].

***Netherlands***

College van Beroep voor het bedrijfsleven [Trade and Industry Appeals Tribunal], 20 June 2013, ECLI:NL:CBB:2013:CA3716 (Dollarrevenue/Autoriteit Consument en Markt).
Voorzieningenrechter Rechtbank Amsterdam, 12 February 2004, ECLI:NL:RBAMS:2004:AO3649 (Broadcast Press).
Hoge Raad [Dutch Supreme Court], 9 September 2011, ECLI:NL:HR:2011:BQ8097 (Santander).

* * *

# Short summary

To protect privacy in the area of behavioural targeting, the EU lawmaker mainly relies on the consent requirement for the use of tracking technologies in the e-Privacy Directive, and on general data protection law. With informed consent requirements, the law aims to empower people to make choices in their best interests. But behavioural studies cast doubt on the effectiveness of the empowerment approach as a privacy protection measure. Many people click “I agree” to any statement that is presented to them. Therefore, to mitigate privacy problems such as chilling effects and the lack of individual control over personal information, this study argues for a combined approach of protecting and empowering the individual. Compared to the current approach, the lawmaker should focus more on protecting people.

Chapter 1 introduces the research question: how could European law improve privacy protection in the area of behavioural targeting, without being unduly prescriptive?

Chapter 2 explains what behavioural targeting is, by distinguishing five phases. During the first phase of behavioural targeting, firms track people’s online behaviour. Second, firms store data about individuals. Third, firms analyse the data. Fourth, firms disclose data to other parties. In the fifth phase, data are used to target ads to specific individuals.

Chapter 3 discusses the right to privacy in European law, and the privacy implications of behavioural targeting. Three privacy perspectives are distinguished in this study: privacy as limited access, privacy as control, and privacy as identity construction. The chapter discusses three main privacy problems of behavioural targeting. First, the massive collection of information on user behaviour can have a chilling effect. Second, people lack control over their information. Third, behavioural targeting enables social sorting and discriminatory practices. Also, some fear that personalised ads and other content could be manipulative, or could narrow people's horizons.

Chapter 4 gives an overview of the data protection principles. Data protection law is Europe's main legal tool to protect information privacy, and aims to ensure that personal data processing happens fairly and transparently. The chapter shows that there's a tension within data protection law between empowering and protecting the individual. This tension is a recurring theme in this study.

Chapter 5 concerns the material scope of data protection law. Many behavioural targeting firms say data protection law doesn't apply to them, because they only process "anonymous" data. The chapter makes two points. First, an analysis of current law shows that data protection law generally applies to behavioural targeting. Data protection law also applies if firms don't tie a name to individual profiles. Second, from a normative perspective, data protection law *should* apply.

Chapter 6 discusses the role of informed consent in the regulation of behavioural targeting. Current law regarding behavioural targeting places a good deal of emphasis on informed consent. The e-Privacy Directive requires firms to obtain informed consent for the use of most tracking technologies, such as cookies. Furthermore, in general data protection law, consent is one of the legal bases that a firm can rely on for personal data processing.

Chapter 7 analyses practical problems with informed consent in the area of behavioural targeting. The chapter reviews law and economics literature, behavioural economics literature, and empirical research on how people make privacy choices. The potential of data protection law's informed consent requirement as a privacy protection measure is very limited. People generally ignore privacy policies, and click "I agree" to almost any online request.

Chapter 8 discusses measures to improve individual *empowerment*. Strictly enforcing and tightening data protection law would be a good start. For example, firms shouldn't be allowed to infer consent from mere inactivity from the individual, and long unreadable privacy policies shouldn't be accepted. User-friendly mechanisms should be developed to foster transparency and to enable people to express their choices. This study doesn't suggest that data subject control over personal information can be fully achieved. Nevertheless, some improvement must be possible, as now people's data are generally accumulated and used without meaningful transparency or consent.

Chapter 9 discusses measures to improve individual *protection*. Certain data protection principles could protect people, even if they consent to data processing. While the role of informed consent in data protection law is important, it's at the same time limited. People can't waive data protection law's safeguards, or contract around the rules. The protective data protection principles should be enforced more strictly; but this won't be enough. In addition to general data protection law, more specific rules regarding behavioural targeting are needed. And if society is better off if certain behavioural targeting practices don't happen, the lawmaker should consider banning them.

Chapter 10 summarises the main findings and answers the research question. There's no easy solution, but legal privacy protection can be improved in the area of behavioural targeting. While current regulation emphasises empowerment, without much reflection on practical issues, this study argues for a combined approach of protecting and empowering people. To improve privacy protection, the data protection principles should be more strictly enforced. But the limited potential of informed consent as a privacy protection measure should be taken into account. Therefore, the lawmaker should give more attention to rules that protect, rather than empower, people.

* * *

# Short summary in Dutch

## Betere privacybescherming op het gebied van behavioural targeting

Behavioural targeting is een vorm van marketing waarbij mensen op internet worden gevolgd, en er op basis van afgeleide interesses gerichte advertenties worden getoond aan mensen. Deze praktijk wordt door veel mensen ervaren als een aantasting van privacy. Behavioural targeting is al deels gereguleerd in Europese wetgeving. In dit verband zijn de belangrijkste Europese regels om online privacy te beschermen het toestemmingsvereiste in de e-Privacyrichtlijn voor tracking cookies en vergelijkbare volgtechnieken, en de regels in de algemene Richtlijn Bescherming Persoonsgegevens. In Nederland zijn deze regels geïmplementeerd in artikel 11.7a van de Telecommunicatiewet, respectievelijk in de Wet bescherming persoonsgegevens.

Door bedrijven te verplichten geïnformeerde toestemming te vragen voor behavioural targeting, probeert de wetgever mensen in staat te stellen keuzes te maken in hun eigen belang. Het idee is dat mensen zo zelf kunnen beslissen of, en in welke gevallen, zij een deel van hun privacy opgeven. Kortom, via geïnformeerde toestemming streeft de wetgever naar *empowerment* van het individu.

Inzichten uit *behavioural economics* (gedragseconomie) trekken de effectiviteit van deze *empowerment*-aanpak in twijfel. In de praktijk klikken veel mensen OK op elk verzoek dat zij tegenkomen op het internet. De wetgever zou zich daarom meer moeten richten op *protection*, het beschermen van mensen. In dit proefschrift wordt gepleit voor een gecombineerde aanpak van *empowerment* en *protection*.

In hoofdstuk 1 wordt de onderzoeksvraag toegelicht: welke maatregelen zou de EU wetgever kunnen nemen om de privacy van internetgebruikers beter te beschermen als het gaat om behavioural targeting, zonder daarbij onnodige lasten en regels op te leggen?

In hoofdstuk 2 wordt uitgelegd hoe behavioural targeting werkt. Deze studie onderscheidt vijf fasen in het proces van behavioural targeting. In fase 1 verzamelen bedrijven informatie over wat mensen doen op internet. Dit gebeurt vaak door middel van tracking cookies. Een cookie is een klein tekstbestand dat op de computer van een internetgebruiker geplaatst kan worden. Met behulp van tracking cookies kan een bedrijf iemands surfgedrag in kaart brengen. In fase 2 slaan bedrijven de informatie op. De informatie over een persoon is gekoppeld aan unieke identificatiecode, die in onder meer in een cookie kan worden opgenomen. In fase 3 worden de gegevens geanalyseerd. In fase 4 stellen bedrijven de gegevens ter beschikking aan adverteerders of aan andere bedrijven. In fase 5 tonen bedrijven gerichte, op vermeende individuele interesses gebaseerde, advertenties aan specifieke personen.

In hoofdstuk 3 worden de privacyproblemen geanalyseerd die het gevolg zijn van behavioural targeting. Privacy is moeilijk te definiëren. In deze studie worden drie perspectieven op privacy onderscheiden: privacy als beperkte toegang, privacy als zeggenschap of controle over persoonlijke informatie, en privacy als de vrijheid van onredelijke beperkingen op identiteitsvorming. Vanuit elk van de drie privacy-perspectieven is behavioural targeting problematisch. Drie van de belangrijkste privacyproblemen veroorzaakt door behavioural targeting zijn (i) *chilling effects*, (ii) een gebrek aan controle over persoonlijke informatie, en (iii) het risico op discriminatie en manipulatie. Een *chilling effect* kan optreden als gevolg van grootschalige gegevensverzameling: mensen passen hun gedrag aan als zij weten dat hun activiteiten worden gevolgd. Het tweede privacyprobleem is dat mensen niet weten welke informatie over hen wordt verzameld, hoe deze informatie gebruikt wordt, en met wie deze wordt gedeeld. Hierdoor verliezen zij zeggenschap over de hen betreffende gegevens. Ten derde maakt behavioural targeting discriminatie mogelijk. Sommigen vrezen daarnaast dat behavioural targeting kan worden gebruikt om mensen te manipuleren. Gepersonaliseerde reclame zou zo effectief kunnen worden dat adverteerders een oneerlijk voordeel verkrijgen ten opzichte van consumenten.

In hoofdstuk 4 wordt een overzicht gegeven van het Europese juridische kader voor de verwerking van persoonsgegevens. Deze regels hebben als hoofddoel te bevorderen dat de verwerking van persoonsgegevens eerlijk en transparant gebeurt. Het hoofdstuk laat zien dat er een spanning bestaat in het gegevensbeschermingsrecht tussen *empowerment* en *protection* van mensen. Deze spanning is een terugkerend thema in dit onderzoek.

In hoofdstuk 5 wordt besproken of behavioural targeting binnen de werkingssfeer van het gegevensbeschermingsrecht valt. Veel bedrijven die aan behavioural targeting doen, zeggen dat het gegevensbeschermingsrecht niet van toepassing is op hun praktijken, omdat ze alleen "anonieme" gegevens verwerken. Europese gegevensbeschermingsautoriteiten (zoals het Nederlandse College Bescherming Persoonsgegevens), samenwerkend in de Artikel 29 Werkgroep, zeggen echter dat behavioural targeting doorgaans de verwerking van persoonsgegevens met zich meebrengt, ook als een bedrijf geen naam kan koppelen aan de gegevens over een individu. Als een bedrijf gegevens gebruikt om iemand te individualiseren of iemand te onderscheiden binnen een groep, dan zijn die gegevens persoonsgegevens volgens de Werkgroep. In deze studie wordt dit standpunt onderschreven.

In hoofdstuk 6 staat het concept van geïnformeerde toestemming centraal. Sinds 2009 volgt uit de e-Privacyrichtlijn, kort gezegd, dat tracking cookies slechts geplaatst mogen worden als de betrokkene toestemming heeft verleend, na te zijn voorzien van duidelijke en volledige informatie. Bovendien staat de Richtlijn Bescherming Persoonsgegevens bedrijven slechts toe om persoonsgegevens te verwerken, als zij de verwerking op toestemming of op een andere wettelijke grondslag kunnen baseren.

In hoofdstuk 7 worden de praktische problemen bij het geven van geïnformeerde toestemming voor behavioural targeting geanalyseerd. Uit onderzoek uit op het gebied van de rechtseconomie (*law and economics*) en de gedragseconomie (*behavioural economics*), en uit empirisch onderzoek naar hoe mensen keuzes maken over privacy, blijkt dat er in de praktijk vrijwel onoplosbare problemen zijn met geïnformeerde toestemming. Vrijwel niemand leest privacyverklaringen of toestemmingsverzoeken. Veel mensen klikken OK op vrijwel elk verzoek dat zij tegenkomen op het internet. Eigenlijk kan ook niet van mensen verwacht worden dat zij elk verzoek zouden lezen. Onderzoek toont aan dat het mensen enkele weken per jaar zou kosten om elke privacyverklaring die zij tegenkomen op het internet te lezen. Bovendien: zelfs als iemand een toestemmingsverzoek zou lezen en begrijpen, dan nog is er een grote kans dat hij of zij toch op OK klikt bij een privacy-onvriendelijke verzoek. De wet staat website-houders in veel gevallen toe om mensen een *take-it-or-leave-it* keuze te bieden. Zo installeren veel websites tracking-muren of cookie-muren – barrières waar mensen alleen langs komen als zij op toestaan dat er via de website tracking cookies worden geplaatst.

Er is daarom voor de wetgever reden tot ingrijpen. Gezien de beperkte mogelijkheden van geïnformeerde toestemming als privacybeschermingsmaatregel, wordt in deze studie gepleit voor een gecombineerde aanpak van *empowerment* en *protection* van mensen.

Overigens is het is onduidelijk of, vanuit een economisch perspectief, de maatschappij als geheel beter of slechter wordt van behavioural targeting. Ook is omstreden of behavioural targeting nodig is om “gratis” websites te financieren. Advertenties die niet gebaseerd zijn op behavioural targeting zijn ook mogelijk, zoals contextuele reclame: advertenties voor auto’s op websites over auto’s.

Hoofdstuk 8 bespreekt mogelijke maatregelen om mensen beter in staat te stellen om voor hun eigen belangen op te komen (*empowerment*). Om de informatieasymmetrie in de context van behavioural targeting te verminderen, zou het transparantiebeginsel beter gehandhaafd moeten worden. De wetgever zou moeten afdwingen dat toestemmingsverzoeken simpel, kort, en gemakkelijk te begrijpen zijn. Privacyverklaringen en toestemmingsverzoeken kunnen veel duidelijker en begrijpelijker worden geformuleerd. De bestaande regels over toestemming moeten strenger gehandhaafd worden. "Wie zwijgt stemt toe" zou niet geaccepteerd mogen worden.

In hoofdstuk 9 worden maatregelen toegelicht om het individu te beschermen (*protection*). Als de wetgeving voor de bescherming van persoonsgegevens volledig nageleefd zou worden, dan zouden mensen redelijke bescherming genieten, ook als zij OK klikken op elk toestemmingsverzoek. Hoewel toestemming een belangrijke rol speelt in het gegevensbeschermingsrecht, geeft toestemming bedrijven geen vrijbrief om met persoonsgegevens te doen wat zij willen. Ook als iemand toestemming heeft gegeven, dient het bedrijf nog te voldoen aan de overige eisen uit het gegevensbeschermingsrecht. Het gaat immers om dwingend recht. Zo eist de wet dat bedrijven persoonsgegevens beveiligen, en verbiedt de wet het gebruik van persoonsgegevens voor doelen die onverenigbaar zijn met het verzameldoel. Verder mogen bedrijven geen disproportionele hoeveelheden gegevens verzamelen en verwerken – ook niet na toestemming van het individu. Met betere handhaving en explicitering van de huidige normen, kan de wetgever een deel van de omvangrijke privacy-problemen adresseren. Maar dit is waarschijnlijk niet voldoende. Als de samenleving beter af is als bepaalde behavioural targeting praktijken niet plaatsvinden, dan zou de wetgever een verbod van dergelijke praktijken moeten overwegen.

Zo zouden tracking-muren verboden moeten worden voor publieke omroepen en voor overheidswebsites. In Nederland ligt nu een wetsvoorstel voor met een vergelijkbare regel. De wetgever zou ook een stap verder kunnen gaan, door alle commerciële dataverzameling voor behavoural targeting en vergelijkbare doelen te verbieden op overheidswebsites.

Hoofdstuk 10 bevat de conclusie. Er bestaat geen wondermiddel voor privacybescherming als het gaat om behavioural targeting. Terwijl de huidige regelgeving veel nadruk legt op *empowerment,* zonder veel reflectie op de praktijk, zou een gecombineerde aanpak van *protection* en *empowerment* effectiever zijn. Om de privacy van mensen beter te beschermen, moet het gegevensbeschermingsrecht strikter worden gehandhaafd. Maar omdat geïnformeerde toestemming als privacybeschermingsmaatregel tekort schiet, moet de wetgever niet al te hoge verwachtingen hebben van empowerment. Er moet ook voldoende aandacht gegeven worden aan het beschermen van mensen.

* * *